%% file: main.tex
\documentclass[aps,prd,article]{revtex4-2}
\pdfoutput=1
\usepackage[dvipsnames]{xcolor}
\usepackage{amsmath}
\usepackage{amssymb}
\usepackage{amsfonts}
\usepackage{enumerate}
\usepackage{float}
\usepackage{graphicx}
\usepackage{hyperref}
\usepackage{mathrsfs}
\usepackage{mathtools}
\usepackage{bm}
\usepackage{array}
\usepackage[normalem]{ulem}
\usepackage{comment}
\usepackage{appendix}
\usepackage{lineno}

\allowdisplaybreaks

\graphicspath{{./figures/}}

\def\pp#1{\left( #1 \right)}
\def\bb#1{\left[ #1 \right]}

\def\vv#1{\left\vert #1 \right\vert}

\def\del{\partial}

\def\vec#1{\bm{\mathrm{#1}}}

\def\d{\mathop{}\!\mathrm{d}}
\def\D#1{\mathop{}\!\mathrm{d^#1}}

\def\Ma{\mathscr M}

\def\Ceu{C_{eu}}
\def\Ced{C_{ed}}
\def\Clqi{C_{\ell q}^{(1)}}
\def\Clqiii{C_{\ell q}^{(3)}}
\def\Clu{C_{\ell u}}
\def\Cld{C_{\ell d}}
\def\Cqe{C_{qe}}

\def\APVe{A_{\rm PV}^{(e)}}
\def\APVp{A_{\rm PV}^{(p)}}

\def\APVH{A_{\rm PV}^{(H)}}
\def\APVpD{A_{\rm PV}^{(p(D))}}
\def\ALC{A_{\rm LC}}

\def\ALCH{A_{\rm LC}^{(H)}}
\def\ALCpD{A_{\rm LC}^{(p(D))}}

\def\adet{a_{\rm det}}

\begin{document}
%	\linenumbers
	
	\title{Neutral-Current Electroweak Physics and SMEFT Studies at the EIC}

\input{authorlist}

    \begin{abstract}
		We study the potential for precision electroweak (EW) measurements and beyond-the-Standard Model (BSM) searches using cross-section asymmetries in neutral-current (NC) deep inelastic scattering at the electron-ion collider (EIC). Our analysis uses a complete and realistic accounting of systematic errors from both theory and experiment and considers the potential of both proton and deuteron beams for a wide range of energies and luminosities. We also consider what can be learned from a possible future positron beam and a potential ten-fold luminosity upgrade of the EIC beyond its initial decade of running. We use the SM effective field theory (SMEFT) framework to parameterize BSM effects and focus on semi-leptonic four-fermion operators, whereas for our precision EW study, we determine how well the EIC can measure the weak mixing angle. New features of our study include the use of an up-to-date detector design of EIC Comprehensive Chromodynamics Experiment (ECCE) and accurate running conditions of the EIC, the simultaneous fitting of beam polarization uncertainties and Wilson coefficients to improve the sensitivity to SMEFT operators, and the inclusion of the weak mixing angle running in our fit template. We find that the EIC can probe BSM operators at scales competitive with and in many cases exceeding LHC Drell-Yan bounds while simultaneously not suffering from degeneracies between Wilson coefficients.
    \end{abstract}
  
    \maketitle

	\tableofcontents
	
	\section{Introduction \label{sec:introduction}}
	\input{./sections/1-introduction}
	
	\section{Neutral-current DIS Measurements at the EIC\label{sec:nc-dis-at-eic}}
	\input{./sections/2a-dis_and_smeft_formalism}
	\input{./sections/2b-measurement_of_pv_asymmetries_at_eic}
	\input{./sections/2c-measurement_of_lc_asymmetries_at_eic}
	
	\section{Projection of Parity-Violation and Lepton-Charge Asymmetry Data\label{sec:proj}}
	\input{./sections/3a-ecce_detector_config_for_inclusive_nc}
	\input{./sections/3b-simulation_with_fast_smearing}

	\input{./sections/3c-event_selection}
	\input{./sections/3d-integrated_luminosity}
	\input{./sections/3e-stat_uncertainty_projection_for_pv_asymmetries}
	\input{./sections/3f-stat_and_qed_uncertainty_projections_for_lc_asymmetries}
	\input{./sections/3g-projection_for_hl_eic}
	
	\section{Pseudodata Generation and the Uncertainty Matrix\label{sec:pseudodata-errmat}}
	\input{./sections/4a-pseudodata_for_pv_asymmetries}
	\input{./sections/4b-pseudodata_for_lc_asymmetries}
	\input{./sections/4c-uncertainty_matrix}
	\input{./sections/4d-comparison_of_uncertainty_components}

	\section{Extraction of the SM Weak Mixing Angle\label{sec:sinth}}
	\input{./sections/5-extraction_of_weak_mixing_angle}

	\section{Framework for the SMEFT Analysis\label{sec:analysis_framework}}
	\input{./sections/6a-data_generation_and_selection}
	\input{./sections/6b-structure_of_smeft_asymmetry_corrections}
	\input{./sections/6c-best-fit_analysis_of_wilson_coefficients}

	\section{SMEFT fit results\label{sec:smeft-fits}}
	\input{./sections/7a-fits_of_single_wilson_coefficients}
	\input{./sections/7b-fits_of_two_wilson_coefficients}

	\section{Conclusions\label{sec:conclusions}}	
	\input{./sections/8-conclusions}

	%%%%%%%%%%%%%%%%%
	
	\appendix
	
	\section{Additional fits \label{app:additional_fits}}
	\input{./appendices/a1-luminosity_difference_fits}
	\input{./appendices/a2-beam_polarization_fits}

	\section{Complete set of fitted results on Wilson coefficients \label{app:complete_results}}
	\input{./appendices/b1-fits_of_single_wilson_coefficients}
	\input{./appendices/b2-fits_of_two_wilson_coefficients}
	\input{./appendices/c1-fits_of_six_wilson_coefficients}

	%%%%%%%%%%%%%%%%%	
	
	\noindent{\bf Acknowledgment:} 
	R.~B. is supported by the US Department of Energy (DOE) contract DE-AC02-06CH11357. T.~K. is supported by the DOE grant DE-SC0020240 and the Zuckerman STEM Leadership Program. F.~P., K.~\c{S}., and D.~W. are supported by the DOE grants DE-FG02-91ER40684 and DE-AC02-06CH11357. X.~Z. and M.~N. are supported by the DOE grant  DE-SC0014434. The authors would like to thank the ECCE Consortium for performing a full simulation of their detector design, for providing up-to-date information on EIC run conditions, and for suggestions and comments on the manuscript. X. Z. would like to thank H.~Spiesberger for suggestions on the use of the {\tt Djangoh} generator and useful discussions related to the analysis.
		
	\bibliography{refs}
	
\end{document}

%% file: authorlist.tex
\author{Radja Boughezal$^1$, Alexander Emmert$^2$, Tyler Kutz$^3$, Sonny Mantry$^4$, Michael Nycz$^2$, Frank Petriello$^{1,5}$, Ka{\u{g}}an \c{S}im\c{s}ek$^5$, Daniel Wiegand$^5$, Xiaochao Zheng$^2$
\\ \vspace{0.1cm}
{\sl $^1$ Argonne National Laboratory, Lemont, IL, USA}   \\
{\sl $^2$ University of Virginia, Charlottesville, VA, USA} \\
{\sl $^3$ Massachusetts Institute of Technology, Cambridge, MA, USA} \\
{\sl $^4$ University of North Georgia, Dahlonega, GA, USA} \\
{\sl $^5$ Northwestern University, Evanston, IL, USA} \\
}

%% file: sections/1-introduction.tex
The Standard Model (SM) of particle physics currently describes all known laboratory phenomena. All particles predicted by the SM have now been found after the discovery of the Higgs boson at the Large Hadron Collider (LHC). No new particles beyond those present in the SM have been discovered and no appreciable deviation from SM predictions has been conclusively observed. Despite the enormous success of this theory, it contains numerous shortcomings. It does not contain an explanation of the dark matter observed in the universe or of the baryon-antibaryon asymmetry and it does not describe neutrino masses. It additionally suffers from several aesthetic issues, such as the hierarchy problem and an extreme hierarchy of fermion Yukawa couplings. Even the sectors of the theory that have been experimentally successful still contain unsatisfying and poorly understood features. For example, the exact composition of the proton spin in terms of the spin and orbital angular momentum of its constituent quarks and gluons is still poorly known. 

Numerous experimental programs that attempt to address these residual issues in our understanding of Nature are either running or under design. Our focus in this manuscript will be on the Electron-Ion Collider (EIC) to be built at Brookhaven National Laboratory in Upton, New York. The EIC will be a particle accelerator that collides electrons with protons and nuclei in the intermediate-energy range between fixed-target scattering facilities and high-energy colliders. It will provide luminosity orders of magnitude higher than HERA, the only electron-proton collider operated to date. It will also be the first lepton-ion collider with the ability to polarize both the electron and the proton (ion) beams and the first collider with a fast spin-flip capacity. These unique design features will allow direct extraction of parity-violating (PV) asymmetries in the electroweak neutral-current (NC) scattering cross section associated with either the electron, $\APVe$,  or the proton (ion) spin-flip, $\APVpD$. Experimental uncertainties from effects such as luminosity measurement and detector acceptance or efficiency will be substantially reduced due to these capabilities.

Although the EIC was designed primarily to explore outstanding issues in QCD such as the proton spin mentioned above, it also has strong potential to probe several aspects of precision electroweak (EW) and beyond-the-SM (BSM) physics. It can measure the value of the weak mixing angle over a wide range of momentum transfer complementary to $Z$-pole measurements and low-energy determinations. The possibility of polarizing both electron and proton/ion beams gives it unique handles on BSM physics. Our goal in this manuscript is to provide a detailed accounting of the EW and BSM potential of the EIC with a realistic simulation of anticipated experimental uncertainties. We explore the use of the asymmetries $\APVe$ and $\APVpD$. In addition to determining the BSM reach of PV observables, we consider the reach of the lepton-charge asymmetry $\ALCpD$ at the EIC for the first time, assuming a positron beam will become available in the future. 

Since no new particles beyond the SM have so far been discovered, we adopt the Standard Model Effective Field Theory (SMEFT) for our BSM studies (for a review of the SMEFT, see Ref.~\cite{Brivio:2017vri}). The SMEFT contains higher-dimensional operators formed by using SM fields, assuming all new physics is heavier than both SM states and the accessible collider energy. The leading dimension-6 operator basis of SMEFT for on-shell fields has been completely classified (there is a dimension-5 operator that violates lepton number, which we do not consider here)~\cite{Buchmuller:1985jz,Arzt:1994gp,Grzadkowski:2010es}. We find that the EIC can probe the full spectrum of SMEFT operators to the few-TeV level or beyond. The wide variety of observables possible at the EIC, which include several asymmetries with either proton or ion beams, ensure that no flat directions remain in the Wilson coefficient parameter space, unlike at the LHC in the neutral-current Drell-Yan process~\cite{Alte:2018xgc,RBoughezal2020,Panico:2021vav}. Our analysis of the determination of the weak mixing angle, assuming a realistic annual luminosity and accounting for experimental and theoretical uncertainties to the best level that can be reached at pre-EIC running stage, finds good precision for this fundamental SM parameter in a kinematic region not explored before. The precision will continue to improve as data are accumulated from decades-long running of the EIC.

Our paper is organized as follows. In Section~\ref{sec:nc-dis-at-eic}, first we provide a complete description of deep inelastic scattering (DIS) that includes both SM contributions and the SMEFT extensions. The DIS cross sections that account for both electron and hadron polarizations are provided in both structure-function and parton-model languages. We follow this theoretical framework by presenting a basic strategy to measure different polarization components of the cross sections and to form PV asymmetries at the EIC. Measurement of the lepton-charge (LC) asymmetry is also discussed. In Section~\ref{sec:proj}, we present data simulation based on the design of the ECCE Detector (recently endorsed as the reference design for EIC detector 1 by the EIC Detector Proposal Advisory Panel~\cite{DPAPreport2022}) using a fast-smearing method and event-selection criteria, followed by projections of statistical precision for PV and LC asymmetries based on the planned annual luminosity of the EIC. Generation of pseudodata, as well as the uncertainty matrix, are presented in Section~\ref{sec:pseudodata-errmat}, followed by extractions of the EW mixing angle in Section~\ref{sec:sinth}.  In Section~\ref{sec:analysis_framework}, we provide an extensive description of our SMEFT analysis framework, with representative results on the fits of single and two Wilson coefficients given in Section~\ref{sec:smeft-fits}. We also show an example fit in which six Wilson coefficients are turned on simultaneously, in order to demonstrate that EIC data is capable of removing all degeneracies in the semi-leptonic four-fermion operator parameter space. We conclude in Section~\ref{sec:conclusions}.  In Appendix~\ref{app:additional_fits}, we present novel analysis methods to simultaneously fit PV asymmetries and the beam polarization or LC asymmetries and the luminosity difference between $e^+$ and $e^-$ runs. A complete collection of all the SMEFT fit results of single and two Wilson coefficients from this study is given in Appendix~\ref{app:complete_results}.

%% file: sections/2a-dis_and_smeft_formalism.tex
\subsection{Deep Inelastic Scattering and the SMEFT Formalism}
In this section, we give a brief overview of the formalism of DIS and the SMEFT. In particular, we generalize the SM DIS cross-section and asymmetry formulae to include contributions from SMEFT operators, which encode new physics at an energy level $\Lambda$ that lies well beyond the electroweak scale. We denote a lepton scattering off a nucleus as:
\begin{eqnarray}
	\label{eq:DISprocesses}
	\ell (k) + H (P) \longrightarrow \ell (k') + X~,
\end{eqnarray}
where $\ell$ stands for an electron or positron, the hadron $H$ stands for either the proton ($p$) or the deuteron ($D$), and $X$ denotes the final-state hadronic system. The four-momenta of the initial and final leptons and the initial hadron are denoted as $k$, $k'$, and $P$, respectively. Using the momenta of the initial- and final-state leptons and the initial-state hadron, one can define the following Lorentz-invariant kinematic variables: 
\begin{eqnarray}
	s &=&(P+k)^2 \label{eq:s}~, \\
	Q^2 &=&  -(k-k')^2\label{eq:Q2}~, \\
	x &=&\frac{Q^2}{2P\cdot(k-k')} \label{eq:x}~, \\
	y &=&\frac{P\cdot(k-k')}{P\cdot k} \label{eq:y}~, \\
	W^2 &=& (P+k-k')^2 \label{eq:W}~,
\end{eqnarray}
where $s$ is the center-of-mass energy squared, $Q^2$ is the negative of the lepton four-momentum transfer squared, the Bjorken-$x$ variable is the longitudinal hadron momentum fraction carried by the struck parton, the inelasticity parameter $y$ gives the fractional energy loss of the lepton in the hadron rest frame, and $W$ gives the invariant mass of the final-state hadronic system $X$. The kinematic variables $x$, $y$, $s$, and $Q^2$ are related to each other via $Q^2=xy(s-M^2)$, where $M$ is the mass of the proton or deuteron. 

\begin{figure}
	[h]\centering
	\includegraphics[width=.4\textwidth]{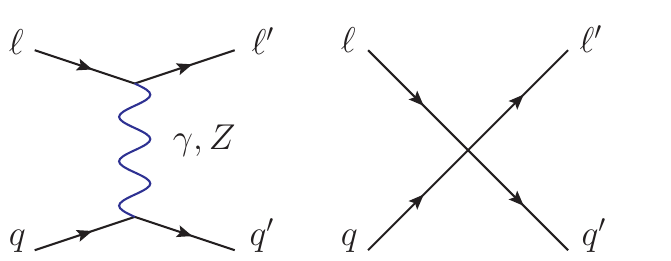}
	\caption{The Feynman diagrams for $ \ell + H \to \ell + X $ at the parton level from one-boson exchange (left) and SMEFT contact interactions (right).}
	\label{fig_1}
\end{figure}

The diagrams in Fig.~\ref{fig_1} show the partonic tree-level processes that contribute to Eq.~(\ref{eq:DISprocesses}). These are the contributions to the total tree-level amplitude from single-photon exchange, single-$Z$-boson exchange, and the SMEFT contact interactions. The SMEFT Lagrangian that describes these contact interactions has the form:
\begin{eqnarray}
	\label{eq:SMEFTLag}
	{\cal L}_\mathrm{SMEFT}  = \frac{1}{\Lambda^2}\sum_{r} C_r {\cal O}_r + \cdots~,
\end{eqnarray}
where the summation index $r$ runs over the set of dimension-6 SMEFT operators and the ellipsis denotes SMEFT operators of mass-dimension greater than 6. We restrict our analysis to include only the effects of dimension-6 SMEFT operators since the higher-dimensional operators are formally suppressed by additional powers of $E^2/\Lambda^2$, where $E$ is the typical energy scale of the scattering process. Although these effects can be important for Drell-Yan production at the LHC~\cite{Alioli:2020kez,Boughezal:2021tih}, the low energy of the EIC renders them negligible in this analysis. ${\cal O}_{r}$ denotes the $r^{\rm th}$ dimension-6 operator and $C_r$ is the corresponding (dimensionless) Wilson coefficient arising from integrating out the new-physics degrees of freedom at the scale $\Lambda$. These Wilson coefficients can be constrained through a comparison of SMEFT predictions with precision measurements of various processes studied in a variety of experiments across a wide range of energy scales.

The subset of dimension-6 operators that we consider in our analysis of DIS is given in Table~\ref{tab:SMEFTops}. We note that there are additional SMEFT operators but they are known to be far better bounded through other data sets such as precision $ Z $-pole observables~\cite{Dawson:2019clf,DeBlas:2019qco,Corbett:2021eux} and we neglect them here. The above assumptions leave us with the seven Wilson coefficients associated with the listed operators which enter the predictions for DIS cross sections and asymmetries. 

As seen in Table~\ref{tab:SMEFTops}, the SMEFT operators ${\cal O}_r$ are expressed in terms of the basis of SM fields before electroweak symmetry breaking. For the purposes of DIS phenomenology below the electroweak scale, it is useful to rewrite these  SMEFT operators in the vector and axial-vector basis using Dirac fields that describe the massive electrons ($e$) and quarks ($q_f$) after electroweak symmetry breaking, which is a customary parameterization (see e.g.~\cite{PDG:2020}):
\begin{eqnarray}
	\label{eq:SMEFTLagDIS}
	{\cal L}_\mathrm{SMEFT}  = \frac{1}{\Lambda^2}\sum_r \tilde{C}_r \Big \{ \sum_f \bar{e} \gamma^\mu (c_{V_r}^e - c_{A_r}^e \gamma_5) e\> \bar{q}_f \gamma^\mu (c_{V_r}^f - c_{A_r}^f\gamma_5) q_f \Big \} +\cdots~,
\end{eqnarray}
where the specific values of the vector and axial-vector couplings---$c_{V_k}^{e,q}$ and $c_{A_k}^{e,q}$, respectively---for the $r^{\rm th}$ SMEFT operator follow from the corresponding chiral and flavor structure of the SMEFT operators. The coefficients $\tilde{C}_r$ are related to the $C_r$ by an overall factor and can be fixed by comparing Eqs.~(\ref{eq:SMEFTLag}) and (\ref{eq:SMEFTLagDIS}). There is freedom to always redefine the $\tilde{C}_r$ by absorbing an overall factor into the couplings $c_{V_r}^{e,q}$, $c_{A_r}^{e,q}$. We specify in Table~\ref{tab:SMEFTops} the exact definitions that we use. These couplings are analogous to the vector and axial-vector couplings, $g_{V}^{e,q}$ and $g_{A}^{e,q}$, of the $Z$-boson but are instead generated from integrating out UV physics associated with the scale $\Lambda$.
\begin{table}[b]
	\centering
	\begin{tabular}{|c|c|c|c|c|c|c|c|c|}
		\hline
		$C_r$ & ${\cal O}_r$ & $\tilde{C}_r$ & $c_{V_r}^e$ & $c_{A_r}^e$ & $c_{V_r}^{u}$ & $c_{A_r}^{u}$ & $c_{V_r}^{d,s}$ & $c_{A_r}^{d,s}$  \\
		\hline
		\hline
		$C_{\ell q}^{(1)}$  & ${\cal O}_{\ell q}^{(1)}=(\bar{L}_L \gamma^\mu L_L)( \bar{Q}_L \gamma_\mu Q_L)$ & $C_{\ell q}^{(1)}$/4 & 1 & 1 & 1 & 1 & 1 & 1 \\
		$C_{\ell q}^{(3)}$  & ${\cal O}_{\ell q}^{(3)}=(\bar{L}_L \gamma^\mu \tau^I L_L)( \bar{Q}_L \gamma_\mu \tau^I Q_L)$ & $C_{\ell q}^{(3)}$/4  & 1 & 1  & -1 & -1 & 1 & 1\\      
		$C_{eu}$  & ${\cal O}_{eu}=(\bar{e}_R \gamma^\mu e_R)( \bar{u}_R \gamma_\mu u_R)$ &  $C_{eu}$/4 & 1 & -1 & 1  & -1 & 0 & 0 \\     
		$C_{ed}$  & ${\cal O}_{ed}=(\bar{e}_R \gamma^\mu e_R)( \bar{d}_R \gamma_\mu d_R)$ & $C_{ed}$/4 & 1 & -1 & 0 & 0 & 1 & -1\\ 
		$C_{\ell u}$  & ${\cal O}_{\ell u}=(\bar{L}_L \gamma^\mu L_L)( \bar{u}_R \gamma_\mu u_R)$ & $C_{\ell u}$/4 & 1 & 1 & 1 & -1 & 0 & 0 \\            
		$C_{\ell d}$  & ${\cal O}_{\ell d}=(\bar{L}_L \gamma^\mu L_L)( \bar{d}_R \gamma_\mu d_R)$ &  $C_{\ell d}$/4 & 1 & 1 & 0 & 0 & 1 & -1 \\            
		$C_{qe}$  & ${\cal O}_{qe}=(\bar{Q}_L \gamma^\mu Q_L)( \bar{e}_R \gamma_\mu e_R)$ & $C_{qe}$/4 & 1  & -1 & 1 & 1 & 1 & 1 \\             
		\hline
	\end{tabular}
	\caption{List of SMEFT operators relevant to DIS in the basis of SM fields before electroweak symmetry breaking and rexpressed in the vector and axial-vector current basis after electroweak symmetry breaking:  $C_r {\cal O}_r = \tilde{C}_r \sum_f \bar{e} \gamma^\mu (c_{V_r}^e - c_{A_r}^e \gamma_5) e\> \bar{q}_f \gamma^\mu (c_{V_r}^f - c_{A_r}^f\gamma_5) q_f+ \cdots$. The coefficients $c_{V,A}^f$ give the chiral structure of each operator.}
	\label{tab:SMEFTops}
\end{table}

As seen in Fig.~\ref{fig_1}, the total tree-level amplitude can be decomposed into three contributions:
\begin{equation}
	\Ma = \Ma_\gamma + \Ma_Z + \Ma_r~,
\end{equation}
where $\Ma_\gamma, \Ma_Z$, and $\Ma_r$ denote the contributions from single-photon exchange, single-$Z$-boson exchange, and the SMEFT operators, respectively. In particular, $\Ma_r = \sum_i \Ma_i$, where the summation index $i$ runs over the amplitudes arising from the SMEFT operators listed in Table~\ref{tab:SMEFTops}. Up to leading order in the SMEFT power counting, where only dimension-6 SMEFT operators that scale as $1/\Lambda^2$ are kept, the total amplitude squared can be written as:
\begin{eqnarray}
	\vv{\Ma}^2 &=& \Ma_{\gamma\gamma}+ 2\Ma_{\gamma Z} + \Ma_{ZZ} + 2\Ma_{\gamma r} + 2\Ma_{Zr}~,
\end{eqnarray} 
where $\Ma_{\gamma \gamma}=\vv{\Ma_\gamma}^2$, $\Ma_{ZZ}=\vv{\Ma_Z}^2$, $2 \Ma_{\gamma Z}= \Ma_\gamma^* \Ma_Z + \Ma_\gamma \Ma_Z^*$, $2 \Ma_{\gamma r}= \Ma_\gamma^* \Ma_r + \Ma_r \Ma_\gamma^*$, and $2 \Ma_{Z r}= \Ma_Z^* \Ma_r + \Ma_r \Ma_Z^*$. These denote the amplitudes of the single-photon exchange, single-$Z$-boson exchange, the interference between the single-photon and single-$Z$-boson exhange, the interference between the single-photon exchange and the SMEFT, and the interference between the single-$Z$-boson exchange and SMEFT, respectively. Here, we ignore the $\vv{\Ma_r}^2$ contribution since it scales as $1/\Lambda^4$, formally the same size as contributions from dimension-8 SMEFT operators interfering with the SM.

For the hadron-level cross sections and asymmetries, these different contributions will give rise to corresponding structure functions. In particular, in addition to the usual structure functions encountered in the SM DIS, new structure functions corresponding to SMEFT contributions arise. Thus, including the SMEFT contributions, the DIS differential cross section takes the general form:
\begin{eqnarray}
	\label{eq:DIS_cross_sec}
	\frac{\D2\sigma}{\d x \d y} = \frac{2 \pi y \alpha^2}{Q^4 } \Big\{\> \eta^\gamma L_{\mu \nu}^\gamma W_\gamma^{\mu \nu} + \eta^{\gamma Z} L_{\mu \nu}^{\gamma Z} W_{\gamma Z}^{\mu \nu} + \eta^Z L_{\mu \nu}^{Z} W_Z^{\mu \nu} +\sum_r \xi^{\gamma r} L_{\mu \nu}^{\gamma r} W_{\gamma r}^{\mu \nu} + \sum_r \xi^{Z r} L_{\mu \nu}^{Z r} W_{Z r}^{\mu \nu}  \>\Big \}~,
\end{eqnarray}
where $\alpha$ is the electromagnetic fine structure constant and $L_{\mu \nu}^{\gamma,\gamma Z,Z,\gamma r, Zr}$ and $W_{\mu \nu}^{\gamma,\gamma Z,Z,\gamma r, Zr}$ are the leptonic and hadronic tensors, respectively. The first three terms on the right-hand side (RHS) correspond to the SM contributions from $\Ma_{\gamma \gamma}$, $2 \Ma_{\gamma Z}$, and $\Ma_{ZZ}$, respectively, and the last two sets of terms correspond to the contributions from the SMEFT operators, i.e. $2 \Ma_{\gamma r}$ and $2 \Ma_{Z r}$,  respectively. For completeness, below we collect some useful results to make the form of the cross section explicit. The dimensionless coefficients $\eta^{\gamma, \gamma Z, Z}$, $\xi^{\gamma r}$, and $\xi^{Z r}$ are given by:
\begin{eqnarray}
	\eta^{\gamma } &=& 1~, \nonumber \\
	\eta^{\gamma Z} &=& \frac{G_{\rm F} M_Z^2}{2\sqrt{2} \pi \alpha} \frac{Q^2}{Q^2 + M_Z^2}~, \nonumber \\
	\eta^{Z} &=& \left (\eta^{\gamma Z} \right)^2~,  \\
	\xi^{\gamma r} &=& \frac{\tilde{C}_r}{4 \pi \alpha}\frac{Q^2}{\Lambda^2}~, \nonumber \\
	\xi^{Z r} &=& \eta^{\gamma Z} \frac{\tilde{C}_r}{4 \pi \alpha}\frac{Q^2}{\Lambda^2}~, \nonumber
\end{eqnarray}
where $G_{\rm F}=1.1663787(6) \times 10^{-5}$~GeV$^{-2}$ is the Fermi constant and $M_Z=91.1876 \pm 0.0021$~GeV~\cite{PDG:2020} is the mass of the $Z$ boson. The leptonic tensors in Eq.~(\ref{eq:DIS_cross_sec}) are:
\begin{eqnarray}
	L_{\mu \nu}^{\gamma} &=& 2\big [\> k_\mu k_\nu ' + k_\mu ' k_\nu - k\cdot k ' g_{\mu \nu} - i \lambda_e \epsilon_{\mu \nu \alpha \beta} k^\alpha (k ')^{\beta } \> \big ]~, \nonumber \\
	L_{\mu \nu}^{\gamma Z} &=& -(g_V^e - \lambda_e g_A^e)L_{\mu \nu}^{\gamma}~, \nonumber\\
	L_{\mu \nu}^{Z} &=& (g_V^e - \lambda_e g_A^e)^2L_{\mu \nu}^{\gamma}~,  \\
	L_{\mu \nu}^{\gamma r} &=& (c_{V_r}^e - \lambda_e c_{A_r}^e)L_{\mu \nu}^{\gamma}~,  \nonumber\\
	L_{\mu \nu}^{Zr} &=& -(c_{V_r}^e - \lambda_e c_{A_r}^e)(g_V^e - \lambda_e g_A^e)L_{\mu \nu}^{\gamma}~,  \nonumber
\end{eqnarray}
where $\lambda_e=\pm 1$ denotes the lepton helicity. For positrons, one flips the sign of all the $g_A^e$ and $c_{Ar}^e$ terms and the overall sign of $L^{\gamma Z}$ and $L^{\gamma r}$ above. Using these identities for the leptonic tensors, Eq.~(\ref{eq:DIS_cross_sec}) can be written more explicitly as:
\begin{eqnarray}
	\label{eq:DIS_cross_sec_explicit}
	\frac{\D2\sigma}{\d x \d y} &=& \frac{2 \pi y \alpha^2}{Q^4 } L_{\mu \nu}^\gamma \Big\{\> \eta^\gamma  W_\gamma^{\mu \nu} - \eta^{\gamma Z} (g_V^e - \lambda_e g_A^e) W_{\gamma Z}^{\mu \nu} + \eta^{Z} (g_V^e - \lambda_e g_A^e)^2  W_Z^{\mu \nu} \nonumber \\
	&& +\sum_r \xi^{\gamma r} (c_{V_r}^e - \lambda_e c_{A_r}^e)  W_{\gamma r}^{\mu \nu} - \sum_r \xi^{Z r}  (c_{V_r}^e - \lambda_e c_{A_r}^e)(g_V^e - \lambda_e g_A^e) W_{Z r}^{\mu \nu}  \>\Big \}~.
\end{eqnarray}
Based on the general Lorentz-tensor structure, the available four-momenta, and the nucleus spin vector, $S^\mu$, numerous hadronic tensors are parameterized in terms of structure functions as
\begin{eqnarray}
	\label{eq:HadTensor}
	W_{\mu \nu}^{j} &=& \Big(-{g_{\mu \nu}} + \frac{q_\mu q_\nu}{q^2}\Big) F_1^{j} + \frac{\hat{P}_\mu \hat{P}_\nu}{(P\cdot q)} F_2^j + \frac{i\epsilon_{\mu \nu \alpha \beta}}{2 (P\cdot q)} \Big [ {P^\alpha q^\beta} F_3^j + 2 q^\alpha S^\beta g_1^j  \Big ] \nonumber \\
	&-& \frac{S\cdot q}{(P\cdot q)} \Big [\frac{\hat{P}_\mu \hat{P}_\nu}{P\cdot q} g_4^j + \left(g_{\mu \nu} - \frac{q_\mu q_\nu}{q^2}\right) g_5^j \Big ]~,
\end{eqnarray}
where $\hat{P}_\mu \equiv P_\mu - q_\mu (P\cdot q)/q^2$. The index $j$ denotes the possibilities $\{\gamma, \gamma Z, Z, \gamma r, Z r \}$  and $F_{1,2,3}^j$ and $g_{1,4,5}^j$ denote various unpolarized and polarized nuclear structure functions, respectively. We omit two additional possible Lorentz structures in the hadronic tensor, typically denoted as the polarized structure functions $g_2$ and $g_3$, since these terms give a contribution to the cross section that is suppressed by $M^2/Q^2$ when contracted with the leptonic tensor. Therefore, we do not consider the structure functions $g_{2,3}$ in the rest of our analysis. The nucleus spin vector $S^\mu$ satisfies the constraints $S^2 = -M^2$ and $S\cdot P = 0$. For longitudinal polarization, it takes the canonical form $S^\mu = \lambda_H \> (|\vec{p}|, \> E \frac{\vec{p}}{|\vec{p}|})$, where $\lambda_H = \pm 1$ is the nucleon helicity and $P^\mu = (E, \vec{p})$ is the nucleon four-momentum.

Based on the structure of the cross section in Eq.~(\ref{eq:DIS_cross_sec_explicit}), in conjunction with the form of the hadronic tensor in Eq.~(\ref{eq:HadTensor}), it becomes useful to define the following combinations of structure functions that also include the SMEFT contributions:
\begin{eqnarray}
	\label{eq:GenStrucFunc}
	F_i &=& F_i^{\rm SM, NC} + F_i^{\rm SMEFT}~, \nonumber \\ 
	g_i &=& g_i^{\rm SM, NC} + g_i^{\rm SMEFT}~, 
\end{eqnarray}
where the SM contributions are given by the commonly known NC structure functions 
\begin{eqnarray}
	\label{eq:GenStrucFunc_SM}
	F_i^{\rm SM,NC} &=& F_i^{\gamma} - \eta^{\gamma Z} (g_V^e - \lambda_e g_A^e) F_i^{\gamma Z} + \eta^{Z}  (g_V^e - \lambda_e g_A^e)^2 F_i^Z~,  \nonumber \\
	g_i^{\rm SM,NC} &=& g_i^{\gamma} - \eta^{\gamma Z} (g_V^e - \lambda_e g_A^e) g_i^{\gamma Z} + \eta^{ Z}  (g_V^e - \lambda_e g_A^e)^2 g_i^Z~,  
\end{eqnarray}
and the SMEFT contributions are given by:
\begin{eqnarray}
	\label{eq:GenStrucFunc_SMEFT}
	F_i^{\rm SMEFT} &=&\sum_r \xi^{\gamma r} (c_{V_r}^e - \lambda_e c_{A_r}^e)  F_i^{\gamma r} - \sum_r \xi^{Z r}  (c_{V_r}^e - \lambda_e c_{A_r}^e)(g_V^e - \lambda_e g_A^e) F_i^{Z r}~, \nonumber \\
	g_i^{\rm SMEFT} &=&\sum_r \xi^{\gamma r} (c_{V_r}^e - \lambda_e c_{A_r}^e)  g_i^{\gamma r} - \sum_r \xi^{Z r}  (c_{V_r}^e - \lambda_e c_{A_r}^e)(g_V^e - \lambda_e g_A^e) g_i^{Z r}~.
\end{eqnarray}
The parton-model expressions for the SM structure functions
are summarized below. We also provide the corresponding expressions for the structure functions arising from the interference of the SM with the SMEFT operators:
\begin{eqnarray}
	\left[F_2^\gamma, F_2^{\gamma Z}, F_2^Z, F_2^{\gamma r}, F_2^{Z r}\right] &=& x\sum_f \left[Q_f^2, 2Q_f g_V^f, {g_V^f}^2+{g_A^f}^2, 2Q_f c_{V_r}^f, 2 (g_V^f c^f_{V_r}+g_A^f c^f_{A_r}) \right] (q_f+\bar q_f)~,\label{eq:SF_F2} \nonumber  \\
	\left[F_3^\gamma, F_3^{\gamma Z}, F_3^Z,F_3^{\gamma r},F_3^{Zr}\right] &=& \sum_f \left[0, 2Q_f g_A^f, 2{g_V^f}{g_A^f},2Q_f c_{Ar}^f, 2(g_V^fc_{A_r}^f + g_A^f c_{V_r}^f ) \right] (q_f-\bar q_f)~, \nonumber \\
	\left[g_1^\gamma, g_1^{\gamma Z}, g_1^Z,g_1^{\gamma r},g_1^{Z r}\right] &=& \frac{1}{2} \sum_f \left[Q_f^2, 2Q_f g_V^f, {g_V^f}^2+{g_A^f}^2, 2Q_f c_{V_r}^f, 2 (g_V^f c^f_{V_r}+g_A^f c^f_{A_r})\right] (\Delta q_f+\Delta\bar q_f)~,\label{eq:SF_g1} \nonumber \\
	\left[g_5^\gamma, g_5^{\gamma Z}, g_5^Z, g_5^{\gamma r}, g_5^{Z r} \right] &=& \sum_f \left[0, Q_f g_A^f, g_V^f g_A^f, Q_f c_{A_r}^f, g_V^fc_{A_r}^f + g_A^f c_{V_r}^f\right] (\Delta q_f-\Delta \bar q_f)~,\label{eq:SF_g5}
\end{eqnarray}
where $q_f(x,Q^2)$ and $\Delta q_f(x,Q^2)$ are unpolarized and polarized parton distribution functions (PDFs) of quark flavor $f$, respectively, and $Q_f$ denotes the electric charge in units of the proton charge $e$. In the parton model, at leading order (LO), one has for the structure functions the Callan-Gross relations $F_2^i=2xF_1^i$ and $g_4^i=2xg_5^i$ for  $i=\gamma, \gamma Z, Z, \gamma r, Z r$. For an ion beam (or nuclear target), the neutron PDFs can be related to the proton PDFs by assuming isospin symmetry for the valence quarks: 
\begin{eqnarray}
	\label{eq:pdf_isospin1}
	q_{u/n}(x,Q^2) &=& q_{d/p}(x,Q^2)~,\nonumber \\ q_{d/n}(x,Q^2) &=& q_{u/p}(x,Q^2)~,\\
	\Delta q_{u/n}(x,Q^2)&=&\Delta q_{d/p}(x,Q^2)~, \nonumber \\
	\Delta q_{d/n}(x,Q^2) &=&\Delta q_{u/p}(x,Q^2)~, \nonumber
\end{eqnarray}
while the charm and strange sea quark PDFs are assumed to be identical for the proton and the neutron:
\begin{eqnarray}
	\label{eq:pdf_isospin2}
	q_{s/n}(x,Q^2) &=& q_{s/p}(x,Q^2)~,\nonumber \\ q_{c/n}(x,Q^2) &=& q_{c/p}(x,Q^2)~,\\
	\Delta q_{s/n}(x,Q^2)&=&\Delta q_{s/p}(x,Q^2)~, \nonumber \\
	\Delta q_{c/n}(x,Q^2) &=&\Delta q_{c/p}(x,Q^2)~. \nonumber
\end{eqnarray}
For the deuteron, an isoscalar bound state of a proton and a neutron, the PDFs can be constructed from the proton and neutron PDFs as:
\begin{eqnarray}
	q_{f/D}(x,Q^2) &=& \frac 12 [q_{f/p}(x,Q^2) + q_{f/n}(x,Q^2)]~,  \\
	\Delta q_{f/D}(x,Q^2) &=& \frac 12 [\Delta q_{f/p}(x,Q^2) + \Delta q_{f/n}(x,Q^2)]~, \nonumber
\end{eqnarray}
for quark flavor $f$. 

In terms of the generalized structure functions in Eq.~(\ref{eq:GenStrucFunc}), which include dependence on the electron helicity, $\lambda_e$, as seen in Eqs.~(\ref{eq:GenStrucFunc_SM}) and (\ref{eq:GenStrucFunc_SMEFT}), one can write the cross section for given electron and nucleon helicities,  including SMEFT operator contributions, as: 
\begin{eqnarray}
	\label{eq:spin-dep-cross-section}
	\frac{\D2\sigma (\lambda_e, \lambda_H)}{\d x \d y} &=& \frac{4 \pi \alpha^2}{x y Q^2 }\Bigg \{ xy^2 F_1 + (1-y) F_2 - \lambda_e \frac{y}{2} (2-y)\> xF_3 + \lambda_e \lambda_H (2-y)\> xy \> g_1 \nonumber \\
	&-&  \lambda_H  (1-y) \>g_4 - \lambda_H  \> x y^2 \>g_5 \Bigg \}~,\label{eq:DISxsec_spin}
\end{eqnarray}
where we ignore the electron mass and all target-mass correction terms that are proportional to $M^2/Q^2$.

To connect to experimentally measured observables, it is convenient to write the scattering cross section of Eq.~(\ref{eq:DISxsec_spin}) as the sum of four components that depend on the spin direction of the initial electron and hadron: $\d\sigma_0, \d\sigma_e, \d\sigma_H$, and $\d\sigma_{eH}$, where each $\d\sigma$ represents the differential cross section as $\D2\sigma/(\d x\d y)$. The quantity $\d\sigma_0$ is the unpolarized cross section, $\d\sigma_e$ and $\d\sigma_H$ denote the cross-section differences between initial electron and hadron states of opposite helicity, respectively, and $\d\sigma_{eH}$ is the cross-section difference between initial electron and hadron states with the same and opposite helicities defined in the center-of-mass frame. These quantities can be formed by using Eq.~(\ref{eq:DISxsec_spin}) as:
\begin{eqnarray}
	\label{eq:cross_section_quantities}
	\d\sigma_0 &=&  \frac{1}{4} \Big [\> \d\sigma |_{\lambda_e=+1, \lambda_H=+1} + \d\sigma |_{\lambda_e=+1, \lambda_H=-1} + \d\sigma |_{\lambda_e=-1, \lambda_H=+1} + \d\sigma |_{\lambda_e=-1, \lambda_H=-1}\>\Big ]~, \nonumber \\
	\d\sigma_e &=& \frac{1}{4} \Big [\> \d\sigma |_{\lambda_e=+1, \lambda_H=+1} + \d\sigma |_{\lambda_e=+1, \lambda_H=-1} - \d\sigma |_{\lambda_e=-1, \lambda_H=+1} - \d\sigma |_{\lambda_e=-1, \lambda_H=-1}\>\Big ]~, \nonumber \\
	\d\sigma_H &=&\frac{1}{4} \Big [\> \d\sigma |_{\lambda_e=+1, \lambda_H=+1} - \d\sigma |_{\lambda_e=+1, \lambda_H=-1} + \d\sigma |_{\lambda_e=-1, \lambda_H=+1} - \d\sigma |_{\lambda_e=-1, \lambda_H=-1}\>\Big ]~,\nonumber \\
	\d\sigma_{eH} &=& \frac{1}{4} \Big [\> \d\sigma |_{\lambda_e=+1, \lambda_H=+1} - \d\sigma |_{\lambda_e=+1, \lambda_H=-1} - \d\sigma |_{\lambda_e=-1, \lambda_H=+1} + \d\sigma |_{\lambda_e=-1, \lambda_H=-1}\>\Big ]~, 
\end{eqnarray}
and can be computed in conjunction with Eqs.~(\ref{eq:GenStrucFunc}), (\ref{eq:GenStrucFunc_SM}), and (\ref{eq:GenStrucFunc_SMEFT}). 

The SM contributions to the DIS cross sections with the target-mass terms omitted are: 
\begin{eqnarray}
	\frac{\D2\sigma_0}{\d x \d y} &=& \frac{4\pi\alpha^2}{xyQ^2}\left\{ xy^2 \left[F_1^\gamma - g_V^e\eta_{\gamma Z}F_1^{\gamma Z}+({g_V^e}^2+{g_A^e}^2) \eta_ZF_1^Z\right]
	\right.\nonumber\\
	&& \left.+(1-y)\left[F_2^\gamma - g_V^e\eta_{\gamma Z}F_2^{\gamma Z} + ({g_V^e}^2+{g_A^e}^2)\eta_Z F_2^Z\right]\right.\nonumber\\
	&&\left. -\frac{xy}{2}(2-y)\left[g_A^e\eta_{\gamma Z} F_3^{\gamma Z}-2g_V^eg_A^e\eta_Z F_3^Z\right]\right\}~, \nonumber \label{eq:sig0_shortcut}\\
	\frac{\D2\sigma_e}{\d x \d y} &=& \frac{4\pi\alpha^2}{xyQ^2}\left\{ xy^2\left[g_A^e\eta_{\gamma Z}F_1^{\gamma Z}-2g_V^eg_A^e\eta_Z F_1^Z\right]+
	(1-y)\left[g_A^e\eta_{\gamma Z}F_2^{\gamma Z}-2g_V^eg_A^e\eta_Z F_2^Z\right]\right.\nonumber\\
	&&\left.+\frac{xy}{2}(2-y)\left[g_V^e\eta_{\gamma Z} F_3^{\gamma Z}-({g_V^e}^2+{g_A^e}^2) \eta_Z F_3^Z\right]
	\right\}~,\label{eq:sige_shortcut}
	\\
	\frac{\D2\sigma_H}{\d x \d y} &=& \frac{4\pi\alpha^2}{xyQ^2}\left\{
	\left(2-y\right)xy\left[g_A^e\eta_{\gamma Z}g_1^{\gamma Z}-2g_V^eg_A^e\eta_Z g_1^Z\right] \right.\nonumber\\
	&&\left.+ (1-y)\left[g_V^e\eta_{\gamma Z}g_4^{\gamma Z}-({g_V^e}^2+{g_A^e}^2) \eta_Z g_4^Z\right]\right.\nonumber\\
	&&\left.-xy^2\left[g_V^e\eta_{\gamma Z} g_5^{\gamma Z}-({g_V^e}^2+{g_A^e}^2) \eta_Z g_5^Z\right]\right\}~, \nonumber\label{eq:sigp_shortcut}
	\\
	\frac{\D2\sigma_{eH}}{\d x \d y} &=& \frac{4\pi\alpha^2}{xyQ^2}\left\{
	\left(2-y\right)xy\left[g_1^\gamma-g_V^e\eta_{\gamma Z}g_1^{\gamma Z}+({g_V^e}^2+{g_A^e}^2)\eta_Z g_1^Z\right]\right.\nonumber\\
	&& \left.- (1-y)\left[g_A^e\eta_{\gamma Z}g_4^{\gamma Z}-2g_V^e g_A^e \eta_Z g_4^Z\right]-xy^2\left[g_A^e\eta_{\gamma Z} g_5^{\gamma Z}-2 g_V^e g_A^e \eta_Z g_5^Z\right]
	\right\}~. \nonumber\label{eq:sigep_shortcut}
\end{eqnarray}
The SMEFT contributions are:
\begin{eqnarray} 
	\frac{\D2\sigma_0^\textrm{SMEFT}}{\d x \d y} &=&\frac{4\pi\alpha^2}{xyQ^2}\sum_r\left[xy^2(c_{V_r}^e\xi_{\gamma r}F_1^{\gamma r} - (c_{V_r}^eg_{V}^e + c_{A_r}^eg_A^e)\xi_{Zr}F_1^{Zr})\right.\nonumber\\
	&&+(1-y)(c_{V_r}^e\xi_{\gamma r}F_2^{\gamma r} - (c_{V_r}^eg_V^e + c_{A_r}^eg_A^e)\xi_{Zr}F_2^{Zr})\nonumber\\
	&&+\left.\frac{xy}{2}(2-y)(c_{A_r}^e\xi_{\gamma r}F_3^{\gamma r} - (c_{V_r}^eg_A^e + c_{A_r}^eg_V^e)\xi_{Zr}F_3^{Zr})\right]~,\nonumber\\
	\frac{\D2\sigma_e^\textrm{SMEFT}}{\d x \d y} &=&-\frac{4\pi\alpha^2}{xyQ^2}\sum_r\left[xy^2(c_{A_r}^e\xi_{\gamma r}F_1^{\gamma r} - (c_{V_r}^eg_{A}^e + c_{A_r}^eg_V^e)\xi_{Zr}F_1^{Zr})\right.\nonumber\\
	&&+(1-y)(c_{A_r}^e\xi_{\gamma r}F_2^{\gamma r} - (c_{V_r}^eg_A^e + c_{A_r}^eg_V^e)\xi_{Zr}F_2^{Zr})\nonumber\\
	&&\left.+\frac{xy}{2}(2-y)(c_{V_r}^e\xi_{\gamma r}F_3^{\gamma r} - (c_{A_r}^eg_A^e + c_{V_r}^eg_V^e)\xi_{Zr}F_3^{Zr})\right]~,\nonumber\\
	\frac{\D2\sigma_H^\textrm{SMEFT}}{\d x \d y} &=&-\frac{4\pi\alpha^2}{xyQ^2}\sum_r\left[xy(2-y)(c_{A_r}^e \xi_{\gamma r}g_1^{\gamma r} -(c_{V_r}^eg_A^e+c_{A_r}^eg_V^e)\xi_{Zr}g_1^{Zr})\right.\nonumber\\
	&&+(1-y)( c_{V_r}^e \xi_{\gamma r}g_4^{\gamma r} -(c_{A_r}^eg_A^e+c_{V_r}^eg_V^e)\xi_{Zr}g_4^{Zr})\nonumber\\
	&&\left.+xy^2( c_{V_r}^e\xi_{\gamma r} g_5^{\gamma r} -(c_{A_r}^eg_A^e+c_{V_r}^eg_V^e)\xi_{Zr}g_5^{Zr})\right]~,\nonumber\\
	\frac{\D2\sigma_{eH}^\textrm{SMEFT}}{\d x \d y} &=&\frac{4\pi\alpha^2}{xyQ^2}\sum_r\left[xy(2-y)( c_{V_r}^e\xi_{\gamma r} g_1^{\gamma r} -(c_{A_r}^eg_A^e+c_{V_r}^eg_V^e)\xi_{Zr}g_1^{Zr})\right.\nonumber\\
	&&+(1-y)( c_{A_r}^e\xi_{\gamma r} g_4^{\gamma r} -(c_{V_r}^eg_A^e+c_{A_r}^eg_V^e)\xi_{Zr}g_4^{Zr})\nonumber\\
	&&\left.+xy^2(c_{A_r}^e\xi_{\gamma r} g_5^{\gamma r} -(c_{V_r}^eg_A^e+c_{A_r}^eg_V^e)\xi_{Zr}g_5^{Zr})\right]~.
\end{eqnarray}

If a positron beam becomes available at the EIC, one can measure both $e^+H$ and $e^-H$ cross sections and study the differences. Neglecting target-mass terms again and writing the SM and SMEFT contributions all together, we have: 
\begin{eqnarray}
	\frac{\D2\sigma_0^{e^+}}{\d x \d y} - \frac{\D2\sigma_0^{e^-}}{\d x \d y} &=& \frac{4\pi\alpha^2}{xyQ^2}g_A^e \left[
	xy(2-y)\left(\eta_{\gamma Z}F_3^{\gamma Z}-2g_V^e\eta_{Z} F_3^Z\right)\right] \nonumber\\
	&& -\frac{8\pi\alpha^2}{xyQ^2}\sum_r\frac{xy}{2} (2 - y)  c_{A_r}^e (\xi_{\gamma r} F_3^{\gamma r} + 2g_V^e \xi_{Zr}F_3^{Zr} )~,\nonumber\\
	\frac{\D2\sigma_e^{e^+}}{\d x \d y} - \frac{\D2\sigma_e^{e^-}}{\d x \d y} &=& -\frac{4\pi\alpha^2}{xyQ^2}g_A^e \left[
	2(1-y)\left(\eta_{\gamma Z}F_2^{\gamma Z}-2g_V^e\eta_{Z} F_2^Z\right)
	+ 2xy^2\left(\eta_{\gamma Z}F_1^{\gamma Z}-2g_V^e\eta_Z F_1^Z\right)\right] \nonumber\\
	&& +\frac{8\pi\alpha^2}{xyQ^2}\sum_r\left[(1-y) c_{A_r}^e(\xi_{\gamma r}F_2^{\gamma r} + 2g_V^e \xi_{Zr} F_2^{Zr}) + xy^2 c_{A_r}^e(\xi_{\gamma r}F_1^{\gamma r} + 2g_V^e \xi_{Zr} F_1^{Zr}) \right]~,\nonumber\\
	\frac{\D2\sigma_H^{e^+}}{\d x \d y} - \frac{\D2\sigma_H^{e^-}}{\d x \d y} &=& -\frac{4\pi\alpha^2}{xyQ^2}g_A^e \left[
	2xy\left(2-y\right)\left(\eta_{\gamma Z}g_1^{\gamma Z}-2g_V^e\eta_{Z} g_1^Z\right)\right] \nonumber\\
	&& +\frac{8\pi\alpha^2}{xyQ^2}\sum_r xy (2 - y)c_{A_r}^e (\xi_{\gamma r}g_1^{\gamma r} -2 g_V^e\xi_{Zr} g_1^{Zr}) ~,\nonumber\\
	\frac{\D2\sigma_{eH}^{e^+}}{\d x \d y} - \frac{\D2\sigma_{eH}^{e^-}}{\d x \d y} &=& \frac{4\pi\alpha^2}{xyQ^2}g_A^e\left[2(1-y)\left(\eta_{\gamma Z}g_4^{\gamma Z}-2g_V^e \eta_{Z} g_4^Z\right)+2xy^2\left(\eta_{\gamma Z} g_5^{\gamma Z}-2 g_V^e \eta_{Z} g_5^Z\right)\right] \nonumber\\
	&& -\frac{8\pi\alpha^2}{xyQ^2}\sum_r\left[ (1 - y) c_{A_r}^e (\xi_{\gamma r}g_4^{\gamma r} +2g_V^e \xi_{Zr} g_4^{Zr})+xy^2 c_{A_r}^e (\xi_{\gamma r}g_5^{\gamma r} + 2g_V^e \xi_{Zr} g_5^{Zr}) \right]~.
\end{eqnarray}

In this study, we focus on measurements of both parity-violating and lepton-charge asymmetries. The parity-violating asymmetry can be formed either by comparing right-handed and left-handed electron scattering from unpolarized hadrons, referred to as the ``unpolarized PV asymmetry'':
\begin{eqnarray}
	\APVe &\equiv&\frac{\d\sigma_e}{\d\sigma_0}~,
\end{eqnarray}
or by comparing unpolarized electron scattering off right-handed and left-handed hadrons, referred to as the``polarized PV asymmetry'':
\begin{eqnarray}
	\APVH &\equiv&\frac{\d\sigma_H}{\d\sigma_0}~.
\end{eqnarray}
If a positron beam becomes available in the future, the lepton-charge asymmetry, defined as the unpolarized DIS cross-section asymmetry between electron and positron beams: 
\begin{eqnarray}
	\ALCH &=& \frac{\d\sigma_0^{e^+} - \d\sigma_0^{e^-}}{\d\sigma_0^{e^+} + \d\sigma_0^{e^-}}~, \label{eq:Alc}
\end{eqnarray}
will provide additional constraints on SMEFT interactions. On the other hand, the double-spin asymmetry, $A_{\rm PV}^{(eH)} \equiv{\d\sigma_{eH}}/{\d\sigma_0}$, is the primary observable to study the nucleon spin structure but is not within the scope of this work. Similarly, a complete list of lepton-charge asymmetries that includes lepton-polarization dependence can be found in~\cite{Zheng:2021hcf}, but they provide similar constraints to SM and SMEFT studies as the unpolarized asymmetry defined in Eq.~\ref{eq:Alc} and are not discussed in this work.

\vspace{-.3cm}

%% file: sections/2b-measurement_of_pv_asymmetries_at_eic.tex
\subsection{Measurement of Parity-Violating Asymmetries at the EIC}
In DIS experiments utilizing an electron beam of polarization $P_e$ and a hadron beam of polarization $P_H$, the measured differential cross section is:
\begin{eqnarray}
	\label{eq:crossPePH}
	\d\sigma &=& \d\sigma_{0} + P_e \d\sigma_{e} + P_H \d\sigma_{H} + P_e P_H \d\sigma_{eH}~,
\end{eqnarray}
where $P_e$ and $P_H$ have the same sign as the respective beam helicities, $\lambda_e$ and $\lambda_H$, and can take the values $-1\leqslant P_e,P_H\leqslant 1$. The various cross-section components in Eq.~(\ref{eq:crossPePH}) are given in Eqs.~(\ref{eq:sig0_shortcut}). 

The PVDIS asymmetry can be formed by flipping the spin direction of either the electron beam or the ion beam. For the EIC, beams of opposite polarizations will be injected into the storage rings alternately, and thus each of the signs of both electron and ion polarizations is flipped periodically on a short time scale. This is in contrast to HERA, where data were taken with positive then negative electron polarization, with such long time intervals in between that runs with opposite electron polarizations are essentially two independent experiments. 

We express the measured DIS event counts during a certain beam-helicity state as:
\begin{eqnarray}
	N^{++} &=& \adet L^{++}\left(\d\sigma_0 + \vert P_e^{++}\vert  \d\sigma_e + \vert P_H^{++}\vert  \d\sigma_H + \vert P_e^{++}\vert \vert P_H^{++}\vert  \d\sigma_{eH}\right)~,\\
	N^{+-} &=& \adet L^{+-}\left(\d\sigma_0 + \vert P_e^{+-}\vert  \d\sigma_e - \vert P_H^{+-}\vert  \d\sigma_H - \vert P_e^{+-}\vert \vert P_H^{+-}\vert  \d\sigma_{eH}\right)~,\\
	N^{-+} &=& \adet L^{-+}\left(\d\sigma_0 - \vert P_e^{-+}\vert  \d\sigma_e + \vert P_H^{-+}\vert  \d\sigma_H - \vert P_e^{-+}\vert \vert P_H^{-+}\vert  \d\sigma_{eH}\right)~,\\
	N^{--} &=& \adet L^{--}\left(\d\sigma_0 - \vert P_e^{--}\vert  \d\sigma_e - \vert P_H^{--}\vert  \d\sigma_H + \vert P_e^{--}\vert \vert P_H^{--}\vert  \d\sigma_{eH}\right)~,
\end{eqnarray}
%where $ij=++,+-,-+,--$ represents the electron and proton helicity states with their time sequence depending on the helicity pattern of the beam injection, $L^{ij}$ stands for the integrated luminosity, and $P_e^{ij}$ and $P_H^{ij}$ are the electron and proton (or ion) beam polarizations, respectively, during the corresponding helicity bunch $ij$. The factor $\adet$ represents the detector phase space, acceptance, and efficiency.  
where $L^{ij}$ stands for the integrated luminosity, and $P_e^{ij}$ and $P_H^{ij}$ are the electron and the proton (or ion) beam polarizations during the corresponding helicity bunch $ij$. The superscripts $ij=++,+-,-+,--$ represent the electron and the proton helicity states with their time sequence depending on the helicity pattern of the beam injection. The $a_{det}$ factor represents the detector phase space, acceptance and efficiency. In the simplest case, if we assume both beam polarizations, the luminosity, and detector efficiency and acceptance do not vary with time, then: 
\begin{eqnarray}
	\label{eq:cross_pol}
	\d\sigma_0 &=& \frac{1}{4}\left(\d\sigma^{++}+\d\sigma^{+-}+\d\sigma^{-+}+\d\sigma^{--}\right)~,\\
	\d\sigma_e &=& \frac{1}{4\vert P_e\vert}\left(\d\sigma^{++}+\d\sigma^{+-}-\d\sigma^{-+}-\d\sigma^{--}\right)~,\\
	\d\sigma_H &=& \frac{1}{4\vert P_H\vert}\left(\d\sigma^{++}-\d\sigma^{+-}+\d\sigma^{-+}-\d\sigma^{--}\right)~,\\
	\d\sigma_{eH} &=& \frac{1}{4\vert P_e\vert \vert P_H\vert}\left(\d\sigma^{++}-\d\sigma^{+-}-\d\sigma^{-+}+\d\sigma^{--}\right)~, 
\end{eqnarray} 
where we define the experimentally measured cross section by $\d\sigma^{ij}\equiv N^{ij}/L^{ij}/\adet$. The PVDIS asymmetry due to the electron spin-flip can be extracted from data by taking the ratio of the cross sections. Because spin-flips of both electron and hadron beams will be carried out at very short time-scale, the factor $\adet$ can be assumed constant and cancels out when forming the asymmetry, and we can extract the asymmetry from experimentally measured yields, defined by $Y^{ij}\equiv N^{ij}/L^{ij}$, as:
\begin{eqnarray}
	\APVe &\equiv&
	\frac{\d\sigma_e}{\d\sigma_0}=\frac{1}{\vert P_e\vert}\frac{Y^{++}+Y^{+-}-Y^{-+}-Y^{--}}{Y^{++}+Y^{+-}+Y^{-+}+Y^{--}}~, \label{eq:Apv_e}
\end{eqnarray}
and that due to proton (ion) spin flip can be similarly extracted as:
\begin{eqnarray}
	\APVH &\equiv&\frac{\d\sigma_H}{\d\sigma_0}=\frac{1}{\vert P_H\vert}\frac{Y^{++}-Y^{+-}+Y^{-+}-Y^{--}}{Y^{++}+Y^{+-}+Y^{-+}+Y^{--}}~. \label{eq:Apv_H}
\end{eqnarray}
The design of the EIC requires that the point-to-point luminosity uncertainty be at $10^{-4}$ level. Therefore, the dominant experimental uncertainty would come from electron and proton (ion) polarimetry for $\APVe$ and $\APVH$, respectively.

\vspace{-.3cm}

%% file: sections/2c-measurement_of_lc_asymmetries_at_eic.tex
\subsection{Measurement of Lepton-Charge Asymmetries at the EIC\label{sec:Alc_meas}}
Unlike PV asymmetries, which can be formed by comparing scattering yields of right-handed vs. left-handed electron or hadron scattering on a short time-scale, the measurement of the LC asymmetry requires comparison between electron and positron runs and thus relies on two independent cross-section measurements.  To reduce the uncertainty in the measurement of $\ALCH$, we can reverse the polarity of the magnet to minimize the systematic uncertainty due to differences in $e^-$ and $e^+$ detection.  In this case, the main experimental systematic uncertainty will come from the luminosity difference between $e^-$ and $e^+$ runs, which is assumed to be 2\% (relative in luminosity, absolute in $\ALCH$) in this analysis.

%% file: sections/3a-ecce_detector_config_for_inclusive_nc.tex
\subsection{ECCE Detector Configuration for Inclusive Neutral-Current Study}\label{sec:ecce_detector}
The EIC Comprehensive Chromodynamics Experiment (ECCE) detector concept~\cite{ecce-detector-proposal} addresses the full EIC science mission as described in the EIC community White Paper~\cite{Accardi:2012qut} and the 2018 National Academies of Science (NAS) Report~\cite{NAP25171}. It is simultaneously fully capable, low-risk, and cost-effective. ECCE strategically repurposes select components of existing experimental equipment to maximize its overall capabilities within the envelope of planned resources. For example, the central barrel of the detector incorporates the storied 1.4-T BaBar superconducting solenoid and the sPHENIX barrel hadronic calorimeter, currently under construction. 

For EW NC physics studied in this work, we focus on the detection and identification of inclusive scattered electrons, provided by ECCE’s tracking system~\cite{ecce-note-det-2021-03} combined with electromagnetic calorimetry~\cite{ecce-note-det-2021-02} in a nearly hermetic coverage. ECCE features a hybrid tracking detector design using three state-of-the-art technologies to achieve high-precision primary and decay-vertex determination, fine momentum-tracking, and distance-of-closest-approach resolution in the region $\vert\eta\vert\leqslant 3.5$ with full azimuth coverage. The ECCE tracking detector consists of the Monolithic Active Pixel Sensor (MAPS)-based silicon vertex/tracking subsystem, the µRWELL tracking subsystem, and the AC-LGAD outer tracker, which also serves as the time-of-flight detector, all optimized by artificial intelligence. For the electromagnetic  calorimeter, the system employed by ECCE consists of the PbWO$_4$-based Electron Endcap EM Calorimeter (EEMC) for the region $-3.7<\eta<-1.8$, the SciGlass-based barrel ECal for the region $-1.7<\eta<1.3$, and the Pb-Scintillator shashlik-type forward ECal (FEMC, hadron beam direction) that covers roughly $1.3<\eta<4$. 

For the inclusive DIS kinematics determination, we use single-electron simulations in the full detector to study the measurement of the electron momentum and trajectory, and characterize the difference between detected and true values as smearing in the electron momentum and polar and azimuthal angles. The smearing can then be applied to simulated events without involving the full detector. This is referred to as fast-smearing and is the simulation method adopted here that yield all physics projections provided in this work. On the other hand, other methods that can be used to identify DIS kinematics, such as detecting all hadrons in the final state, or detecting both the scattered electron and all hadrons, are not investigated here. Similarly, the use of the electromagnetic calorimeter can improve track identification in part of the phase space, but is not included in this work.

%% file: sections/3b-simulation_with_fast_smearing.tex
\subsection{Simulation with Fast Smearing}\label{sec:proj_ecce_sim}
We use the {\tt Djangoh} event generator~\cite{Charchula:1994kf} (version 4.6.16~\footnote{\href{https://github.com/spiesber/DJANGOH}{https://github.com/spiesber/DJANGOH}}) that includes full electromagnetic and electroweak radiative effects to generate Monte-Carlo (MC) events for each of the four beam-energy and two beam-type combinations: $18\times 275(137)$, $10\times 275(137)$, $10\times 100$, and $5\times 100$~GeV for $ep$ ($eD$) collisions, respectively. For the deuterium ion beam, the energy specified is per nucleon. The fast-smearing method is applied to inclusive electron events in the {\tt Djangoh} output, and the physics cross section and parity-violating asymmetries are calculated event-by-event using a modified user routine of {\tt Djangoh}. The number of scattered DIS electrons is then calculated using the cross-section information and the expected integrated luminosity after correcting for bin migration.

\begin{figure}[h]
	\includegraphics[width=0.48\textwidth]{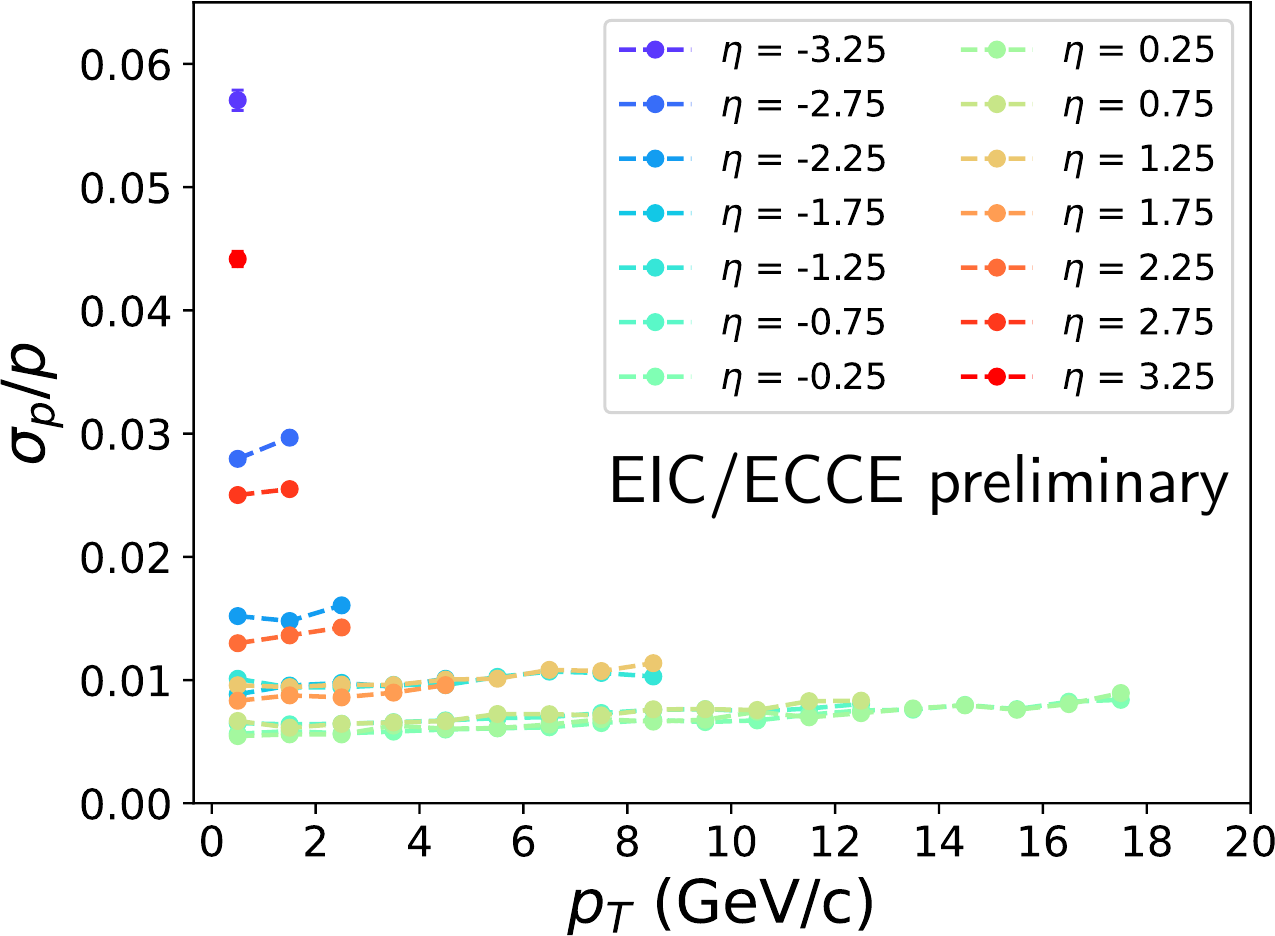}\\
	\includegraphics[width=0.48\textwidth]{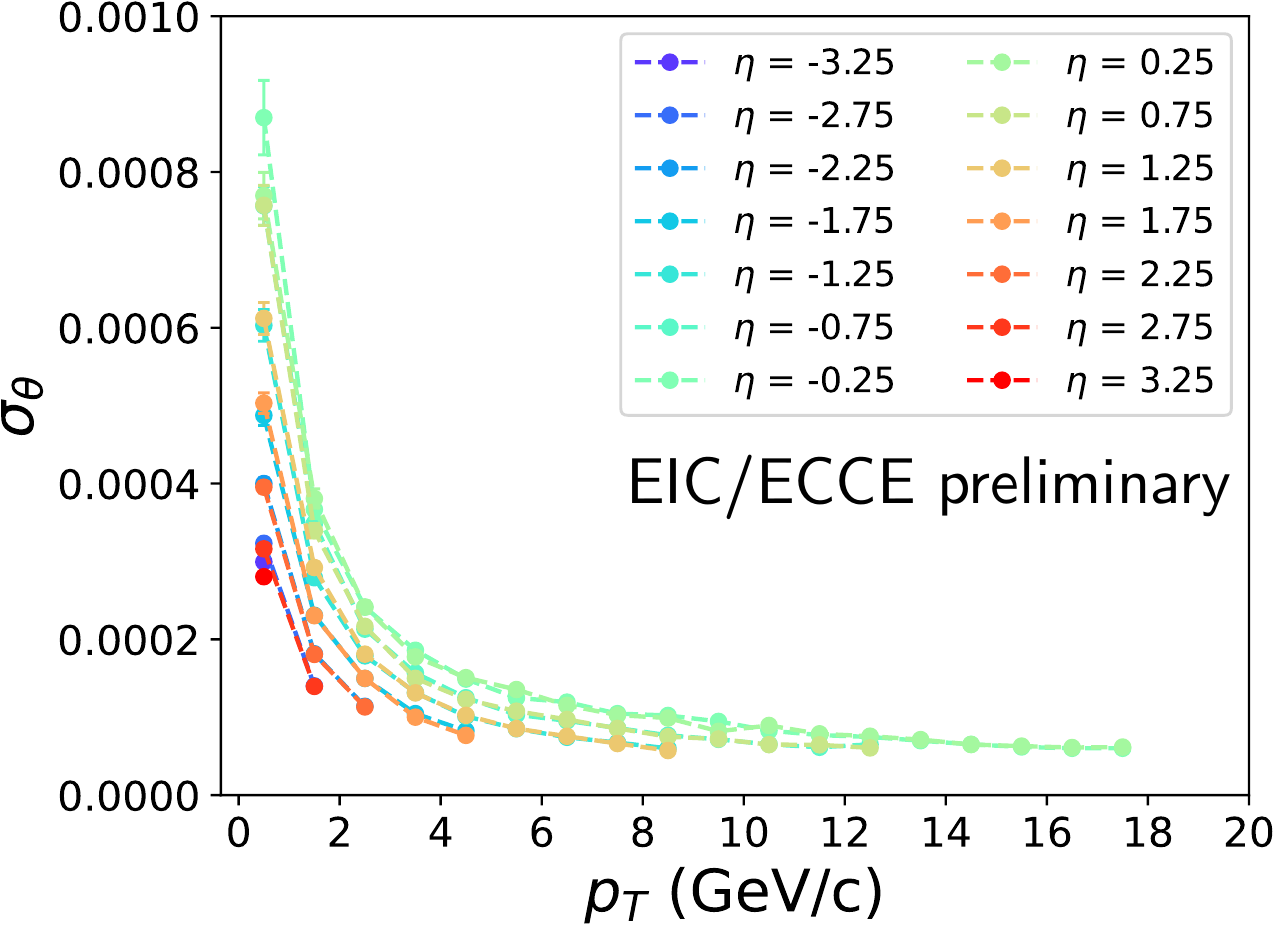}
	\includegraphics[width=0.48\textwidth]{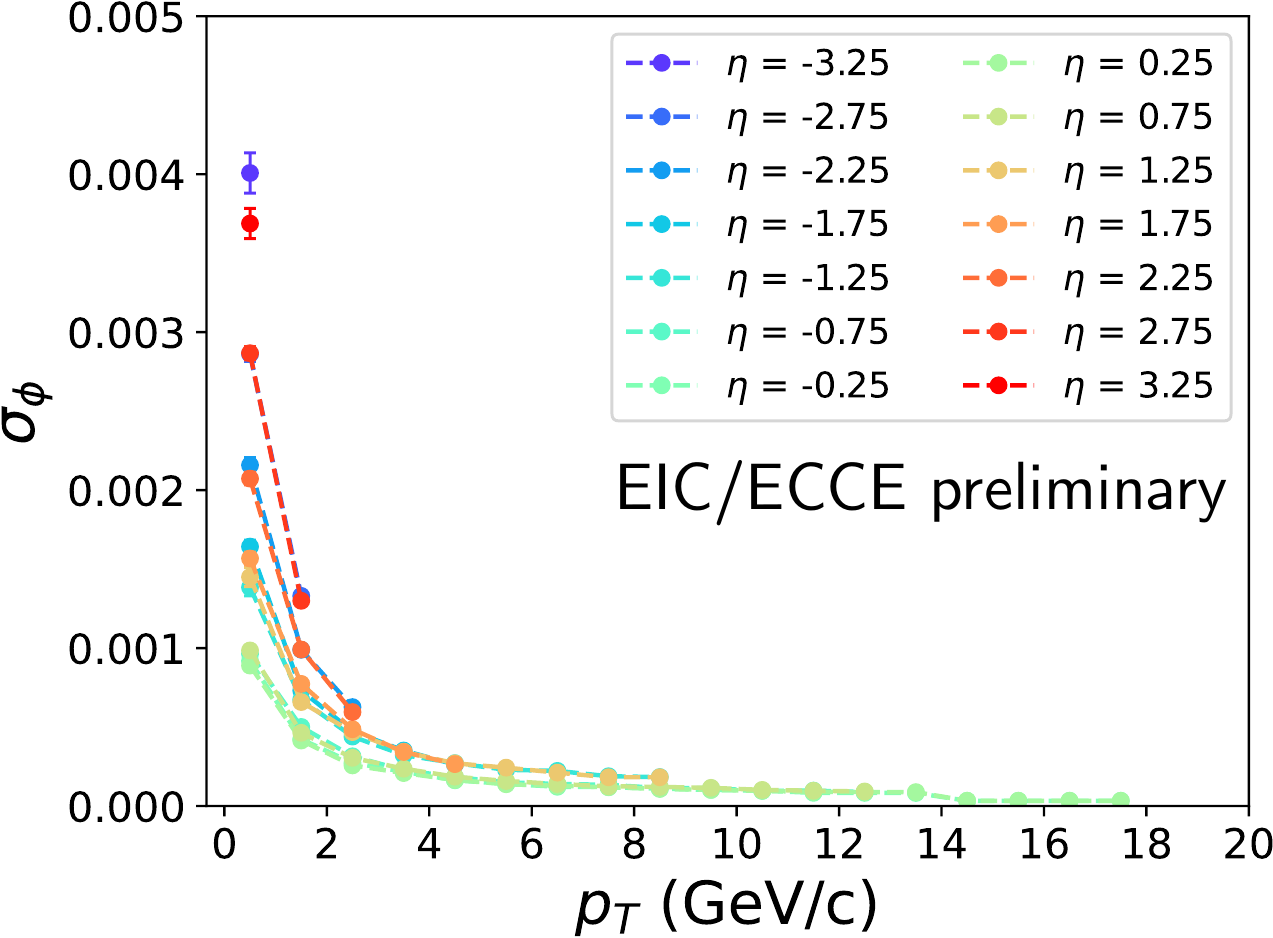}
	\caption{RMS values for fast-smearing spectra obtained from single electron-gun simulation of July 2021 concept of ECCE. The unit for $\sigma_\theta$ and $\sigma_\phi$ is radians. }
	\label{fig:fast_smearing_res}
\end{figure}

The detector fast-smearing is obtained from a single-electron gun simulation.  Resolution spectra are determined for 57 evenly-spaced bins for the pseudo-rapidity range $\eta=(-3.5625,$ $3.5625)$ and 1~GeV-wide bins in the transverse momentum, $p_T$. For each {\tt Djangoh}-simulated event, smearing in the electron momentum, $p$, and polar and azimuthal angles $\theta$ and $\phi$ are randomly picked from the corresponding spectrum and applied to the event, which are used to determine the detected kinematics of the event. While the smearing spectra are not exactly Gaussian-shaped, they are fitted with a Gaussian function. The fitted root-mean-square (RMS) values extracted for illustration purposes are displayed in Fig.~\ref{fig:fast_smearing_res}.

Using the fast-smearing method, we generate 20~M total MC events for each of the beam-energy combinations.Of these 20~M, 10~M events are generated to study the kinematic coverage over the full phase space. The remaining 10~M events are generated with $Q^2_\mathrm{min}=50$~GeV$^2$ for which DIS events have the most impact on the extraction of the weak mixing angle. The drawback of the fast-smearing method is that no selection of the hadronic state is implemented. Methods utilizing hadronic final states such as the double-angle method may provide better DIS event identification for certain kinematic range and thus improve precision of the analysis.

Bin migration of inclusive scattering electrons due to internal and external radiative effects is studied with fast-smearing simulation and treated using the ``R matrix" unfolding method~\cite{Schmitt:2016orm}. Background reactions are studied using the hadronic final state generated by {\tt Djangoh} (with $Q^2_{\rm min}=1.0$~GeV$^2$), and another Monte-Carlo simulation of photoproduction events are generated by {\tt Pythia} (version 6.428, with $Q^2_{\rm min}=0$). All events are passed through the full ECCE simulation. We find that the highest background events occur at high $y$ values. These events are rejected at the event-selection stage; see the next section.

We have also studied how our results change if a simple ``theory-only" simulation without a detailed detector simulation is performed. We find two major differences with respect to the current analysis:
\begin{itemize}

\item As mentioned in the next section, we use the inelasticity constraint $0.1<y<0.9$ in our current simulation. We find that the regions $0.1<y<0.2$ and $0.8<y<0.9$ are not reliably modeled without a detailed detector simulation. Our theory-only simulation cannot accurately reproduce the expected event counts in this region due to missing detector response effects. We therefore must remove these regions, leading to an effective reduction of statistics for the theory-only simulation.

\item Secondly, in the $0.2<y<0.8$ region considered, the total error is relatively 10 to 30\% more optimistic in each bin compared to the full detector simulation, with the 30\% differences occurring near the boundaries of the $y$ region.

\end{itemize}
The net result of these two competing effects is that theory-only bounds are up to 10\% more optimistic than those found with a full detector simulation.

%% file: sections/3c-event_selection.tex
\subsection{Event Selection}\label{sec:ecce_event_selection}
\begin{figure}[h]
	\includegraphics[width=0.48\textwidth]{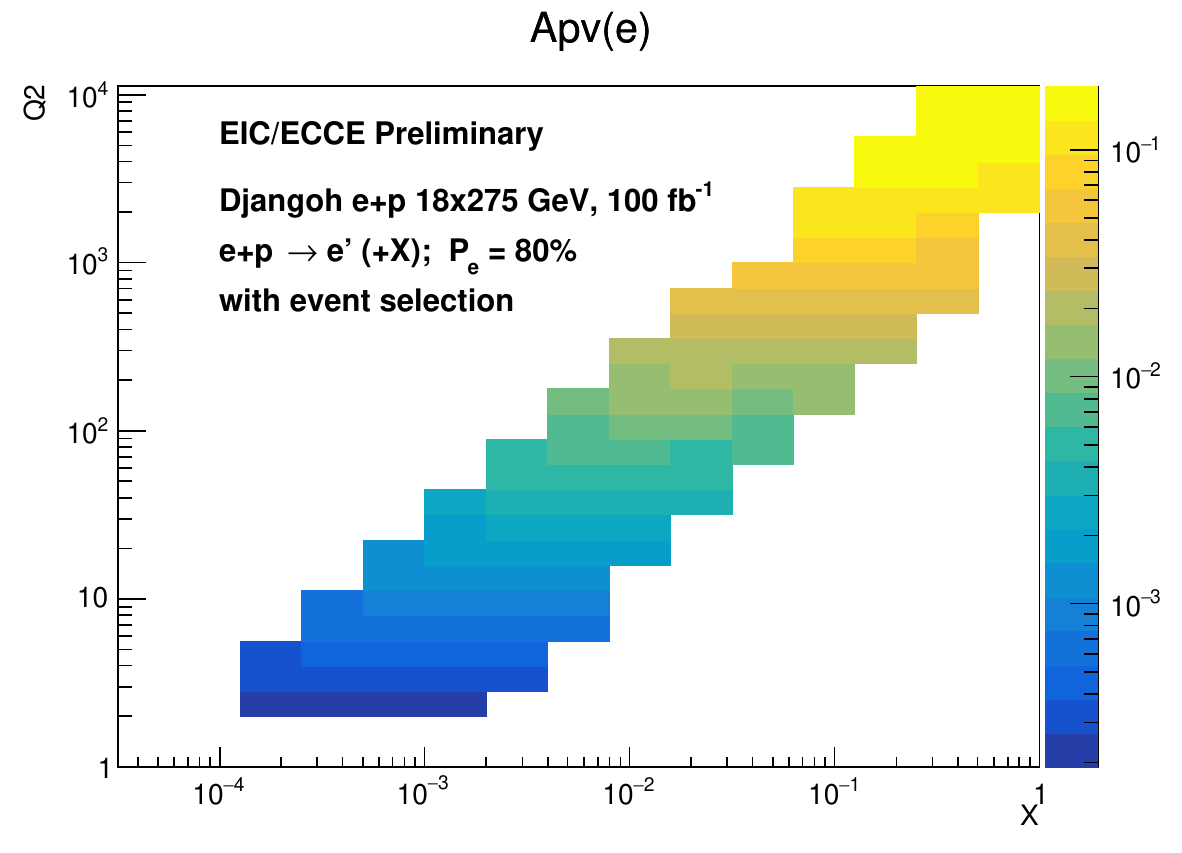}
	\includegraphics[width=0.48\textwidth]{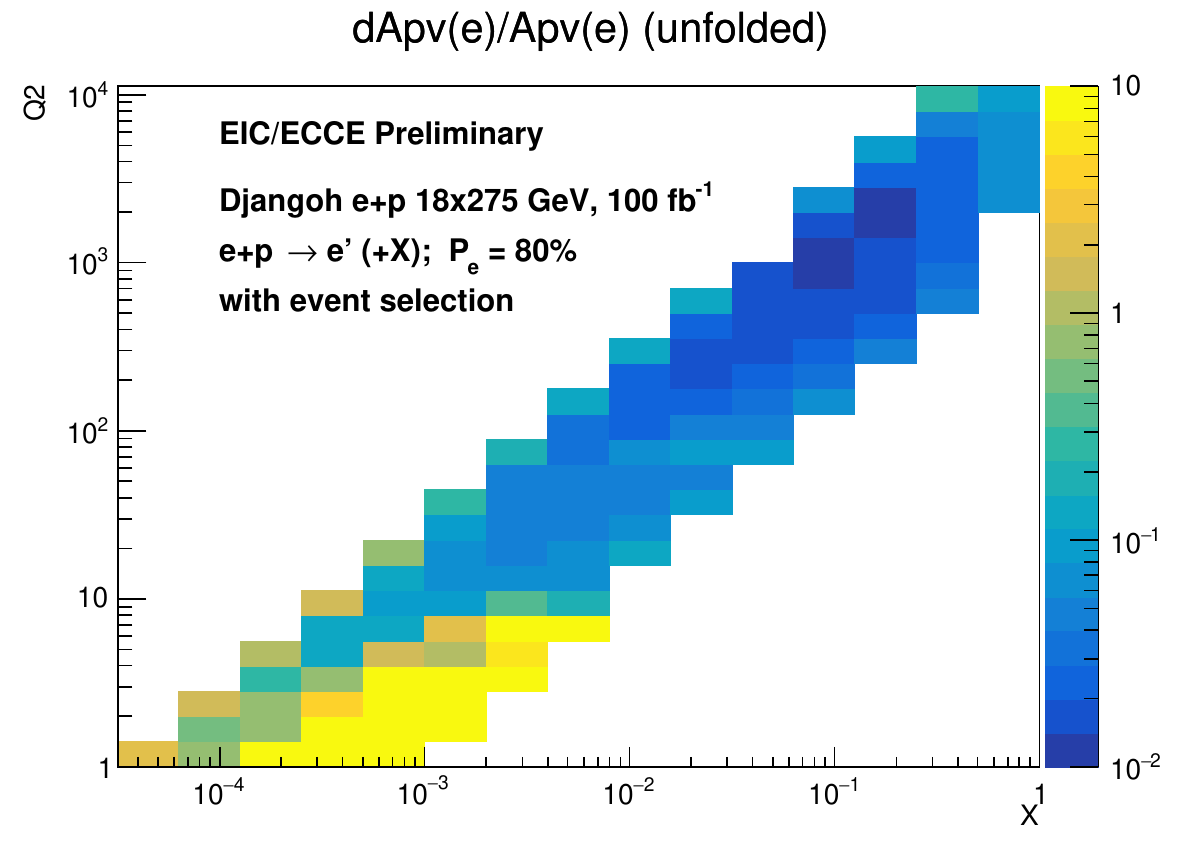}
	\caption{Projection for $\APVe$ (left), and $\d A_{\rm PV, stat}^{(e)}/\APVe$ after unfolding (right) for $18\times 275$~GeV $ep$ collisions, with event-selection criteria applied. An integrated luminosity of 100~fb$^{-1}$ and an electron polarization of 80\% are assumed.}
	\label{fig:dApvE_ep_18_275_ana}
\end{figure}

\begin{figure}[h]
	\includegraphics[width=0.48\textwidth]{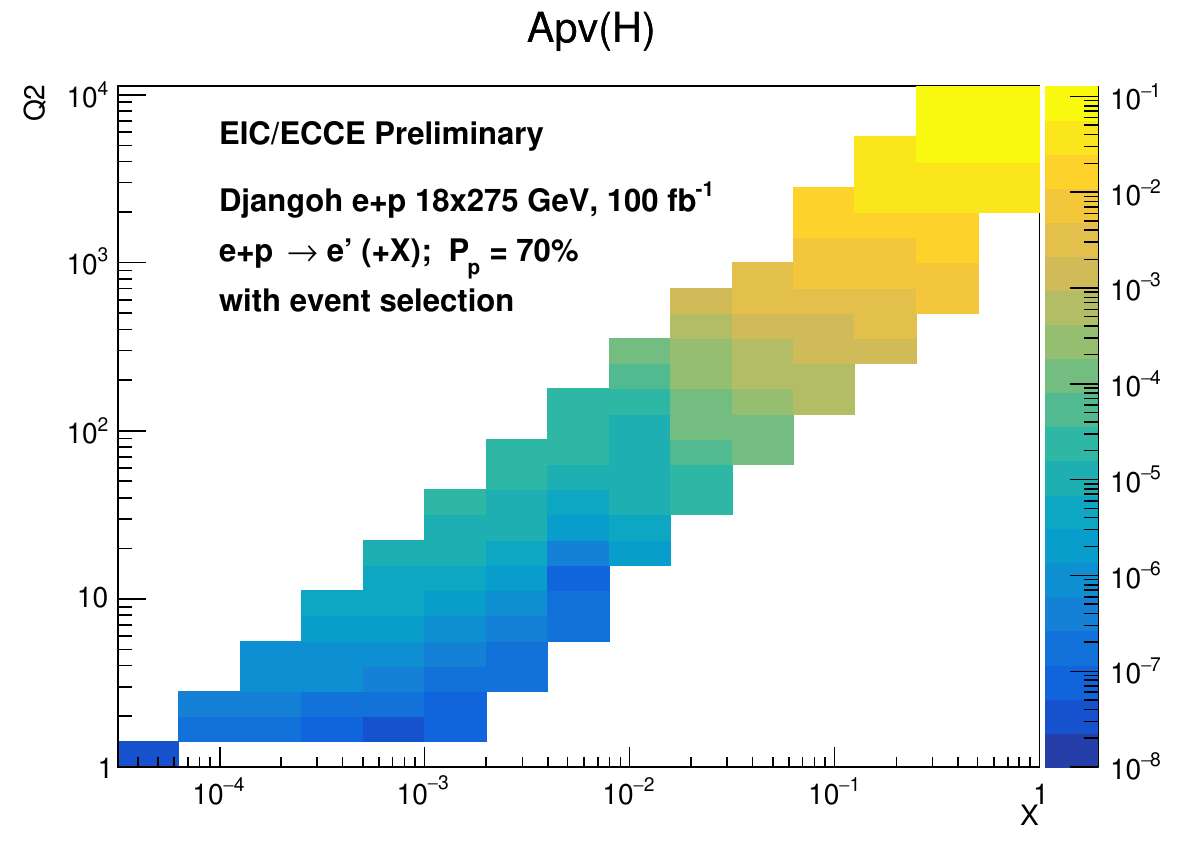}
	\includegraphics[width=0.48\textwidth]{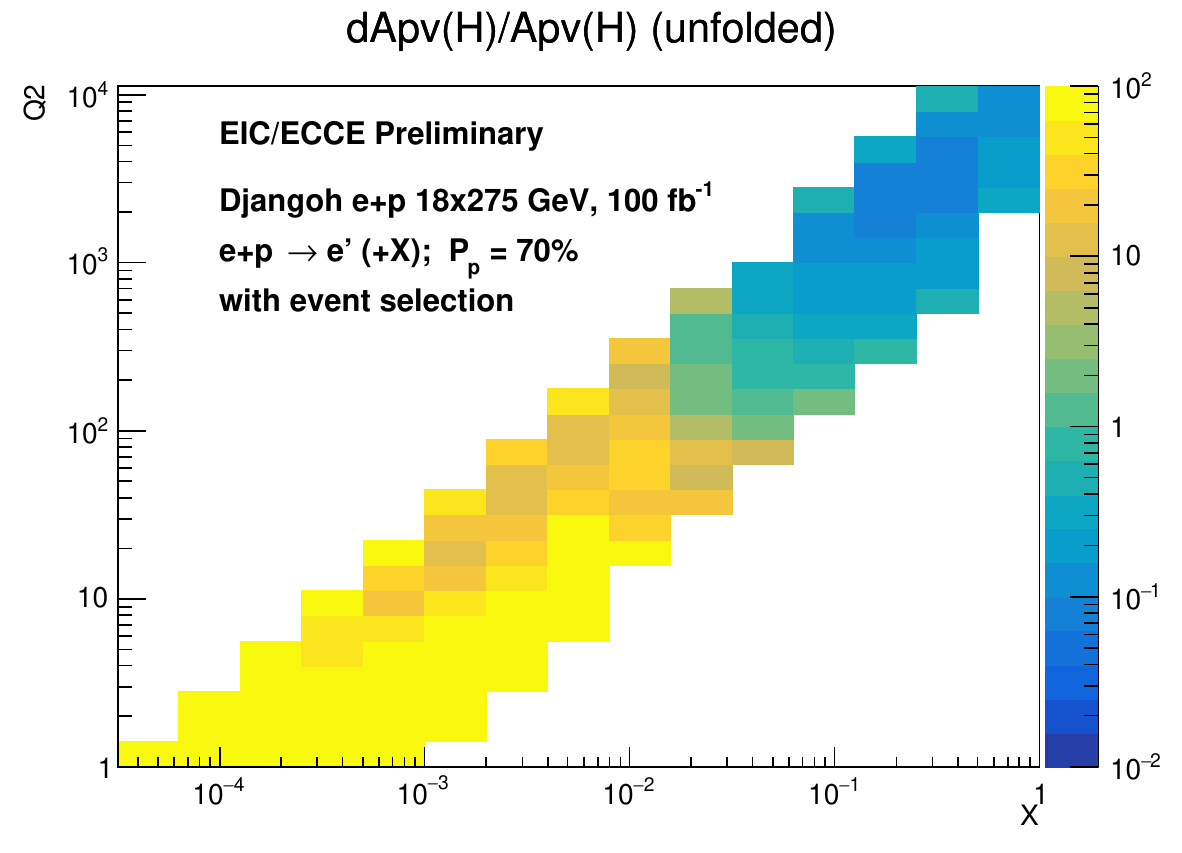}
	\caption{Projection for $\APVp$ (left), and $\d A_{\rm PV, stat}^{(p)}/\APVp$ after unfolding (right) for $18\times 275$~GeV $ep$ collisions, with event-selection criteria applied. An integrated luminosity of 100~fb$^{-1}$ and an proton polarization of 70\% are assumed.}
	\label{fig:dApvH_ep_18_275_ana}
\end{figure}

For the 20~M fast-smearing events, event-selection criteria are applied to choose DIS events ($Q^2_\mathrm{det}>1.0$~GeV$^2$) in order to avoid regions with severe bin migration and unfolding uncertainty ($y_\mathrm{det}>0.1$), to avoid regions with high photoproduction background ($y_\mathrm{det}<0.90$), to restrict events in the main acceptance of the ECCE detector where the fast-smearing method is applicable ($\eta_\mathrm{det}>-3.5$ and  $\eta_\mathrm{det}<3.5625$), and to ensure high purity of electron samples ($E'>2.0$~GeV). Here, the subscript ``det" implies the variables are calculated using the detected information of the electron. The projected values and statistical uncertainties for $\APVe$ and $\APVH$ after unfolding are shown in Figs.~\ref{fig:dApvE_ep_18_275_ana} and ~\ref{fig:dApvH_ep_18_275_ana}, respectively, for $18\times 275$~GeV $ep$ collisions with an integrated luminosity of $100$~fb$^{-1}$. 

\vspace{-.3cm}

%% file: sections/3d-integrated_luminosity.tex
\subsection{Integrated Luminosity}
To account for realistic running conditions, the annual luminosity---the ``high-divergence configuration'' value as shown in Table 10.1 of the EIC Yellow Report (YR)~\cite{AbdulKhalek:2021gbh}, multiplied by $10^7$~s---are used. These values are shown in Table~\ref{tab:expconfig} and will be referred to as ``Nominal Luminosity (NL)'' hereafter. As a comparison with the weak mixing angle extraction presented in the YR, we also carry out projections for 100~fb$^{-1}$ $18\times 275$~GeV $ep$ and 10~fb$^{-1}$ $18\times 137$~GeV $eD$ collisions as the ``YR reference point''. We abbreviate the $ep$ pseudodata sets as P1, P2, P3, P4, and P5 and the $eD$ pseudodata sets as D1, D2, D3, D4, and D5; see Table~\ref{tab:expconfig}. The YR reference point is denoted by P6. Simulated pseudodata sets with polarized hadrons are indicated as $\Delta$D1--5 and $\Delta$P1--6, while positron data sets are referred to as LD1--5 and LP1--6 (with ``L'' for lepton charge).

As an exercise, we consider the additional statistical power that could be obtained by a hypothetical future high-luminosity upgrade to the EIC (HL-EIC) that delivers a ten-fold increase in the integrated luminosity ($10\times$ higher than those in Table~\ref{tab:expconfig}) for these measurements. As the EIC is not yet built, there is no technical basis to assume that such an upgrade is possible. We choose the factor of $10\times$ luminosity increase to explore the sensitivity of the measurements we study in this paper, without making a comment as to the feasibility of such an upgrade. These projections will be denoted with an ``High Luminosity (HL)" label hereafter. 
\begin{table}
	[H]\centering
	\begin{tabular}
		{|l|l||l|l|}
		\hline
		\hline
		D1 & $ 5 {\rm\ GeV} \times 41 {\rm\ GeV} $ $ eD $, $ 4.4 {\rm\ fb^{-1}} $ & P1 & $ 5 {\rm\ GeV} \times 41 {\rm\ GeV} $ $ ep $, $ 4.4 {\rm\ fb^{-1}} $ \\ 
		\hline
		D2 & $ 5 {\rm\ GeV} \times100 {\rm\ GeV} $ $ eD $, $36.8 {\rm\ fb^{-1}} $ & P2 & $ 5 {\rm\ GeV} \times100 {\rm\ GeV} $ $ ep $, $36.8 {\rm\ fb^{-1}} $ \\
		\hline
		D3 & $10 {\rm\ GeV} \times100 {\rm\ GeV} $ $ eD $, $44.8 {\rm\ fb^{-1}} $ & P3 & $10 {\rm\ GeV} \times100 {\rm\ GeV} $ $ ep $, $44.8 {\rm\ fb^{-1}} $ \\
		\hline
		D4 & $10 {\rm\ GeV} \times137 {\rm\ GeV} $ $ eD $, $ 100 {\rm\ fb^{-1}} $ & P4 & $10 {\rm\ GeV} \times275 {\rm\ GeV} $ $ ep $, $ 100 {\rm\ fb^{-1}} $ \\
		\hline
		D5 & $18 {\rm\ GeV} \times137 {\rm\ GeV} $ $ eD $, $15.4 {\rm\ fb^{-1}} $ & P5 & $18 {\rm\ GeV} \times275 {\rm\ GeV} $ $ ep $, $15.4 {\rm\ fb^{-1}} $ \\
		\hline
		& & P6 & $18 {\rm\ GeV} \times275 {\rm\ GeV} $ $ ep $, $ 100 {\rm\ fb^{-1}} $ \\
		\hline
		\hline
	\end{tabular}
	\caption{Beam energy, beam type, and the corresponding nominal annual luminosity assumed for the EIC in our analysis. P6 is the YR reference setting.}
	\label{tab:expconfig}
\end{table}

%% file: sections/3e-stat_uncertainty_projection_for_pv_asymmetries.tex
\subsection{Statistical Uncertainty Projection for Parity-Violating Asymmetries}\label{sec:proj_apv}
For a given value of integrated luminosity, the statistical uncertainty of an asymmetry measurement is:
\begin{eqnarray}
	\d A_\mathrm{stat, measured} = \frac{1}{\sqrt{N}}~,
\end{eqnarray}
where $N$ is the total number of events detected, assumed to be approximately equally divided between the two scattering types---between left- and right-handed electron beam, between left- and right-handed proton (ion) beam, or between positron and electron runs. The unfolding process increases the statistical precision only slightly for the region where the relative statistical uncertainty on the asymmetry is most precise. 

If the asymmetry originates from polarization (as for the case of PV asymmetries), one must correct for the beam polarization:
\begin{eqnarray}
	\d A_\mathrm{PV, stat}^{(e)} = \frac{1}{\vert P_e\vert }\frac{1}{\sqrt{N}}~, ~~{\rm and}~~~
	\d A_\mathrm{PV, stat}^{(H)} = \frac{1}{\vert P_H\vert }\frac{1}{\sqrt{N}}~. 
\end{eqnarray}
For $\APVe$ projections, an electron beam polarization of $P_e=80\%$ with relative 1\% systematic uncertainty from the electron polarimetry is assumed. Similarly, for $\APVH$ projections, a proton (ion) beam polarization of $P_H=70\%$ with relative 2\% systematic uncertainty from the proton (ion) polarimetry is used. An illustration of the relative precision of PV asymmetries is provided in Figs.~\ref{fig:dApvE_ep_18_275_ana} and ~\ref{fig:dApvH_ep_18_275_ana}. The statistical uncertainty of $\APVH$ is rather large because of the much smaller size of $\APVH$ than $\APVe$.

%% file: sections/3f-stat_and_qed_uncertainty_projections_for_lc_asymmetries.tex
\subsection{Statistical and QED Uncertainty Projection for Lepton-Charge Asymmetries}\label{sec:proj_alc}
As described in Section~\ref{sec:Alc_meas}, to measure the lepton-charge asymmetry $\ALCH$, one can reverse the polarity of the magnet to minimize the systematic uncertainty due to differences in $e^-$ and $e^+$ detection.  In this case, the main experimental systematic uncertainty would come from the luminosity difference between $e^-$ and $e^+$ runs, which is assumed to be 2\% (relative in luminosity, absolute in $\ALC$) in this analysis.  If the detector magnet polarity is reversed, then the detection of DIS positrons would be very similar to that of DIS electrons and all the data simulations, event selections, unfolding, etc. described in Section~\ref{sec:proj_ecce_sim} apply.  The statistical uncertainty in $\ALC$ is thus determined by the luminosity of $e^+$ run, which we assume to be one-tenth of that of the electron beam. Note that beam polarization and thus polarimetry uncertainties do not affect $\ALC$ measurements.

\begin{figure}[h]
	\includegraphics[width=0.48\textwidth]{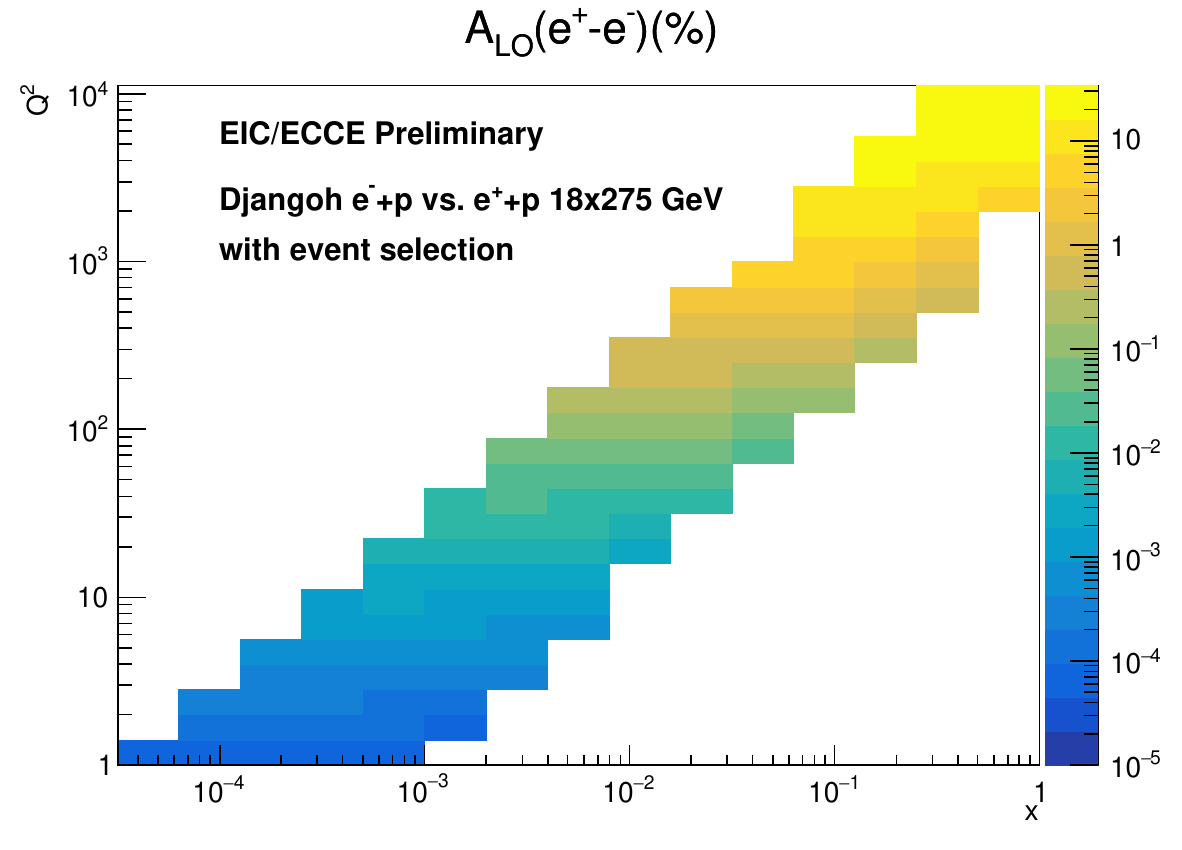}
	\includegraphics[width=0.48\textwidth]{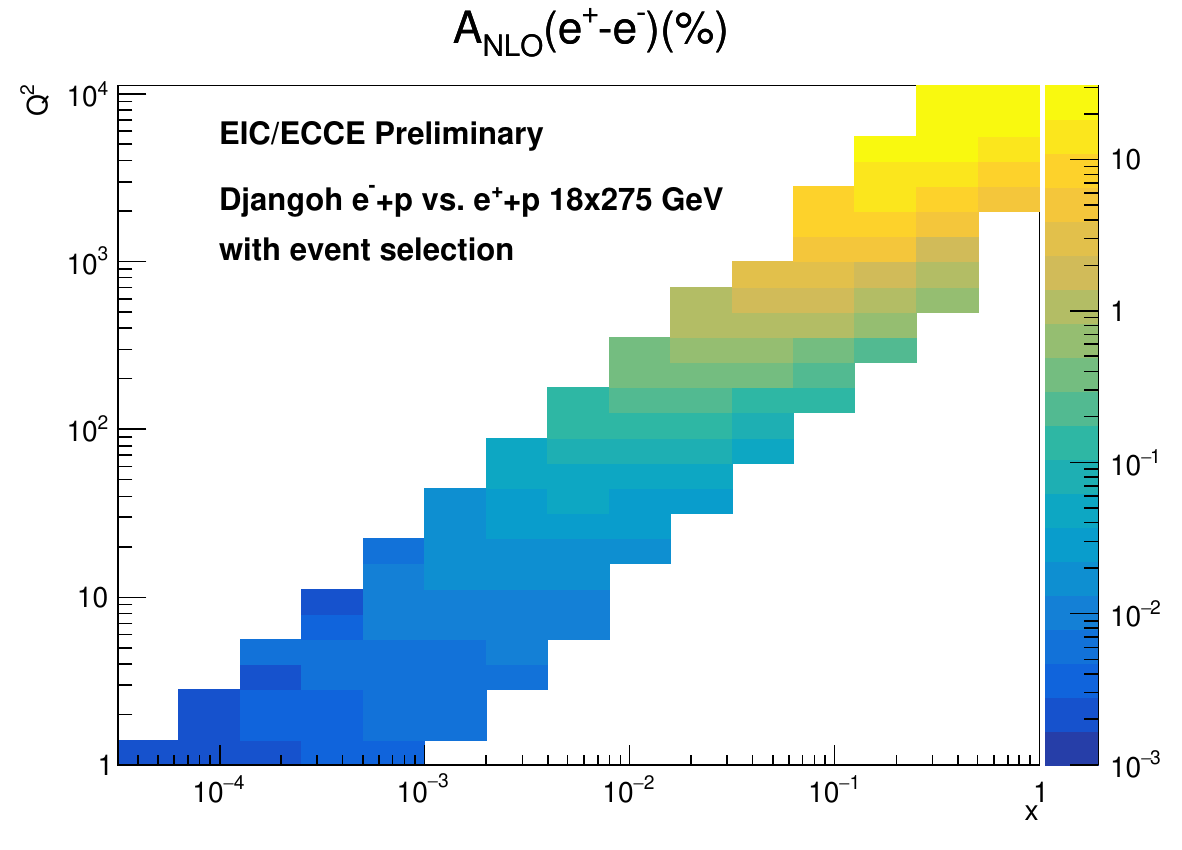}
	\caption{Calculation for $\ALC$ at the Born (LO) (left) and NLO (right) levels for $e^+p$ vs. $e^-p$ collisions at $18\times 275$~GeV. The LO calculation includes only the $\gamma Z$ interference term, which is of main interest of this study. The NLO calculation includes box diagrams, which introduces a large QED effect to the asymmetry and is effectively a background to the EW and SMEFT study presented here.}
	\label{fig:Alc}
\end{figure}

The EW physics reach of $\ALC$ is further clouded by the difference in $e^-$ vs. $e^+$ DIS cross sections due to higher-order QED effects. We calculate the value of $\ALC$ using {\tt Djangoh} version 4.6.19 in both the Born LO (that includes one-boson exchange only) and NLO radiated mode (that includes higher-order EW and QED effects); see Fig.~\ref{fig:Alc}. The difference of NLO minus Born values is taken as an estimate of QED NLO effects and the uncertainty is assumed to be 5\% relative.

Because of the moderate $Q^2$ reach of the EIC, the 2\% absolute uncertainty from luminosity measurement is a dominating systematic effect for the uncertainty of $\ALCH$. In Appendix \ref{app:additional_fits:luminosity_difference_fits}, we present a method to simultaneously fit the luminosity term with SMEFT coefficients; however, we find this method yields 15 to 20\% weaker SMEFT constraints. 

%% file: sections/3g-projection_for_hl_eic.tex
\subsection{Projection for High-Luminosity EIC}\label{sec:proj_hleic}
%
%https://indico.bnl.gov/event/14504/
In addition to the nominal luminosity expected for the EIC, we also carry out projections considering the possibility of an additional ten-fold increase in the annual luminosity beyond EIC's initial phase of running, the so-called high-luminosity EIC (HL-EIC). Assuming all experimental systematic effects remain the same, we scale the projected statistical uncertainty of asymmetry observables described in the previous section by a factor of $1/\sqrt{10}$. For beam energies with lower luminosity (hence larger statistical uncertainty) or asymmetries of smaller sizes such as $\APVH$, the ten-fold increase in luminosity will push the physics reach one step further.  On the other hand, for beam energies with already high luminosity and for observables where systematic effects dominate over the statistical ones, such as $\APVe$ for $10\times 275$~GeV $ep$ and $10\times 137$~GeV $eD$ collisions and $\ALCH$, the impact from the luminosity increase of HL-EIC on the physics reach is marginal. 

%% file: sections/4a-pseudodata_for_pv_asymmetries.tex
\subsection{Pseudodata for Parity-Violating Asymmetries}
We discuss first the case of the two PV asymmetries: polarized electron asymmetries with unpolarized hadrons, $\APVe$, and polarized hadron asymmetries with unpolarized electrons, $\APVH$. The experimental uncertainties are from three sources: statistical, $\sigma_\mathrm{stat}$; experimental systematic, $\sigma_\mathrm{sys}$, which is mainly due to particle background, also including other imperfections of the measurement, and is assumed to be fully uncorrelated; and beam polarimetry, $\sigma_\mathrm{pol}$, which is assumed to be fully correlated within data of the same $\sqrt{s}$ and beam type. 

For the $ b^{\rm th} $ bin, with given $ \sqrt s $, $ x $, and $ Q^2 $ values and using the nominal PDF set under consideration, first we compute the theoretical SM prediction, $ (A_{\rm PV})_{{\rm SM},0,b}^{\rm theo} $. Combining the given uncertainties in quadrature separately for uncorrelated and correlated ones, we obtain a pseudo-experimental asymmetry value by:
\begin{eqnarray}
	(A_{\rm PV})_{b}^{\rm pseudo} = 
	(A_{\rm PV})_{{\rm SM},0,b}^{\rm theo} 
	+ r_b \sqrt{
		\sigma_{{\rm stat},b}^2 
		+ \bb{(A_{\rm PV})_{{\rm SM},0,b}^{\rm theo} \pp{\frac{\sigma_{\rm sys}}A}_b}^2}
	+ r' \sqrt{\bb{(A_{\rm PV})_{{\rm SM},0,b}^{\rm theo} \pp{\frac{\sigma_{\rm pol}}A}_b}^2}~, 
	\label{4}
\end{eqnarray}
where $ r_b $ and $ r' $ are random numbers chosen from a normal distribution of mean 0 and standard deviation 1. Note that the correlated errors are incorporated using a single random number, $r'$, across all the bins. The systematic uncertainties are $\sigma_\mathrm{sys}/A=1\%$, $\sigma_\mathrm{pol}/A=1\%$ for $\APVe$, and $\sigma_\mathrm{pol}/A=2\%$ for $\APVH$.

%% file: sections/4b-pseudodata_for_lc_asymmetries.tex
\subsection{Pseudo Data for Lepton-Charge Asymmetries}
We consider next unpolarized electron-positron asymmetries with unpolarized hadrons, namely the lepton-charge (LC) asymmetries. The uncertainties used in the data generation are from three sources: statistical, $\sigma_\mathrm{stat}$; experimental systematic, $\sigma_\mathrm{sys}$, which is mainly due to background and is assumed to be fully uncorrelated; luminosity difference between $e^+$ and $e^-$ runs, $\sigma_\mathrm{lum}$, which is fully correlated within data of the same $\sqrt{s}$ and ion beam type; and higher-order QED effects, $\sigma_{\rm QED~NLO}$, taken as 5\% of the difference between the calculated NLO and Born (LO) $\ALC$ values. 

In analogy with Eq.~\eqref4, for the LC asymmetries, we write:
\begin{eqnarray}
	(A_{\rm LC})_{b}^{\rm pseudo} =
	(A_{\rm LC})_{{\rm SM},0,b}^{\rm theo} 
	+ r_b \sqrt{\sigma_{\rm stat}^2 + \bb{(A_{\rm LC})_{{\rm SM},0,b} \pp{\frac{\sigma_{\rm sys}}A}_b}^2 + \sigma_{{\rm QED~NLO},b}^2} 
	+ r' \sigma_{{\rm lum},b}~.
	\label{5}
\end{eqnarray}

\vspace{-.5cm}

%% file: sections/4c-uncertainty_matrix.tex
\subsection{Uncertainty Matrix}\label{sec:proj_err_matrix}
The uncertainty matrix, $ \Sigma^2 $, for a given data set with $ N_{\rm bin} $ bins is an $ N_{\rm bin} \times N_{\rm bin} $ symmetric matrix. It consists of two parts, which we call $ \Sigma_0^2 $ and $ \Sigma_{\rm pdf}^2 $:
\begin{eqnarray}
	\left(\Sigma^2\right)_{bb'}= \left(\Sigma_0^2\right)_{bb'} + \left(\Sigma_{\rm pdf}^2\right)_{bb'}~.
\end{eqnarray}

The first part of the matrix, $\Sigma_0^2$, is constructed using all the uncertainty components (statistical, systematic, polarimetry or luminosity, and QED NLO if present) other than the PDF uncertainties. All the uncertainties that enter $ \Sigma_0^2 $ must be absolute; relative uncertainties are converted to absolute ones by multiplying with the theoretical SM prediction, $ A_{{\rm SM},0,b}^{\rm theo} $, computed using the central member of the PDF set taken into account. The first part of the matrix then takes the form:
\begin{eqnarray}
	\Sigma_0^2 = 
	\begin{pmatrix}
		\sigma_1^2 & \rho_{12} \tilde \sigma_1 \tilde \sigma_2 & \cdots & \rho_{1N_{\rm bin}} \tilde \sigma_1 \tilde \sigma_{N_{\rm bin}} \\
		& \sigma_2^2                           & \cdots & \rho_{2N_{\rm bin}} \tilde \sigma_2 \tilde \sigma_{N_{\rm bin}} \\
		&                                      & \ddots & \vdots \\
		&                                      &        & \sigma_{N_{\rm bin}}^2            
	\end{pmatrix}_{\rm sym}~,
\end{eqnarray}
where, for the PV asymmetries, we have for the diagonal elements:
\begin{eqnarray}
	\sigma_b^2 = 
	\sigma_{{\rm stat},b}^2 
	+ \bb{(A_{\rm PV})_{{\rm SM},0,b}^{\rm theo} \pp{\frac{\sigma_{\rm sys}}A}_b}^2 
	+ \bb{(A_{\rm PV})_{{\rm SM},0,b}^{\rm theo} \pp{\frac{\sigma_{\rm pol}}A}_b}^2~,
\end{eqnarray}
and for the off-diagonal elements:
\begin{eqnarray}
	\tilde \sigma_b =
	(A_{\rm PV})_{{\rm SM},0,b}^{\rm theo} \pp{\frac{\sigma_{\rm pol}}A}_b~.
\end{eqnarray}
For the LC asymmetries, we have for the diagonal elements: 
\begin{eqnarray}
	\sigma_b^2 =
	\sigma_{{\rm stat},b}^2 
	+ \bb{(A_{\rm LC})_{{\rm SM},0,b}^{\rm theo} \pp{\frac{\sigma_{\rm sys}}A}_b}^2 
	+ \sigma_{{\rm lum},b}^2
	+ \sigma_{{\rm QED~NLO},b}^2~,
\end{eqnarray}
and for the off-diagonal elements:
\begin{eqnarray}
	\tilde \sigma_b = \sigma_{{\rm lum},b}~.
\end{eqnarray}
Here, $ b $ and $ b' $ are bin numbers and we assume full correlation for uncertainties originating from beam polarimetry or luminosity: $ \rho_{bb'} = 1 $ for all $ b $ and $ b' $.

The second part of the uncertainty matrix, $ \Sigma_{\rm pdf}^2 $, is built using the same procedure for both PV and LC asymmetries by taking into account differences between the theoretical SM asymmetry prediction computed at the nominal PDF member, $ A_{\rm SM,0}^{\rm theo} $, and theoretical SM asymmetry predictions evaluated at all other members of the PDF set under consideration, $ A_{{\rm SM}, m}^{\rm theo} $, where $m=1,2, \ldots, N_{\rm PDF}$ with $N_{\rm PDF}$ the total number of PDF sets or replicas available. For Hessian-based PDF sets, the diagonal and off-diagonal elements can be collectively written as:
\begin{equation}
	\left(\Sigma_{\rm PDF}^2\right)_{bb'}^{\rm Hessian} = \frac{1}{4}
\sum_{m=1}^{N_\mathrm{PDF}/2}(A_{{\rm SM},2m,b}^{\rm theo}- A_{{\rm SM},2m-1,b}^{\rm theo})(A_{{\rm SM},2m,b'}^{\rm theo}- A_{{\rm SM},2m-1,b'}^{\rm theo})~.
\end{equation}
For replica-based PDF sets, this expression becomes:
\begin{eqnarray}
\left(\Sigma_{\rm PDF}^2\right)_{bb'}^{\rm replica} =
\frac 1{N_{\rm PDF}} \sum_{m = 1}^{N_{\rm PDF}} (A_{{\rm SM},m,b}^{\rm theo} - A_{{\rm SM,0},b}^{\rm theo})(A_{{\rm SM},m,b'}^{\rm theo} - A_{{\rm SM,0},b'}^{\rm theo})~.
\end{eqnarray}

%% file: sections/4d-comparison_of_uncertainty_components.tex
\subsection{Comparison of uncertainty components \label{sec:comparison_of_uncertainty_components}}
We present in this section the various uncertainty components that enter the SMEFT analysis. We also investigate the total uncertainties combined in quadrature that contribute to the diagonal entries of the uncertainty matrix. 

\begin{figure}
	[h]\centering
	\includegraphics[width=.45\textwidth]{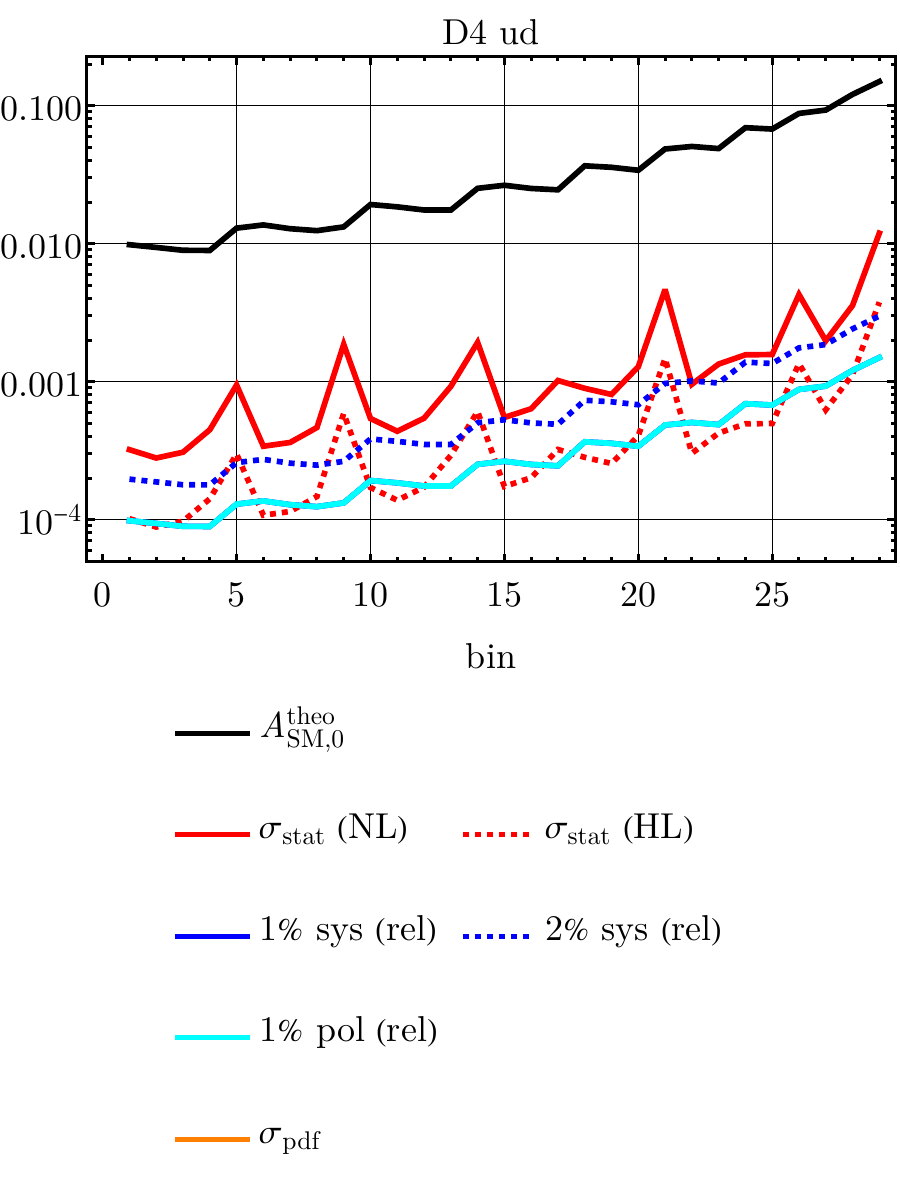}
	\includegraphics[width=.45\textwidth]{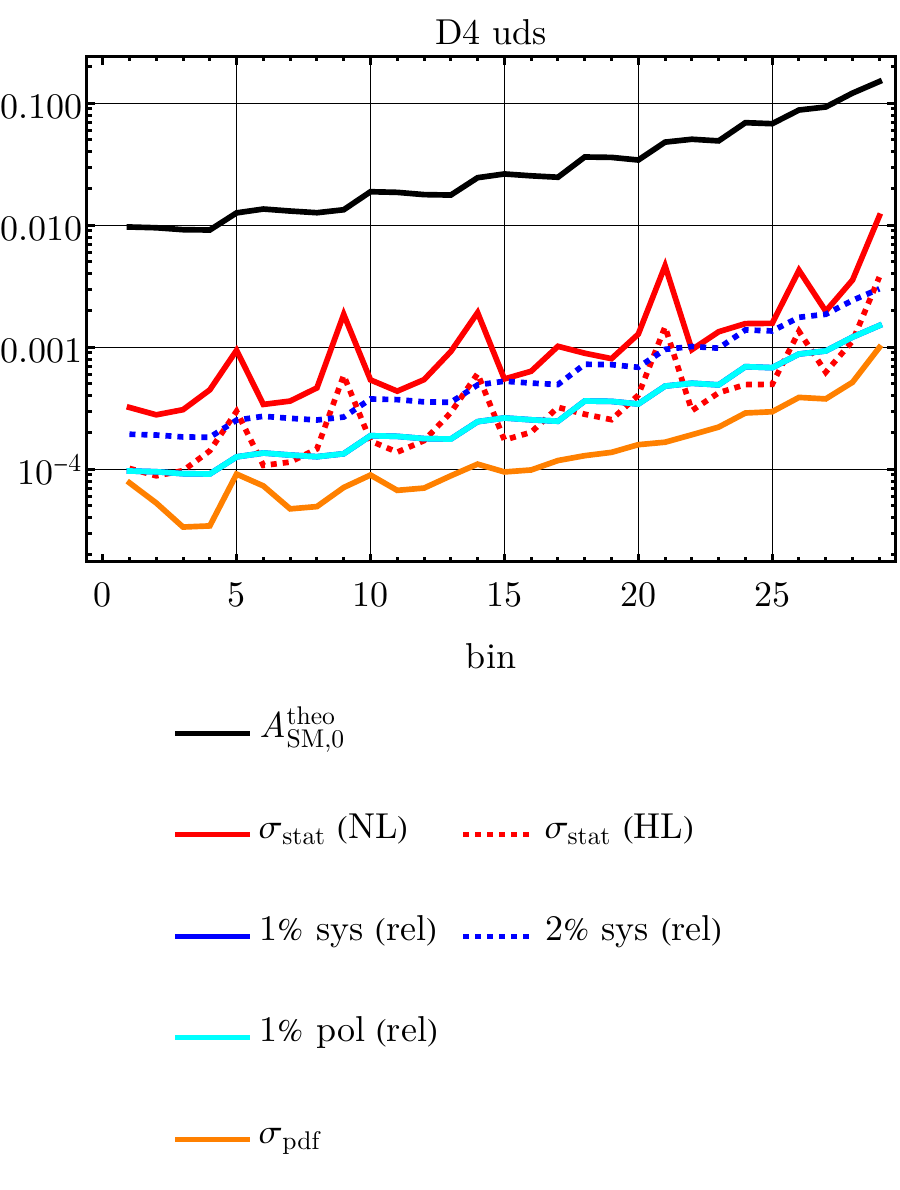}
	\caption{Comparison of the uncertainty components for the data set D4 in the valence-only scenario ({\tt ud}) and with the contributions from the sea quarks ({\tt uds}). Here, ``NL" refers to the currently planned annual luminosity of the EIC, while ``HL" refers to a potential ten-fold luminosity upgrade. }
	\label{fig:D4_ud_uds_err_components}
\end{figure}

\subsubsection{Individual uncertainty components}
We begin by considering the individual components of the uncertainties. We investigate the effects of sea quarks in the analysis by defining a valence-only approximation for the PDFs. The tag {\tt ud} in the plot labels implies the valence-only approximation, in which only up- and down-quark contributions are considered in the hadronic cross section, whereas {\tt uds} indicates that up, down, strange, and their antiquarks are taken into account. Note that for the data sets involving unpolarized deuteron with the {\tt ud} tag, there will be no uncertainty from PDFs since deuteron PDFs, defined in terms of proton and neutron PDFs using isospin symmetry, cancel when analytically forming asymmetries in the valence-only approximation. Note also that for experimental systematic uncertainties other than those from beam polarimetry, both 1\% and 2\% values are shown in all figures of this section, although the 1\% value is used in the results presented. 

Fig.~\ref{fig:D4_ud_uds_err_components} shows the comparison of the uncertainty components for the data set D4 in the {\tt ud} and {\tt uds} scenarios. As for the PDFs, we use {\tt NNPDF3.1 NLO} \cite{NNPDF:2017mvq}  in the unpolarized case and {\tt NNPDFPOL1.1} \cite{ERNocera2014} in the polarized case throughout. Only the $(x,Q^2)$ region relevant for SMEFT analysis is shown, although the full region is used for the extraction of the weak mixing angle. The $x$-axis of these plots is ordered by bin number; these are ordered first from low to high $Q^2$, then from small to large $x$ within each $Q^2$ bin, leading to the observed oscillatory behavior. When we turn on the sea quark contributions, the unpolarized deuteron data sets receive nonzero but highly suppressed PDF uncertainties, indicating that the assumption of deuteron PDFs completely canceling is a reasonably good approximation. The right panel shows that even after including sea quarks, the PDFs are still the smallest uncertainty component. This indicates that potentially poorly determined sea quark and strange quark distributions have little effect on this analysis. The largest single uncertainty component is the statistical uncertainty (shown as a dark red line). This is larger than both the 1\% beam polarization uncertainty (light blue line), and either of the 1\% or 2\% uncorrelated systematic uncertainty assumptions (solid and dotted blue lines, respectively). When we switch to the high-luminosity (HL-EIC) scenario (dotted red line), the statistical uncertainty becomes comparable to the systematic ones. All uncertainties are significantly smaller than the predicted values of the asymmetry, shown as the solid black line in the plots.

\begin{figure}
	[h]\centering
	\includegraphics[width=.45\textwidth]{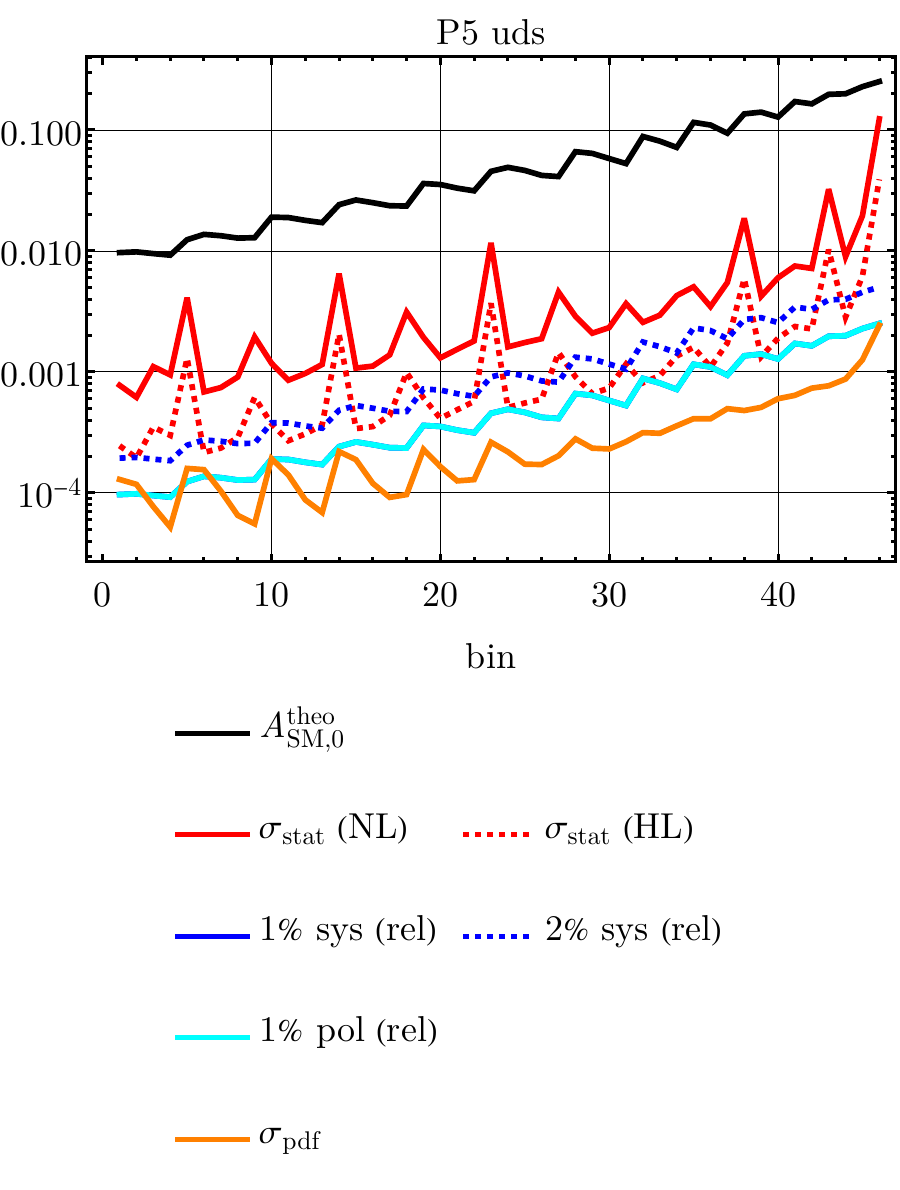}
	\includegraphics[width=.45\textwidth]{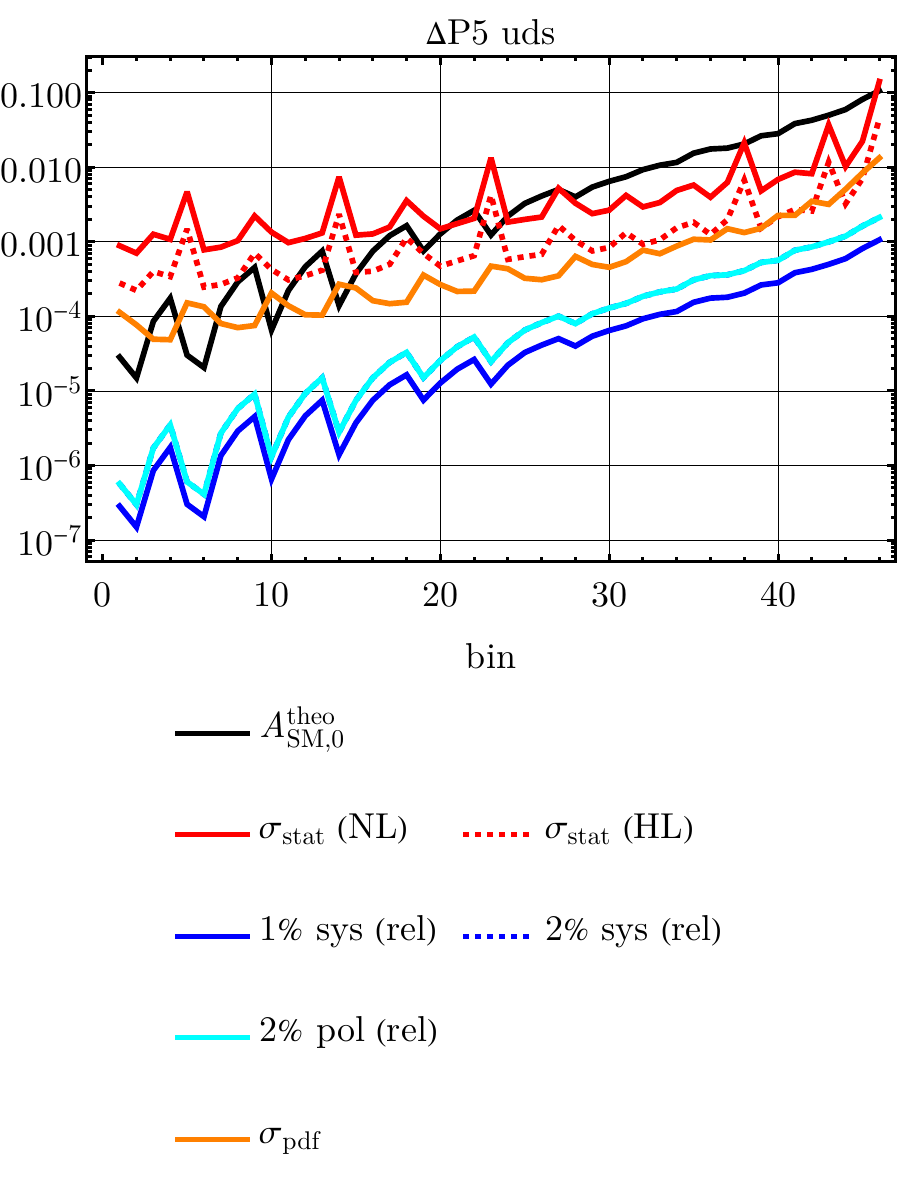}
	\caption{Uncertainty components for the data sets P5 and $ \Delta $P5.}
	\label{fig:P5_err_components}
\end{figure}

In Fig. \ref{fig:P5_err_components}, we display the different contributions to the diagonal entries of the uncertainty matrix of the data sets P5 and $ \Delta $P5. The pattern of uncertainties for P5 is very similar to that observed for D4. The statistical ones are the largest single uncertainty source, while the PDFs are the smallest. Assuming high luminosity, the statistical uncertainties become comparable to the anticipated systematic ones. The pattern is different for $\Delta$P5: the statistical uncertainties are largest for all bins, even assuming high luminosity. The PDF uncertainties are also non-negligible, consistent with the expectation that spin-dependent PDFs are not known as precisely as the spin-independent ones. The anticipated experimental systematic uncertainties are negligible for all bins.

\begin{figure}
	[h]\centering
	\includegraphics[width=.7\textwidth]{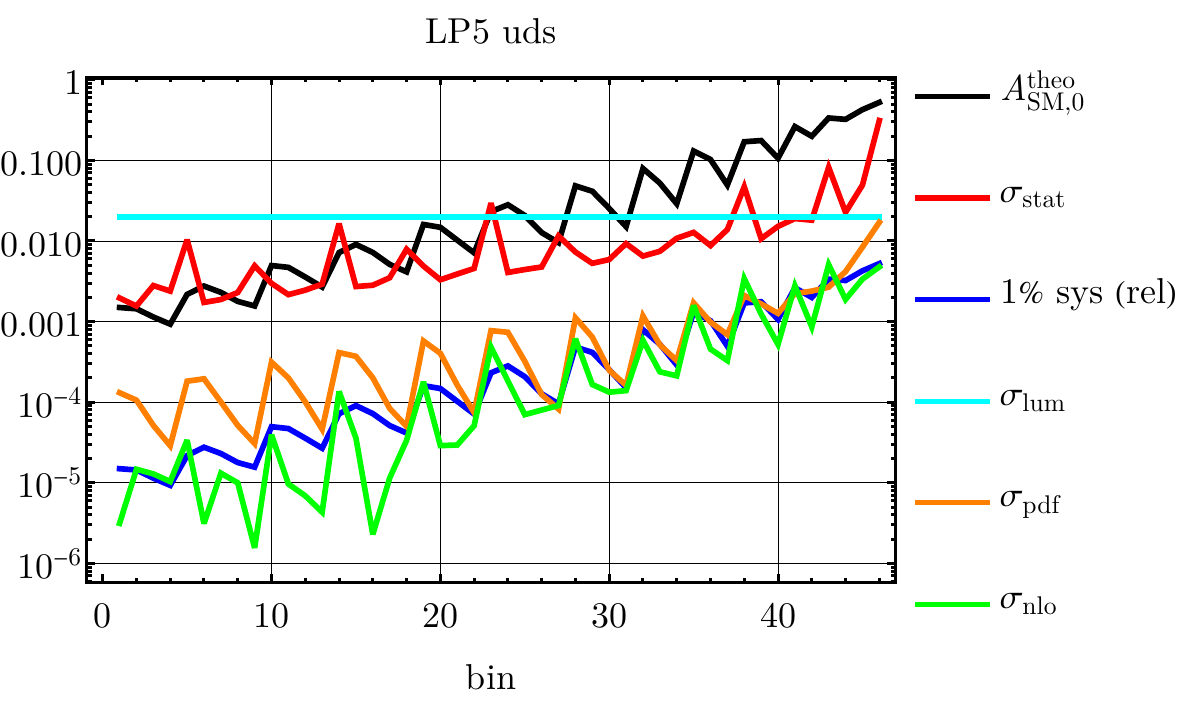}
	\caption{The same as in Fig. \ref{fig:P5_err_components} but for LP5.}
	\label{fig:EP5_err_components}
\end{figure}

Finally, we show in Fig.~\ref{fig:EP5_err_components} the individual uncertainties for the electron-positron asymmetry data set LP5. The error budget is different for this scenario compared to PV asymmetries. Since both beams are unpolarized, there is no uncertainty related to beam polarization. However, since electron and positron runs occur with different beams, there is the possibility of a significant overall luminosity difference between the two runs that can lead to an apparent asymmetry. We assume an absolute 2\% uncertainty, two times the luminosity uncertainty requirement of~\cite{EICCDR}. Finally, we consider the possible errors arising from higher-order QED corrections that may differentiate between electron and positron scattering. We estimate this uncertainty by taking 5\% of the difference between the Born-level and NLO QED results, obtained by using {\tt Djangoh}. The two largest sources of uncertainty throughout the entire kinematic range are the luminosity and statistical uncertainties. PDFs, higher-order QED, and anticipated systematic uncertainties are all significantly smaller.

Summarizing all the figures presented in this section, we can make the following main points:
\begin{itemize}
	\item The expected statistical uncertainties are the dominant ones for the nominal EIC luminosity. If a high-luminosity (HL-EIC) upgrade becomes realistic, they become comparable to experimental systematic uncertainties for PV asymmetries of the unpolarized hadron, $\APVe$.
	\item PDF uncertainties are nearly irrelevant for the asymmetries of unpolarized hadrons, $\APVe$. They become significant, second to statistical uncertainties, for PV asymmetries of polarized hadrons, $\APVH$. 
	\item The luminosity effect dominates over the statistical uncertainty for the majority of the phase space in the case of electron-positron asymmetries, $\ALCH$, particularly at low $ x $ and low $ Q^2 $. On the other hand, uncertainties from missing higher-order QED corrections are expected to be small.
\end{itemize}

\begin{figure}
	[h]\centering
	\includegraphics[width=.45\textwidth]{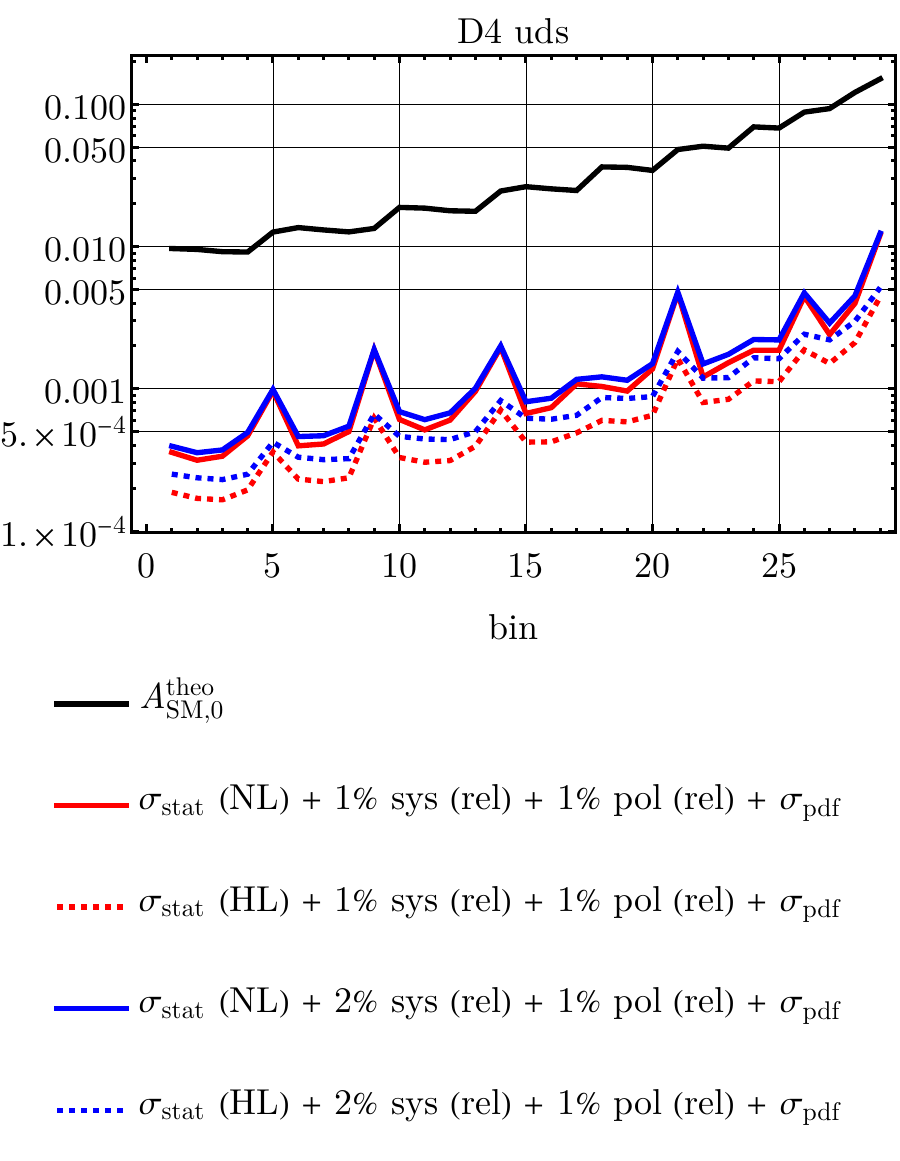}
	\includegraphics[width=.45\textwidth]{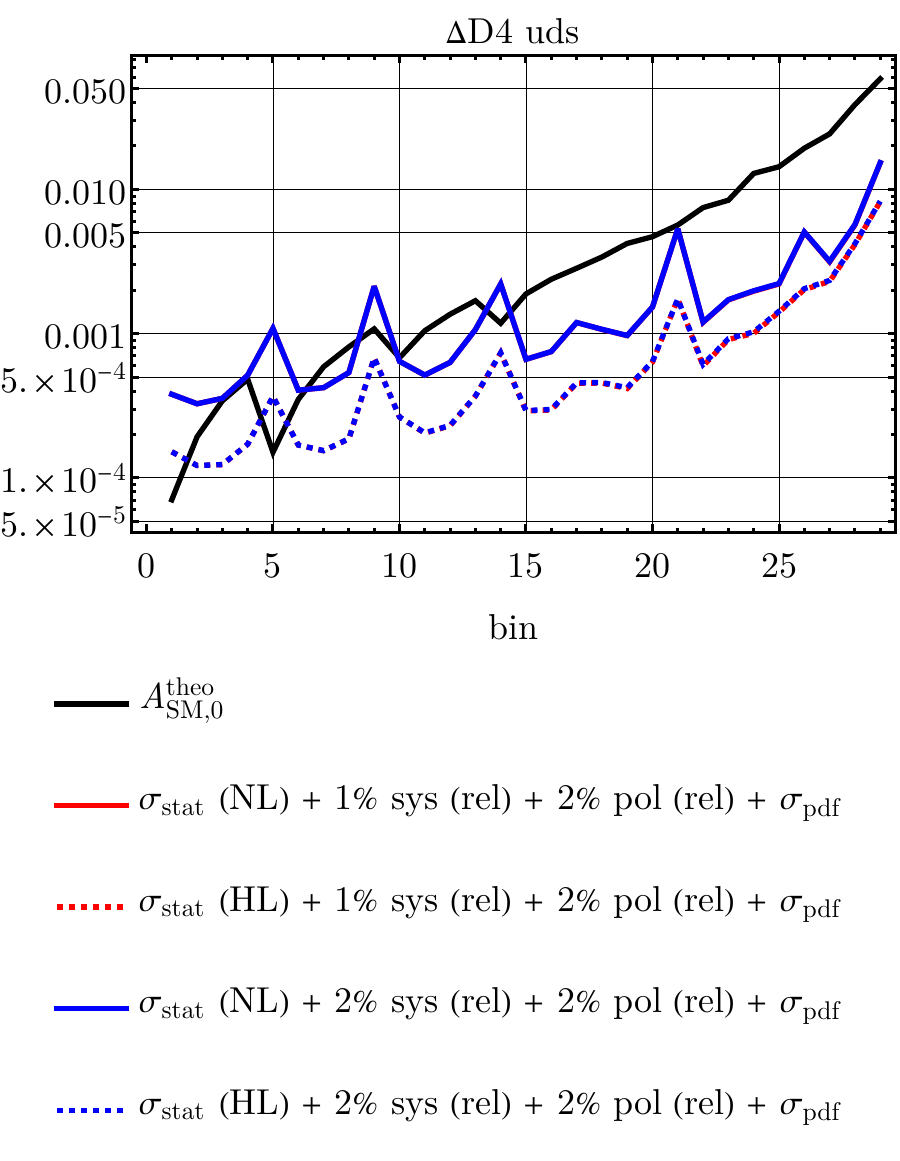}
	\caption{Total uncertainties combined in quadrature for the data sets D4 and $ \Delta $D4 in the {\tt uds} scenario.}
	\label{fig:D4_DeltaD4_total_err}
\end{figure}

\subsubsection{Total uncertainties for nominal luminosity vs. high luminosity}
\begin{figure}
	[h]\centering
	\includegraphics[width=.45\textwidth]{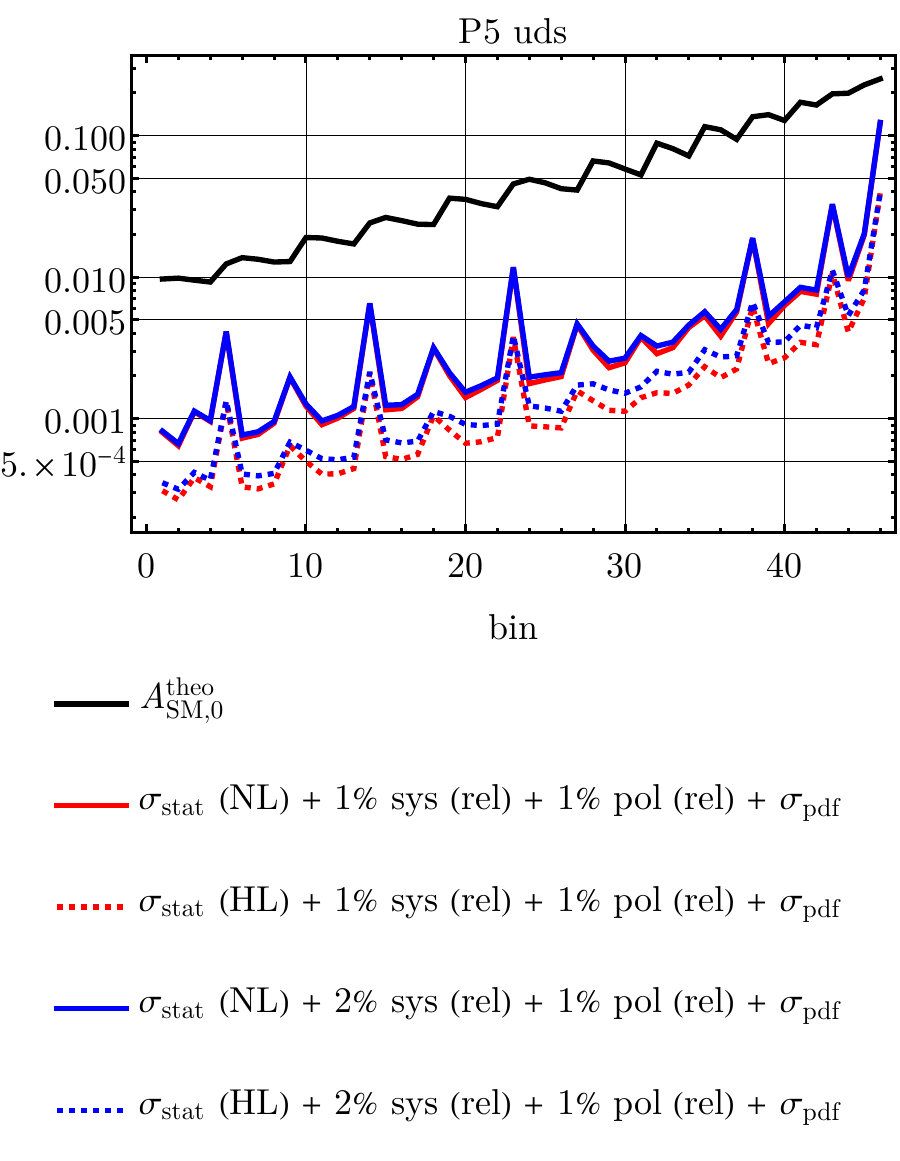}
	\includegraphics[width=.45\textwidth]{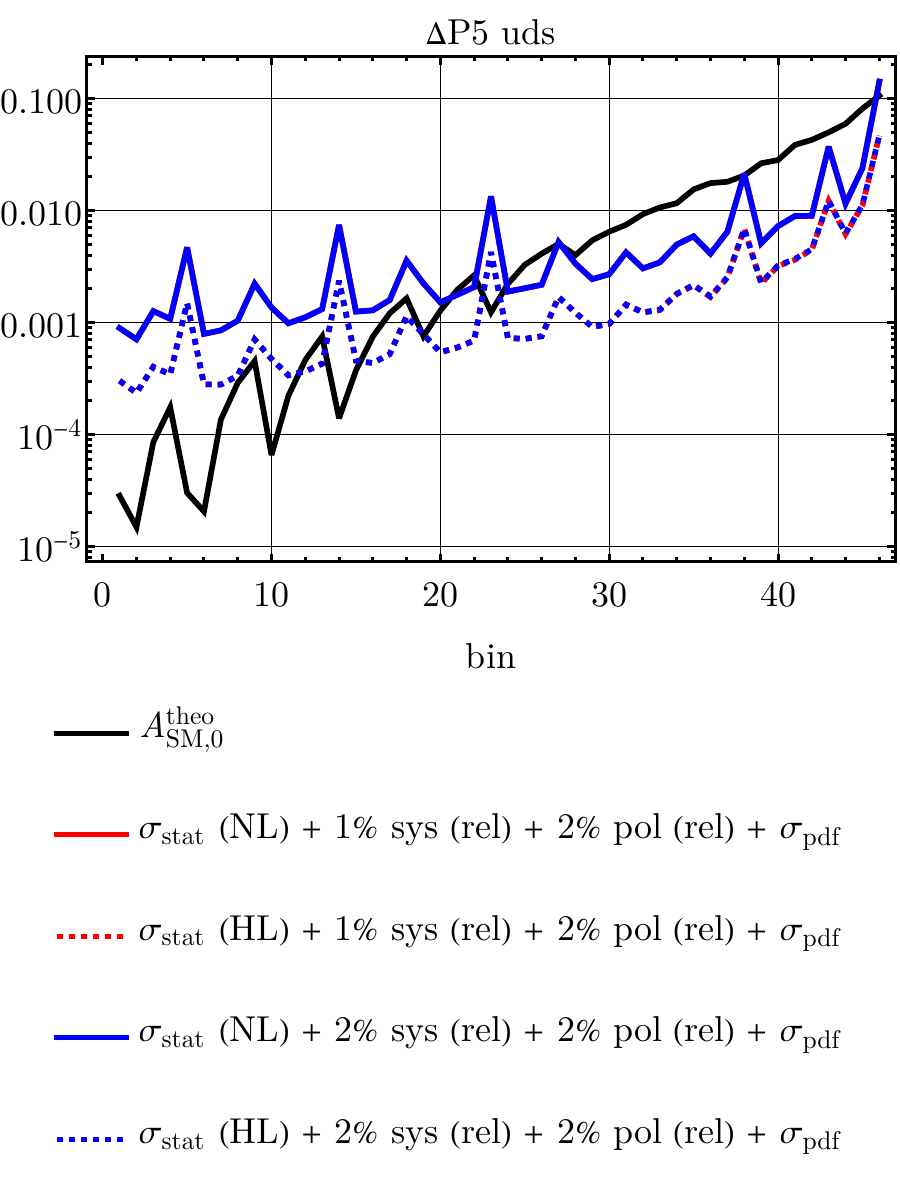}
	\caption{The same as in Fig. \ref{fig:D4_DeltaD4_total_err} but for P5 and $ \Delta $P5.}
	\label{fig:P5_DeltaP5_total_err}
\end{figure}

We now investigate the total uncertainties for the data sets D4, $\Delta$D4, P5, and $\Delta$P5. We consider four different scenarios: the nominal annual luminosity planned for the EIC or a potential high-luminosity upgrade beyond the initial phase of the EIC run, combined with 1\% or 2\% relative experimental systematic uncertainties due to particle background. We show the results in Figs.~\ref{fig:D4_DeltaD4_total_err} and~\ref{fig:P5_DeltaP5_total_err}. We observe first that the dominant uncertainty component in all cases is the statistical one. The four uncertainty scenarios, namely 1\% or 2\% systematic uncertainties combined with nominal or high luminosity, can be in fact reduced to just the luminosity comparison, i.e. nominal vs. high. Next, for both D4 and P5, the asymmetry $\APVe$ is measured to percent-level throughout the considered phase space. This is not the case for the polarized sets $\Delta$D4 and $\Delta$P5. Particularly in the $\Delta$P5 scenario at low $ Q^2 $, the anticipated errors are larger than the asymmetry for all choices of systematic error and luminosity. Only in the very high $Q^2$ bins does a measurement of the asymmetry $\APVH$ become meaningful.

%%%%%%

Our evaluation of the uncertainties indicates that using 1\% or 2\% relative systematic uncertainties makes practically no difference, as the total errors are mostly dominated by the statistical uncertainties for the PV asymmetries or the luminosity difference for the LC asymmetries. We also show that one can take into account the contribution of only the valence quarks to the asymmetries or include the sea quarks up to strange flavor and its antiquark, both of which lead to the same size of PDF errors for the data sets under consideration. In our best-fit analyses, we thus focus on the data sets with 1\% relative systematic uncertainty and nominal luminosity in the {\tt uds} scenario as our main data sets. Comparisons are performed to the ones having high luminosity, keeping the rest of the configuration the same.

An important issue to address is whether a joint fit of PDFs and Wilson coefficients would change the potential of the EIC to probe the SMEFT parameter space that we find in this draft. This issue has been studied for both HERA and LHC dat sets in the literature~\cite{Carrazza:2019sec,Greljo:2021kvv}, where it is found that the interplay between PDFs and Wilson coefficients can become a significant challenge for some future high-luminosity measurements at the LHC. It is beyond the scope of this paper to consider such a joint fit, so we can only speculate regarding the exact answer to this question. However, we can make the following points that are supported by the uncertainty plots in this section of this manuscript.
\begin{itemize}

\item For the unpolarized deuteron data sets, the PDF uncertainties are an order of magnitude smaller than the statistical uncertainties for the nominal luminosity and three to five times smaller than the high-luminosity statistical uncertainties, as shown in Fig.~\ref{fig:D4_ud_uds_err_components}. We therefore expect that a joint fit of the PDFs and Wilson coefficients would not greatly effect the bounds obtained here. The PDFs are already determined sufficiently well from other experiments for the purposes of this analysis,. 

\item The same statement holds for the P5 unpolarized data sets for both nominal and high luminosities, as shown in the left panel of Fig.~\ref{fig:P5_err_components}. This plot indicates that taking ratios to form asymmetries, as we do in this study, greatly reduces the dependence on PDFs.

\item Looking at the right panel of Fig.~\ref{fig:P5_err_components}, we see that the polarized proton PDF error becomes comparable to the statistical error at high $Q^2$ for the high-luminosity data set. In this case, a joint fit of polarized PDFs and Wilson coefficients will be especially important. We note that the bounds from the polarized hadron data sets are generically much weaker than those for the unpolarized hadron sets (see Fig.~\ref{fig:1d_fits_Ceu_Lambda} for example), since the polarized asymmetry is much smaller than the polarized one. We believe that our main point regarding the EIC sensitivity to SMEFT Wilson coefficients is mostly unaffected by this point.

\end{itemize}

%% file: sections/5-extraction_of_weak_mixing_angle.tex
The weak mixing angle, often written as $\sin^2\theta_W$, is a fundamental parameter of the SM and has been measured in experiments ranging from atomic parity violation at eV energy levels to high-energy colliders at the $Z$-pole~\cite{Qweak:2018tjf, CMS:2018ktx, ATLAS:2015ihy}. The EIC will provide constraints on $\sin^2\theta_W$ in the intermediate energy range that resides between the reach of fixed-target and collider facilities. 

For the extraction of the weak mixing angle, we focus on $\APVe$, where $\sin^2\theta_W$ enters through the electron coupling $g_{V,A}^e$ and the corresponding quark couplings in the structure functions. We also include the one-loop renormalization group evolution~\cite{Erler:2004in} of $\sin^2\theta_W$ in the $\overline{\rm MS}$ scheme, including the relevant particle thresholds that arise between $\mu =M_Z$ and $\mu=\sqrt{Q^2}$. Including target-mass correction terms, we can write:
\begin{eqnarray}
	&&\APVe =  \label{eq:Afit}\\
	&&\hspace*{-0.8cm}\frac{\vert P_e\vert \eta_{\gamma Z}\left[g_A^e 2y F_1^{\gamma Z}+g_A^e \left(\frac{2}{xy}-\frac{2}{x}-\frac{2M^2 xy}{Q^2}\right)F_2^{\gamma Z}+g_V^e(2-y)F_3^{\gamma Z}   \right]}
	{2yF_1^\gamma+\left(\frac{2}{xy}-\frac{2}{x}-\frac{2M^2 xy}{Q^2}\right)F_2^\gamma -\eta_{\gamma Z}\left[g_V^e 2y F_1^{\gamma Z}+g_V^e \left(\frac{2}{xy}-\frac{2}{x}-\frac{2M^2 xy}{Q^2}\right)F_2^{\gamma Z}+g_A^e(2-y)F_3^{\gamma Z}   \right]}~, \nonumber
\end{eqnarray}
where $M$ is the nucleon mass. Note that given the moderate $Q^2$ values of the EIC, the pure-$Z$ contribution to the structure functions is omitted for the precision relevant to our analysis. 

A single pseudodata set is generated using a reference value of $\sin^2\theta_W=0.231$ at the $Z$-pole  and the uncertainties in $\APVe$ in each $(x,Q^2)$ bin are obtained from simulation studies. Comparing the theory prediction to the pseudodata, a best-fit value and uncertainty projection for $\sin^2\theta_W$ at the $Z$-pole are obtained by minimizing the $\chi^2$ function defined as:
\begin{eqnarray}
	\chi^2 = [\mathcal A^\mathrm{pseudo-data}-\mathcal A^\mathrm{theory}]^{\rm T}[(\Sigma^{2})^{-1}][\mathcal A^\mathrm{pseudo-data}-\mathcal A^\mathrm{theory}]~,
\end{eqnarray}
where $\mathcal A$ is a dimension-$N_{\rm bin}$ vector with $N_{\rm bin}$ the total number of $(x,Q^2)$ bins, $\Sigma^2$ is the uncertainty matrix of dimension $N_{\rm bin}\times N_{\rm bin}$, described in Section~\ref{sec:proj_err_matrix}, and $\sin^2\theta_W$ to be fitted enters $\mathcal A^\mathrm{theory}$. The PDF portion of the uncertainty matrix is evaluated using the PDF sets {\tt CT18NLO}~\cite{Hou:2019qau} (LHAPDF~\cite{Buckley:2014ana} ID 14400-14458), {\tt MMHT2014nlo\_68cl}~\cite{Harland-Lang:2014zoa} (ID 25100-25150), and {\tt NNPDF31\_nlo\_as\_0118}~\cite{NNPDF:2017mvq} (ID 303400-303500). 

\begin{figure}[!h]
	\begin{center}
		\includegraphics[width=0.7\textwidth]{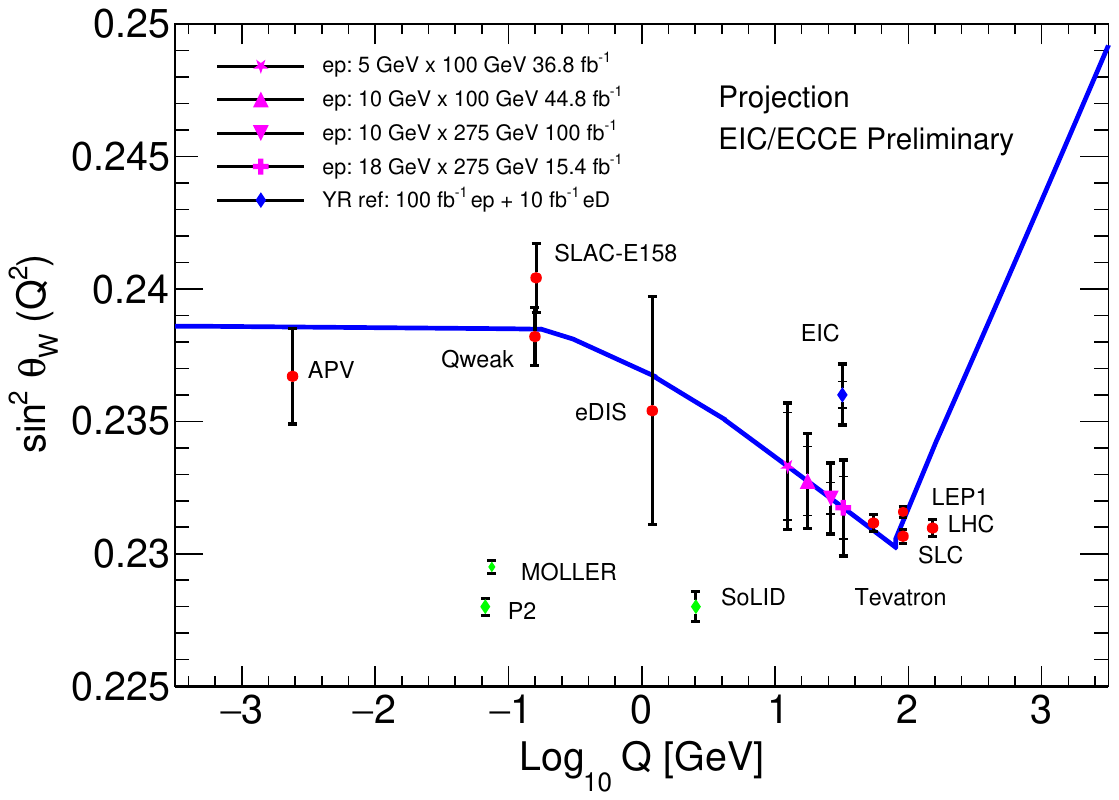}\\
		\includegraphics[width=0.7\textwidth]{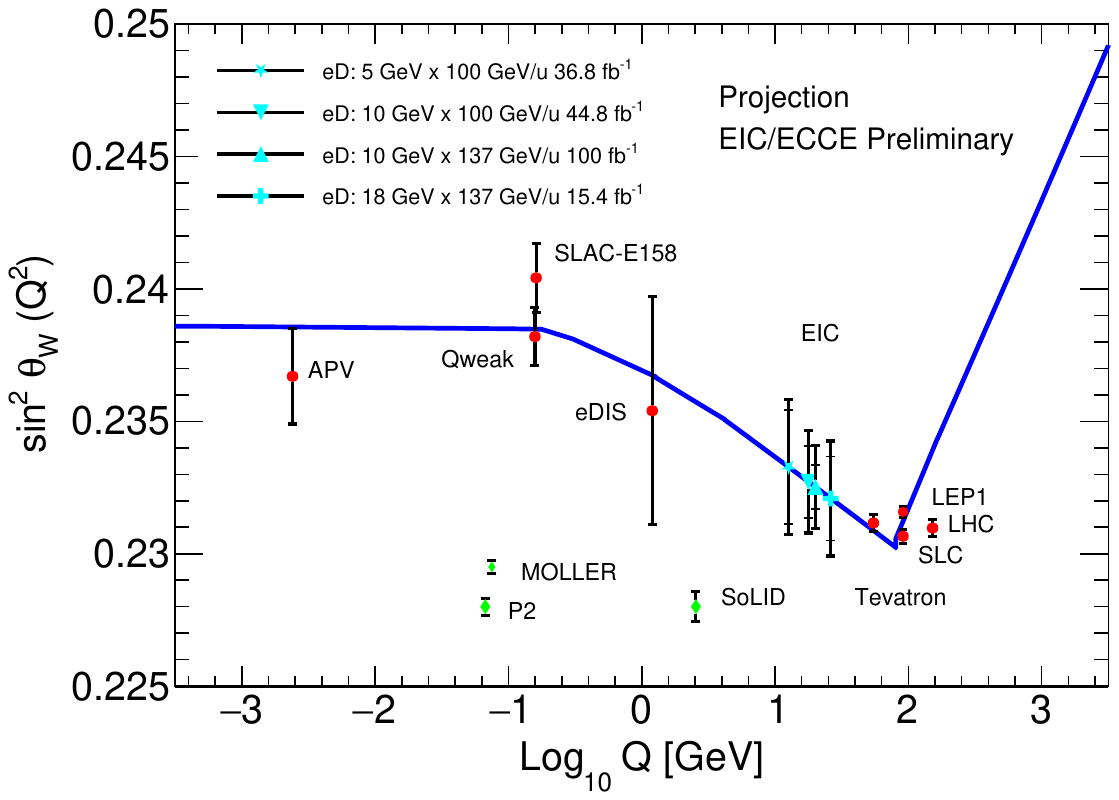}
	\end{center}
	\caption{Projected results for $\sin^2\theta_W$ using $ep$ (top, solid magenta markers) and $eD$ (bottom, solid cyan markers) collision data and the nominal annual luminosity given in Table 10.1 of the Yellow Report~\cite{AbdulKhalek:2021gbh}, along with existing world data (red solid circles) and near-future projections (green diamonds); see text for details. Data points for Tevatron and LHC are shifted horizontally for clarity. The script used to produce this plot is inherited from~\cite{Zhao:2016rfu}. The scale-dependence of the weak mixing angle expected in the SM (blue curve) is defined in the modified minimal subtraction scheme ($\overline{\mathrm{MS}}$ scheme)~\cite{Erler:2004in}.}
	\label{fig:sin2th}
\end{figure}

Our results for $\sin^2\theta_W$ are shown in Tables~\ref{tab:sinth_results_ep} and \ref{tab:sinth_results_eD} for five energy and nominal-annual-luminosity combinations of $ep$ and $eD$ collisions, respectively. These results are illustrated in Fig.~\ref{fig:sin2th}. The inner error bars show the combined uncertainty from statistical and 1\% uncorrelated experimental systematics (due to particle background); the median error bars show the experimental uncertainty that includes statistical, 1\% uncorrelated experimental systematics, and 1\% electron polarimetry. The outermost error bars, which almost coincide with the median error bars, include all the above and the PDF uncertainty evaluated using the set {\tt CT18NLO}. Results evaluated with the sets {\tt MMHT2014} and {\tt NNPDF31NLO} are similar. Along with our projection with the EIC annual nominal luminosity, we show the ``YR reference point'' (blue diamond), obtained from combining 100~fb$^{-1}$ $18\times 275$~GeV $ep$ and 10~fb$^{-1}$ $18\times 137$~GeV $eD$ pseudodata. Also shown are the expected precision from near-future  P2~\cite{Becker:2018ggl}, MOLLER~\cite{MOLLERCDR}, SoLID~\cite{Chen:2014psa}, and PVDIS~\cite{PVDIS,Erler:2014fqa} experiments, respectively, that will dominate the landscape of low to medium energy scales.

We note that our results have larger uncertainties than in the YR~\cite{AbdulKhalek:2021gbh}, which fits PDFs and $\sin^2\theta_W$ simultaneously using the JAM framework~\footnote{https://www.jlab.org/theory/jam}, possibly due to using realistic detector simulation and accurate running conditions.  On the other hand, we find that PDF uncertainties are likely not the dominant ones for the EIC projections, but the electron polarization is, for the settings where the integrated luminosity approaches 100~fb$^{-1}$. Consequently, upgrading the luminosity of the EIC does not bring significant improvement on the uncertainty of  $\sin^2\theta_W$, and therefore we do not show our fitting results for the ten-fold luminosity upgrade. 

\begin{table}[!ht]
	\centering
	\input{tables/EIC_EW_NC_sinth_table_ep.tex}
	\caption{Projected PVDIS asymmetry and fitted results for $\sin^2\theta_W$ using $ep$ collision data and the nominal annual luminosity. Here, $\langle Q^2\rangle$ denotes the value averaged over all $(x,Q^2)$ bins, weighted by $(\d A/A)_\mathrm{stat}^{-2}$ for each bin. The electron beam polarization is assumed to be 80\% with a relative 1\% uncertainty. The total (``tot'') uncertainty is from combining all of statistical, 1\% systematic (background), 1\% beam polarization, and PDF uncertainties evaluated using three different PDF sets. The rightmost column is for comparison with the YR.}
	\label{tab:sinth_results_ep}
\end{table}

\begin{table}[!ht]
	\centering
	\input{tables/EIC_EW_NC_sinth_table_eD.tex}
	\caption{Projected PVDIS asymmetry and fitted results for $\sin^2\theta_W$ using $eD$ collision data and the nominal annual luminosity.  The uncertainty evaluation is the same as Table~\ref{tab:sinth_results_ep}.}
	\label{tab:sinth_results_eD}
\end{table}

Our results show that the EIC will provide a determination of  $\sin^2\theta_W$ at an energy scale that bridges higher-energy colliders with low- to medium-energy SM tests. Additionally, data points of different $\sqrt{s}$ values of the EIC can be combined or the $Q^2$-dependence of the EW parameter can be explored, depending on the runplan of the EIC. Furthermore, one could study the exploratory potential of the EIC beyond the scope of a single SM parameter, and we provide results using the SMEFT framework in the next section.

%% file: tables/EIC_EW_NC_sinth_table_ep.tex
%/home/xz5y/doc/eic/ecce/djangoh.4.6.16/unpolarized/plots/asymmetry_unfolded2_getQ2.xlsx

      \begin{tabular}{|c|c|c|c|c|c|}
      \hline\hline
      %Do NOT remove the following 3 lines. These are used for the official note
          %Beam type & $ep$  & $ep$ & $ep$  & $ep$ & $ep$\\
          %Beam energy (GeV) & $5\times 100$ & $10\times 100$~& $10\times 275$ & $18\times 275$ & $18\times 275$\\
          Beam type and energy & ~$ep$ $5\times 100$~ & ~$ep$ $10\times 100$~ & ~$ep$ $10\times 275$~ & ~$ep$ $18\times 275$~ & ~$ep$ $18\times 275~$~\\
          Label & P2 & P3 & P4 & P5 & P6\\
          \hline
          Luminosity (fb$^{-1}$) & 36.8 & 44.8 & 100 & 15.4 & (100 YR ref)\\
%          Beam polarization & 80\% & 80\% & 80\% & 80\% & 80\% \\
          $\langle Q^2\rangle$ (GeV$^2$) & 154.4 & 308.1 & 687.3 & 1055.1 & 1055.1 \\ 
          $\langle A_{PV}\rangle$ ($P_e=0.8$)& $-0.00854$ & $-0.01617$ & $-0.03254$& $-0.04594$& $-0.04594$\\
          \hline
          $(\mathrm{d}A/A)_\mathrm{stat}$ & 1.54\% & 0.98\% & 0.40\% & 0.80\% & (0.31\%) \\ 
          $(\mathrm{d}A/A)_\mathrm{stat+syst(bg)}$ & 1.55\% & 1.00\% & 0.43\% & 0.81\% & (0.35\%) \\ 
          $(\mathrm{d}A/A)_\mathrm{1\% pol}$ & 1.0\% & 1.0\% & 1.0\% & 1.0\% & (1.0\%) \\ 
          $(\mathrm{d}A/A)_\mathrm{tot}$ & 1.84\% & 1.42\% & 1.09\% & 1.29\% & (1.06\%) \\ \hline 
          Experimental & & & & & \\
%         $\mathrm{d}(\sin^2\theta_W)_\mathrm{stat+bg}$  & 0.001964 & 0.001277 & 0.000594 & 0.001169 & 0.000512 \\
         $\mathrm{d}(\sin^2\theta_W)_\mathrm{stat+syst(bg)}$  & 0.002032 & 0.001299 & 0.000597 & 0.001176 & 0.000516 \\
         $\mathrm{d}(\sin^2\theta_W)_\mathrm{stat+syst+pol}$  & 0.002342 & 0.001759 & 0.001297 & 0.001769 & 0.001244\\\hline
         with PDF & & & & & \\
         $\mathrm{d}(\sin^2\theta_W)_\mathrm{tot,CT18NLO}$  & 0.002388 & 0.001807 & 0.001363 & 0.001823 & 0.001320 \\ 
         $\mathrm{d}(\sin^2\theta_W)_\mathrm{tot,MMHT2014}$  & 0.002353 & 0.001771 & 0.001319 & 0.001781 & 0.001270\\
         $\mathrm{d}(\sin^2\theta_W)_\mathrm{tot,NNPDF31}$  & 0.002351 & 0.001789 & 0.001313 & 0.001801 & 0.001308 \\
         \hline
         \hline
      \end{tabular}

%% file: tables/EIC_EW_NC_sinth_table_eD.tex
%/home/xz5y/doc/eic/ecce/djangoh.4.6.16/unpolarized/plots/asymmetry_unfolded2_getQ2.xlsx

      \begin{tabular}{|c|c|c|c|c|c|}
      \hline\hline
      %Do NOT remove the following 3 lines. These are used for the official note
          %Beam type & $eD$  & $eD$ & $eD$  & $eD$ & $eD$\\
          %Beam energy (GeV) & $5\times 100$ & $10\times 100$~& $10\times 137$ & $18\times 137$ & $18\times 137$\\
          Beam type and energy & ~$eD$ $5\times 100$~ & ~$eD$ $10\times 100$~ & ~$eD$ $10\times 137$~ & ~$eD$ $18\times 137$~ & ~$eD$ $18\times 137~$~\\
          Label & D2 & D3 & D4 & D5 & N/A\\
          \hline
          Luminosity (fb$^{-1}$)  & 36.8 & 44.8 & 100 & 15.4 & (10 YR ref)\\
          $\langle Q^2\rangle$ (GeV$^2$) & 160.0 & 316.9 & 403.5 & 687.2 & 687.2 \\
          $\langle A_{PV}\rangle$ ($P_e=0.8$)& $-0.01028$ & $-0.01923$ & $-0.02366$& $-0.03719$& $-0.03719$\\
          \hline
          $(\mathrm{d}A/A)_\mathrm{stat}$ & 1.46\% & 0.93\% & 0.54\% & 1.05\% & (1.31\%) \\ 
          $(\mathrm{d}A/A)_\mathrm{stat+bg}$ & 1.47\% & 0.95\% & 0.56\% & 1.07\% & (1.32\%) \\ 
          $(\mathrm{d}A/A)_\mathrm{syst, 1\% pol}$ & 1.0\% & 1.0\% & 1.0\% & 1.0\% & (1.0\%)\\ 
          $(\mathrm{d}A/A)_\mathrm{tot}$ & 1.78\% & 1.38\% & 1.15\% & 1.46\% & (1.66\%) \\ \hline
         Experimental & & & & & \\
         $\mathrm{d}(\sin^2\theta_W)_\mathrm{stat+bg}$  & 0.002148 & 0.001359 & 0.000823 & 0.001591 & 0.001963  \\
         $\mathrm{d}(\sin^2\theta_W)_\mathrm{stat+bg+pol}$  & 0.002515 & 0.001904 & 0.001544 & 0.002116 & 0.002414 \\ \hline
         with PDF & & & & & \\
         $\mathrm{d}(\sin^2\theta_W)_\mathrm{tot,CT18}$  & 0.002558 & 0.001936 & 0.001566 & 0.002173 & 0.00247 \\ 
         $\mathrm{d}(\sin^2\theta_W)_\mathrm{tot,MMHT2014}$  & 0.002527 & 0.001917 & 0.001562 & 0.002128 & 0.002424 \\
         $\mathrm{d}(\sin^2\theta_W)_\mathrm{tot,NNPDF31}$  & 0.002526 & 0.001915 & 0.001560 & 0.002127 & 0.002423\\
         \hline
         \hline
      \end{tabular}

%% file: sections/6a-data_generation_and_selection.tex
\subsection{Data Generation and Selection}
We use the procedure described in Section~\ref{sec:proj} to determine the uncertainty of our data projection and the uncertainty matrix. We consider both $ep$ and $eD$ collisions and concentrate on the two highest-energy settings listed in Table~\ref{tab:expconfig}. Because collisions with higher center-of-mass energy are more sensitive to SMEFT operators, we choose four data families with the two highest $\sqrt{s}$ to focus on:
\begin{eqnarray*}
	&&\mbox{$10$ GeV~$\times 137$~GeV $eD$ 100 fb$^{-1}$: D4, $ \Delta $D4, LD4}~,\\
	&&\mbox{$18$ GeV~$\times 137$~GeV $eD$ 15.4 fb$^{-1}$: D5, $ \Delta $D5, LD5}~,\\
	&&\mbox{$10$ GeV~$\times 275$~GeV $ep$ 100 fb$^{-1}$: P4, $ \Delta $P4, LP4}~,\\
	&&\mbox{$18$ GeV~$\times 275$~GeV $ep$ 15.4 fb$^{-1}$: P5, $ \Delta $P5, LP5}~.
\end{eqnarray*}
For the highest $\sqrt{s}$ but lower-luminosity set D5, $\Delta$D5, P5, and $\Delta$P5, we consider two scenarios: the nominal luminosity as indicated above and in Table~\ref{tab:expconfig}, and the high luminosity option denoted with an ``HL" label with ten-fold higher statistics. 

We use Eq. \eqref4 to generate $ N_{\rm exp} = 1000 $ pseudodata sets for each of the data  families. We then impose the following selection criteria on the bin points, $x$ and $ Q^2 $, and the inelasticity, $y$:
\begin{eqnarray}
	x < 0.5~,\quad 
	Q^2 > 100 {\rm\ GeV^2}~,\quad 
	0.1 < y < 0.9 
	\label{cuts_x_Q2_y}~.
\end{eqnarray}
These restrictions are designed to remove large uncertainties from non-perturbative QCD and nuclear dynamics that occur at low $ Q^2 $ and high $ x $, where sensivity to SMEFT effects is anyways expected to be reduced. We note that the condition on $y$ is already applied in the data generation and unfolding stages described in Section~\ref{sec:ecce_event_selection}. 

%% file: sections/6b-structure_of_smeft_asymmetry_corrections.tex
\subsection{Structure of the SMEFT asymmetry corrections \label{sec:description_of_numerical_analysis:structure_of_smeft_asymmetry_corrections}}
In the computation of SMEFT asymmetry values, $ A_{\rm SMEFT} $, we use the central member of the PDF set under consideration. We use the PDF sets {\tt NNPDF31\_nlo\_as\_0118}  \cite{NNPDF:2017mvq} and {\tt NNPDFpol11\_100} \cite{ERNocera2014} for the computation of unpolarized and polarized PV asymmetries, namely $\APVe$ and $\APVH$, respectively. We factor out the UV cut-off scale from all the seven Wilson coefficients, $ C_r \to C_r / \Lambda^2 $, and set $ \Lambda = 1 {\rm\ TeV} $. We turn on only one or two Wilson coefficients at a time and set the remaining ones to zero and linearize the SMEFT expressions with respect to the Wilson coefficient(s) of interest. SMEFT asymmetry expressions then generically take the form:
\begin{eqnarray}
	A_{\rm SMEFT}(x, Q^2, C) = A_{{\rm SM},0}^{\rm theo}(x, Q^2) + C \delta (x, Q^2) \label{6}
\end{eqnarray}
or
\begin{eqnarray}
	A_{\rm SMEFT}(x, Q^2, C_1, C_2) = A_{{\rm SM},0}^{\rm theo}(x, Q^2) + C_1 \delta_1 (x, Q^2) + C_2 \delta_2 (x, Q^2)~.
\end{eqnarray}
Comparing Eq. \eqref6 to \eqref4 or \eqref5, we see that at the end of a multi-run analysis, the distribution of the best-fit values for any single Wilson coefficient should be a Gaussian centered at the origin. 

%% file: sections/6c-best-fit_analysis_of_wilson_coefficients.tex
\subsection{Best-fit analysis of Wilson coefficients \label{sec:description_of_numerical_analysis:chisq_analysis_of_wilson_coefficients}}\label{sec:4}
Generating pseudo-data values, $ A_{\rm SM}^{\rm pseudo} $, and obtaining the SMEFT asymmetry expressions, $ A_{\rm SMEFT} $, we define a $\chi^2$ test statistic as:
\begin{eqnarray}
	\chi^2 =
	\sum_{b=1}^{N_{\rm bin}} \sum_{b'=1}^{N_{\rm bin}} \left[ A_{{\rm SMEFT},b} - A_{{\rm SM},b}^{\rm pseudo} \right] [(\Sigma^2)^{-1}]_{bb'} \left[ A_{{\rm SMEFT},b'} - A_{{\rm SM},b}'^{\rm pseudo} \right]~,
\end{eqnarray}
where $ N_{\rm bin} $ is the number of bins in a given data set. Generically, it looks like:
\begin{eqnarray}
	\chi^2 (C) = k_0 + k_1 C + k_2 C^2
\end{eqnarray}
for a single-parameter fit of Wilson coefficient $ C $, or:
\begin{eqnarray}
	\chi^2 (C_1, C_2) = 
	k_{00} + k_{10} C_1 + k_{01} C_2 + k_{11} C_1C_2 + k_{20} C_1^2 + k_{02} C_2^2
\end{eqnarray}
for a two-parameter fit of Wilson coefficients $ C_1 $ and $ C_2 $. The $ \chi^2 $ function is minimized with respect to $ C $ or to $ C_1 $ and $ C_2 $. This gives us the best-fit values, $ \bar C $ or $ \bar C_1 $ and $ \bar C_2 $. We obtain the inverse square of the error of the single-parameter best-fit value via:
\begin{eqnarray}
	\frac 1{\sigma_C^2} = \frac 12 \frac{\D2 \chi^2}{\d C^2}
\end{eqnarray}
evaluated at $ \bar C $. The inverse covariance matrix, $ V^{-1} $, of the two-parameter fit is constructed in such a way that its $ ij^{\rm th} $ component is given by:
\begin{eqnarray}
	(V^{-1})_{ij} =
	\frac 12 \frac{\del \chi^2}{\del C_i \del C_j}
\end{eqnarray}
for $ i, j = 1, 2 $, evaluated at the best-fit values of $ C_1 $ and $ C_2 $. Inverting $ V^{-1} $, we obtain the individual errors and the correlation of the fit:
\begin{eqnarray}
	V = 
	\begin{pmatrix}
		\sigma_1^2 & \rho_{12} \sigma_1 \sigma_2 \\
		& \sigma_2^2
	\end{pmatrix}_{\rm sym}~.
\end{eqnarray}

\subsubsection{Averaging over multiple pseudodata sets \label{sec:description_of_numerical_analysis:averaging_over_multiple_pseudodata_sets}}\label{sec:5}
When we repeat $ N_{\rm exp} $ times the single-parameter best-fit analysis described in Sec. \ref{sec:4}, we obtain $ N_{\rm exp} $ best-fit values, $ \bar C_e $, with corresponding uncertainties, $ \sigma_{C, e} $, for each pseudo-experiment $ e $. The mean of the best-fit values is obtained by averaging individual best-fit values weighted by the inverse square of the uncertainties:
\begin{eqnarray}
	\bar C = 
	\pp{\sum_{e = 1}^{N_{\rm exp}} \frac 1{\sigma_{C,e}^2}}^{-1} \pp{\sum_{e=1}^{N_{\rm exp}} \frac 1{\sigma_{C,e}^2} \bar C_e}~, \label{Cbest_multirun}
\end{eqnarray}
and the average uncertainty of this mean value is obtained via:
\begin{eqnarray}
	\frac 1{\sigma_C^2} = 
	\frac 1{N_{\rm exp}} \sum_{e=1}^{N_{\rm exp}} \frac 1{\sigma_{C,e}^2}~.\label{best_invV_1d}
\end{eqnarray}
When we repeat $ N_{\rm exp} $ times the two-parameter best-fit analysis on Wilson coefficients described in Sec. \ref{sec:4}, we obtain $ N_{\rm exp} $ pairs of best-fit values, $ \bar C_{1,e} $ and $ \bar C_{2,e} $, and inverse covariance matrices, $ (V^{-1})_e $, for each pseudo-experiment, $ e $. The best-fit values are averaged similarly to the one-dimensional case but with the inverse square of uncertainties replaced by inverse covariance matrices:
\begin{eqnarray}
	\begin{pmatrix}
		\bar C_1 \\
		\bar C_2
	\end{pmatrix} =
	\bb{\sum_{e=1}^{N_{\rm exp}} (V^{-1})_e}^{-1} \bb{\sum_{e=1}^{N_{\rm exp}} (V^{-1})_e \begin{pmatrix}
			\bar C_{1,e}\\
			\bar C_{2,e}
	\end{pmatrix}}~. \label{C1C2best_multirun}
\end{eqnarray}
The average inverse covariance matrix of the resulting best fit is calculated using:
\begin{eqnarray}
	V^{-1} =
	\frac 1{N_{\rm exp}} \sum_{e=1}^{N_{\rm exp}} (V^{-1})_e~.\label{best_invV_2d}
\end{eqnarray}
We note the presence of the factor $ 1/N_{\rm exp} $ in Eqs. \eqref{best_invV_1d} and \eqref{best_invV_2d}. Without it, we would be effectively increasing the luminosity of the corresponding central data set by a factor $N_\mathrm{exp}$. We avoid this by including this factor in computing the average uncertainty or inverse covariance matrix.

\subsubsection{Definition of confidence intervals \label{sec:description_of_numerical_analysis:definition_of_confidence_intervals}}

The result of a single-parameter multi-run fit can be expressed as:
\begin{eqnarray}
	\frac{(C - \bar C)^2}{\sigma_C^2} = \Delta \chi^2~,
\end{eqnarray}
and hence we can express the fitted result and the uncertainty of coefficient $C$ as:
\begin{eqnarray}
	C = C_{\rm best} \pm \sqrt{\Delta \chi^2} \sigma_C~.
\end{eqnarray}
For a two-parameter multi-run fit, the ellipse equation reads:
\begin{eqnarray}
	\begin{pmatrix}
		C_1 - \bar C_1 \\
		C_2 - \bar C_2
	\end{pmatrix}^{\rm T} V^{-1} \begin{pmatrix}
		C_1 - \bar C_1 \\
		C_2 - \bar C_2
	\end{pmatrix} = \Delta \chi^2
\end{eqnarray}
in the $ (C_1, C_2) $ plane. 

The $ \Delta \chi^2 $ values that determine the size of the best-fit region for an arbitrary confidence level are well-known. For 95\% CL, we have $\Delta \chi^2 = 3.841$, $5.991$, and $7.815$ for one-, two-, and three-parameter fits, respectively.

\subsubsection{Combination of best-fits from distinct data sets \label{sec:description_of_numerical_analysis:combination_of_best_fits}}
Suppose we have two data sets, say T1 and T2, from which we obtain the single-parameter best-fit values of Wilson coefficient, $ C $, written as $ \bar C_{\rm T1} $ and $ \bar C_{\rm T2} $, together with the errors $ \sigma_{C, {\rm T1}} $ and $ \sigma_{C, {\rm T2}} $. Assuming these data sets can be treated as uncorrelated to a good approximation, we obtain the combined best-fit value and the corresponding uncertainty by using Eqs. \eqref{Cbest_multirun} and \eqref{best_invV_1d} with slight modifications. First, the summation index $ e $ now runs from 1 to 2, representing the number of data sets combined. Second, the factor $ 1/N_{\rm exp} $ should be removed from Eq.~\eqref{best_invV_1d} because we now have indeed two independent, uncorrelated measurements. This method can be generalized to the combination of the best-fit values from more than two data sets, such as different beam energies, and to the case of multi-parameter fits in a straightforward manner.

\subsubsection{Simultaneous fit of Wilson coefficients and beam polarization or luminosity difference}
We observe in Section~\ref{sec:comparison_of_uncertainty_components} that experimental uncertainties such as the beam polarization and luminosity difference between $e^+$ and $e^-$ runs can be limiting factors for some of the data sets. When the data statistical uncertainty is very precise, there is the possibility that one can use the data themselves to constrain these systematic effects. We present in Appendix~\ref{app:additional_fits:luminosity_difference_fits} a method to simultaneously fit the SMEFT coefficient(s) and the luminosity difference for the LC asymmetries and in Appendix~\ref{app:additional_fits:beam_polarization_fits} a method to simultaneously fit the SMEFT coefficient(s) and the beam polarization for PV asymmetries.

%% file: sections/7a-fits_of_single_wilson_coefficients.tex
\subsection{Fits of single Wilson coefficients}
In this section, we discuss the 95\% CL intervals for the Wilson coefficients in single-parameter fits averaged over 1000 pseudo-experiments. The bounds on the Wilson coefficient $ C_{eu} $ across numerous data sets are representative and exhibit the common features of fits of single Wilson coefficients. We therefore show only the bounds on $ C_{eu} $ to illustrate the main observations and include the remaining Wilson coefficients in Appendix \ref{app:complete_results:1d}. Fig.~\ref{fig:1d_fits_Ceu} displays the 95\% CL intervals of $ C_{eu} $ for the four data families in which we are primarily interested in this paper. The intervals are grouped by asymmetries, namely electron PV asymmetries, $\APVe$, of unpolarized hadrons (``unpolarized $ A_{\rm PV} $''), hadron PV asymmetries, $\APVH$, with unpolarized electrons (``polarized $ A_{\rm PV} $''), and unpolarized electron-positron asymmetries, $\ALCH$, of unpolarized hadrons (``lepton-charge $ A $''). PV asymmetries are then grouped into two, showing the fits in the nominal- and high-luminosity scenarios. In each block of intervals, there are four double lines in the case of PV asymmetries and four single lines in LC asymmetries. These four lines correspond to the data families D4 (black and its shade), D5 (red), P4 (blue), and P5 (orange), respectively. The darker of the two lines indicate the bounds from single-parameter fits with the Wilson coefficient $ C_{eu} $, whereas the lighter ones show the bounds on the Wilson coefficient from simultaneous $(1+1)$-parameter fits with $ C_{eu} $ and the beam polarization. We describe the details of the fits involving the beam polarization as an additional free parameter in Appendix~\ref{app:additional_fits:beam_polarization_fits}. 

\begin{figure}
	[H]\centering
	\includegraphics[width=.85\textwidth]{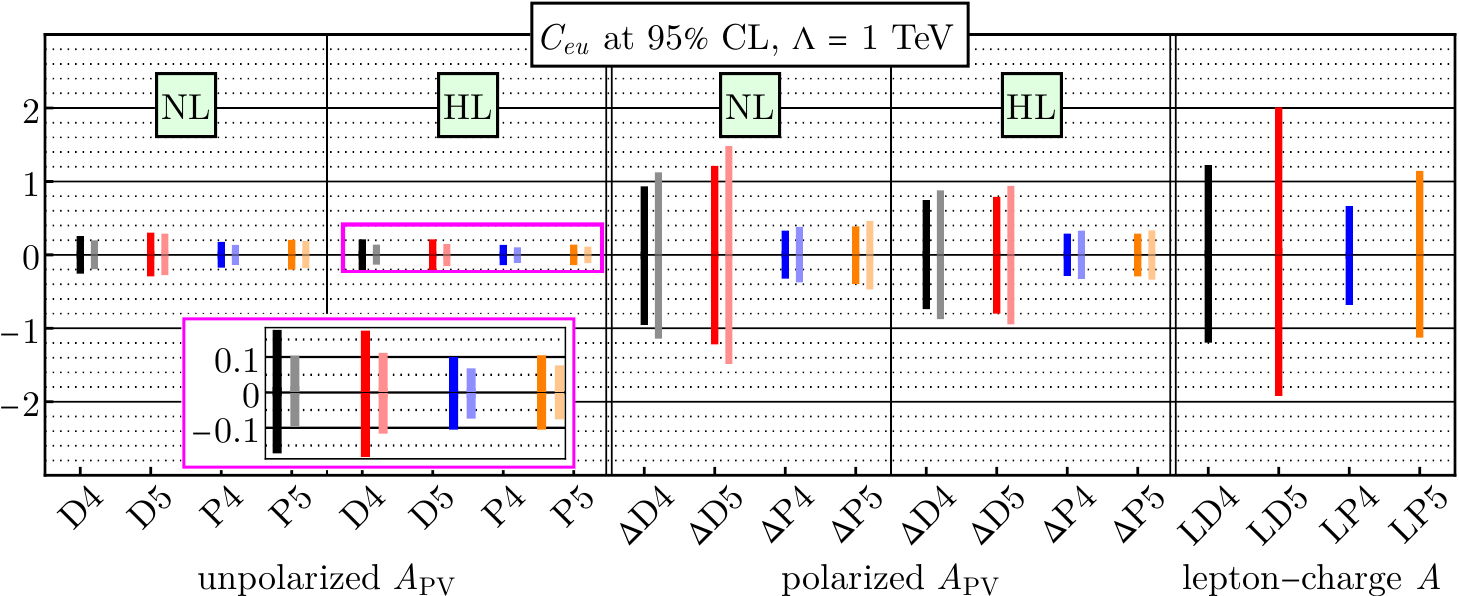}
	\caption{95\% CL bounds of $ C_{eu} $ from single-parameters fits (darker) and from the $(1+1)$-parameter fits with beam polarization as an additional fitting parameter (lighter) using the families of data sets D4, D5, P4, and P5 at $\Lambda = 1 \ {\rm TeV}$.
	}
	\label{fig:1d_fits_Ceu}
\end{figure}

From Fig.~\ref{fig:1d_fits_Ceu}, we can extract the following main points:
\begin{itemize}
	\item Proton asymmetries of all the three types, namely unpolarized and polarized PV asymmetries and LC asymmetries, impose considerably stronger bounds than deuteron.
	\item High-energy low-luminosity data sets D5 and P5 lead to slightly weaker bounds than the less-energetic but higher-luminosity ones, D4 and P4, respectively. 
	\item  Unpolarized PV asymmetries, $\APVe$, offer much stricter bounds than the polarized ones $\APVH$; however, it should be noted that for some Wilson coefficients, unpolarized proton asymmetries yield nearly the same bounds as the corresponding polarized ones.
	\item Data sets in the high-luminosity scenario make a noticeable difference in the size of bounds. The improvement due to increased luminosity is slightly more significant for polarized deuteron asymmetries. 
	\item Bounds from electron-positron asymmetries, $\ALCH$, are comparable to or looser than the ones from polarized hadron asymmetries. They never offer stricter bounds than high-luminosity hadron PV asymmetries.
	\item If the beam polarization is introduced as a new fitting parameter, unpolarized hadron asymmetries give considerably stronger bounds. The improvement is more significant in the high-luminosity scenario. However, the same fitting method yields weaker bounds with polarized hadron asymmetries. We explain this finding in the Appendix~\ref{app:additional_fits:beam_polarization_fits}.
\end{itemize}

Assuming weak correlations, one can also combine the bounds within a given family of data sets, e.g. D4, $ \Delta $D4, and LD4. We find that the resultant bound is never stronger than the strongest one obtained from the individual family members, which is always from the electron PV asymmetry data set. We note that this observation holds for other Wilson coefficient choices and for the two-dimensional fits in the next section, as well.

\begin{figure}
	[h]\centering
	\includegraphics[width=.8\textwidth]{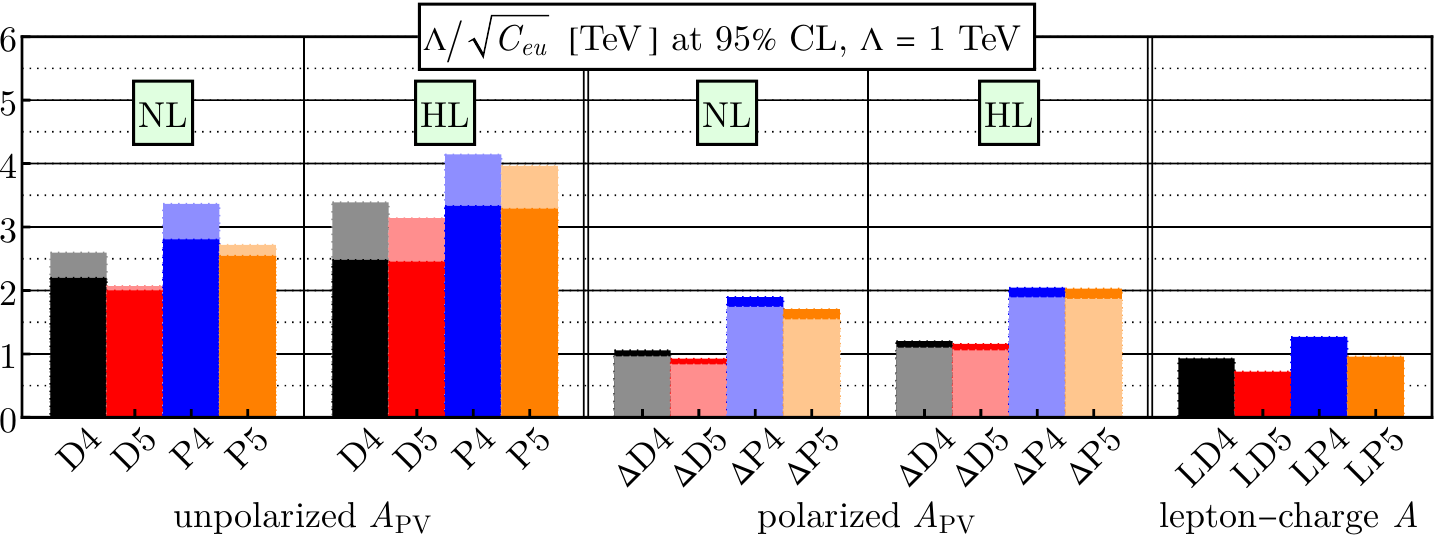}
	\caption{Effective UV cut-off scales, $\Lambda/\sqrt{\Ceu}$, defined in terms of the 95\% CL bounds on the Wilson coefficient $\Ceu$ and with $\Lambda = 1 \ {\rm TeV}$.}
	\label{fig:1d_fits_Ceu_Lambda}
\end{figure}

In Fig.~\ref{fig:1d_fits_Ceu_Lambda}, we present the effective UV cut-off scales, $ \Lambda / \sqrt{C_{eu}} $, with $ \Lambda = 1 \ {\rm TeV} $, corresponding to the bounds shown in Fig.~\ref{fig:1d_fits_Ceu}. The organization of this plot in terms of asymmetries and data sets is the same as in Fig.~\ref{fig:1d_fits_Ceu}. Improved bounds on $ C_{eu} $ with the addition of the beam polarization to the fits are equivalent to higher energy scales in the unpolarized PV asymmetries, which are indicated by the lighter columns in the background; on the other hand, weaker bounds from the fits with beam polarization are depicted by the lighter columns in the foreground for the polarized PV asymmetries.

One can observe that scales reaching 3 TeV can be probed with nominal luminosity, while scales exceeding 4 TeV can be probed for other luminosities. We remark that care must be taken in comparing these mass limits with others found in the literature, which sometimes assume a strong coupling limit, equivalent to setting $ C_r = 4\pi $, and also maximally constructive interference between different quark contributions. For example, converting our results to the notation of \cite{Arrington:2021alx} would yield a bound on $ \Lambda / \sqrt{C_{eu}} $ of 19 TeV, instead of 3 TeV quoted here, which is only very approximate and is calculated by multiplying 3 TeV by $ \sqrt{4\pi} $ and $ \sqrt{\sqrt{5}} $, where the latter is to account for the constructive interference between quark contributions, and by another factor to convert 90\% CL to 95\% CL. 

%% file: sections/7b-fits_of_two_wilson_coefficients.tex
\subsection{Fits of two Wilson coefficients \label{sec:2d_fits}}
In this section, we discuss fits on pairs of Wilson coefficients in order to determine how well the EIC can break degeneracies between parameters that occur in the LHC Drell-Yan data~\cite{RBoughezal2020,RBoughezal2021}. We emphasize that the representative examples shown in this section are the results of the simultaneous fits with beam polarization in light of the significantly improved results of the $(1+1)$-parameter fits in the previous section. The description of the beam-polarization fits is presented in Appendix \ref{app:additional_fits:beam_polarization_fits}. The complete set of plots of confidence ellipses is given in Appendix \ref{app:complete_results:2d}.

In Fig. \ref{fig:ellipses_d4_p4_family_ceu_cqe}, we compare the 95\% CL ellipses for the pair $ (C_{eu}, C_{qe}) $ between the data families D4 and P4. Each asymmetry type gives a distinct correlation pattern, complementary to one another. Electron-positron asymmetries give rise to wide and elongated, band-like ellipses compared to PV asymmetries. As in the case of $(1+1)$-parameter fits, electron PV asymmetries of unpolarized hadrons offer the strongest bounds on the pairs of Wilson coefficients. Comparing deuteron to proton, one can see that proton data are significantly more constraining.

\begin{figure}
	[H]\centering
	\includegraphics[width=.4\textwidth]{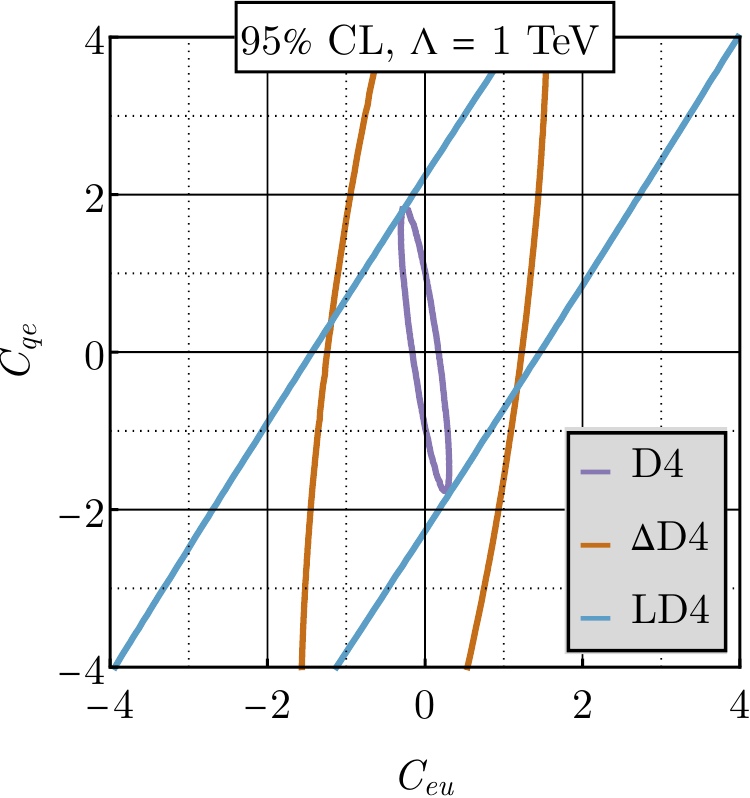}
	\includegraphics[width=.4\textwidth]{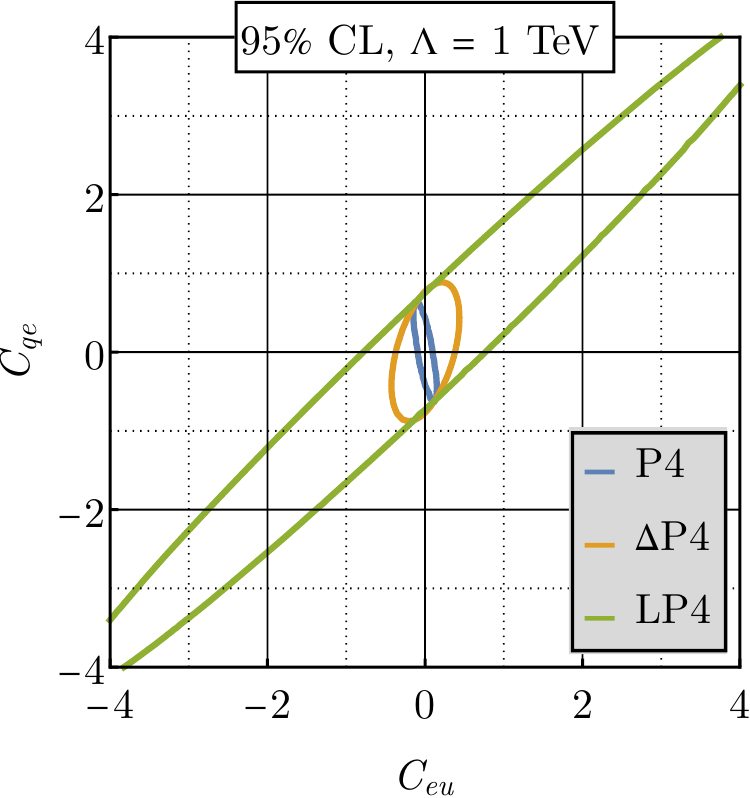}
	\caption{95\% CL ellipses for the Wilson coefficients $ C_{eu} $ and $ C_{qe} $ using the families of data sets D4 and P4 in the simultaneous $(2+1)$-parameter fits that includes the beam polarization as an additional fitting parameter.}
	\label{fig:ellipses_d4_p4_family_ceu_cqe}
\end{figure}

\begin{figure}
	[h]\centering
	\includegraphics[width=.4\textwidth]{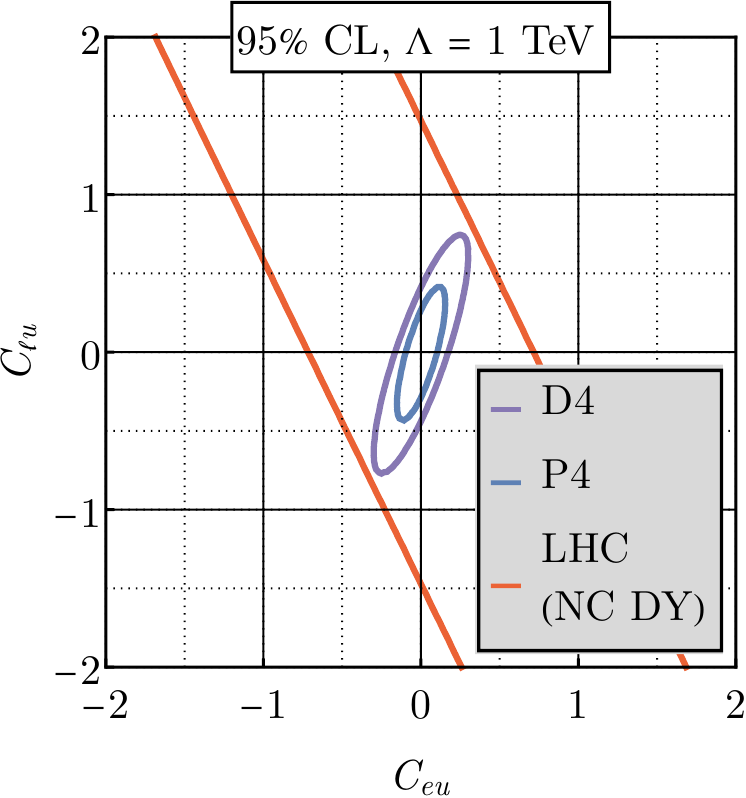}
	\caption{95\% CL ellipses for the Wilson coefficients $ C_{eu} $ and $ C_{qe} $ using the data sets D4 and P4 in the $(2+1)$-parameter fit that includes the beam polarization as an additional fitting parameter, compared with the corresponding two-parameter fit from the LHC data~\cite{RBoughezal2021}.}
	\label{fig:ellipses_p4_nl_hl_lhc_ceu_clu}
\end{figure}

\begin{figure}
	[h]\centering
	\includegraphics[width=.4\textwidth]{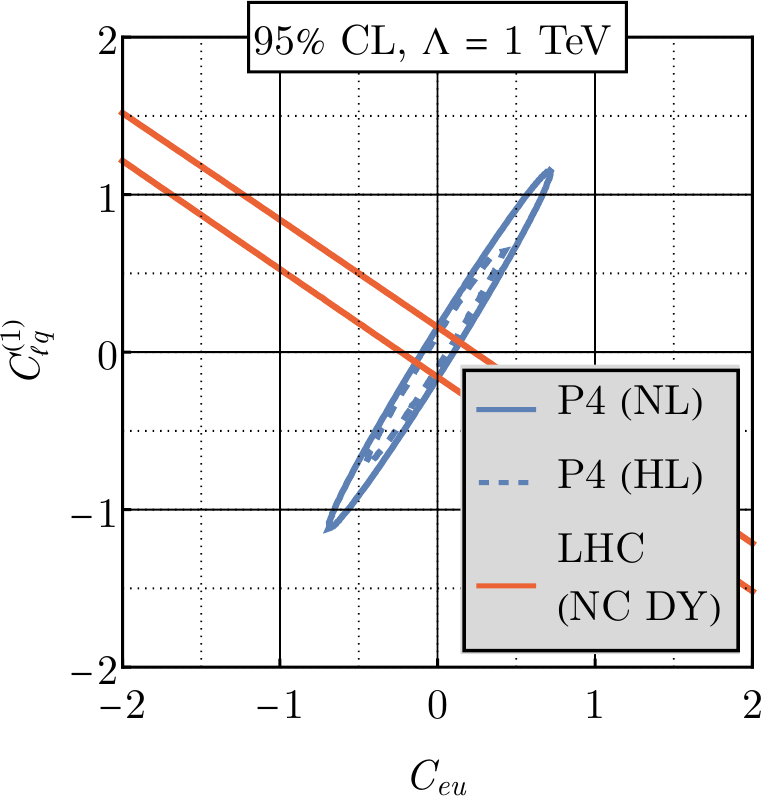}
	\caption{95\% CL ellipses for the Wilson coefficients $ C_{eu} $ and $ C_{\ell q}^{(1)} $ using the nominal- and high-luminosity data set P4 in the $(2+1)$-parameter fit that includes the beam polarization as an additional fitting parameter, compared with the corresponding two-parameter fit from the LHC data \cite{RBoughezal2020}.}
	\label{fig:ellipses_p4_nl_hl_lhc_ceu_clq1}
\end{figure}

Fig.~\ref{fig:ellipses_p4_nl_hl_lhc_ceu_clu} shows the comparison of the simultaneous fit of the Wilson coefficients $ (C_{eu}, C_{\ell u}) $ projected for the EIC to the corresponding fit with the LHC data adapted from \cite{RBoughezal2021}. The LHC fit exhibits a flat direction, i.e. a particular linear combination of the two coefficients cannot be determined. A similar comparison is given in Fig.~\ref{fig:ellipses_p4_nl_hl_lhc_ceu_clq1} for the pair $ (C_{eu}, C_{\ell q}^{(1)}) $, using the nominal- and high-luminosity P4 set of the EIC.  We observe that in both figures, projected EIC fits have different correlation patterns from the LHC. More importantly, the EIC projected data show the capability of resolving flat directions and significantly constraining the aforementioned pairs of Wilson coefficients.

\begin{figure}
	[h]\centering
	\includegraphics[width=.4\textwidth]{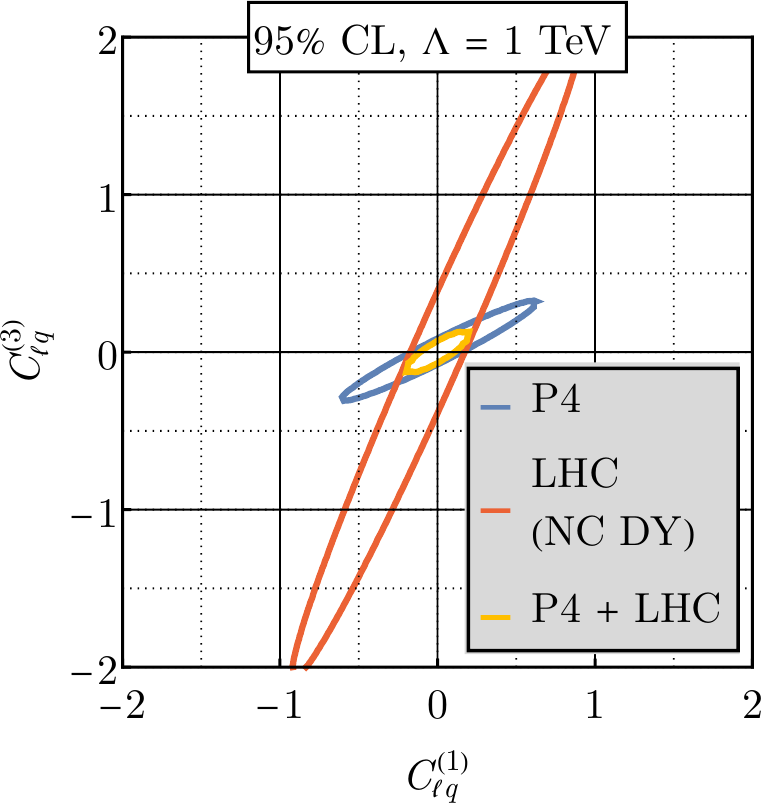}
	\caption{95\% CL ellipses for the Wilson coefficients $ C_{\ell q}^{(1)} $ and $ C_{\ell q}^{(3)} $ using the nominal-luminosity data set P4 in the $(2+1)$-parameter fit that includes the beam polarization as an additional fitting parameter, compared with the corresponding fit from the LHC data~\cite{RBoughezal2020} and the combined fit of the two.}
	\label{fig:ellipses_p4_lhc_p4+lhc_clq1_clq3}
\end{figure}

Finally, in Fig. \ref{fig:ellipses_p4_lhc_p4+lhc_clq1_clq3}, we present the fits from the P4 data set and the LHC adapted from \cite{RBoughezal2020} for the pair $ (C_{\ell q}^{(1)}, C_{\ell q}^{(3)}) $. This figure shows that when the LHC data imposes tight bounds on a pair of Wilson coefficients, the EIC preliminary data can introduce far stronger bounds on the same pair of Wilson coefficients. Moreover, the fits from EIC and LHC have distinct correlations, which indicates the complementarity of the EIC to the LHC as a future collider. Treating the projected EIC and the LHC data to be uncorrelated, we also plot the combined fit of the two, which turns out to even more strongly constrain the chosen pair of Wilson coefficients. We remark that the effective UV scales probed with the combined data set exceed 2 TeV. 

It should be noted that there appear flat directions in the fits of certain pairs of Wilson coefficients with the projected EIC data that utilize the deuteron beam. Examples include $ (C_{eu}, C_{ed}) $ and $ (C_{\ell u}, C_{\ell d}) $. We can explain these observations analytically. We find that these pairs always appear in a specific way in asymmetry expressions, for example $ 2C_{eu} - C_{ed} $ for electron PV asymmetries with unpolarized deuteron. In all such cases, only one of the data families exhibits this behavior, with the degeneracy broken by another data family. 

Our results on the bounds from Wilson coefficients in simultaneous $(2+1)$-parameter fits with the beam polarization as an additional parameter can be summarized as follows: 
\begin{itemize}
	\item Proton asymmetries impose much stricter bounds than deuteron.
	\item Unpolarized hadron asymmetries lead to stronger correlations than the polarized ones.
	\item The three types of asymmetries of deuteron and proton considered in this work, together with the LHC data, are complementary to each other in the sense that they offer distinct correlation patterns.
	\item The projected EIC data are capable of resolving all flat directions that appear in the LHC Drell-Yan data.
	\item The bounds from the projected EIC data can be much stronger than the LHC data, indicating that the EIC has an important role 
	to play in future probes of the SMEFT.
\end{itemize}
We can ask what happens when more than two Wilson coefficients are turned on simultaneously. We study this in Section~\ref{app:sixd_fits}, where we turn on six Wilson coefficients. The resulting bounds are 20 to 30\% weaker than the ones found here in the $2d$ case.  We note that no flat directions appear in these fits, indicating that the EIC can fully probe this parameter space without degeneracies.

%% file: sections/8-conclusions.tex
In this manuscript, we have analyzed the potential of testing the electroweak SM and exploring BSM physics with the future EIC. We have focused on the precision determination of the weak mixing angle over a wide range of momentum transfer and on probes of heavy new physics. We have provided all the formulae for neutral-current DIS and simulation details that will be needed for future studies of these areas. Our BSM analysis utilizes the model-independent SMEFT framework and focuses on the semi-leptonic four-fermion operator sector of the theory. We translate our formalism into the DIS language in terms of parity-violating couplings and structure functions to facilitate cross-talk between the high-energy-physics and nuclear communities. We provide a detailed accounting of uncertainties from statistics, experimental systematic effects, beam polarimetry for parity-violating asymmetries, higher-order QED corrections for lepton-charge asymmetries, and finally PDFs. Additionally, we explore simultaneously fitting the beam polarization with the anticipated high-precision parity-violating asymmetry data as a possible analysis technique to improve upon the experimental limitation from beam polarimetry.

Our BSM analysis finds that UV scales in excess of 3 TeV can be probed with the currently planned (nominal) annual luminosity of the EIC, with scales above 4 TeV possible if a ten-fold luminosity upgrade becomes available beyond EIC's initial decade of running. For the latter, we focus on studying the physics reach and limitations from sources other than statistical, without comment as to the feasibility of such an upgrade. The most stringent bounds come from polarized electron scattering off of unpolarized protons. Constraints from polarized hadrons, deuterons, and from a possible future positron beam provide important complementary probes. Our complete study of correlations between Wilson coefficients finds that no degeneracies remain upon combining all EIC data sets. This is not the case with LHC Drell-Yan measurements, in which numerous degeneracies exist, and will continue to occur even after LHC's high luminosity running.

This demonstrates that the EIC polarization provides a powerful probe of BSM effects. Although the EIC is primarily thought of as a QCD machine, it is also a powerful probe of potential new physics, in many ways complementary to the higher-energy LHC. We note that current  global fits of LHC data, for example to top-quark and Higgs data, probe
orthogonal sets of Wilson coefficients~\cite{Ethier:2021bye}. However, the strongest LHC constraints on the semi-leptonic four-fermion
sector of the SMEFT come from the Drell-Yan data. Given that LHC Drell-Yan cross sections are blind to certain combinations of Wilson coefficients, we envision that high-precision EIC data will help remove these degeneracies in global fits, both at the present moment and when high-luminosity LHC Drell-Yan data become available. We hope that our work motivates future studies of the unexpected power of the EIC for new-physics searches.

%% file: appendices/a1-luminosity_difference_fits.tex
\subsection{Luminosity difference fits \label{app:additional_fits:luminosity_difference_fits}}
Since electron and positron data would be taken at different times with different beam configurations, there is the possibility of a significant offset between the absolute luminosities of the two data sets. In the main text, we include this uncertainty in the error matrix as the luminosity error, $ \sigma_{\rm lum} = 0.02 $, which is assumed absolute. We study here the possibility of simultaneously fitting this luminosity difference together with the Wilson coefficients. 

We fit the pseudodata for the LC asymmetries with an overall shift, $ A_{\rm lum} $, added to the pseudodata. Then, we define the $ \chi^2 $ test statistics as:
\begin{align} 
	\chi^2 = \sum_{b=1}^{N_{\rm bin}} \sum_{b'=1}^{N_{\rm bin}} [A_{{\rm SMEFT}, b} - A_{{\rm SM},b}^{\rm pseudo}] [(\tilde \Sigma^2)^{-1}]_{bb'}  [A_{{\rm SMEFT}, b'} - A_{{\rm SM},b'}^{\rm pseudo}]~,
\end{align}
where we omit the uncertainty in the luminosity difference between $e^+$ and $e^-$ runs from the uncertainty matrix:
\begin{align}
	\tilde \Sigma^2 = \Sigma^2 \big|_{\sigma_{\rm lum} \to 0}~.
\end{align}
However, we keep the luminosity uncertainty in the pseudodata generation. By introducing the luminosity difference, $ A_{\rm lum} $, as a new parameter, we extend our one-parameter and two-parameter Wilson-coefficient fits to $ (1+1)$- and $ (2+1) $-parameter fits. 

\begin{figure}
	[h]\centering
	\includegraphics[width=.5\textwidth]{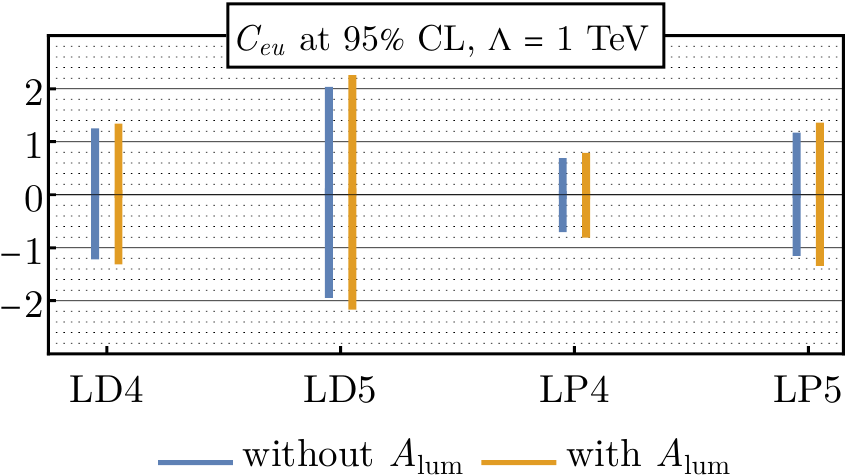}
	\caption{Comparison of the bounds on the Wilson coefficient $ C_{eu} $ with all the LC asymmetry data sets of interest in the absence and presence of the luminosity difference as a new free fitting parameter.}
	\label{fig:lum_Ceu_all}
\end{figure}

\begin{figure}
	[h]\centering
	\includegraphics[width=.4\textwidth]{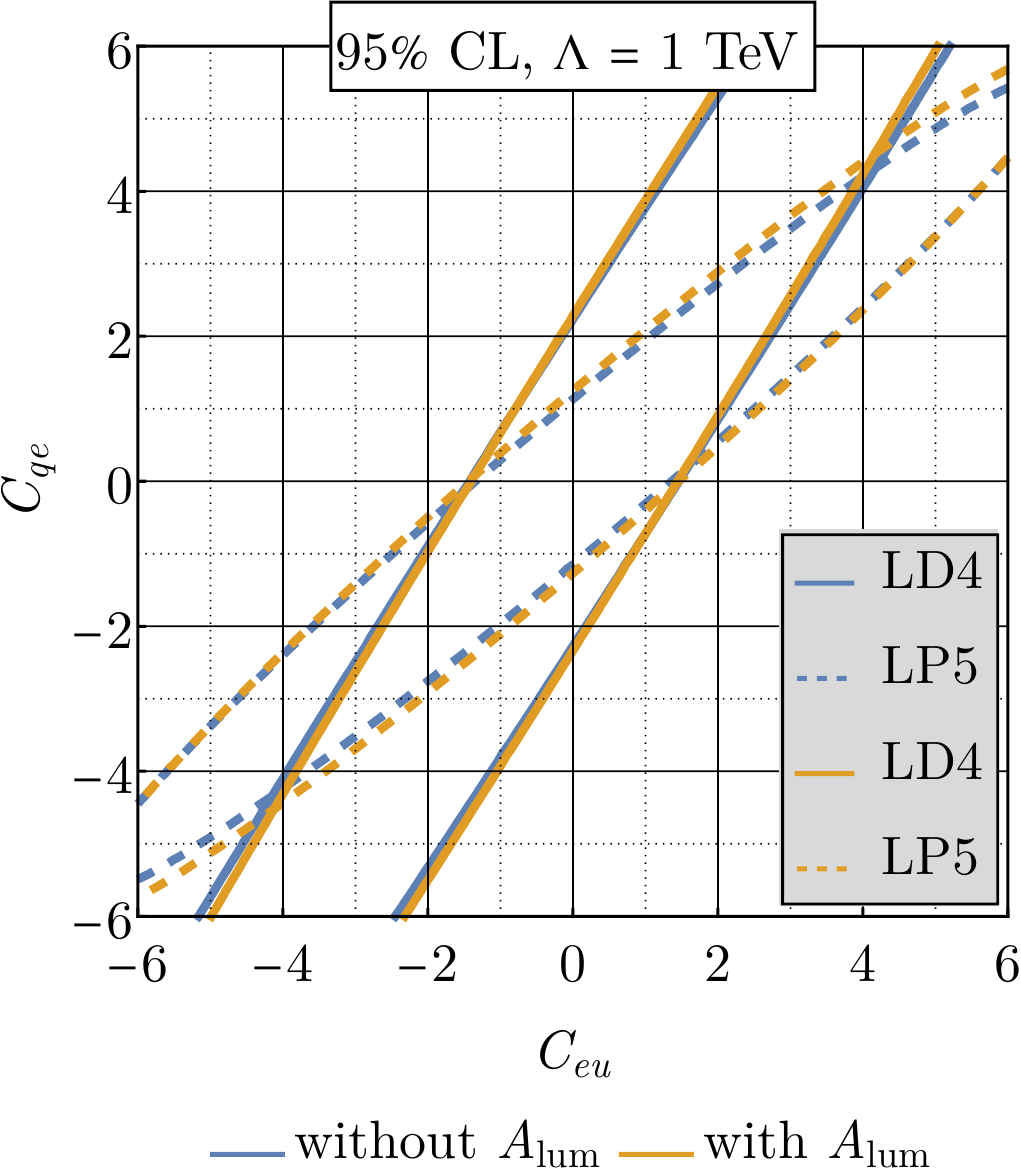}
	\caption{Comparison of the 95\% CL ellipses for the Wilson coefficients $ (C_{eu}, C_{qe}) $ with the data sets LD4 and LP5 in the absence and presence of the luminosity difference as an additional fitting parameter.}
	\label{fig:lum_2w_ed4_ep5}
\end{figure}

We find that there are mild correlations, $ |\rho_r| \lesssim 0.4 $, between $ A_{\rm lum} $ and any $ C_r $ in the $ (1+1) $- and $ (2+1) $-parameter fits. In addition, the fitted results for Wilson coefficients have slightly larger uncertainties when the luminosity difference is treated as a fitting parameter. In Fig. \ref{fig:lum_Ceu_all}, we show the 95\% CL intervals with and without $ A_{\rm lum} $ for the Wilson coefficient $ C_{eu} $ in all the four LC asymmetry data sets of interest. In Fig. \ref{fig:lum_2w_ed4_ep5}, we compare the 95\% CL ellipses of the Wilson coefficients $ (C_{eu}, C_{qe}) $ between the data sets LD4 and LP5 with and without the luminosity difference as a fitted parameter. From these figures, we see that the 95\% CL bounds on $ C_{eu} $ become 15 to 20\% weaker. The difference is less noticeable in the confidence ellipses.

%% file: appendices/a2-beam_polarization_fits.tex
\subsection{Beam polarization fits \label{app:additional_fits:beam_polarization_fits}}
In the same spirit as the previous section, we now consider fitting the beam polarization simultaneously with the Wilson coefficients in an attempt to reduce the uncertainty associated with the experimental limitation from beam polarimetry. We fit the pseudodata for the PV asymmetries by including a factor of $ P $ in the SMEFT asymmetries. We then define a $ \chi^2 $ test statistics as:
\begin{align}
	\chi^2 = \sum_{b=1}^{N_{\rm bin}} \sum_{b'=1}^{N_{\rm bin}} [P A_{{\rm SMEFT}, b} - A_{{\rm SM},b}^{\rm pseudo}] [(\tilde \Sigma^2)^{-1}]_{bb'}  [P A_{{\rm SMEFT}, b'} - A_{{\rm SM},b'}^{\rm pseudo}] + \frac{(P - \bar P)^2}{\delta P^2}~. \label{beam_pol_chi2}
\end{align}
In this approach, we omit the beam polarization uncertainty, $ \sigma_{\rm pol} $, from the uncertainty matrix because it is now treated as a fitting parameter: 
\begin{align}
	\tilde \Sigma^2 = \Sigma^2 \big|_{\sigma_{\rm pol} \to 0}~,
\end{align}
but not during pseudodata generation. The second term on the RHS of Eq.~\eqref{beam_pol_chi2} is added by hand, where $\bar P$ and $\delta P$ are the beam polarization value and its uncertainty provided by the polarimetry, respectively, presumably uncorrelated with the asymmetry measurements. The logic behind this addition is that experimentally, the polarimetry does provide knowledge on the beam polarization, but we hope to obtain a better determination of the polarizations within the uncertainty provided by the polarimetry by fitting data with high statistical precision. As for the beam polarization itself, we use a normalized value of $ \bar P = 1 $ in this study for simplicity. Treating the new term to be the contribution of a new observable, we increase the degrees of freedom of the $ \chi^2 $ distribution by 1. As in the case of luminosity difference fits, we extend our 1- and 2-parameter fits of Wilson coefficients to $ (1+1)$- and $ (2+1) $-parameter simultaneous fits by including the beam polarization as a new parameter. 

From $ (1+1)$-parameter fits, we find that $ P $ and any $ C_r $ are rather weakly correlated, $ |\rho_r| \lesssim 0.1 $, in the polarized hadron data sets, whereas there are strong correlations, $ |\rho_r| \gtrsim 0.7 $, in the unpolarized hadron asymmetries. We observe similar correlations in the $ (2+1)$-parameter fits.

\begin{figure}
	[h]\centering
	\includegraphics[width=.6\textwidth]{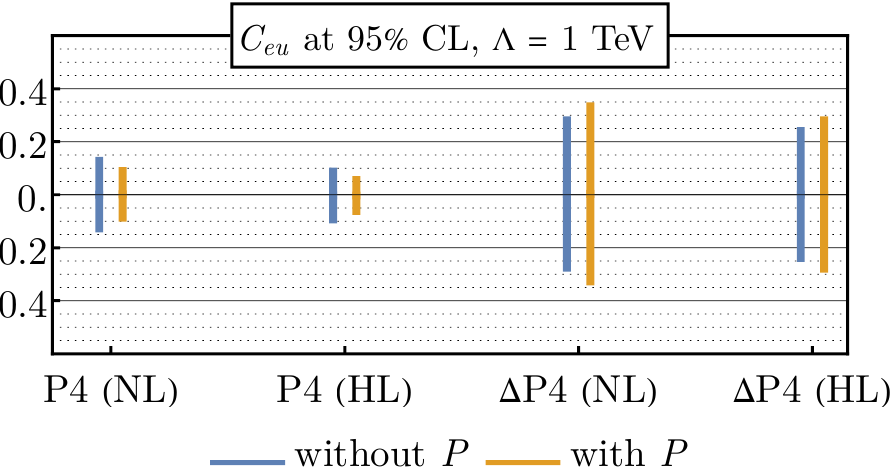}
	\caption{95\% CL bounds on the Wilson coefficient $ C_{eu} $ with the nominal- and high-luminosity data sets P4 and $ \Delta $P4 in the absence and presence of the beam polarization, $ P $, as an additional parameter in the fits.}
	\label{fig:beam_pol_err_bar_p4_deltap4}
\end{figure}

\begin{figure}
	[H]\centering
	\includegraphics[width=.35\textwidth]{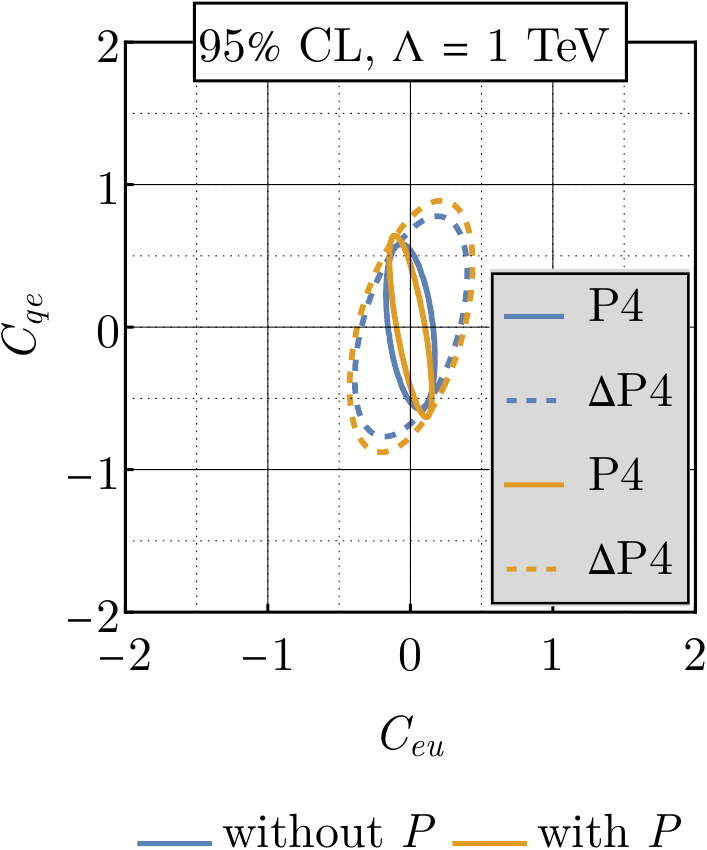}
	\caption{95\% CL ellipse of the Wilson coefficients $ C_{eu} $ and $ C_{qe} $ for the data sets P4 and $ \Delta $P4 in the absence and presence of the beam polarization, $ P $, as a new parameter in the fits.}
	\label{fig:beam_pol_ellipse_ceu_cqe_p4_deltap4}
\end{figure}

In Fig. \ref{fig:beam_pol_err_bar_p4_deltap4}, we present the allowed intervals of the Wilson coefficient $ C_{eu} $ for the nominal- and high-luminosity data sets P4 and $ \Delta $P4, while Fig. \ref{fig:beam_pol_ellipse_ceu_cqe_p4_deltap4} displays the 95\% CL ellipse of the Wilson coefficients $ (C_{eu}, C_{qe}) $ for the same data sets in the nominal-luminosity scenario. We find that bounds from unpolarized hadron data sets become stronger by 30 to 50\%, yet the ones from polarized hadron asymmetries become 15 to 20\% weaker. The improvement is sharper in the high-luminosity unpolarized hadron sets, whereas the worsening is significant for the nominal-luminosity polarized hadron sets.  

One can explain why the bounds become weaker in the polarized hadron sets by referring to the correlations. Since in these data sets, the beam polarization and the Wilson coefficients are found to be weakly correlated, one would naively expect the bounds obtained from single-parameter fits of Wilson coefficients to roughly remain the same on the grounds that $ P $ and $ C_k $ can be thought of almost fully independent so that they will not affect each other in the fits. Thus, any increase in the allowed limits of Wilson coefficient can be attributed to the increase in the number of parameters fitted, which is reflected as the normalization of the uncertainties of the fit.

\vspace{-.3cm}

%% file: appendices/b1-fits_of_single_wilson_coefficients.tex
\subsection{Fits of single Wilson coefficients \label{app:complete_results:1d}}
\begin{figure}
	[H]\centering
	\includegraphics[width=.75\textwidth]{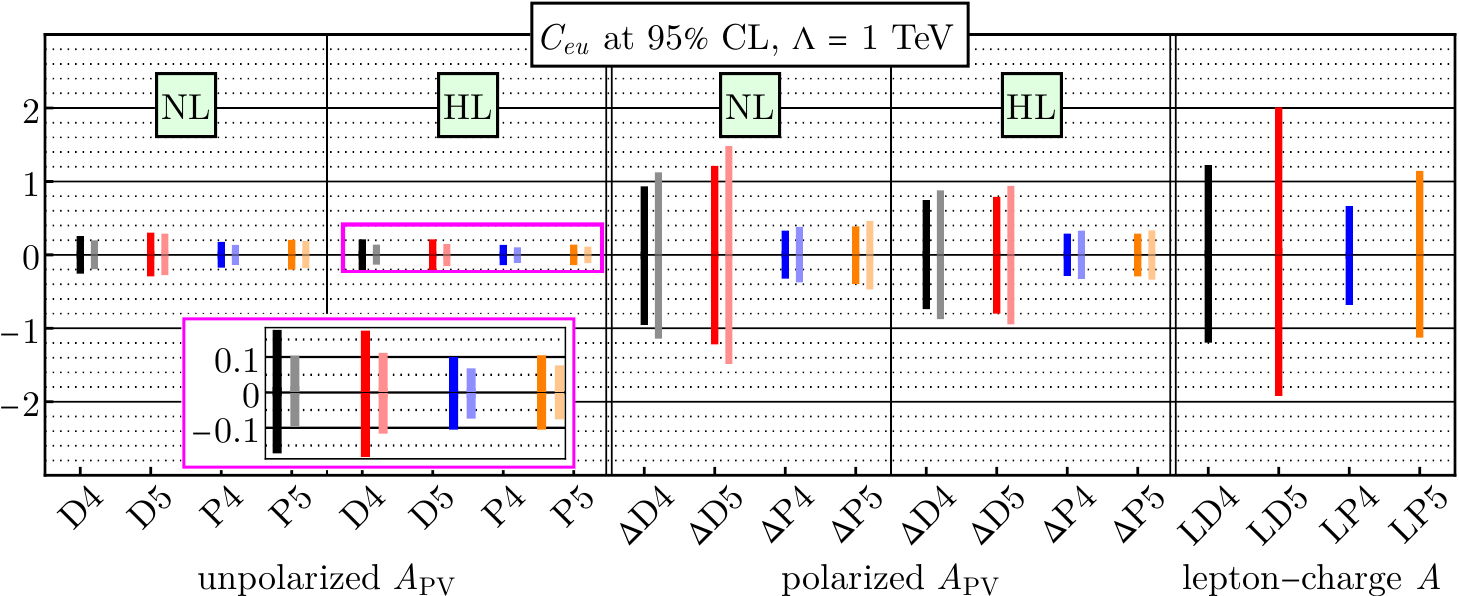}
	\vspace*{-0.3cm}
	\caption{95\% CL bounds of $\Ceu$ from 1-parameter fits (darker) and from simultaneous $(1+1)$-parameter fits with beam polarization (lighter) using the families of data sets D4, D5, P4, and P5.}
	\label{fig:err-bar-Ceu}
\end{figure}

\begin{figure}
	[H]\centering
	\includegraphics[width=.75\textwidth]{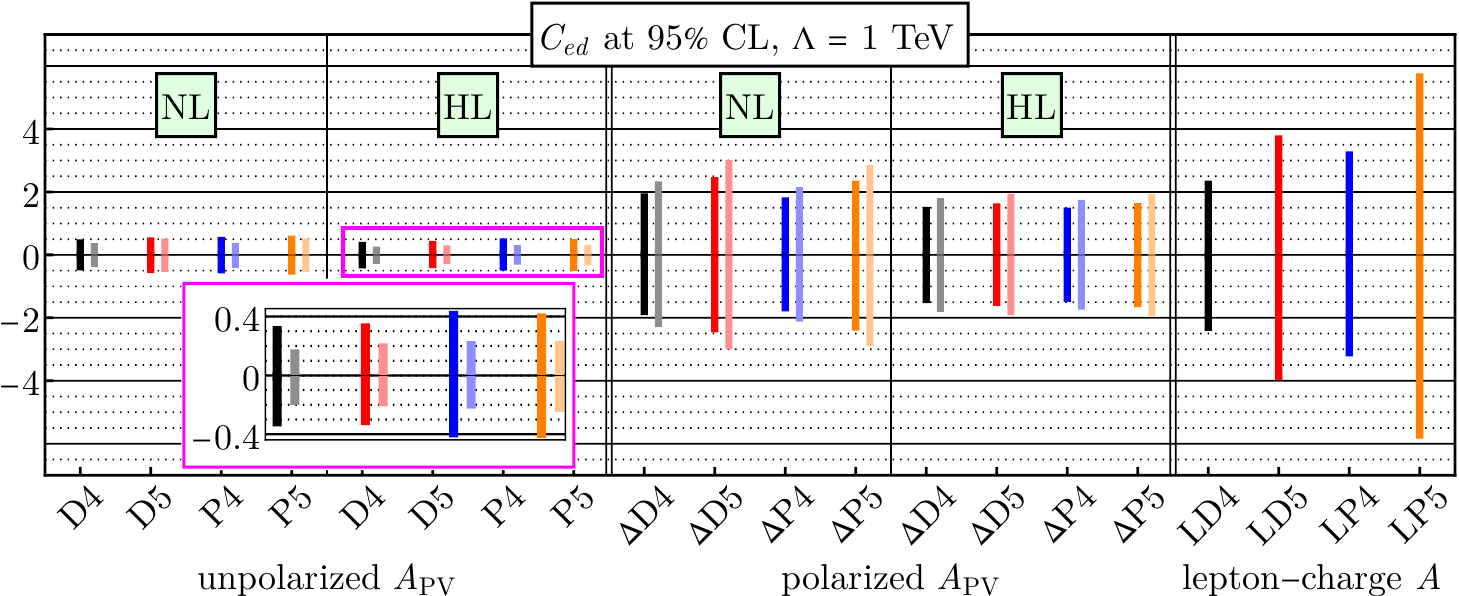}
	\vspace*{-0.3cm}
	\caption{The same as in Fig.~\ref{fig:err-bar-Ceu} but for $\Ced$.}
	\label{fig:err-bar-Ced}
\end{figure}

\begin{figure}
	[H]\centering
	\includegraphics[width=0.75\textwidth]{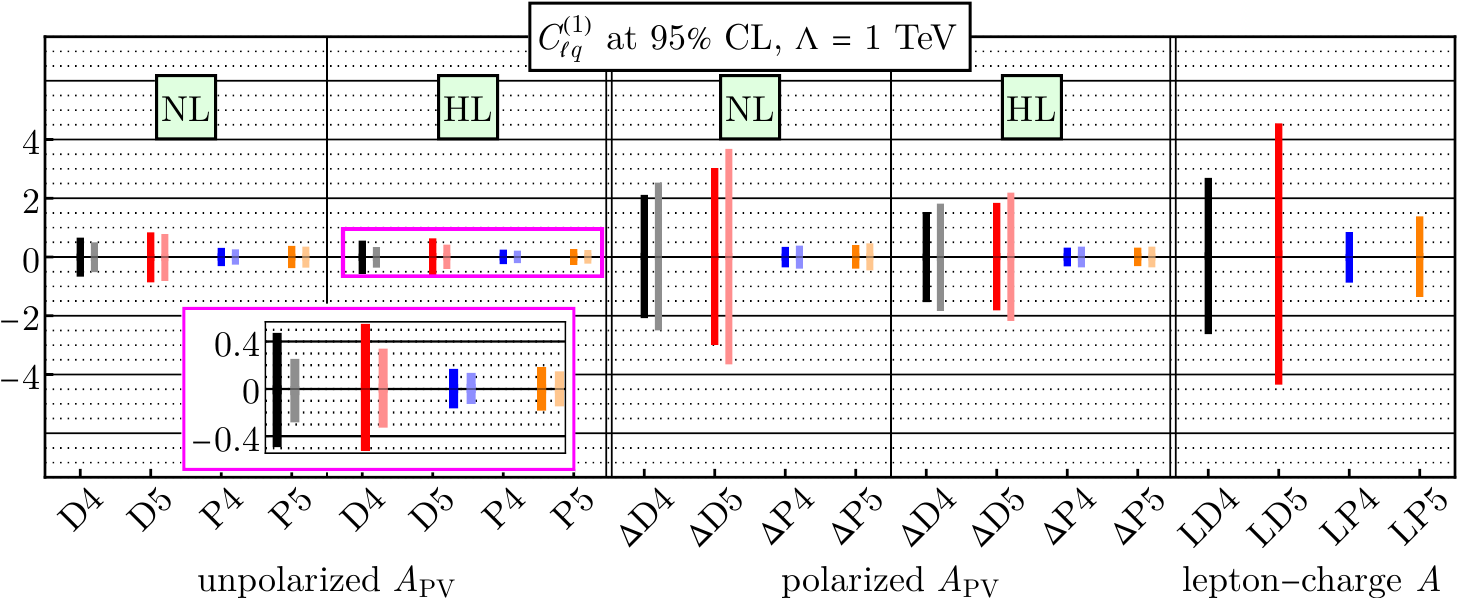}
	\vspace*{-0.3cm}
	\caption{The same as in Fig.~\ref{fig:err-bar-Ceu} but for $\Clqi$.}
	\label{fig:err-bar-Clq1}
\end{figure}

\begin{figure}
	[H]\centering
	\includegraphics[width=0.75\textwidth]{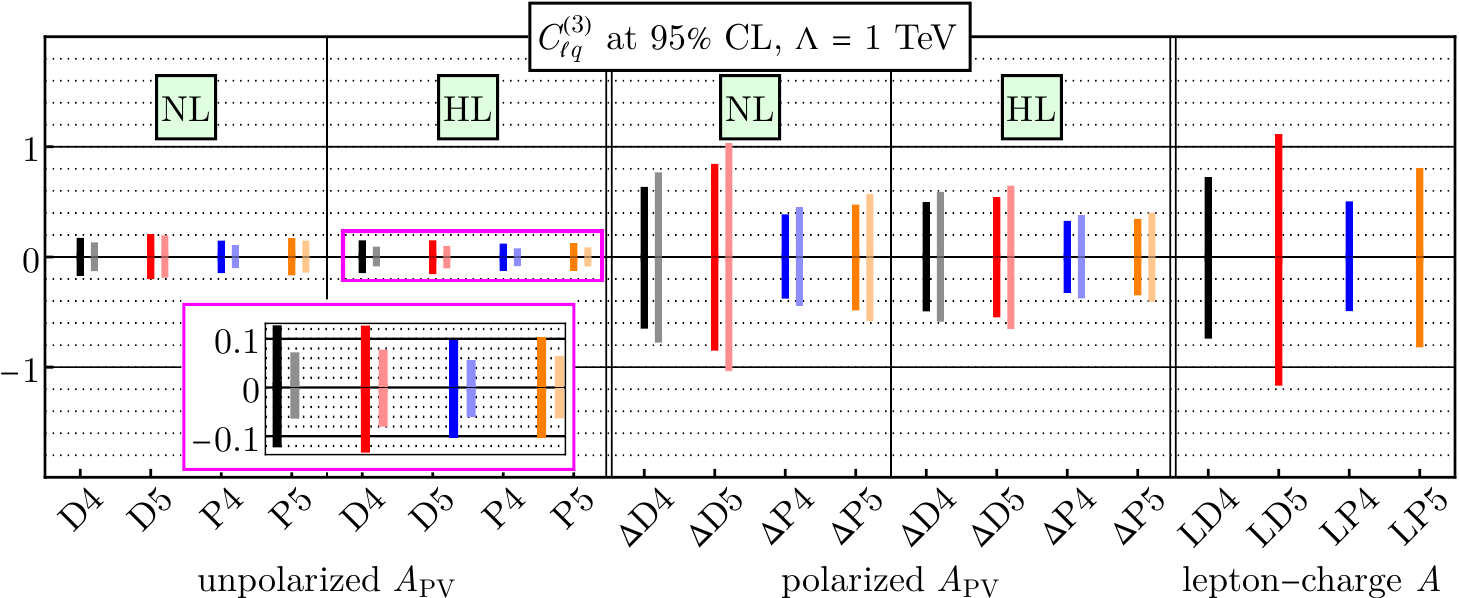}
	\vspace*{-0.3cm}
	\caption{The same as in Fig.~\ref{fig:err-bar-Ceu} but for $\Clqiii$.}
	\label{fig:err-bar-Clq3}
\end{figure}

\begin{figure}
	[H]\centering
	\includegraphics[width=0.75\textwidth]{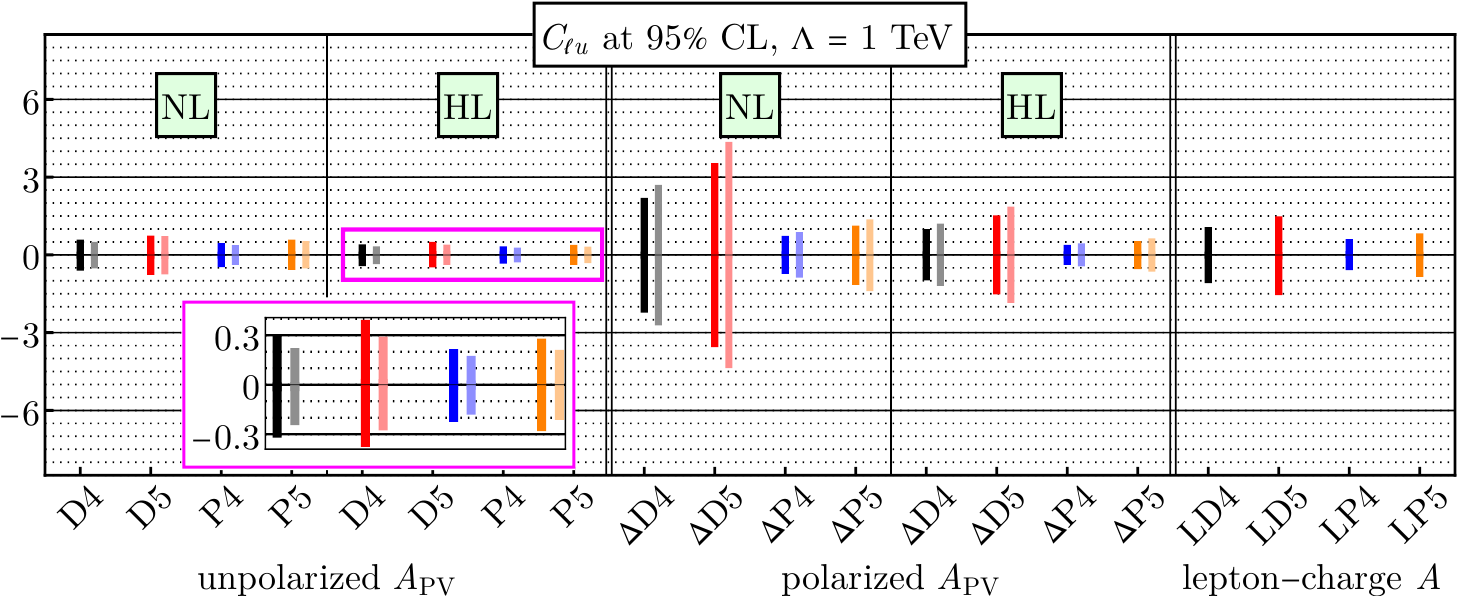}
	\vspace*{-0.3cm}
	\caption{The same as in Fig.~\ref{fig:err-bar-Ceu} but for $\Clu$.}
	\label{fig:err-bar-Clu}
\end{figure}

\begin{figure}
	[H]\centering
	\includegraphics[width=0.82\textwidth]{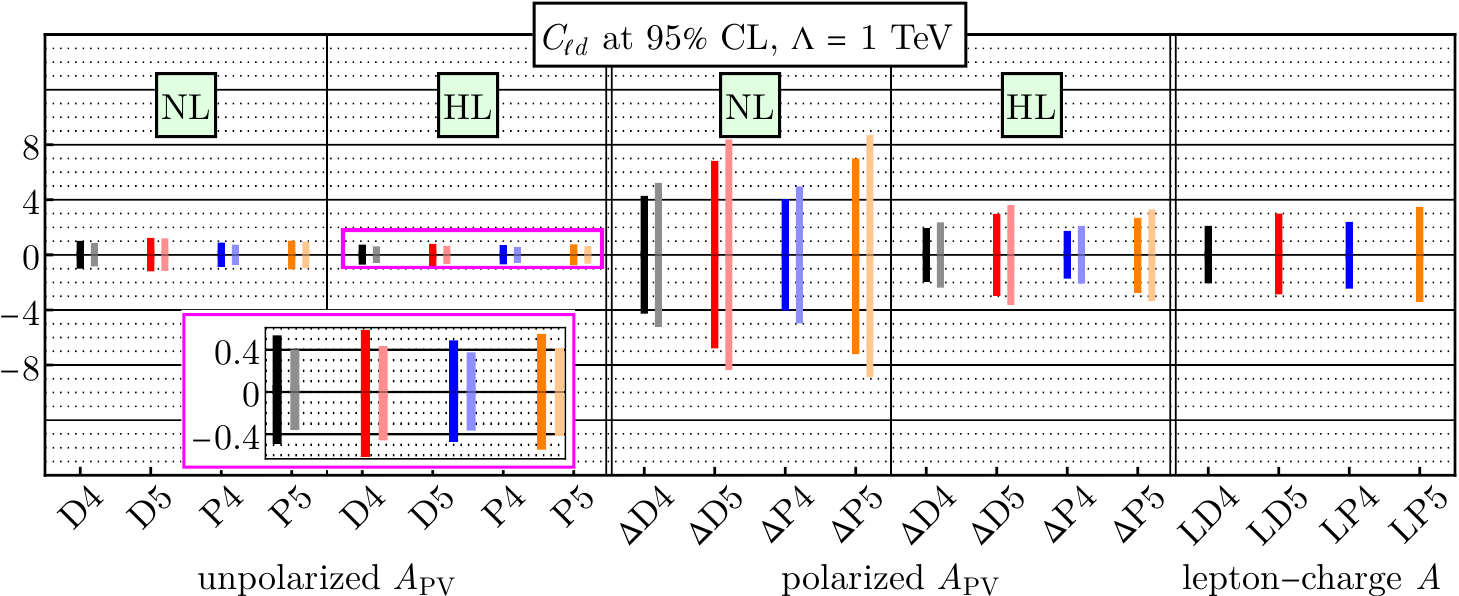}
	\vspace*{-0.3cm}
	\caption{The same as in Fig.~\ref{fig:err-bar-Ceu} but for $\Cld$.}
	\label{fig:err-bar-Cld}
\end{figure}

\begin{figure}
	[H]\centering
	\includegraphics[width=0.82\textwidth]{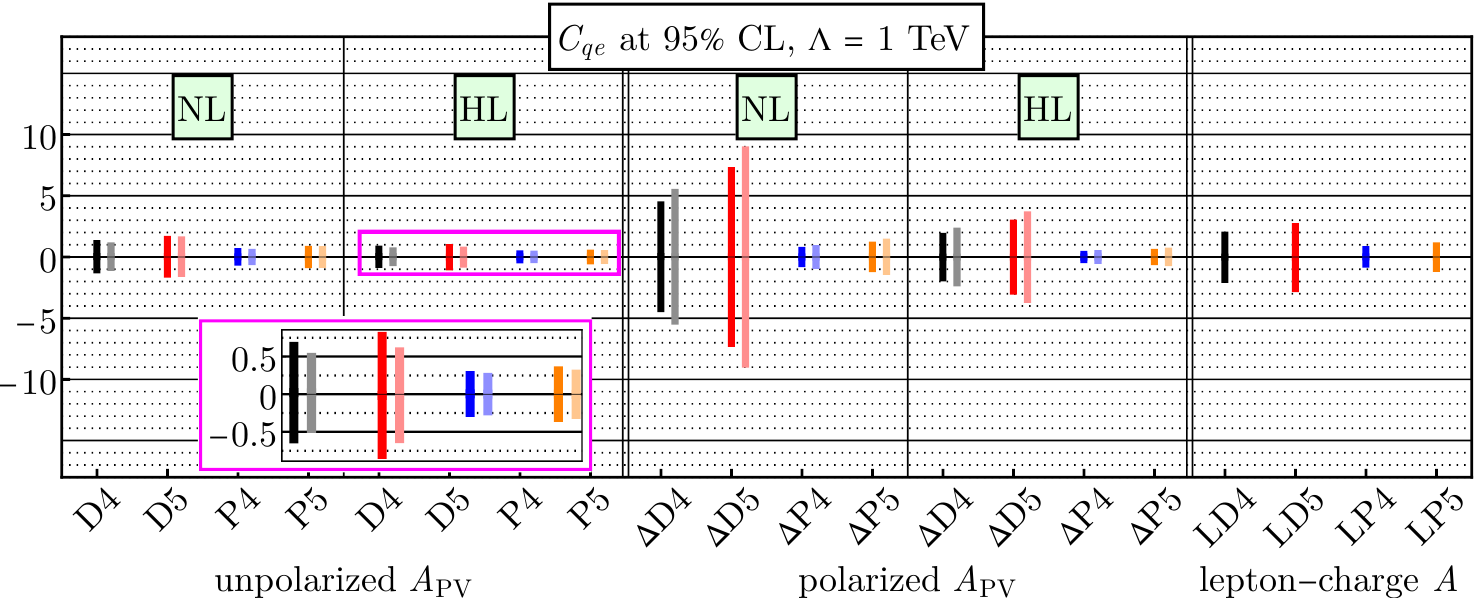}
	\vspace*{-0.3cm}
	\caption{The same as in Fig.~\ref{fig:err-bar-Ceu} but for $\Cqe$.}
	\label{fig:err-bar-Cqe}
\end{figure}

In this section, we present the 95\% CL intervals and the corresponding effective UV cut-off scales for all the seven Wilson coefficients in single-parameter fits averaged over 1000 pseudo-experiments. We recall the following abbreviations for the EIC preliminary data sets:
\vspace{-.3cm}
\begin{itemize}
	\item electron PV asymmetries of unpolarized deuteron, $\APVe$:
	\vspace{-.2cm}
	\begin{itemize}
		\item D4: $ eD $ $ 10 \ {\rm GeV} \times 137 \ {\rm GeV}$, $ 100 \ {\rm fb^{-1}}$
		\item D5: $ eD $ $ 18 \ {\rm GeV} \times 137 \ {\rm GeV}$, $ 15.4 \ {\rm fb^{-1}}$
	\end{itemize} 
	\vspace{-.4cm}
	\item electron PV asymmetries of unpolarized proton, $\APVe$:
	\vspace{-.2cm}
	\begin{itemize}
		\item P4: $ ep $ $ 10 \ {\rm GeV} \times 275 \ {\rm GeV}$, $ 100 \ {\rm fb^{-1}}$
		\item P5: $ ep $ $ 18 \ {\rm GeV} \times 275 \ {\rm GeV}$, $ 15.4 \ {\rm fb^{-1}}$
	\end{itemize}
	\vspace{-.4cm}
	\item hadron PV asymmetries with unpolarized electron, $\APVH$: $\Delta$D4, $\Delta$D5, $\Delta$P4, and $\Delta$P5 with the same energy and luminosity configuration as the corresponding D- and P-sets.
	\vspace{-.2cm}
	\item unpolarized electron-positron asymmetries of unpolarized hadrons, $\ALCH$: LD4, LD5, LP4, and LP5 with the same energy configuration as the corresponding D- and P-sets, but with the luminosity of the positron beam assumed to be 10 times smaller than that of the electron beam. 
\end{itemize}
Figs.~\ref{fig:err-bar-Ceu}--\ref{fig:err-bar-Cqe} display the 95\% CL bounds of each Wilson coefficient for the four primary data families. As in the main part of the manuscript, the intervals are grouped by asymmetries, namely electron PV asymmetries of unpolarized hadrons, $\APVe$ (``unpolarized $ A_{\rm PV} $''), hadron PV asymmetries with unpolarized electrons, $\APVH$ (``polarized $ A_{\rm PV} $''), and unpolarized electron-positron asymmetries of unpolarized hadrons, $\ALCH$ (``lepton-charge $ A $''). PV asymmetries are then grouped into two, showing the fits in the nominal- and high-luminosity scenarios. The nominal luminosity (``NL'') refers to the annual integrated luminosity of Table 10.1 of the YR~\cite{AbdulKhalek:2021gbh}.  The high luminosity (``HL'') is assumed to be 10 times higher than the nominal one and requires a luminosity upgrade of the EIC. In each block of intervals, there are four double lines in the case of PV asymmetries and four single lines in LC asymmetries. These four lines correspond to the data families D4 (black and its shade), D5 (red), P4 (blue), and P5 (orange). The darker of the two lines indicate the bounds from single-parameter fits with the Wilson coefficient $ C_r $, whereas the lighter ones show the bounds on the  Wilson coefficient from simultaneous two-parameter fits with $ C_r $ and the beam polarization. The details of the fits involving the beam polarization as an additional parameter are described in Appendix \ref{app:additional_fits:beam_polarization_fits}.
\vspace{-.6cm}

\begin{figure}
	[H]\centering
	\includegraphics[width=.75\textwidth]{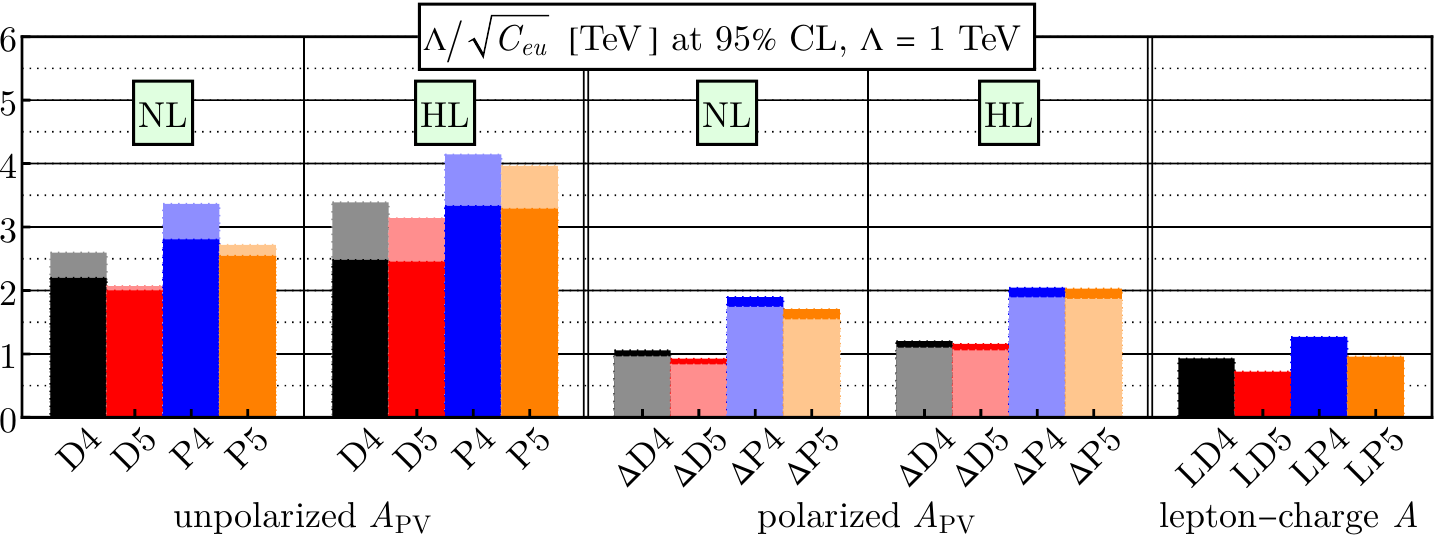}
	\caption{Effective UV cut-off scales, $\Lambda/\sqrt{\Ceu}$, defined in terms of the 95\% CL bounds on the Wilson coefficient $\Ceu$ with $\Lambda = 1 \ {\rm TeV}$. The darker columns in the foreground of unpolarized PV asymmetries and in the background of polarized PV asymmetries indicate the results of single-parameter fits on the Wilson coefficient, $\Ceu$. The lighter columns in the background of unpolarized PV asymmetries and in the foreground of polarized PV asymmetries denote the results of simultaneous $(1+1)$-parameter fits of $\Ceu$ with the beam polarization, $P$.}
	\label{fig:lambda-Ceu}
\end{figure}

\begin{figure}
	[H]\centering
	\includegraphics[width=0.75\textwidth]{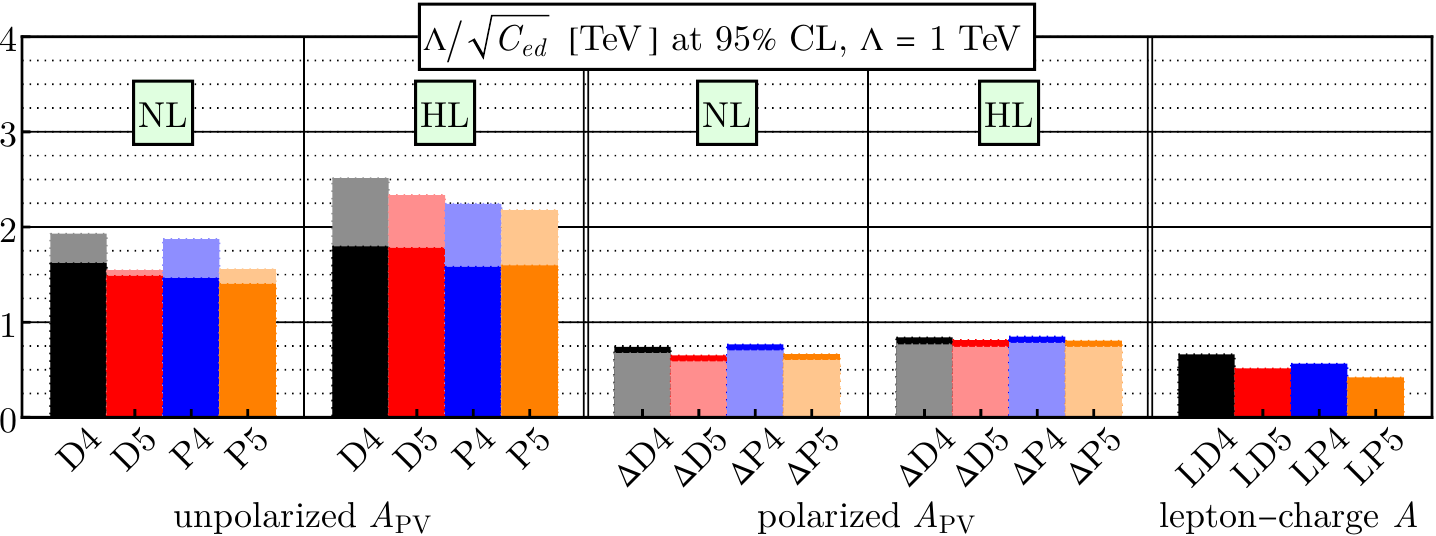}
	\vspace*{-0.3cm}
	\caption{The same as in Fig.~\ref{fig:lambda-Ceu} but for $\Ced$.}
	\label{fig:lambda-Ced}
\end{figure}

\begin{figure}
	[H]\centering
	\includegraphics[width=0.75\textwidth]{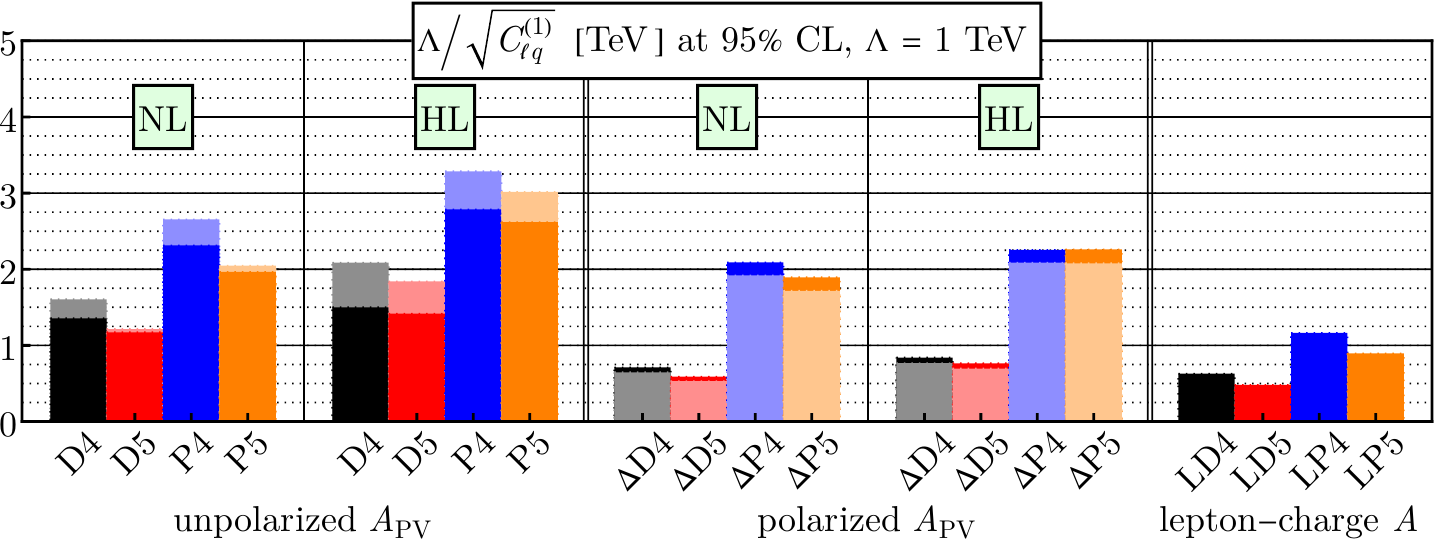}
	\vspace*{-0.3cm}
	\caption{The same as in Fig.~\ref{fig:lambda-Ceu} but for $\Clqi$.}
	\label{fig:lambda-Clq1}
\end{figure}

\begin{figure}
	[H]\centering
	\includegraphics[width=0.75\textwidth]{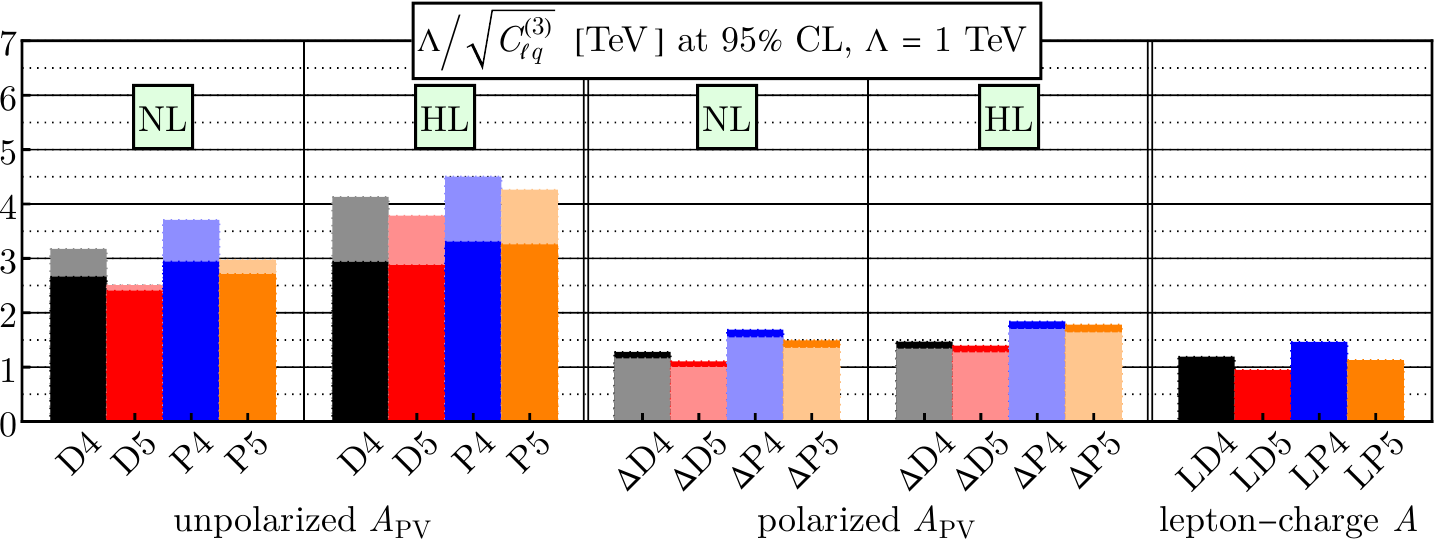}
	\vspace*{-0.3cm}
	\caption{The same as in Fig.~\ref{fig:lambda-Ceu} but for $\Clqiii$.}
	\label{fig:lambda-Clq3}
\end{figure}

\begin{figure}
	[H]\centering
	\includegraphics[width=0.75\textwidth]{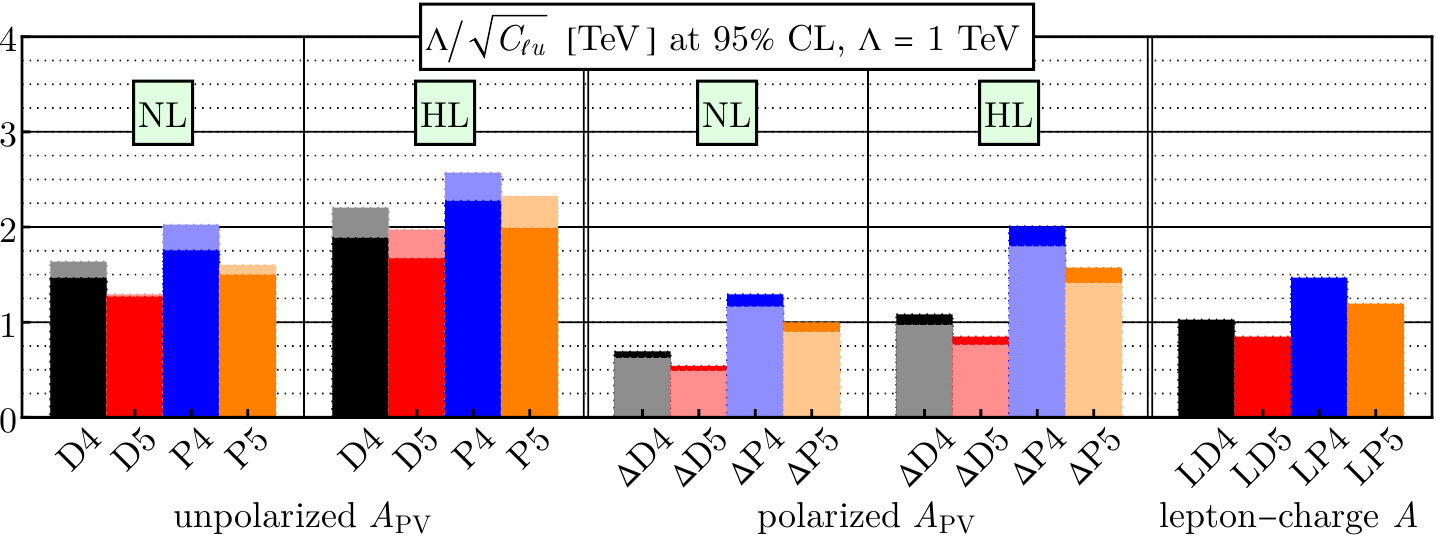}
	\vspace*{-0.3cm}
	\caption{The same as in Fig.~\ref{fig:lambda-Ceu} but for $\Clu$.}
	\label{fig:lambda-Clu}
\end{figure}

\begin{figure}
	[H]\centering
	\includegraphics[width=0.75\textwidth]{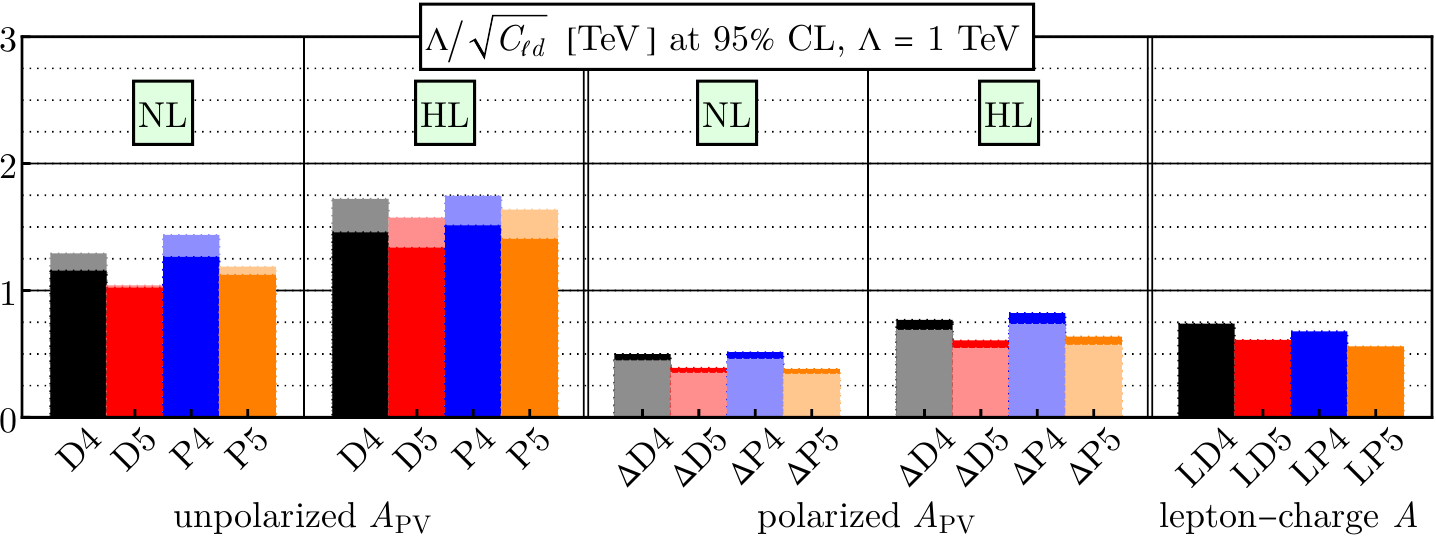}
	\vspace*{-0.3cm}
	\caption{The same as in Fig.~\ref{fig:lambda-Ceu} but for $\Cld$.}
	\label{fig:lambda-Cld}
\end{figure}

\begin{figure}
	[H]\centering
	\includegraphics[width=0.75\textwidth]{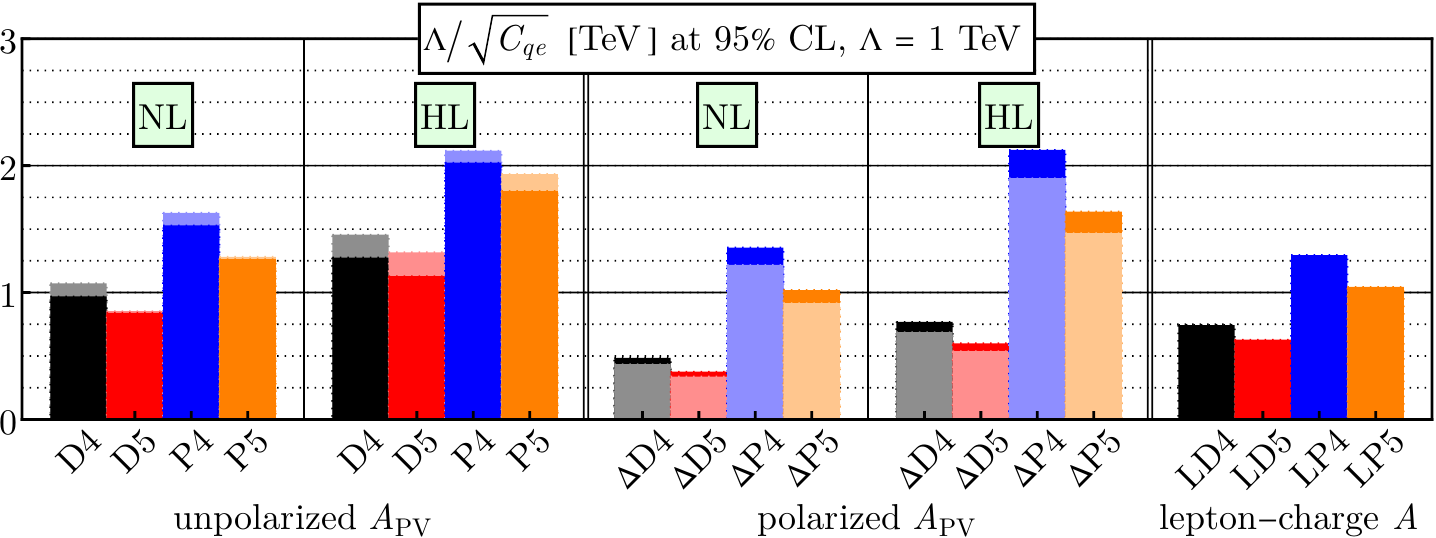}
	\vspace*{-0.3cm}
	\caption{The same as in Fig.~\ref{fig:lambda-Ceu} but for $\Cqe$.}
	\label{fig:lambda-Cqe}
\end{figure}

In Figs.~\ref{fig:lambda-Ceu}--\ref{fig:lambda-Cqe}, we present the effective UV cut-off scales, $ \Lambda / \sqrt{C_r} $, with $ \Lambda = 1 \ {\rm TeV} $, corresponding to the bounds shown in Figs.~\ref{fig:err-bar-Ceu}--\ref{fig:err-bar-Cqe}. The organization of these plots in terms of asymmetries and data sets is the same as in Figs. \ref{fig:err-bar-Ceu}--\ref{fig:err-bar-Cqe}. Improved bounds on $ C_r $ with the addition of the beam polarization to the fits are equivalent to higher energy scales in the unpolarized PV asymmetries, which are indicated by the lighter columns in the background; on the other hand, weaker bounds from the fits with beam polarization are depicted by the lighter columns in the foreground for the polarized PV asymmetries.

\vspace{-.3cm}

\vspace{-.8cm}
\vspace{-.5cm}
\vspace{-.5cm}

For completeness, we summarize in Table \ref{tab:all_1d} the fitting results of all the seven Wilson coefficients with the four families of data sets of interest. The values indicated in this table are the 95\% CL bounds around zero.

\begin{table}
	[H]\centering
	\caption{95\% CL bounds of all the seven Wilson coefficients  around zero at $\Lambda = 1 \ {\rm TeV}$ with the four families of data sets, D4, D5, P4, and P5 in various configurations.}
	\label{tab:all_1d}
	\includegraphics[width=.495\textwidth]{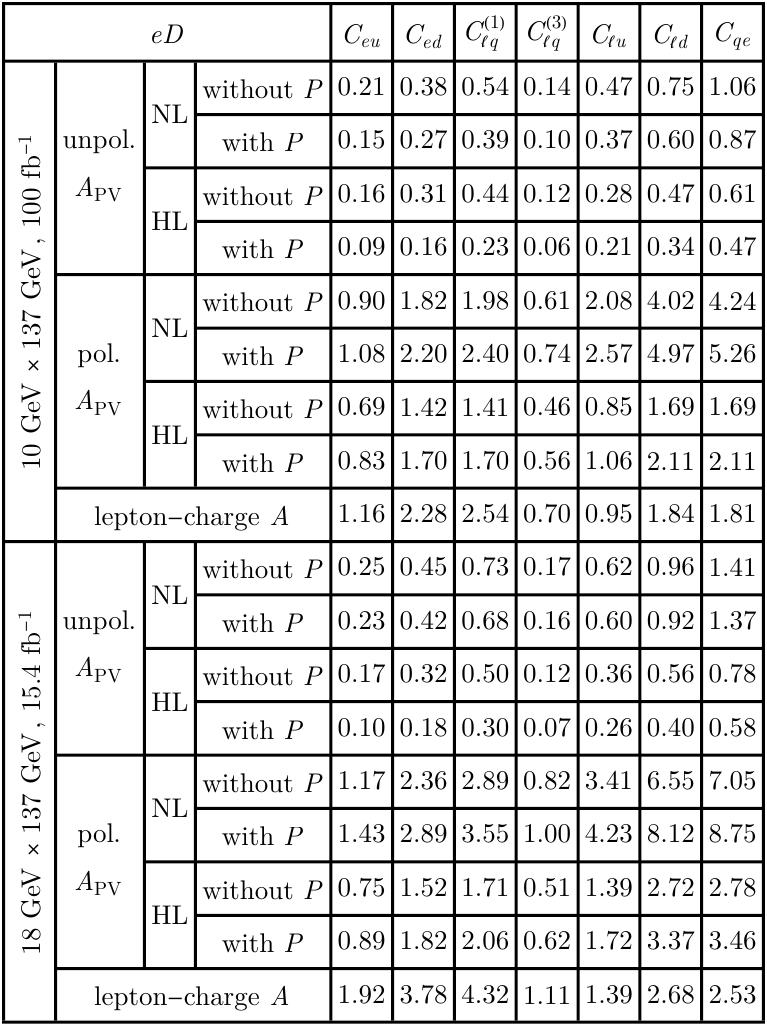}
	\includegraphics[width=.495\textwidth]{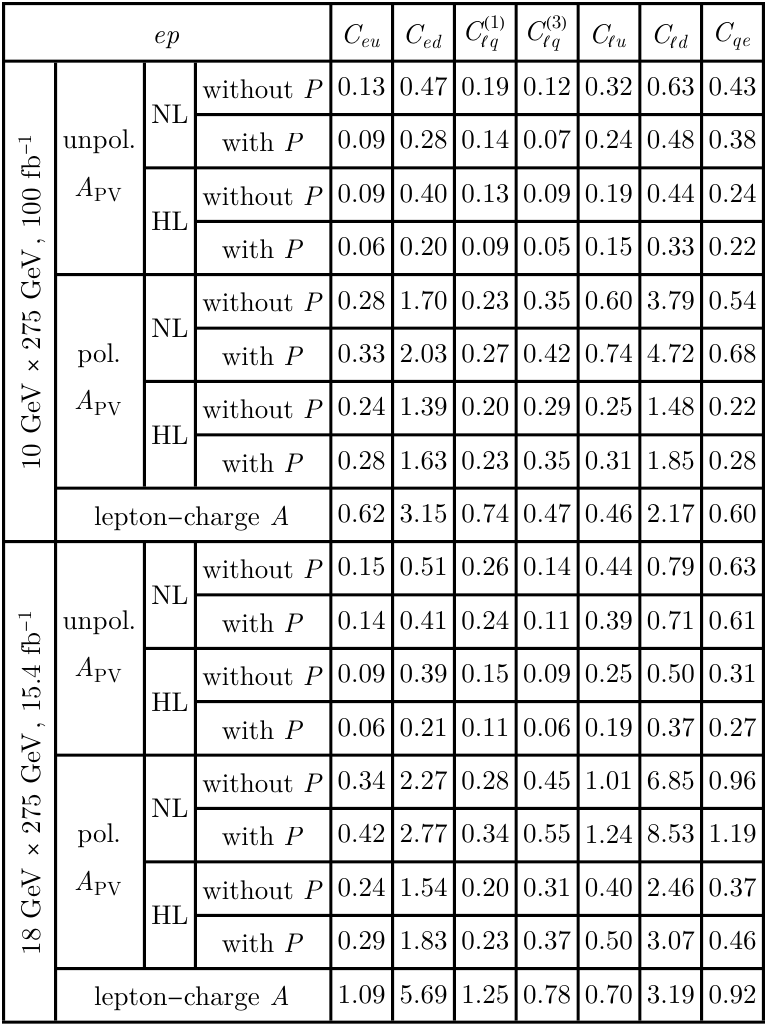}
\end{table}

%% file: appendices/b2-fits_of_two_wilson_coefficients.tex
\subsection{Fits of two Wilson coefficients \label{app:complete_results:2d}}
In this section, we present the complete set of confidence ellipses for all possible pairs of the seven Wilson coefficients that we consider in this work. The ellipses are plotted at 95\% CL and $\Lambda = 1 \ {\rm TeV}$. 

As before, we refer to electron PV asymmetries collectively as unpolarized $A_{\rm PV}$, hadron PV asymmetries as polarized $A_{\rm PV}$, and electron-positron asymmetries as lepton-charge $A$. Data sets with the label NL or HL indicate the luminosity: The nominal luminosity (``NL'') refers to the annual integrated luminosity of Table 10.1 of YR~\cite{AbdulKhalek:2021gbh}.  The high luminosity (``HL'') is assumed to be 10 times higher than the nominal one and requires a luminosity upgrade of the EIC. 

Each figure consists of four panels, containing one of the four families of data sets, namely D4, D5, P4, and P5. We show the fits from polarized and unpolarized PV asymmetry data sets in both nominal- and high-luminosity scenarios for comparison. We remark that the ellipses for the polarized and unpolarized PV asymmetry data sets indicate the results of simultaneous fits on Wilson coefficients with the beam-polarization parameter, $P$, in light of significant improvements in the results with unpolarized PV asymmetries. Moreover, we include for some representative examples fitted results for the LHC Drell-Yan data, adapting from \cite{RBoughezal2020} and \cite{RBoughezal2021}.

\def\elliwi{.47}
\begin{figure}
	[h]\centering
	\includegraphics[width=\elliwi\textwidth]{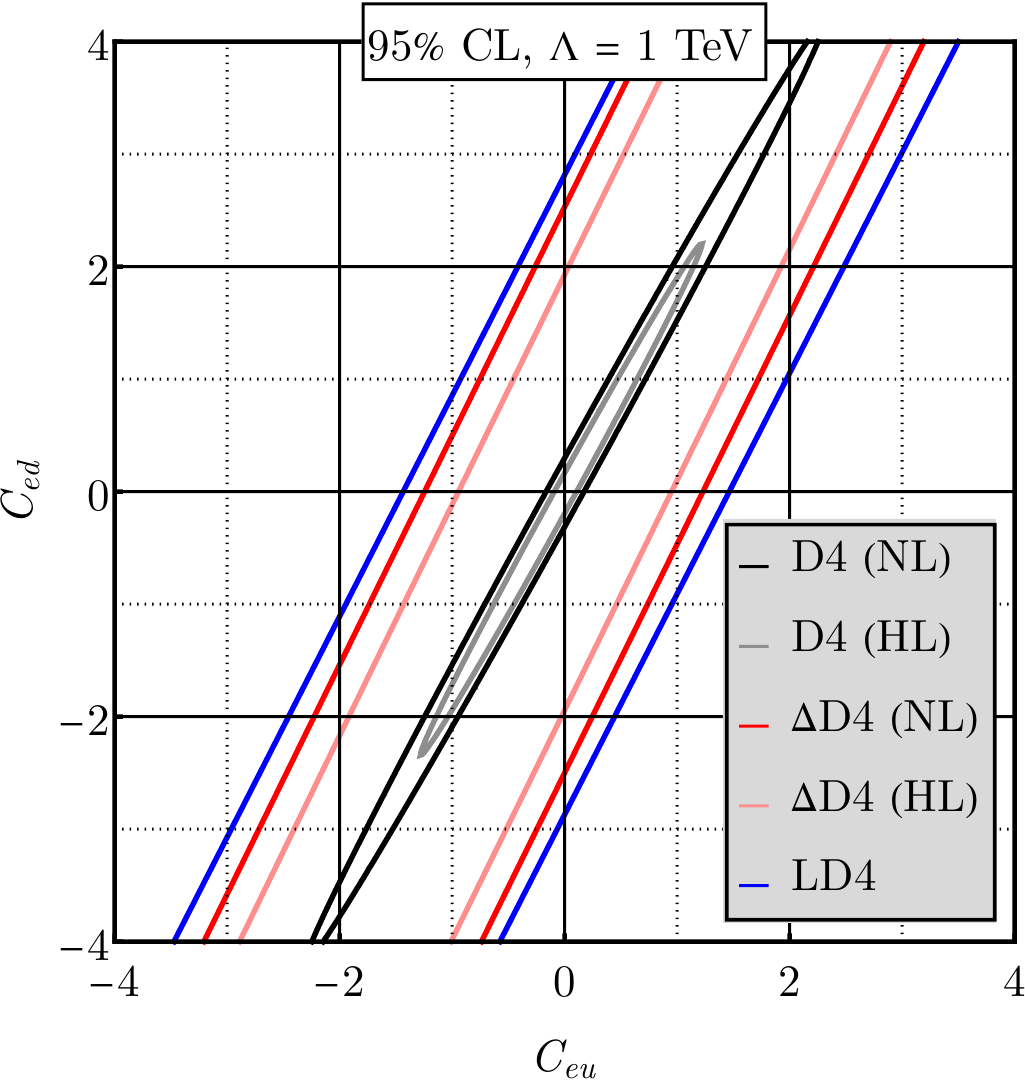}
	\includegraphics[width=\elliwi\textwidth]{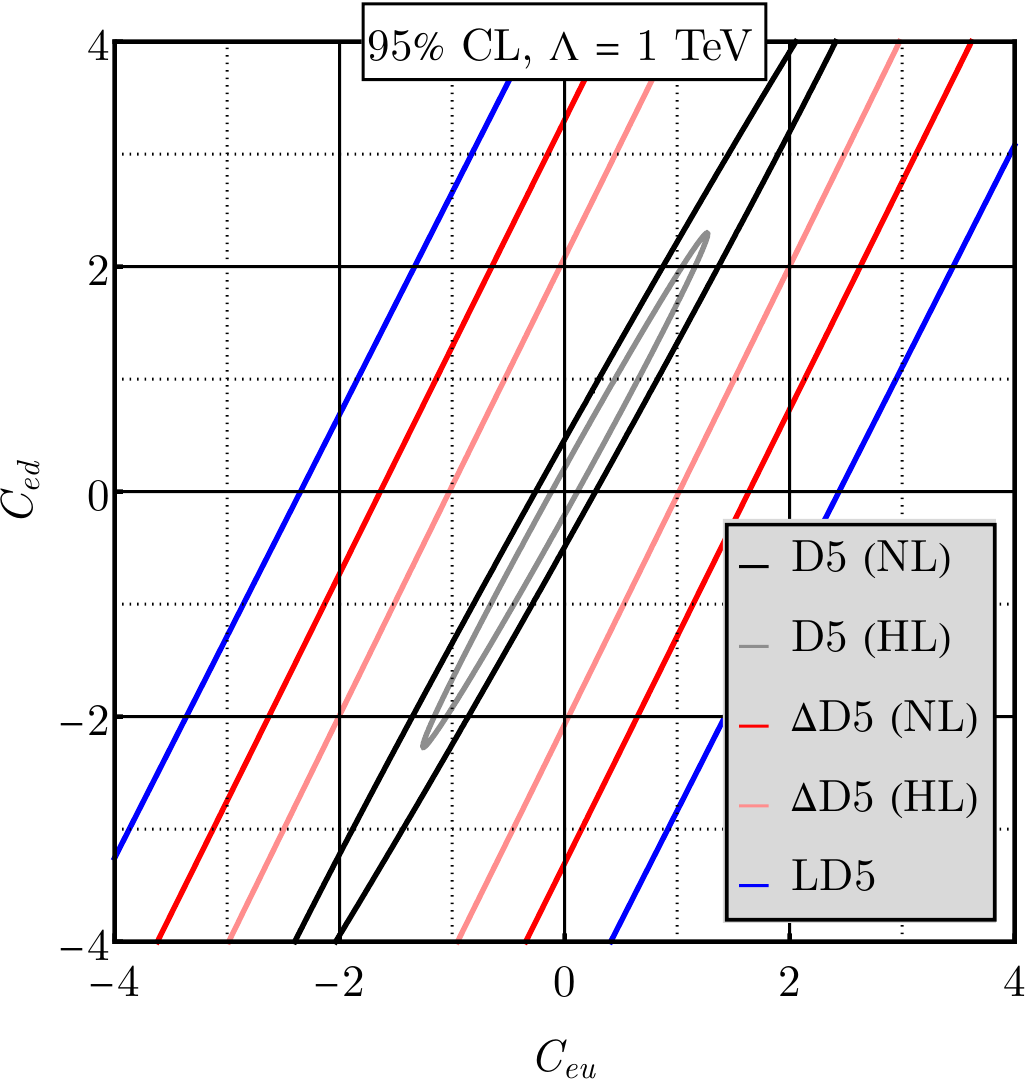}
	\includegraphics[width=\elliwi\textwidth]{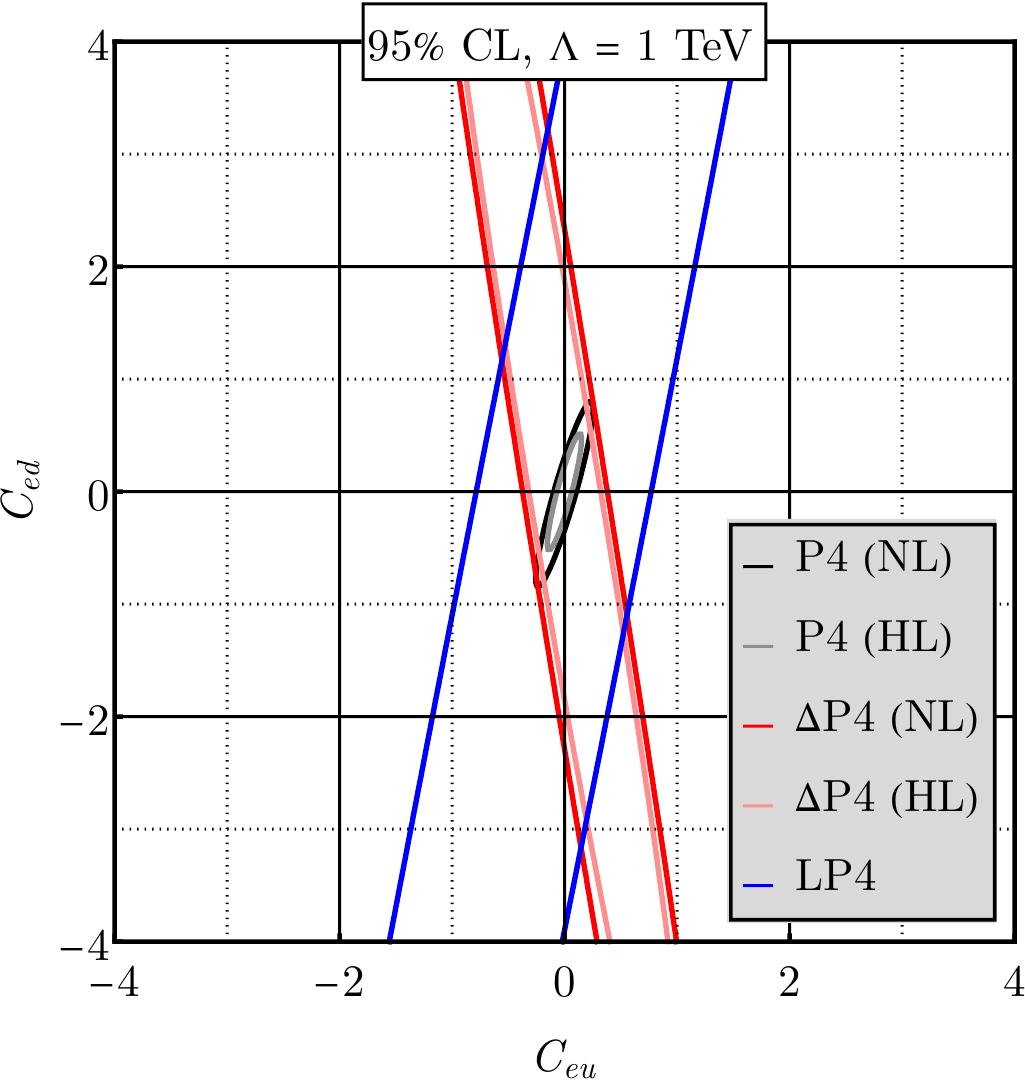}
	\includegraphics[width=\elliwi\textwidth]{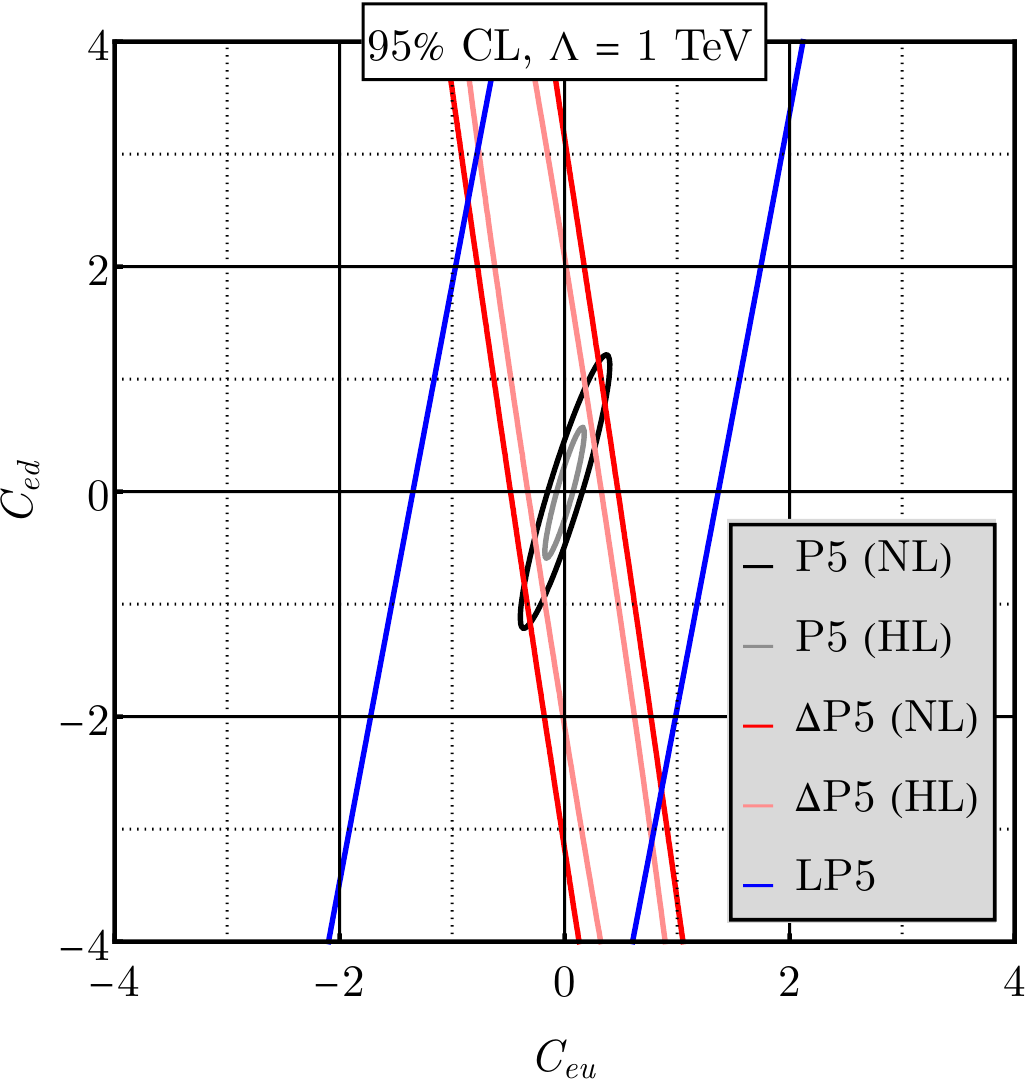}
	\caption{95\% CL ellipses for the Wilson coefficients $\Ceu$ and $\Ced$ using the families of data sets D4, D5, P4, and P5 at $\Lambda = 1 \ {\rm TeV}$. }
	\label{fig:all-ellipses-Ceu-Ced}
\end{figure}

\begin{figure}
	[h]\centering
	\includegraphics[width=\elliwi\textwidth]{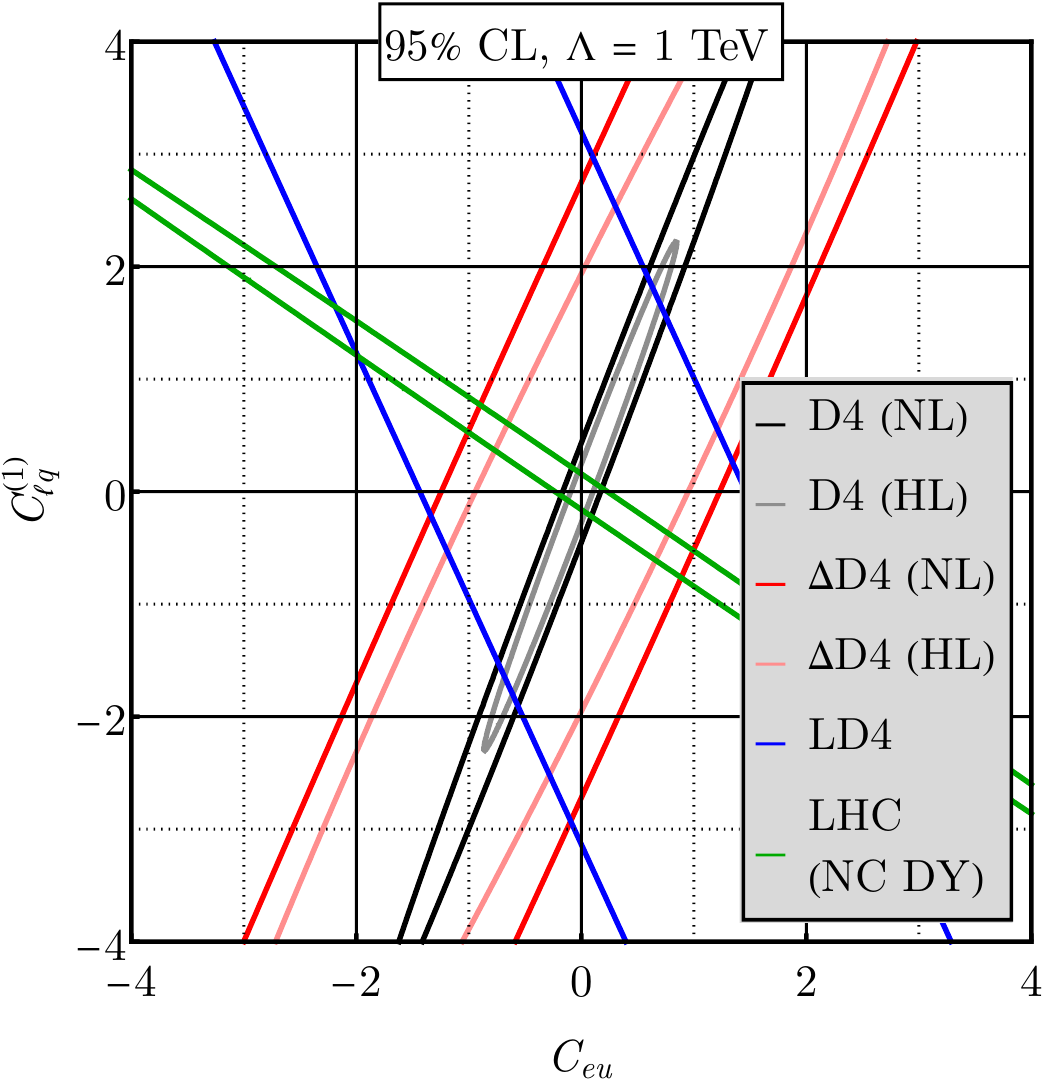}
	\includegraphics[width=\elliwi\textwidth]{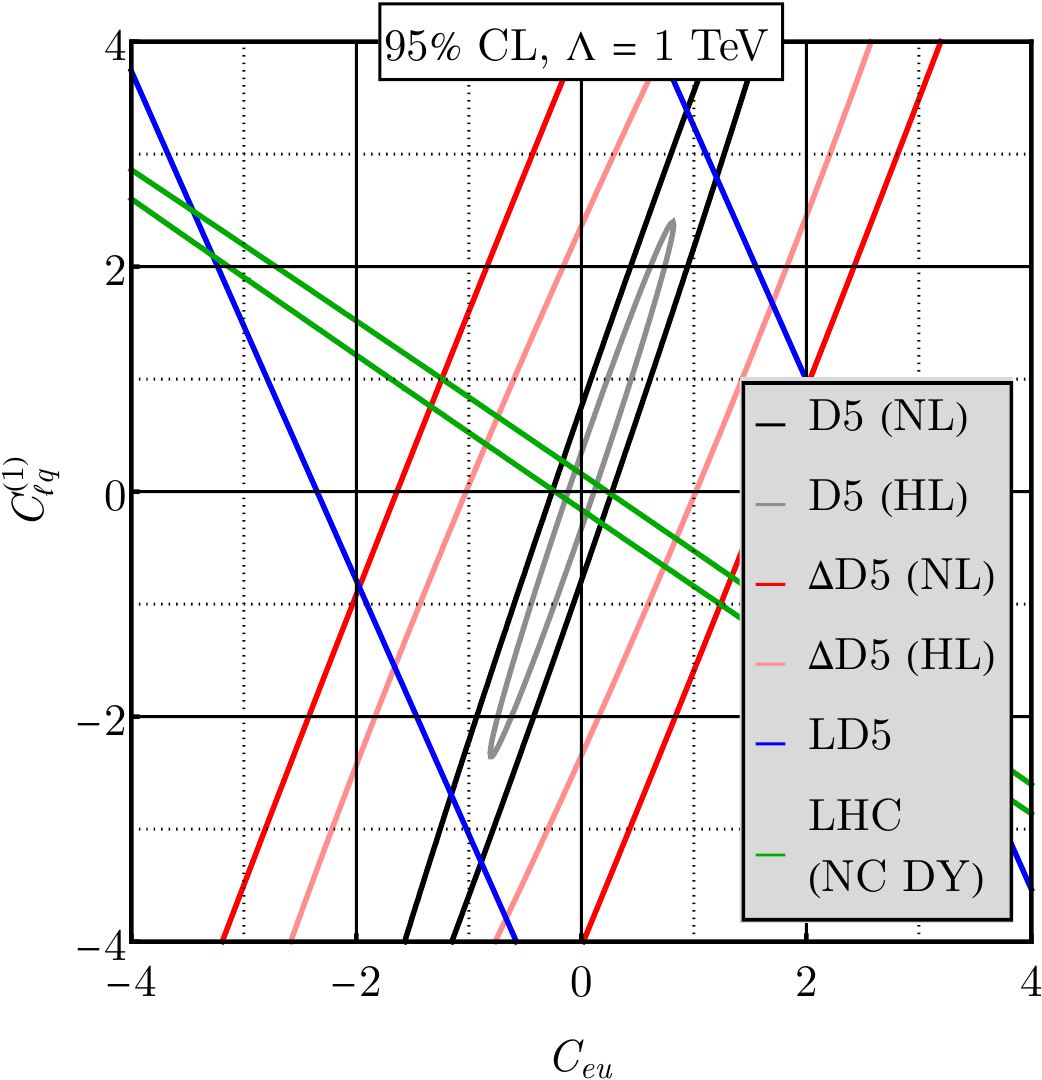}
	\includegraphics[width=\elliwi\textwidth]{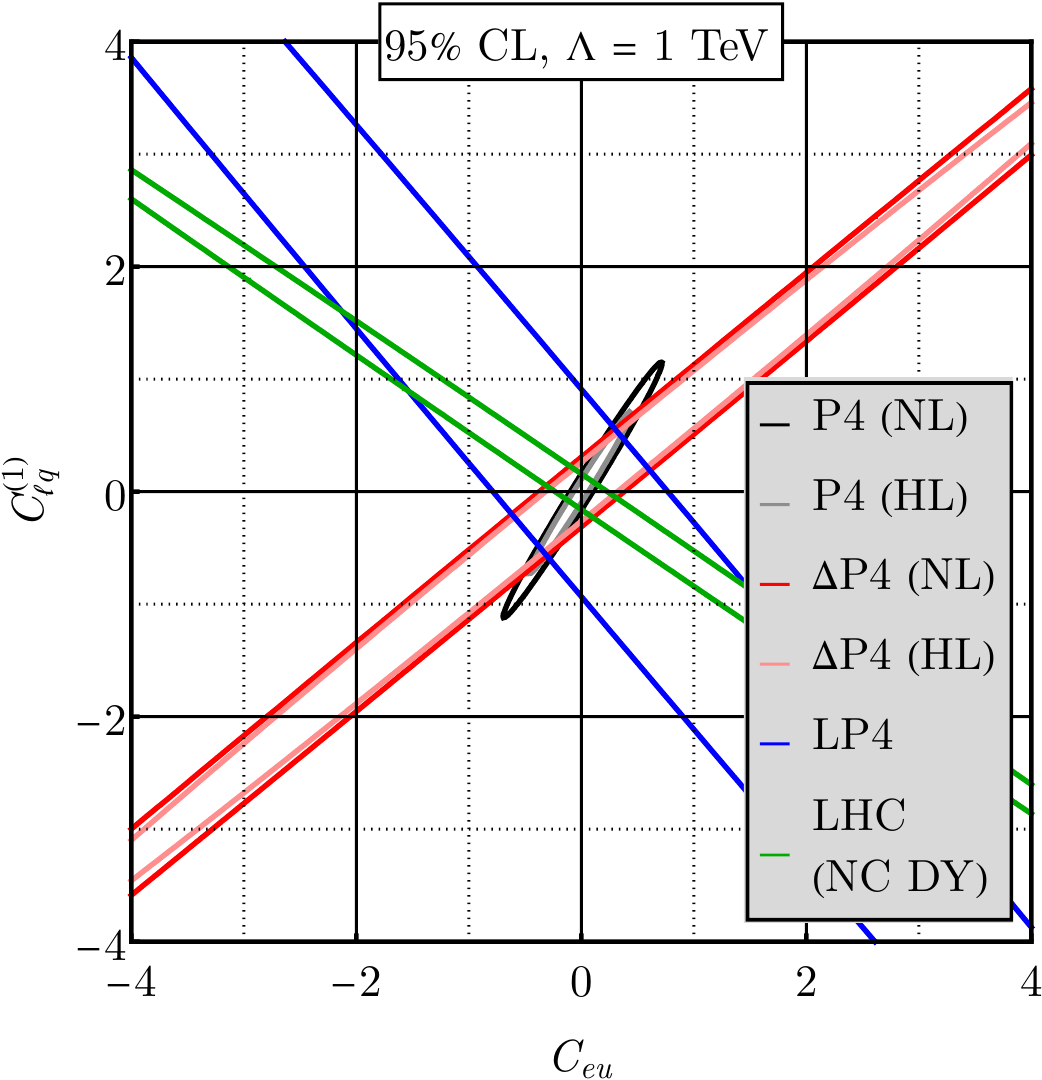}
	\includegraphics[width=\elliwi\textwidth]{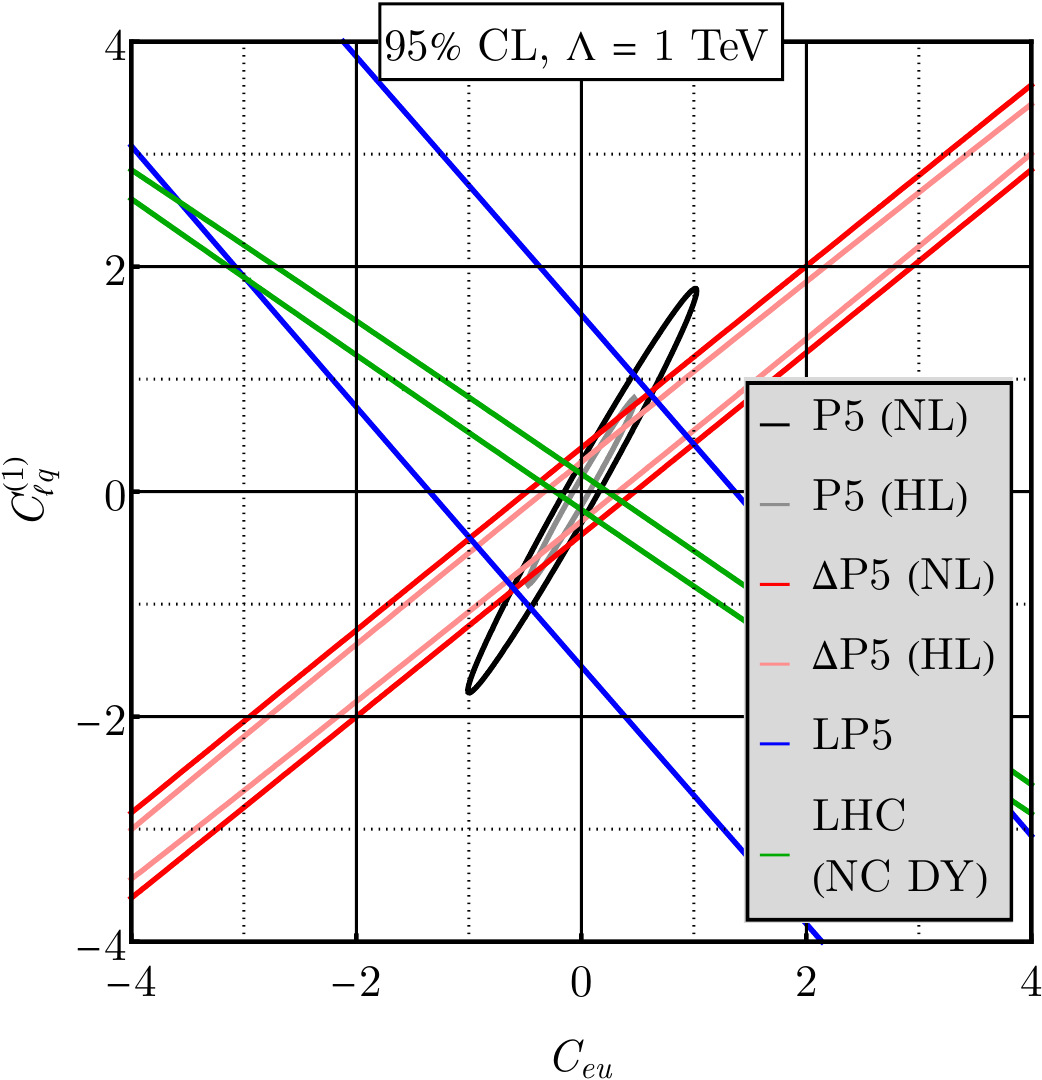}
	\caption{The same as in Fig.~\ref{fig:all-ellipses-Ceu-Ced} but for $\Ceu$ and $\Clqi$.}
	\label{fig:all-ellipses-Ceu-Clq1}
\end{figure}

\begin{figure}
	[h]\centering
	\includegraphics[width=\elliwi\textwidth]{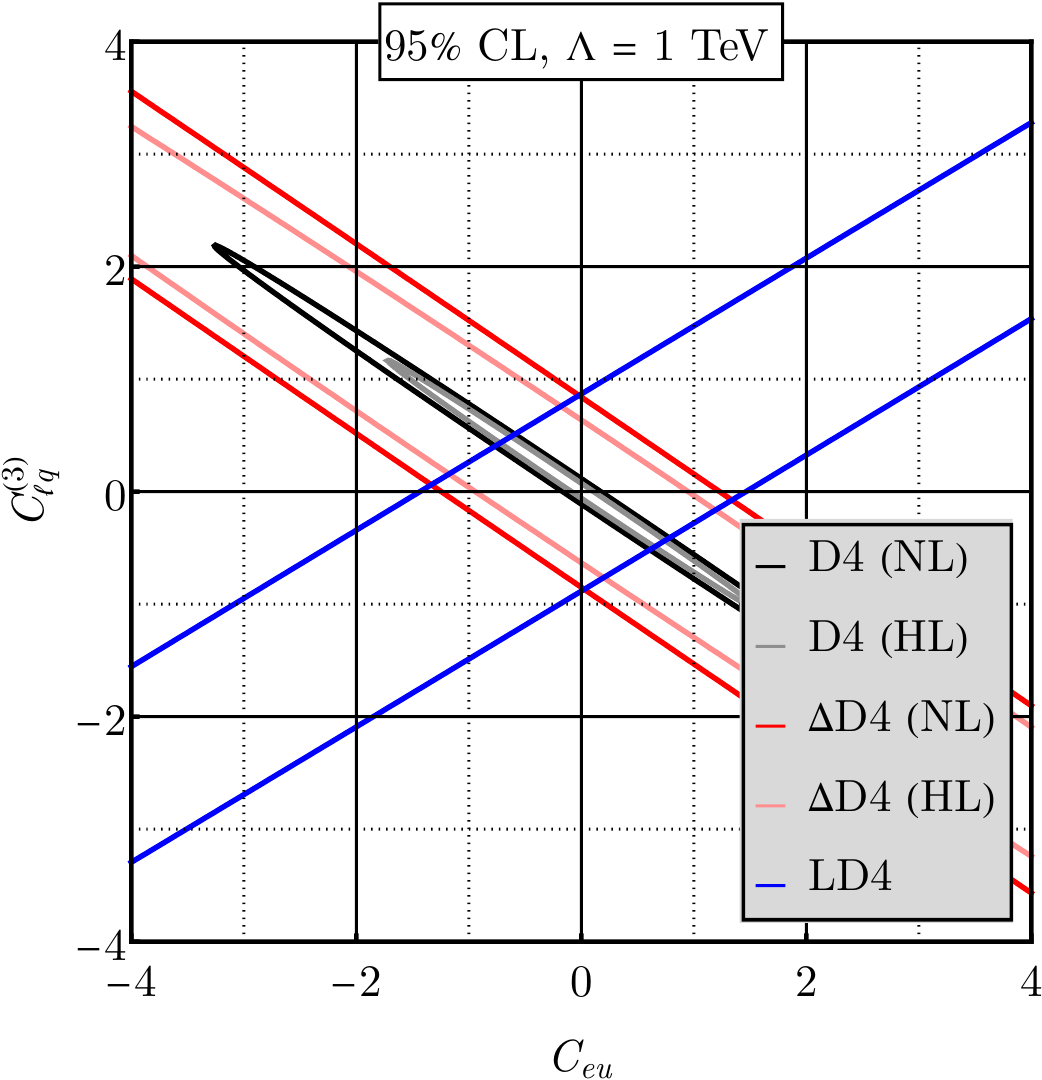}
	\includegraphics[width=\elliwi\textwidth]{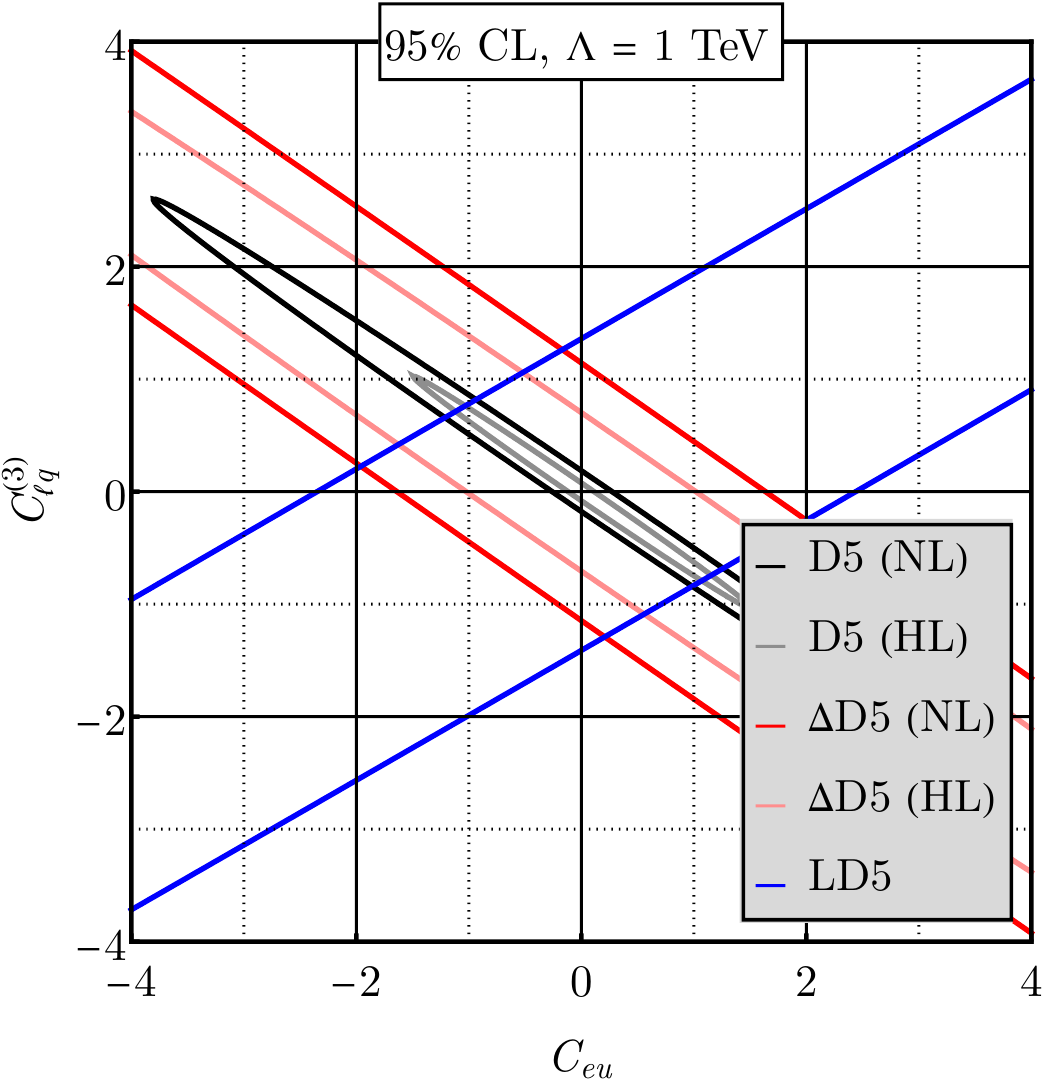}
	\includegraphics[width=\elliwi\textwidth]{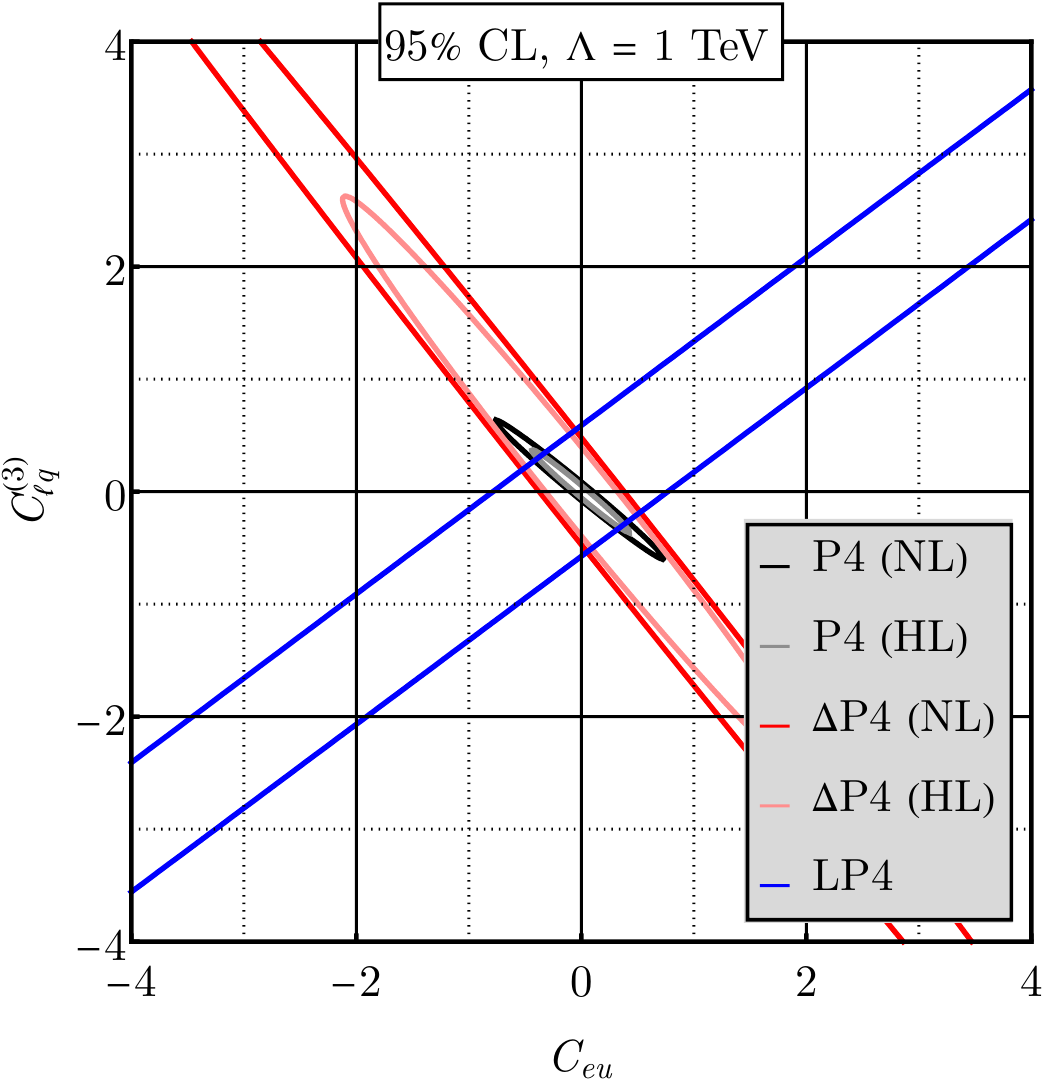}
	\includegraphics[width=\elliwi\textwidth]{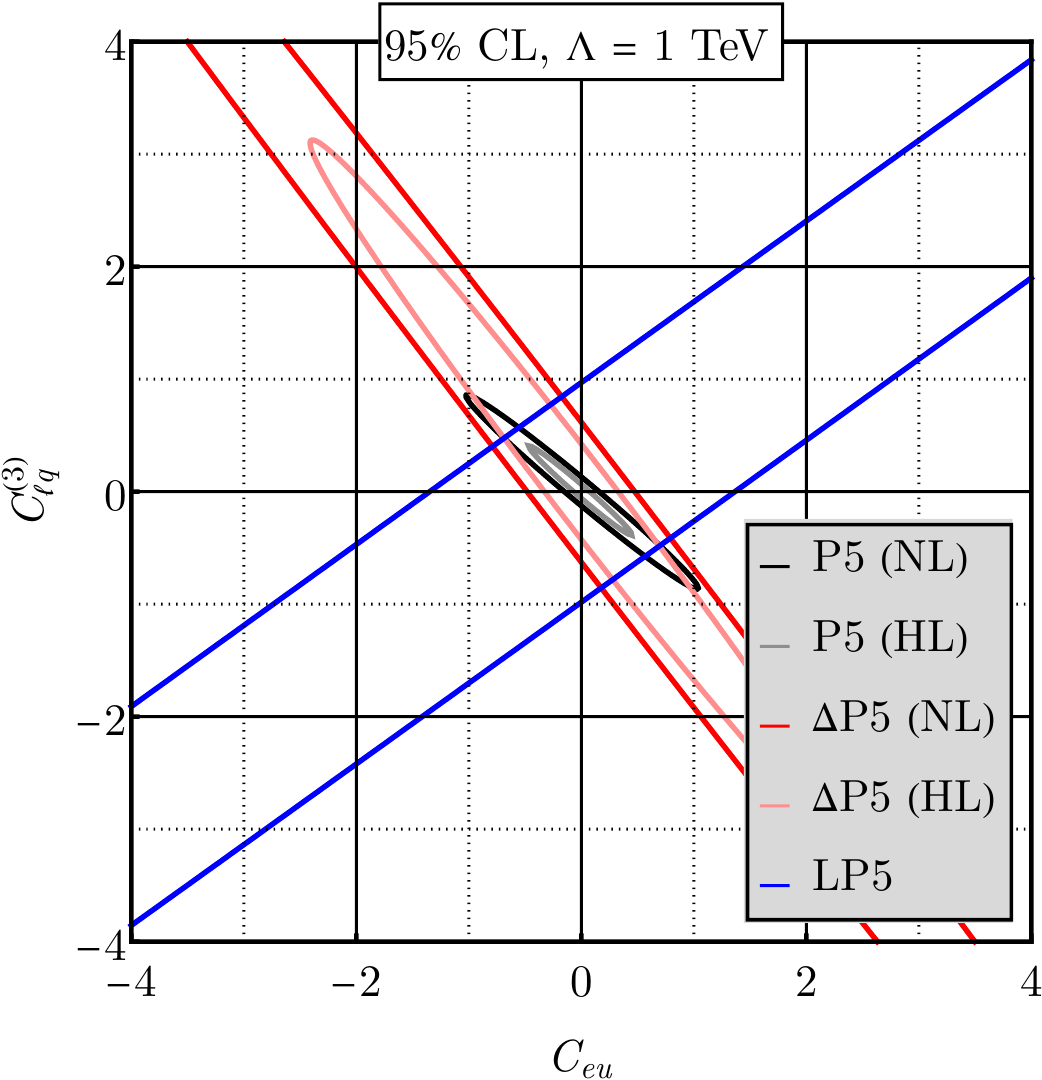}
	\caption{The same as in Fig.~\ref{fig:all-ellipses-Ceu-Ced} but for $\Ceu$ and $\Clqiii$.}
	\label{fig:all-ellipses-Ceu-Clq3}
\end{figure}

\begin{figure}
	[H]\centering
	\includegraphics[width=\elliwi\textwidth]{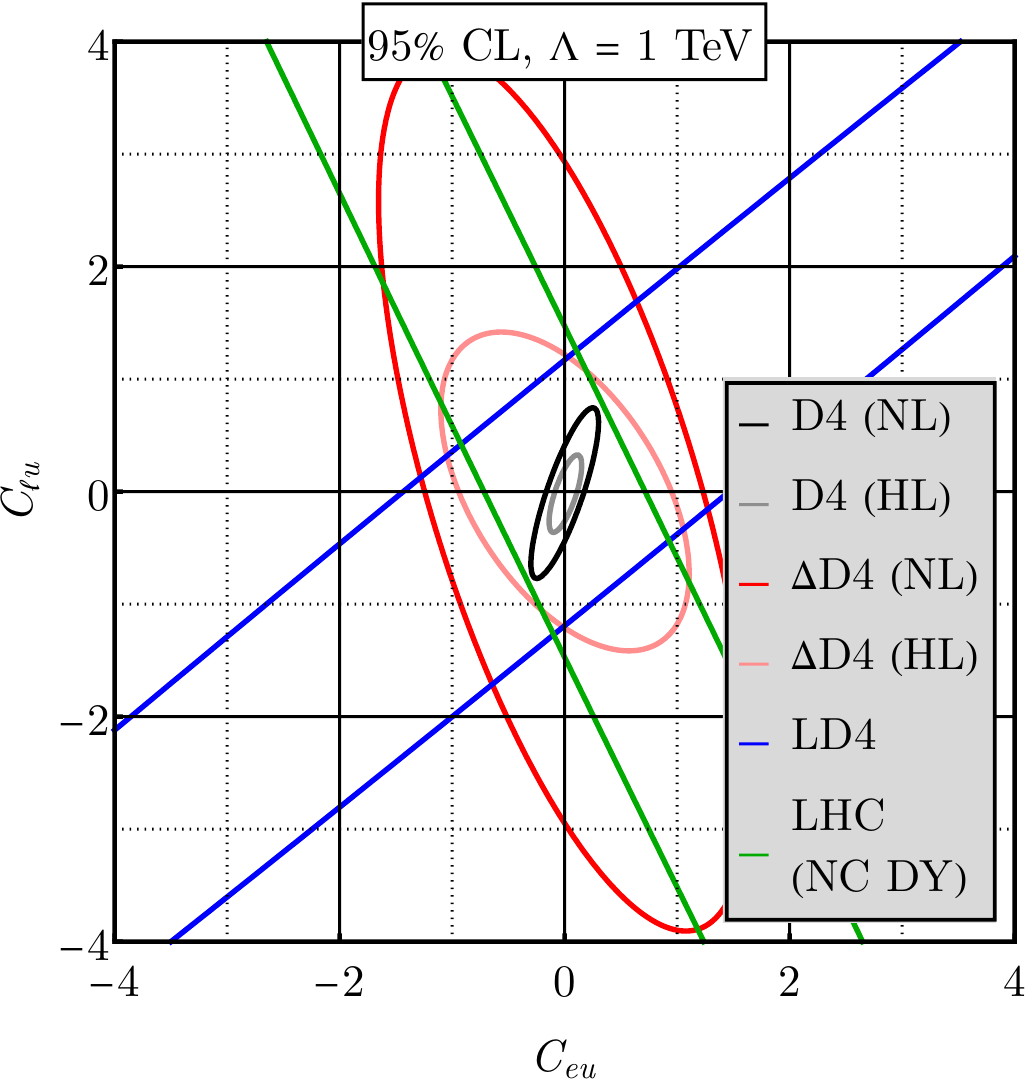}
	\includegraphics[width=\elliwi\textwidth]{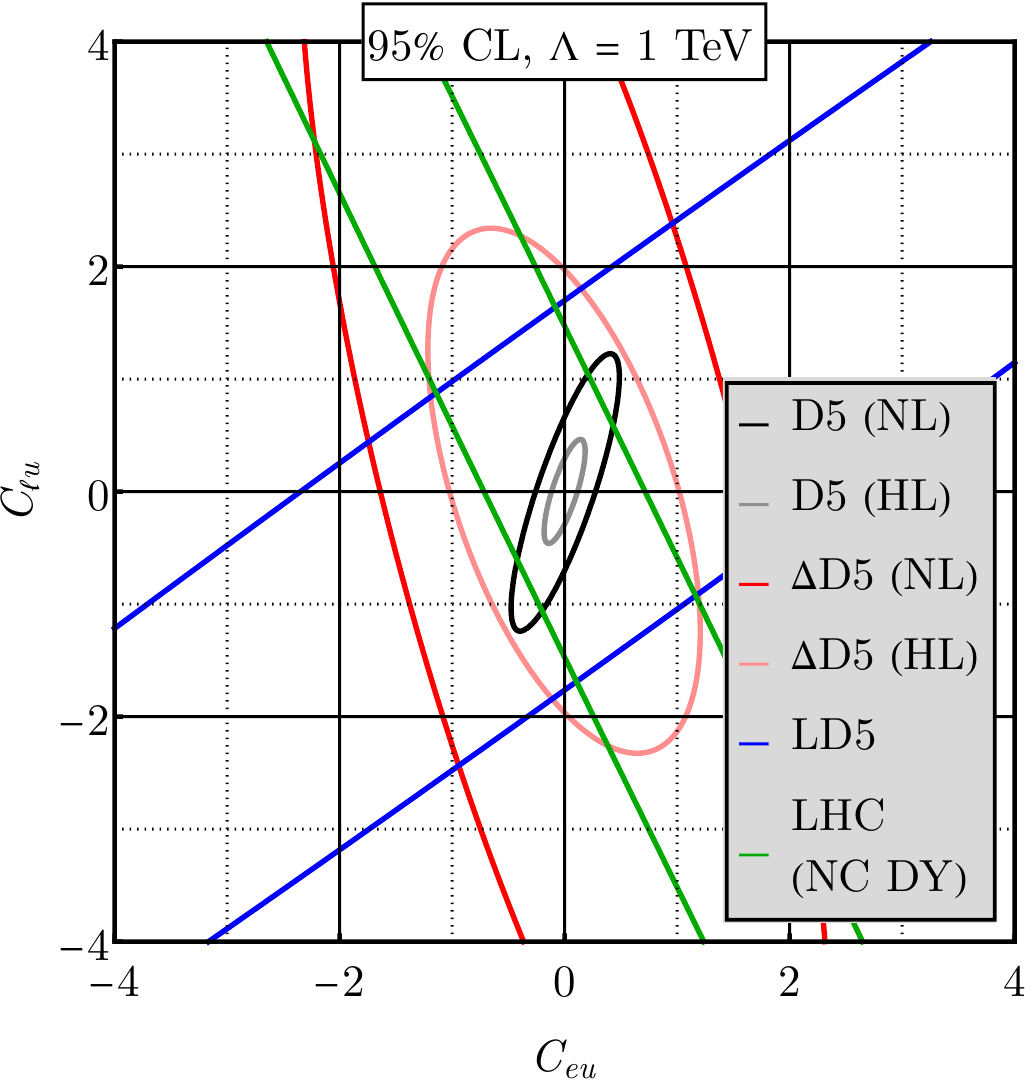}
	\includegraphics[width=\elliwi\textwidth]{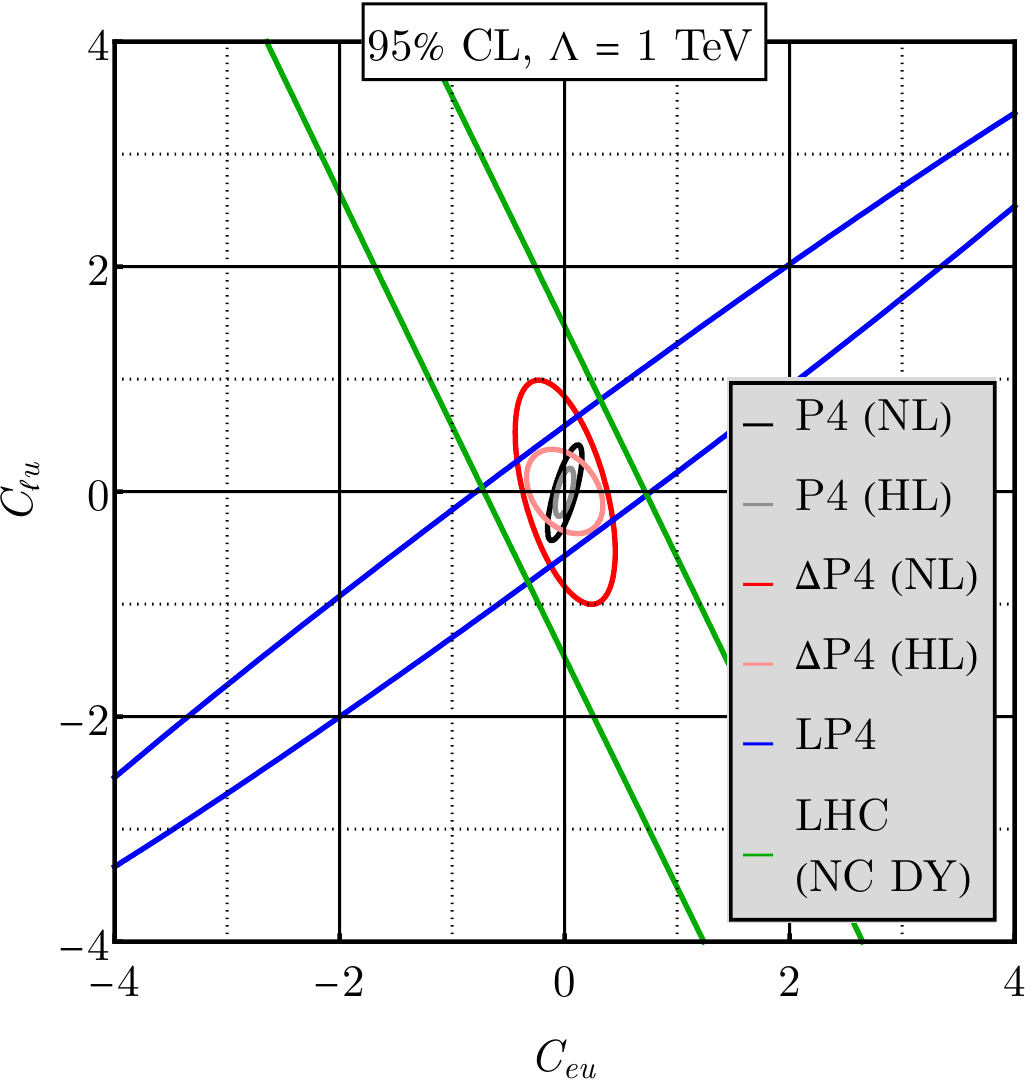}
	\includegraphics[width=\elliwi\textwidth]{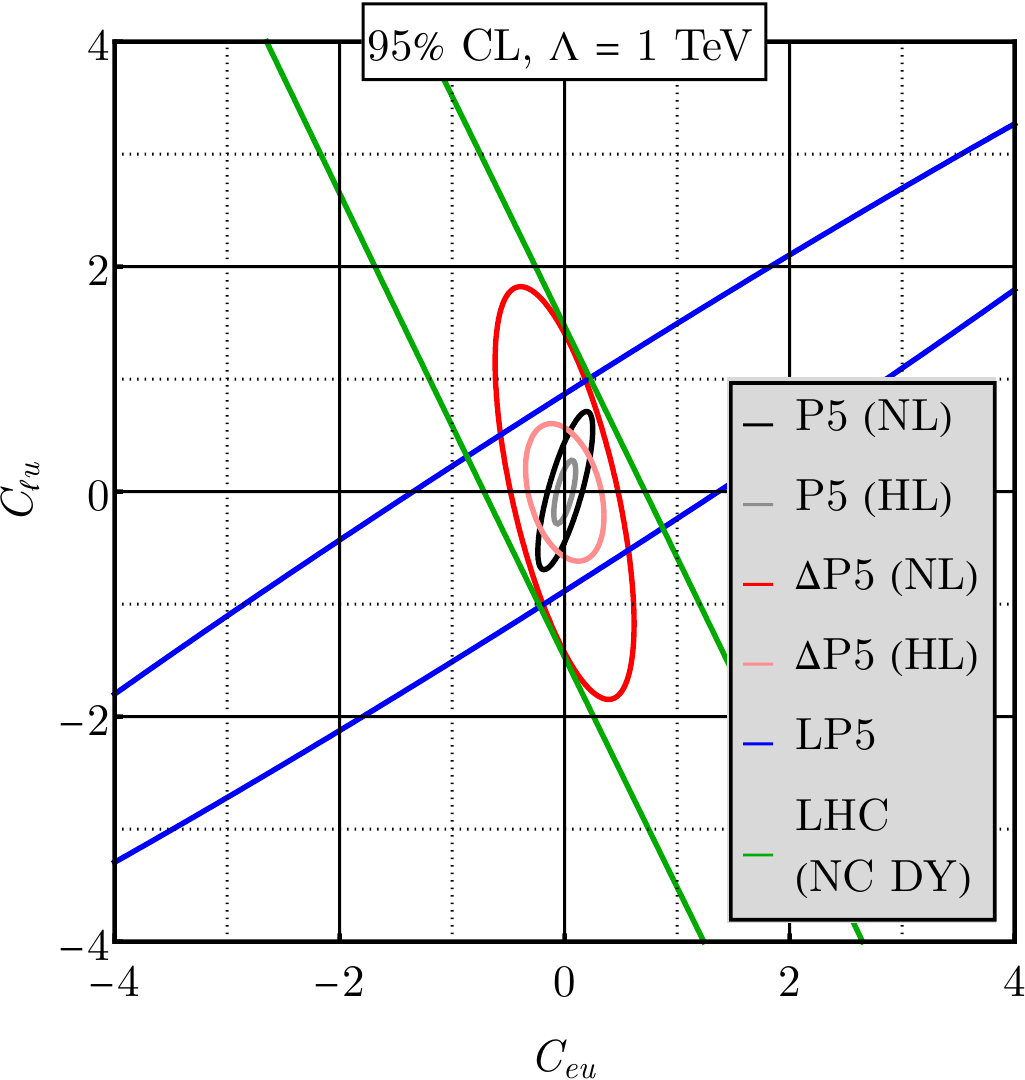}
	\caption{The same as in Fig.~\ref{fig:all-ellipses-Ceu-Ced} but for $\Ceu$ and $\Clu$.}
	\label{fig:all-ellipses-Ceu-Clu}
\end{figure}

\begin{figure}
	[H]\centering
	\includegraphics[width=\elliwi\textwidth]{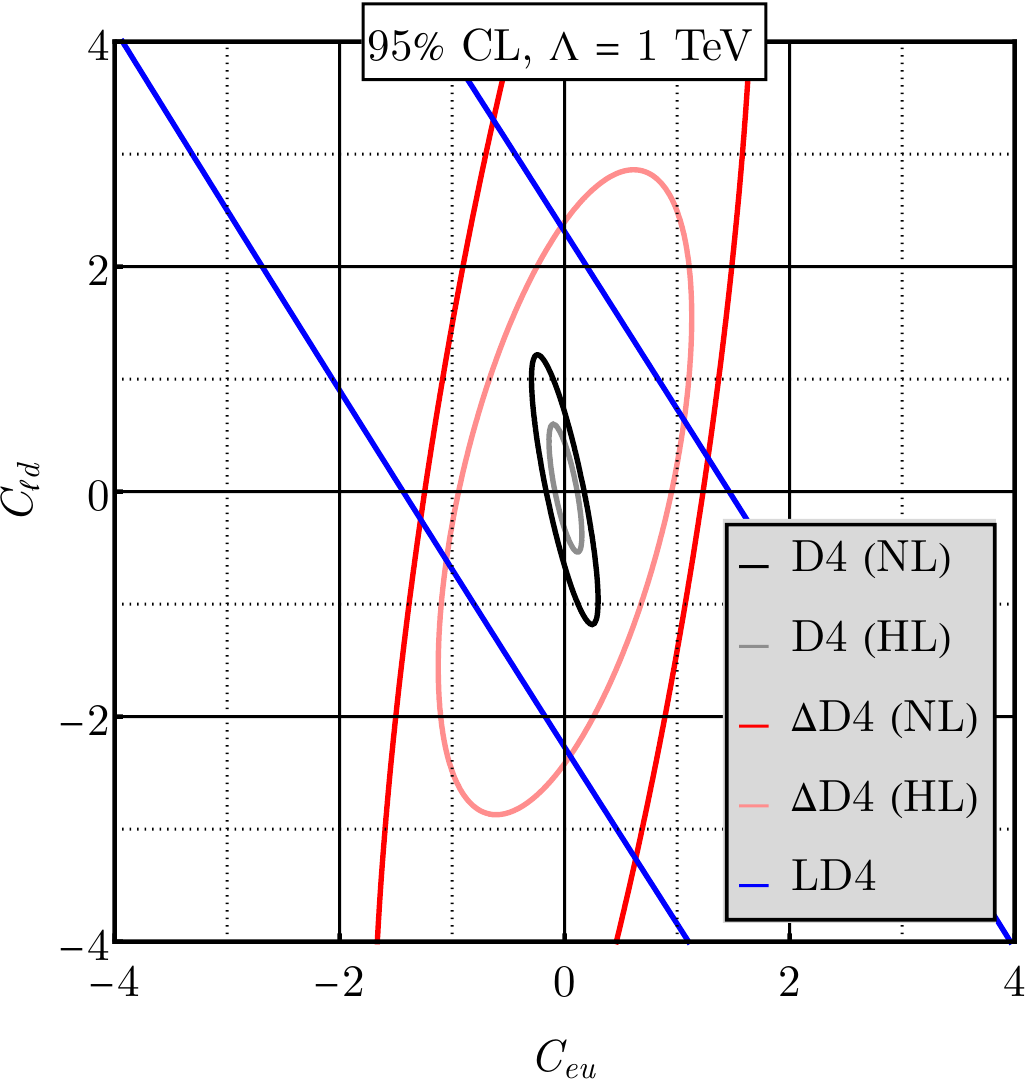}
	\includegraphics[width=\elliwi\textwidth]{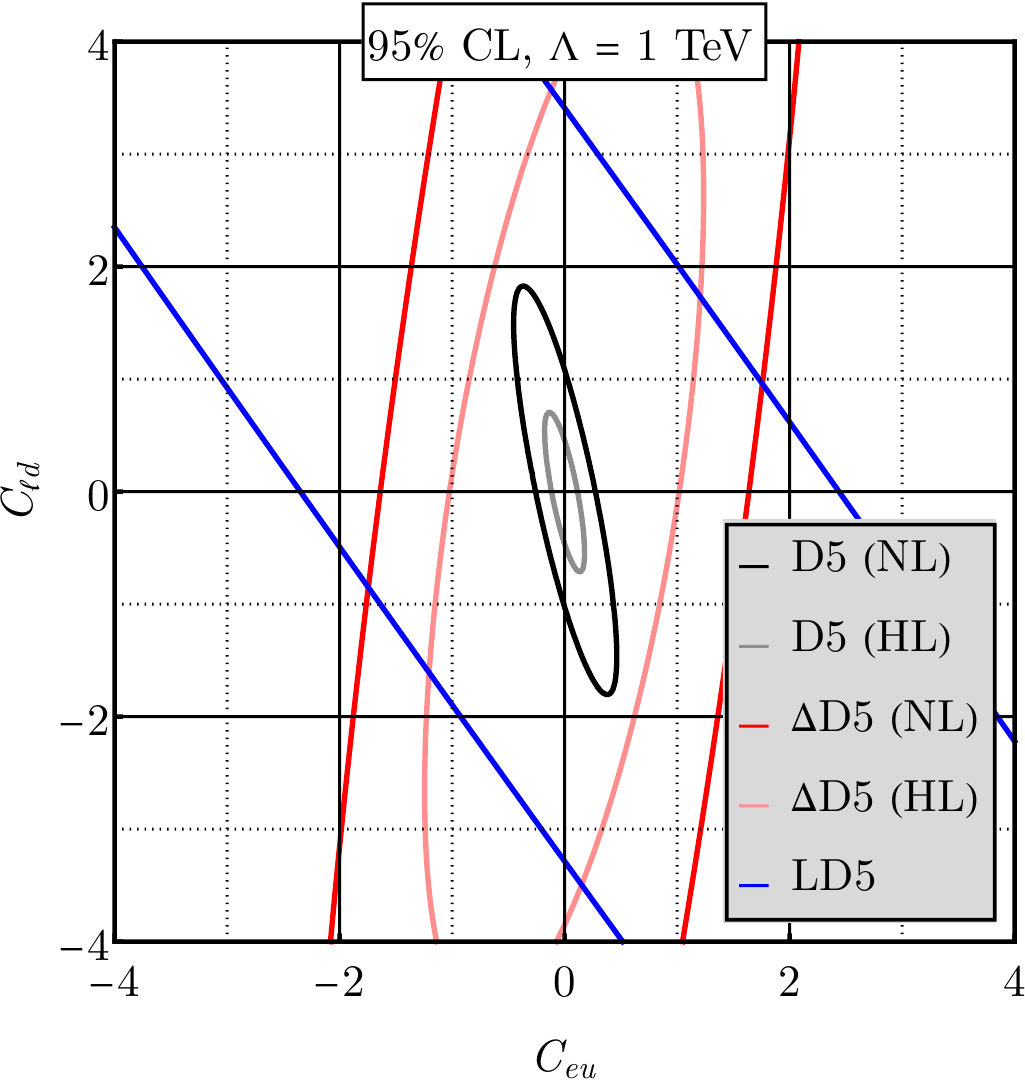}
	\includegraphics[width=\elliwi\textwidth]{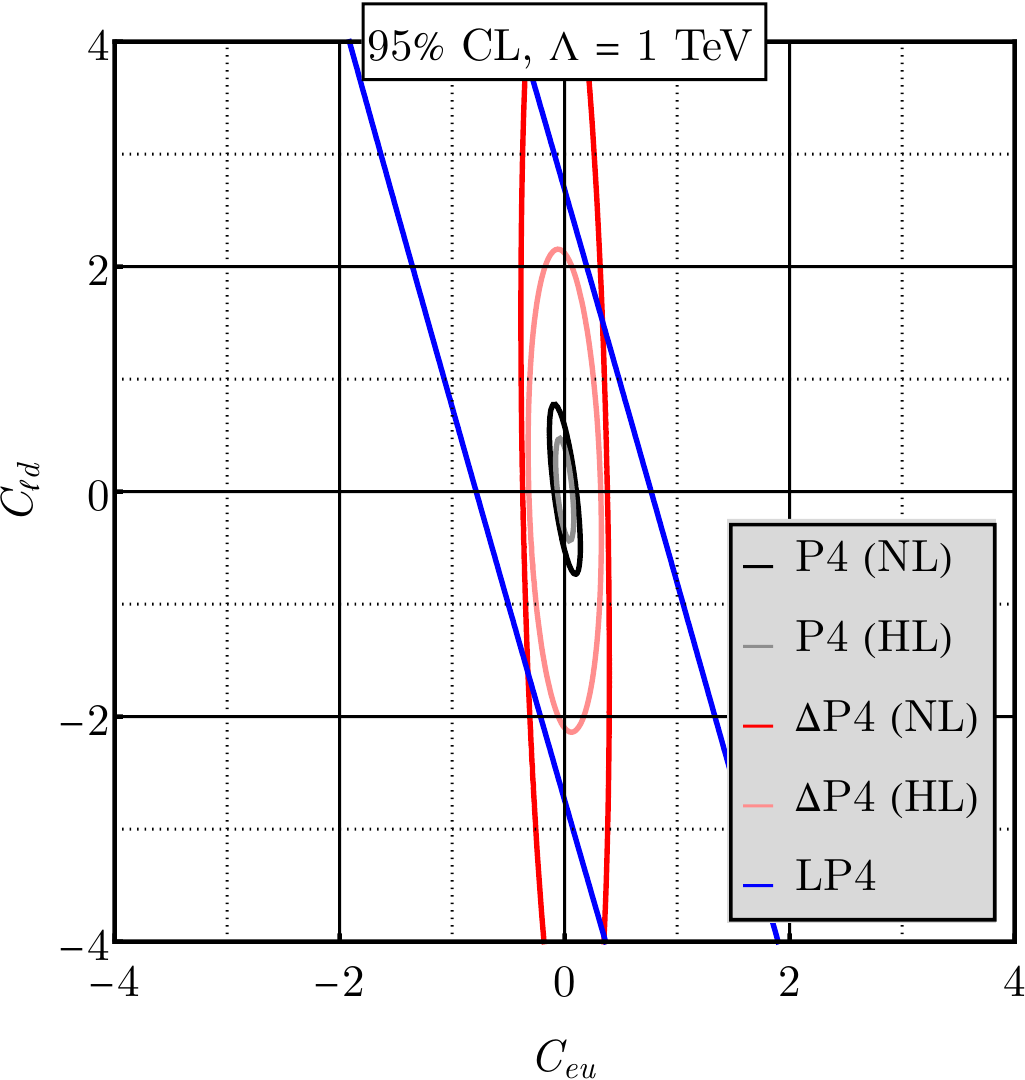}
	\includegraphics[width=\elliwi\textwidth]{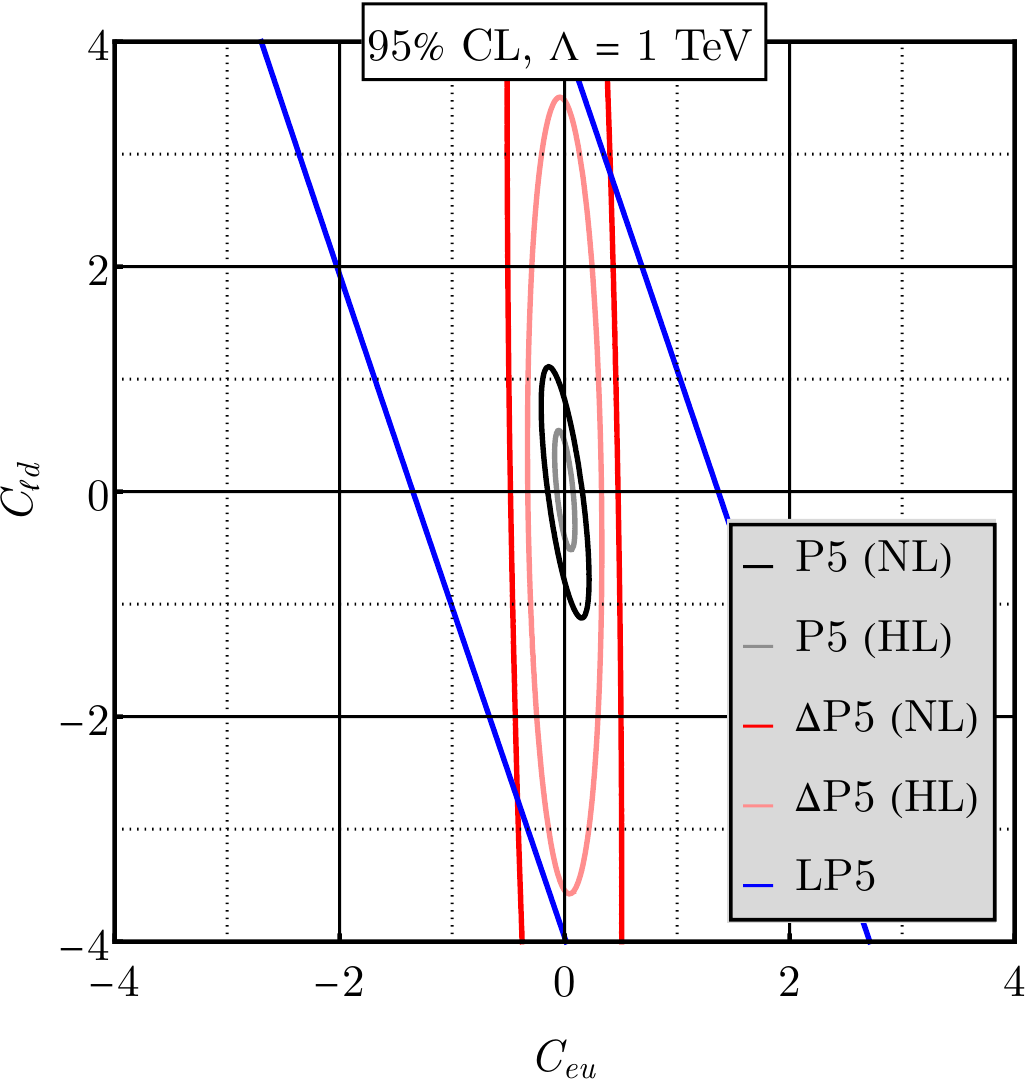}
	\caption{The same as in Fig.~\ref{fig:all-ellipses-Ceu-Ced} but for $\Ceu$ and $\Cld$.}
	\label{fig:all-ellipses-Ceu-Cld}
\end{figure}

\begin{figure}
	[H]\centering
	\includegraphics[width=\elliwi\textwidth]{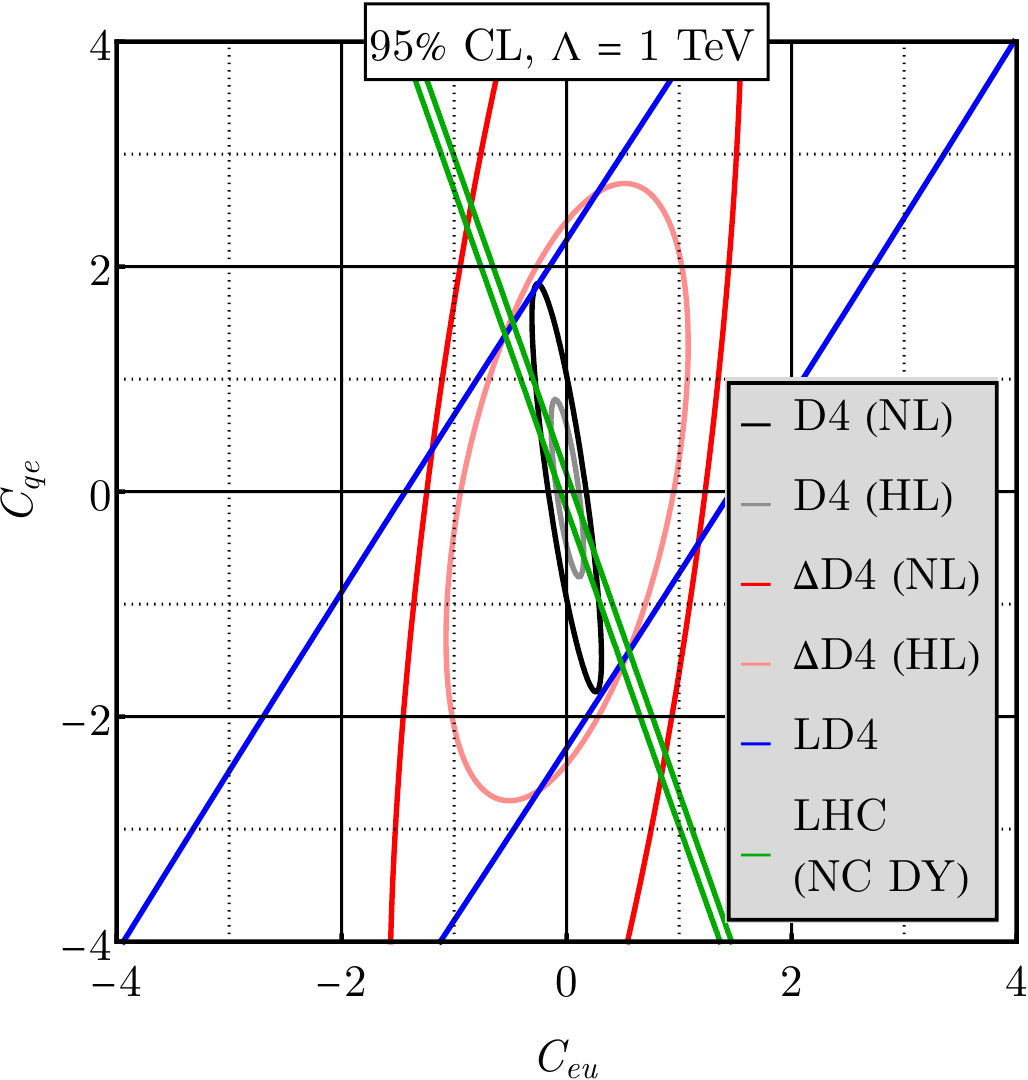}
	\includegraphics[width=\elliwi\textwidth]{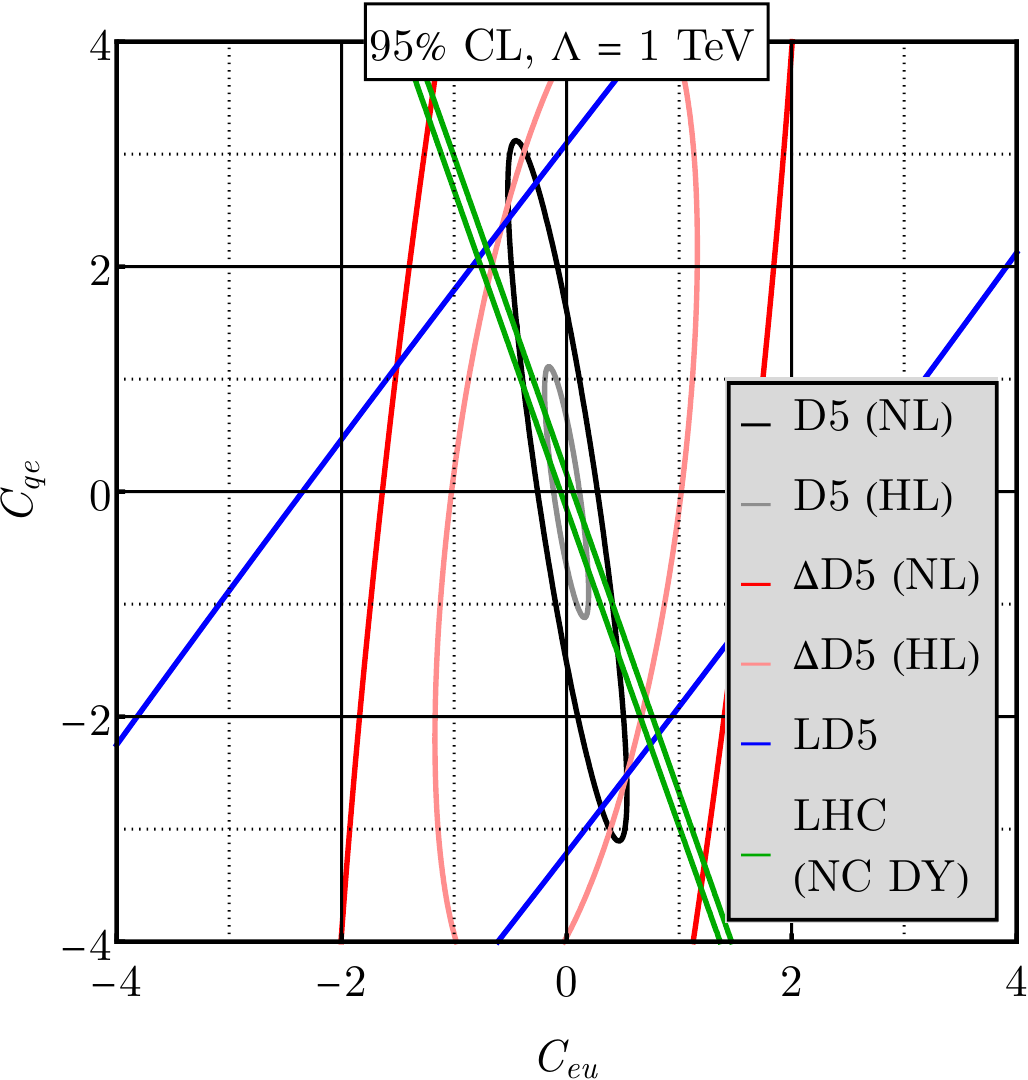}
	\includegraphics[width=\elliwi\textwidth]{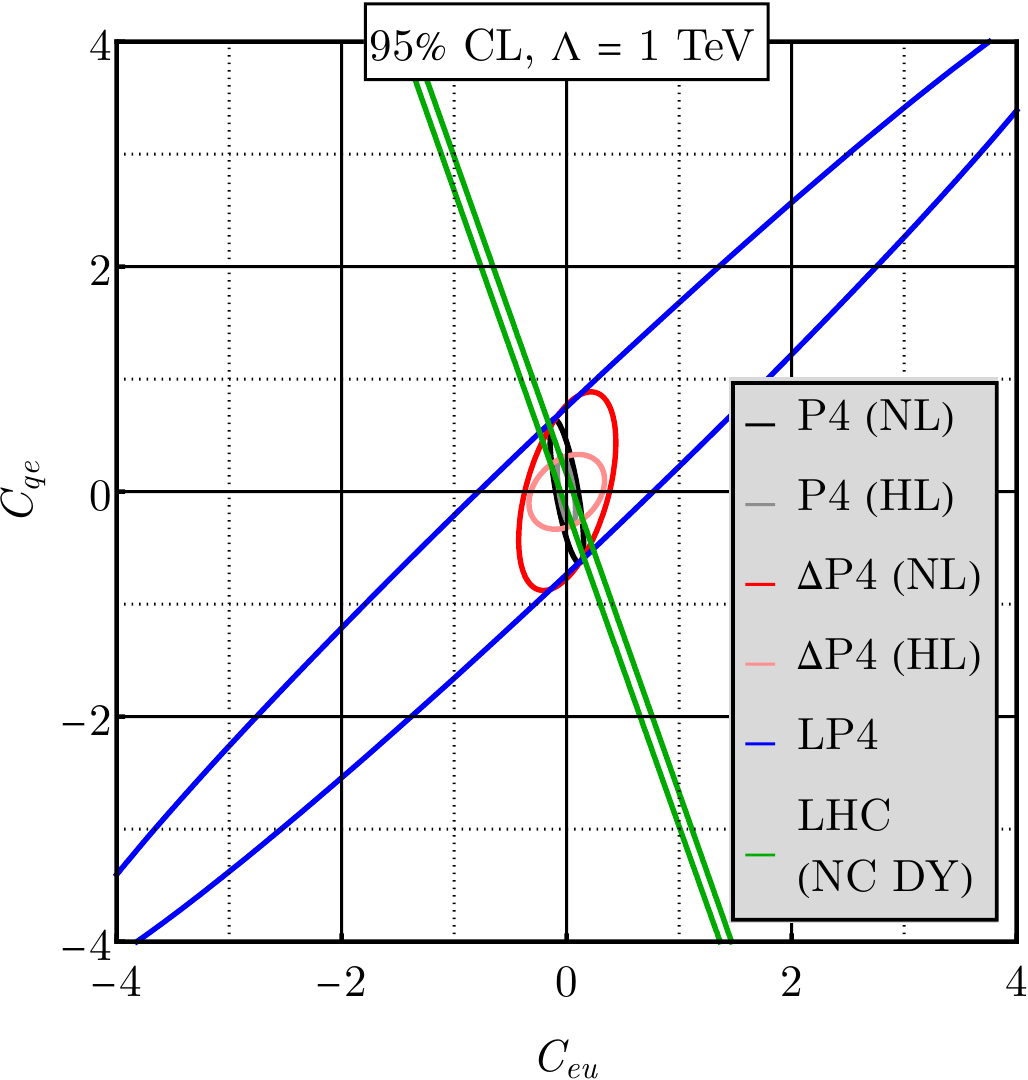}
	\includegraphics[width=\elliwi\textwidth]{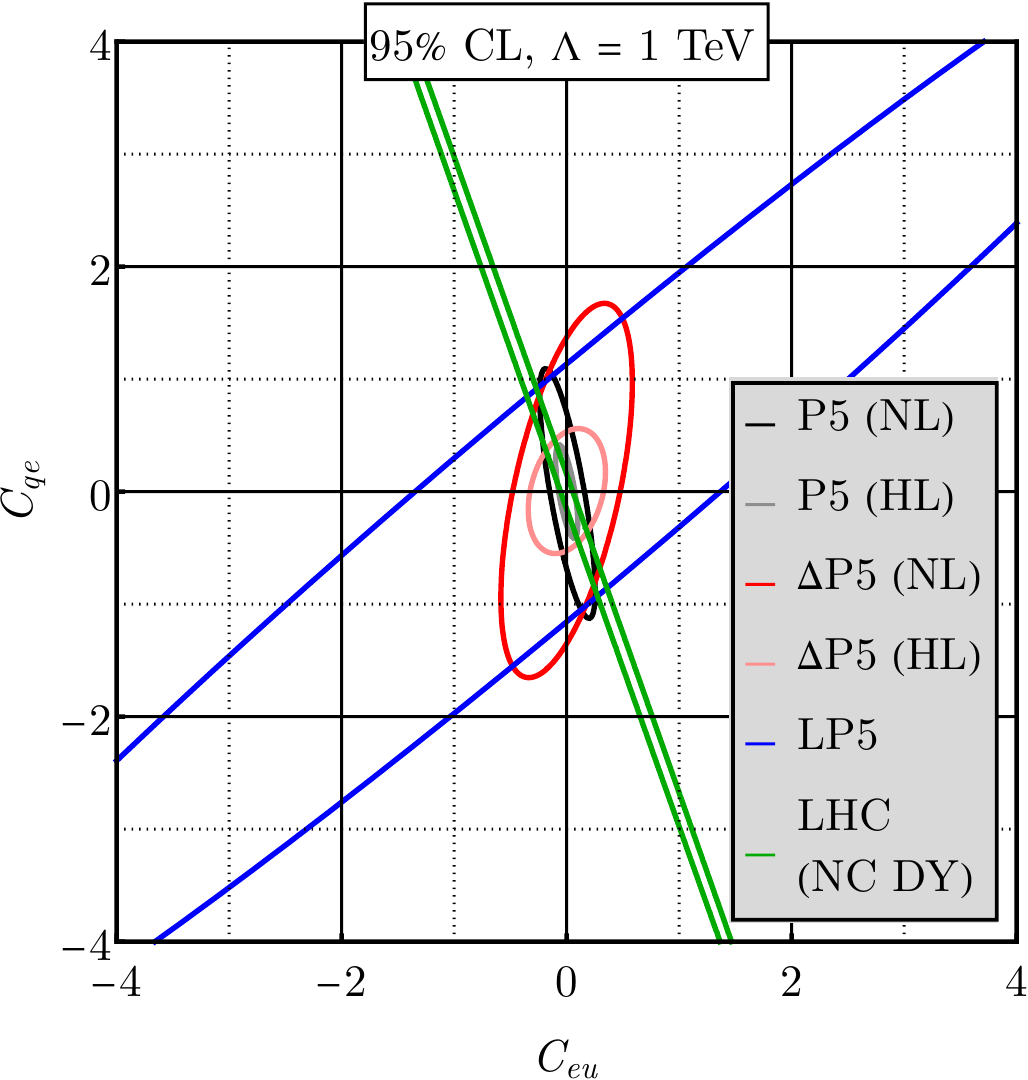}
	\caption{The same as in Fig.~\ref{fig:all-ellipses-Ceu-Ced} but for $\Ceu$ and $\Cqe$.}
	\label{fig:all-ellipses-Ceu-Cqe}
\end{figure}

\begin{figure}
	[H]\centering
	\includegraphics[width=\elliwi\textwidth]{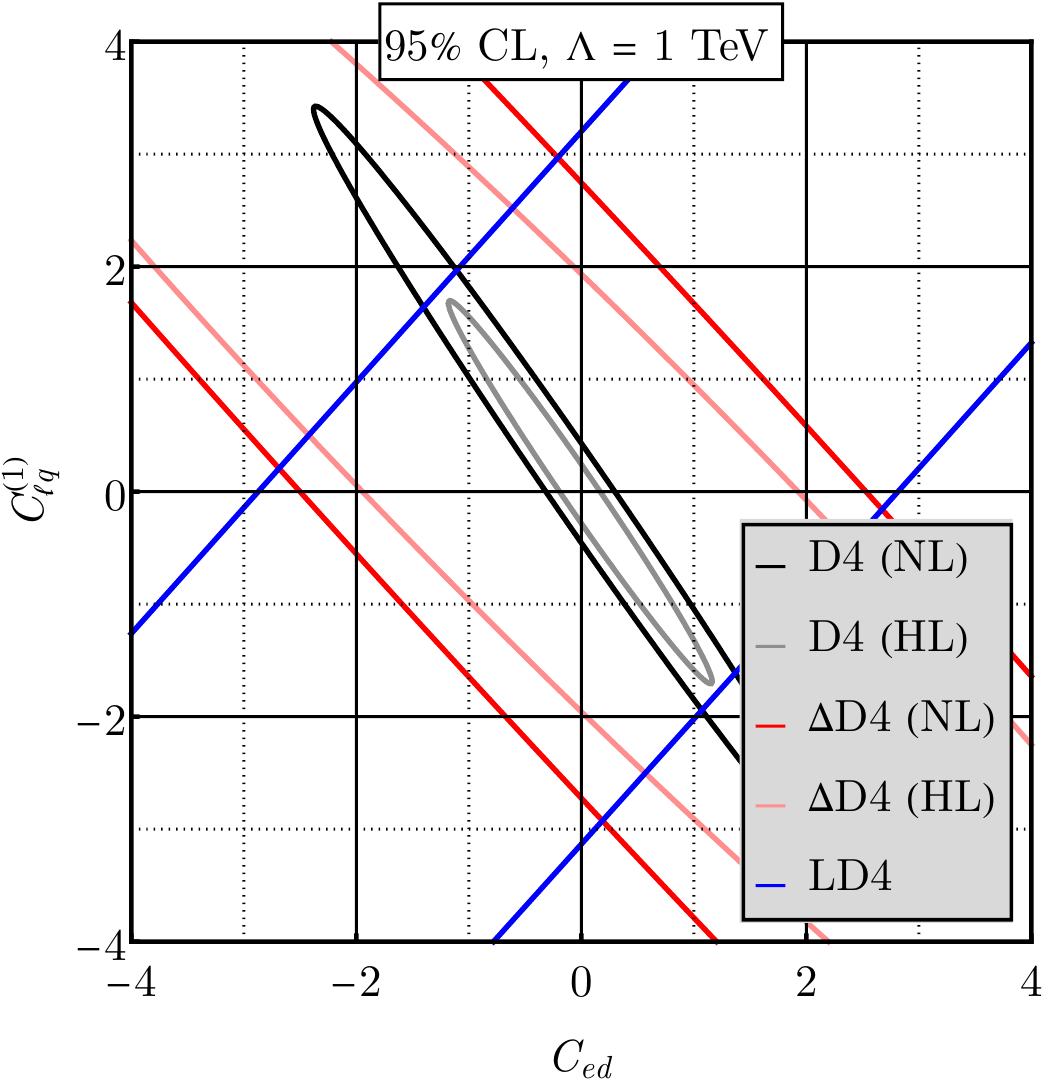}
	\includegraphics[width=\elliwi\textwidth]{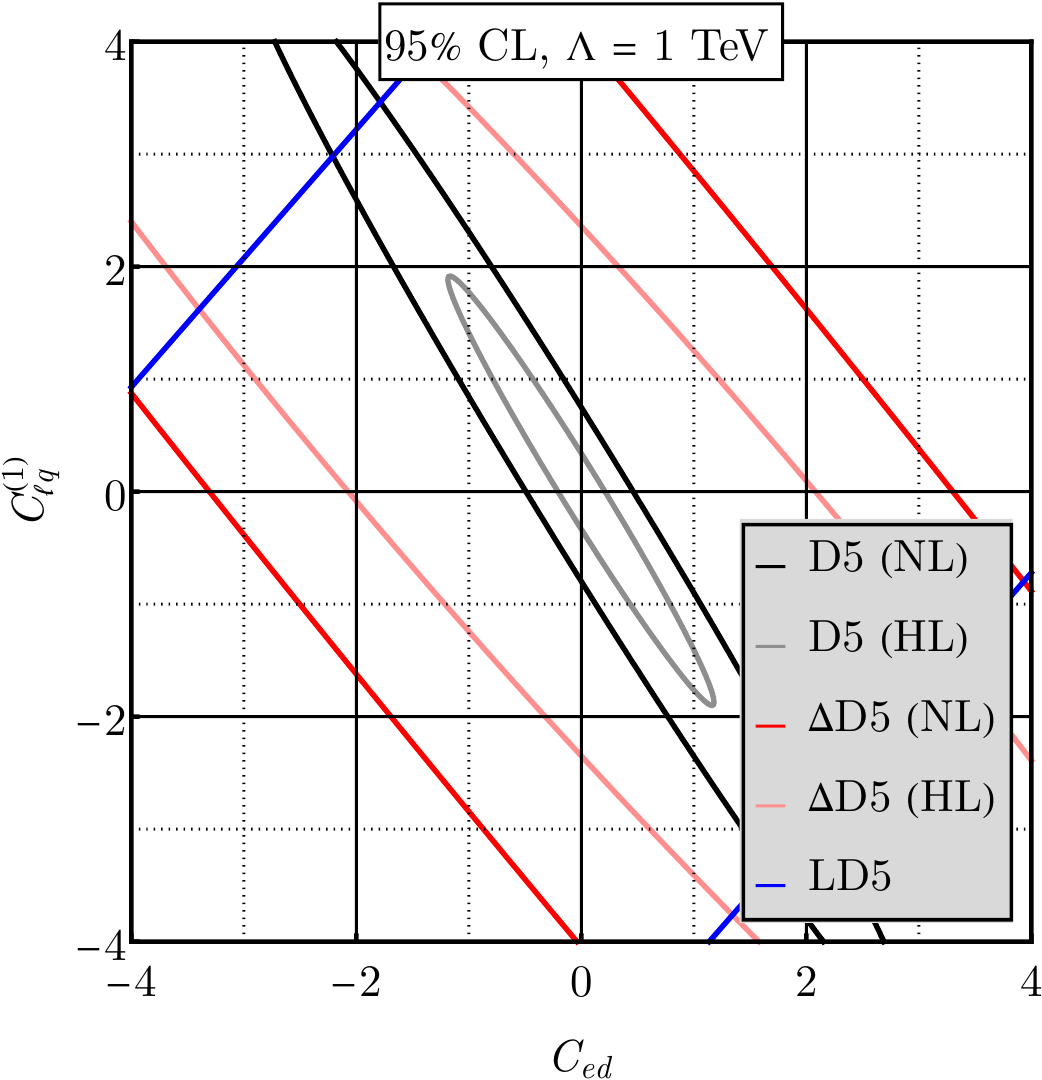}
	\includegraphics[width=\elliwi\textwidth]{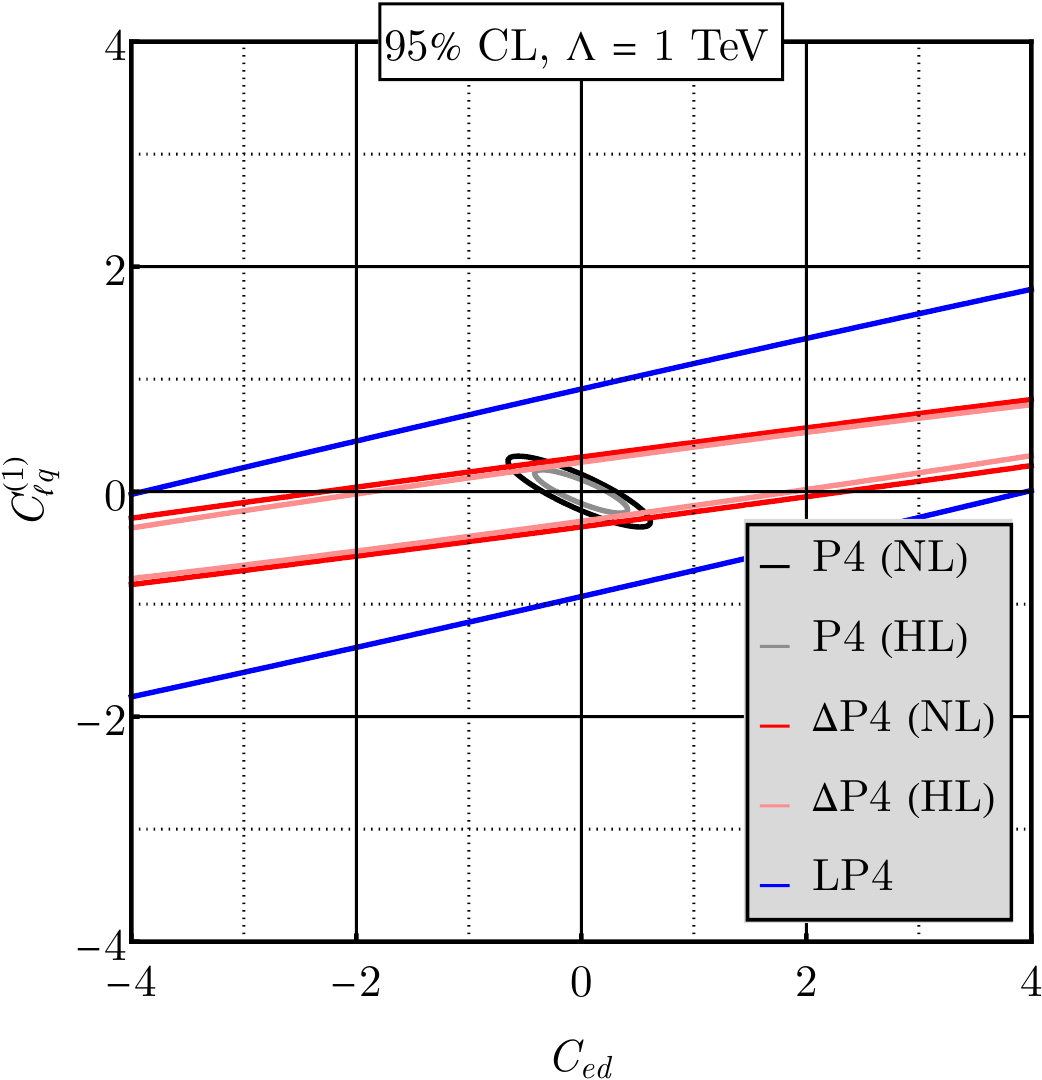}
	\includegraphics[width=\elliwi\textwidth]{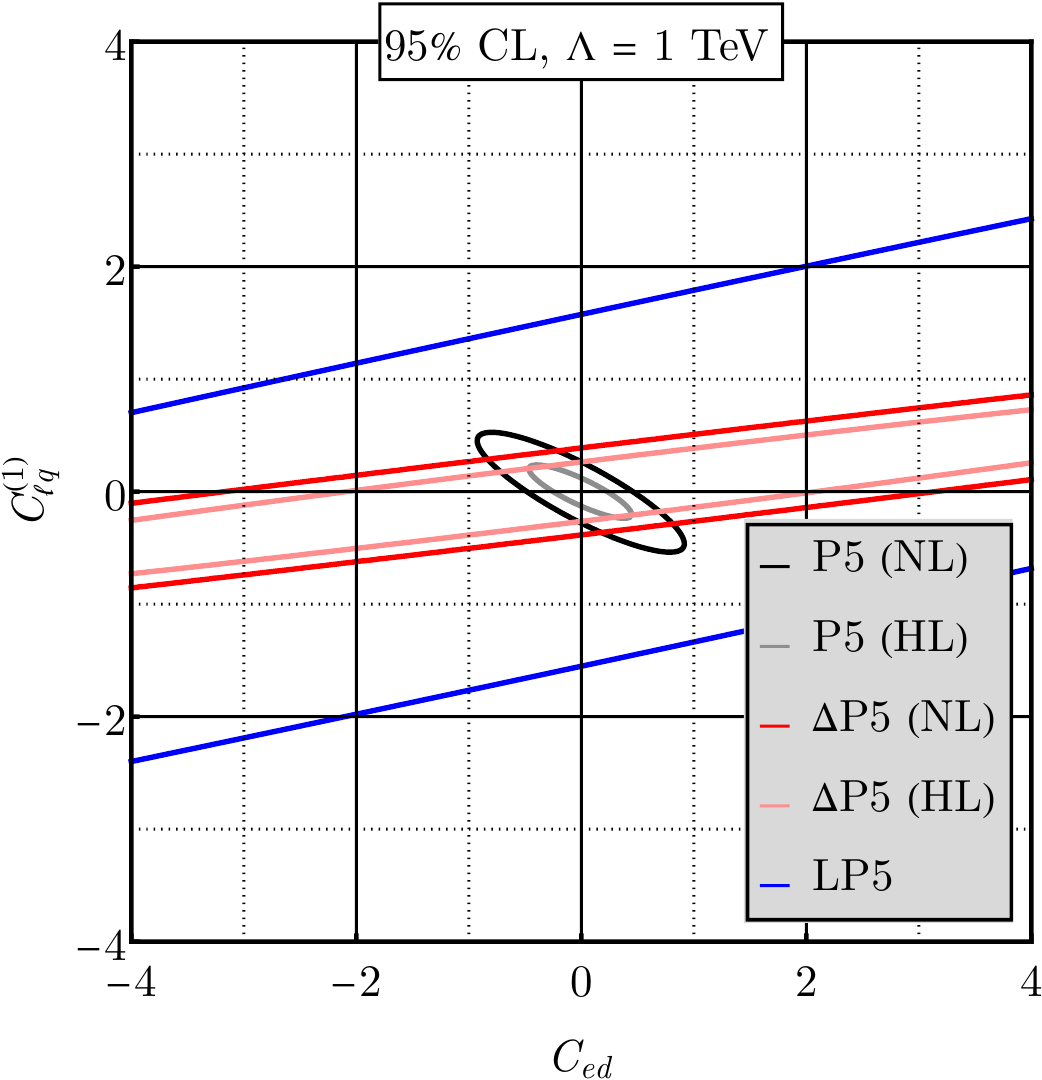}
	\caption{The same as in Fig.~\ref{fig:all-ellipses-Ceu-Ced} but for $\Ced$ and $\Clqi$.}
	\label{fig:all-ellipses-Ced-Clq1}
\end{figure}

\begin{figure}
	[H]\centering
	\includegraphics[width=\elliwi\textwidth]{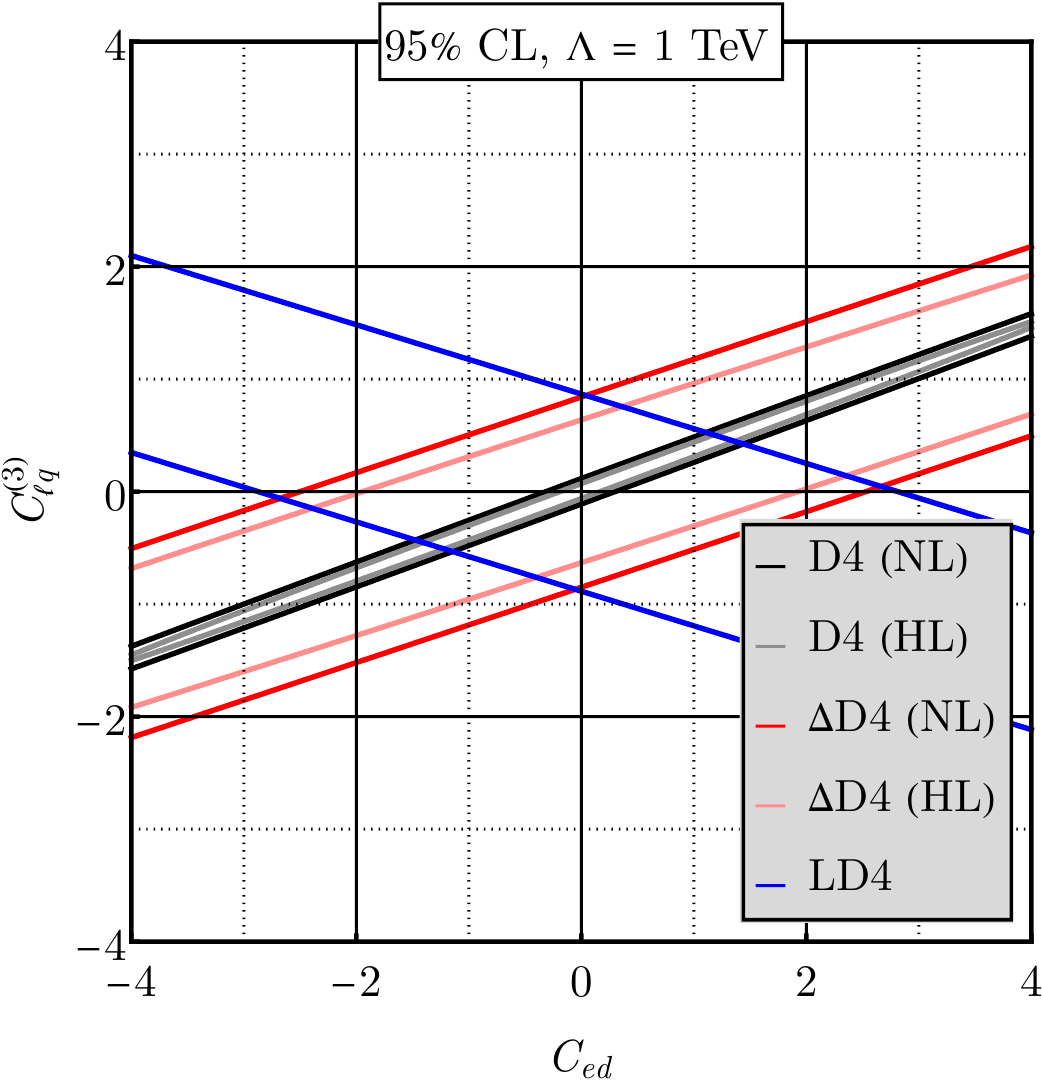}
	\includegraphics[width=\elliwi\textwidth]{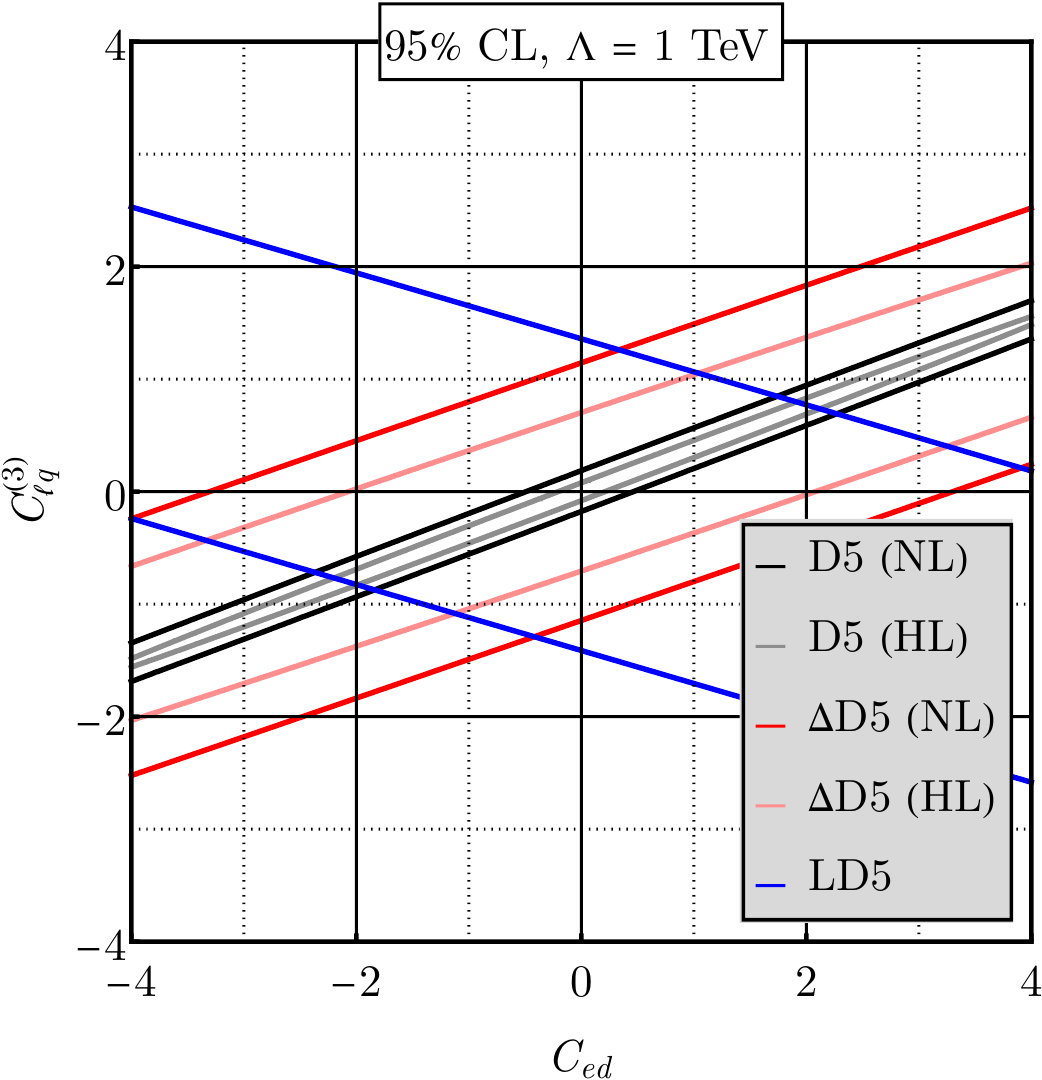}
	\includegraphics[width=\elliwi\textwidth]{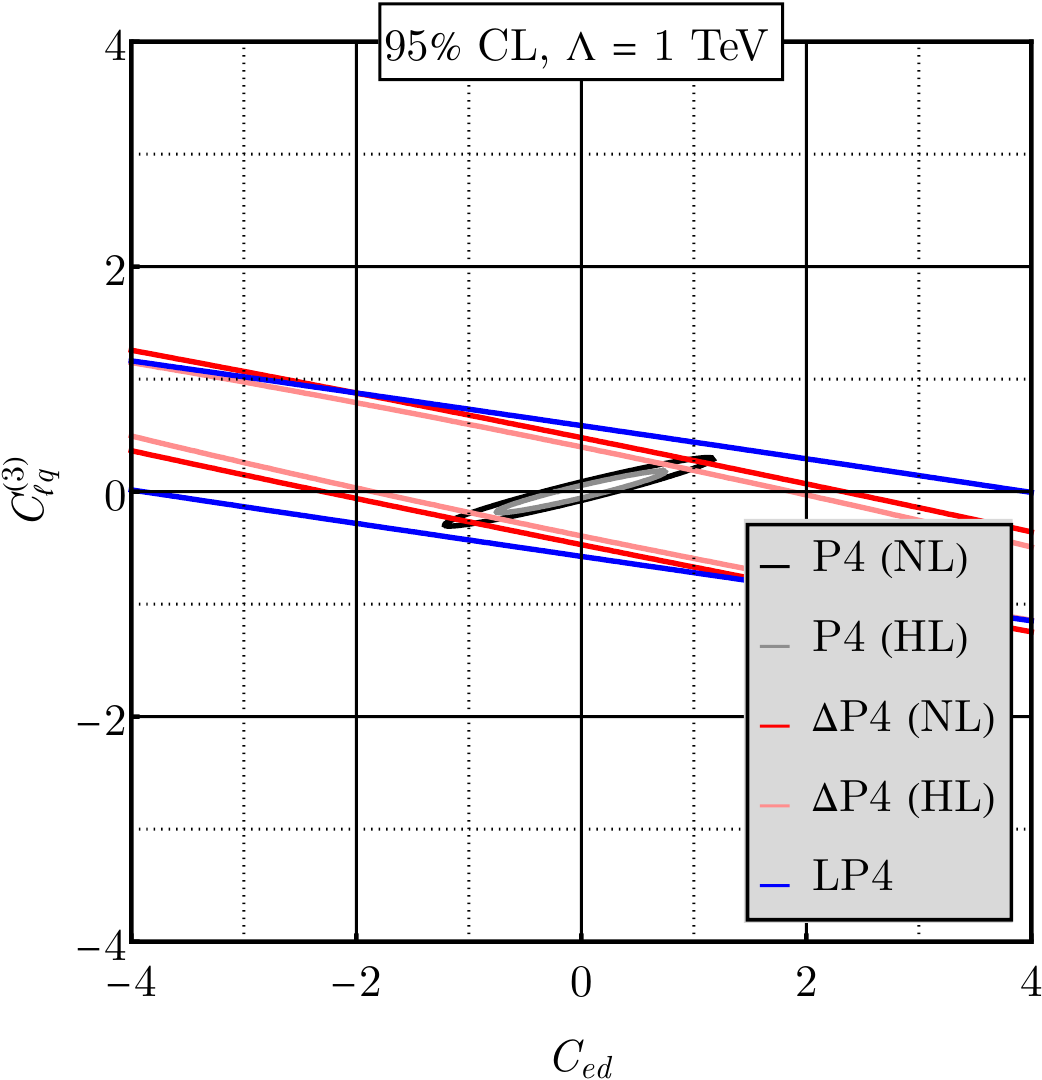}
	\includegraphics[width=\elliwi\textwidth]{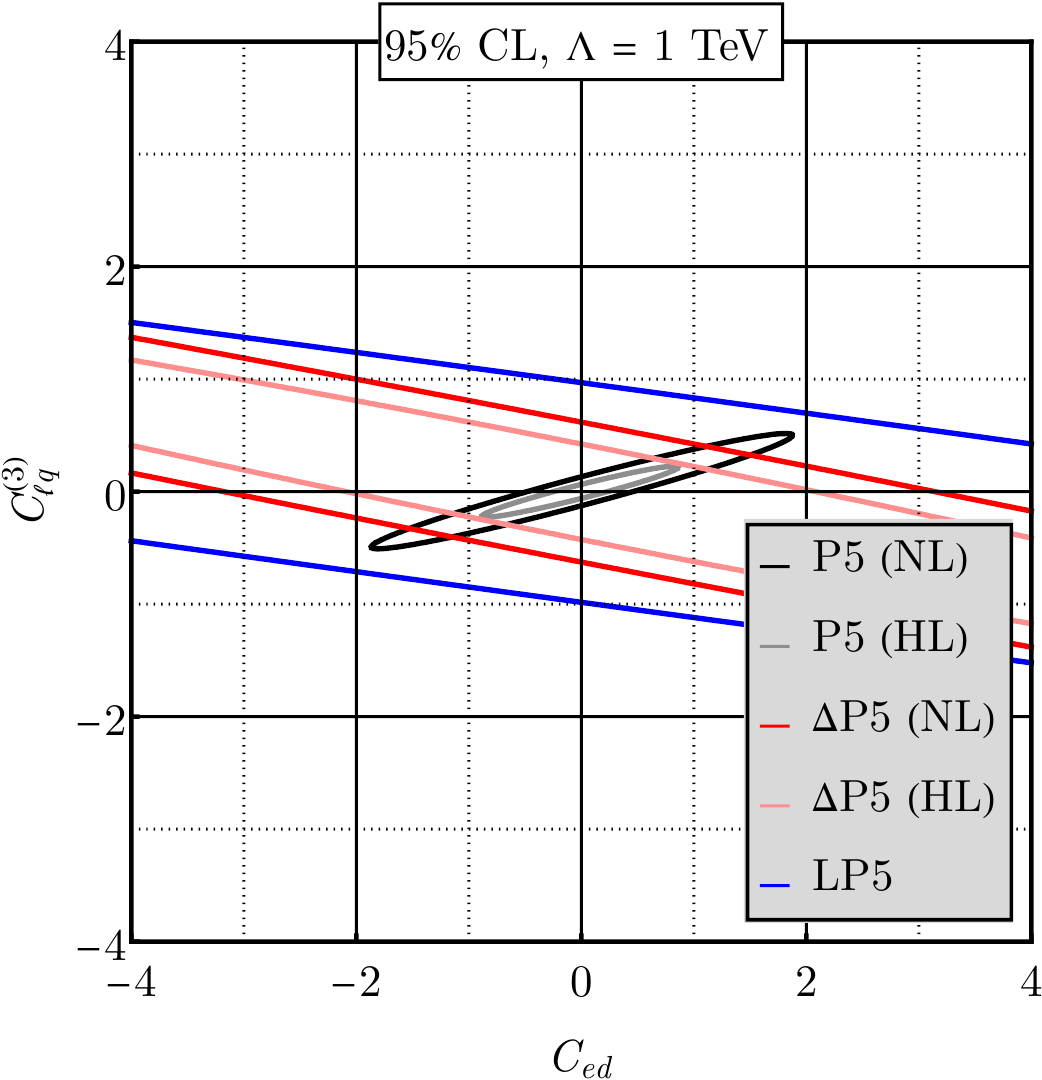}
	\caption{The same as in Fig.~\ref{fig:all-ellipses-Ceu-Ced} but for $\Ced$ and $\Clqiii$.}
	\label{fig:all-ellipses-Ced-Clq3}
\end{figure}

\begin{figure}
	[H]\centering
	\includegraphics[width=\elliwi\textwidth]{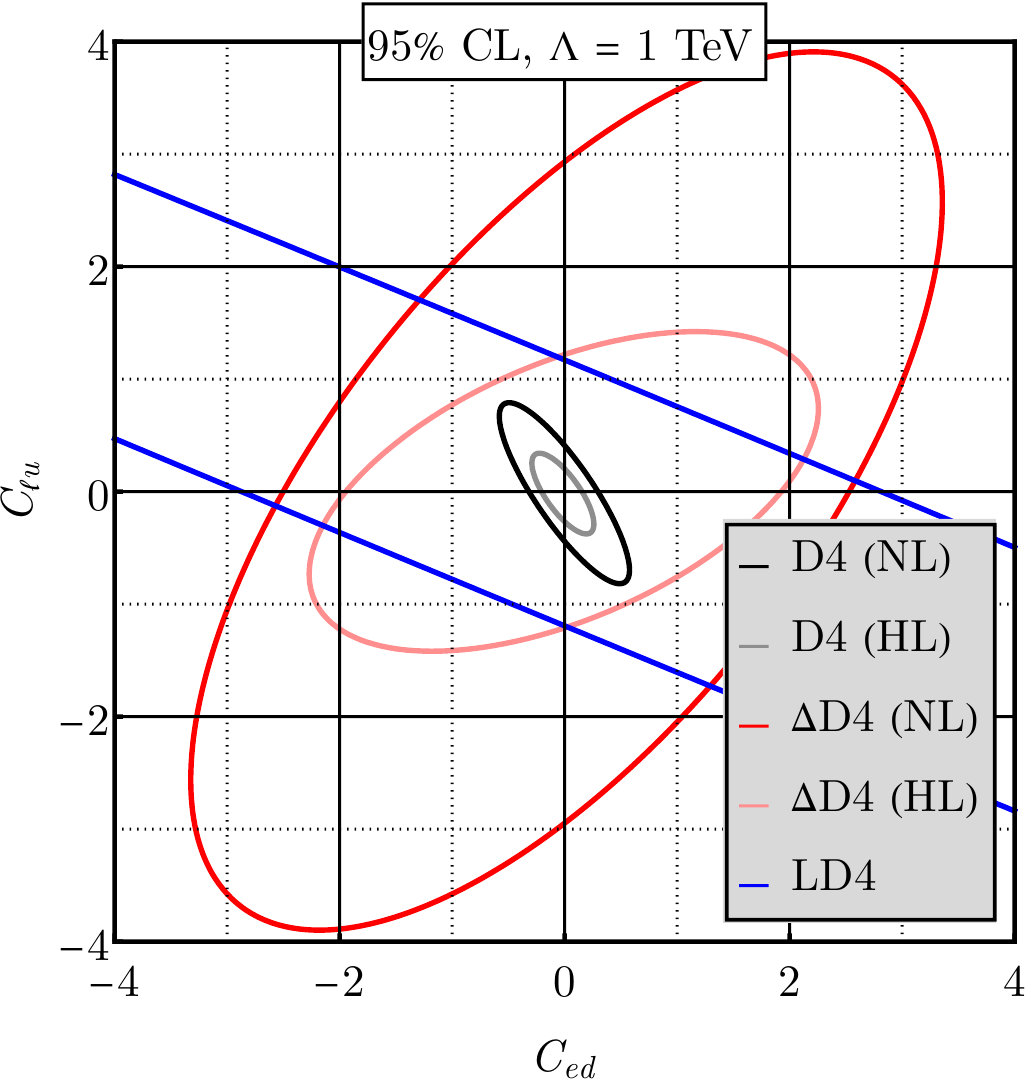}
	\includegraphics[width=\elliwi\textwidth]{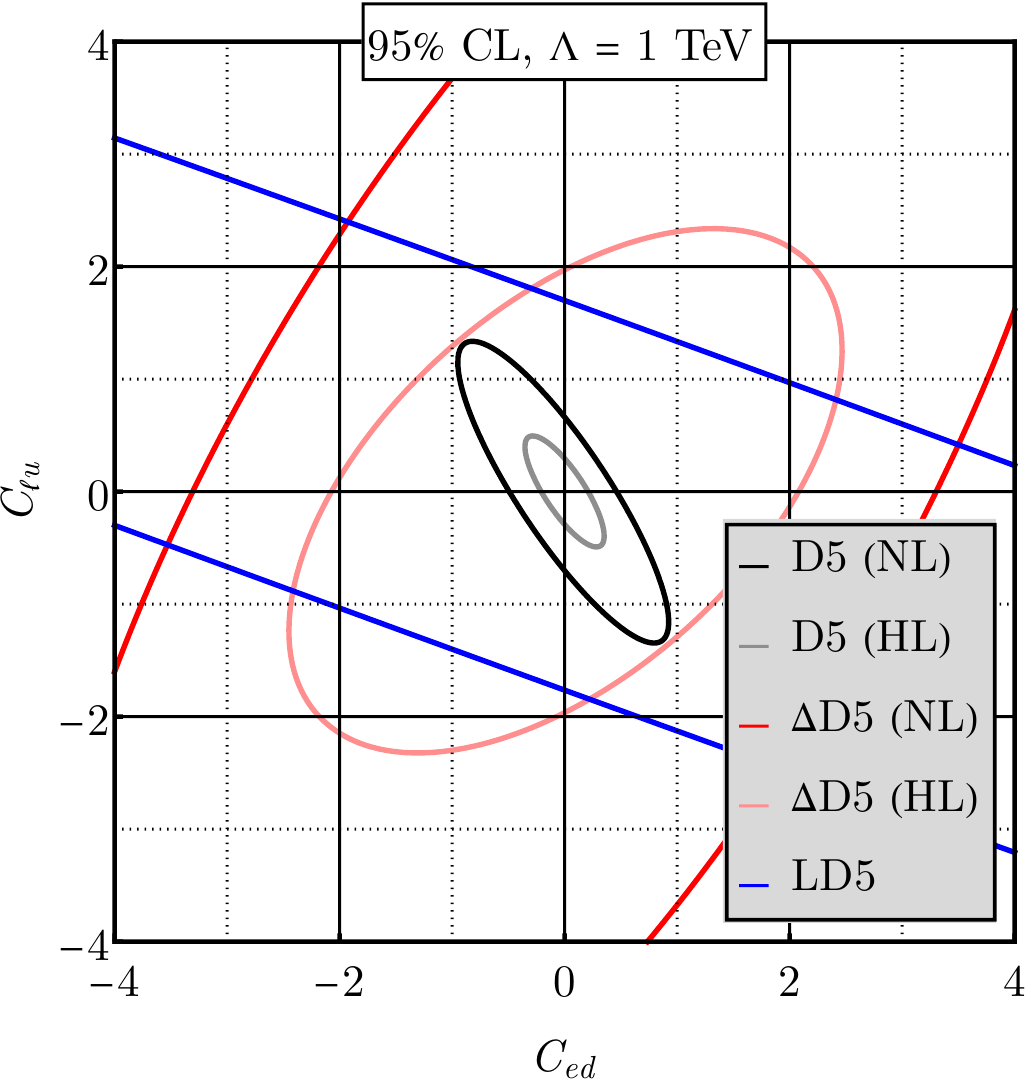}
	\includegraphics[width=\elliwi\textwidth]{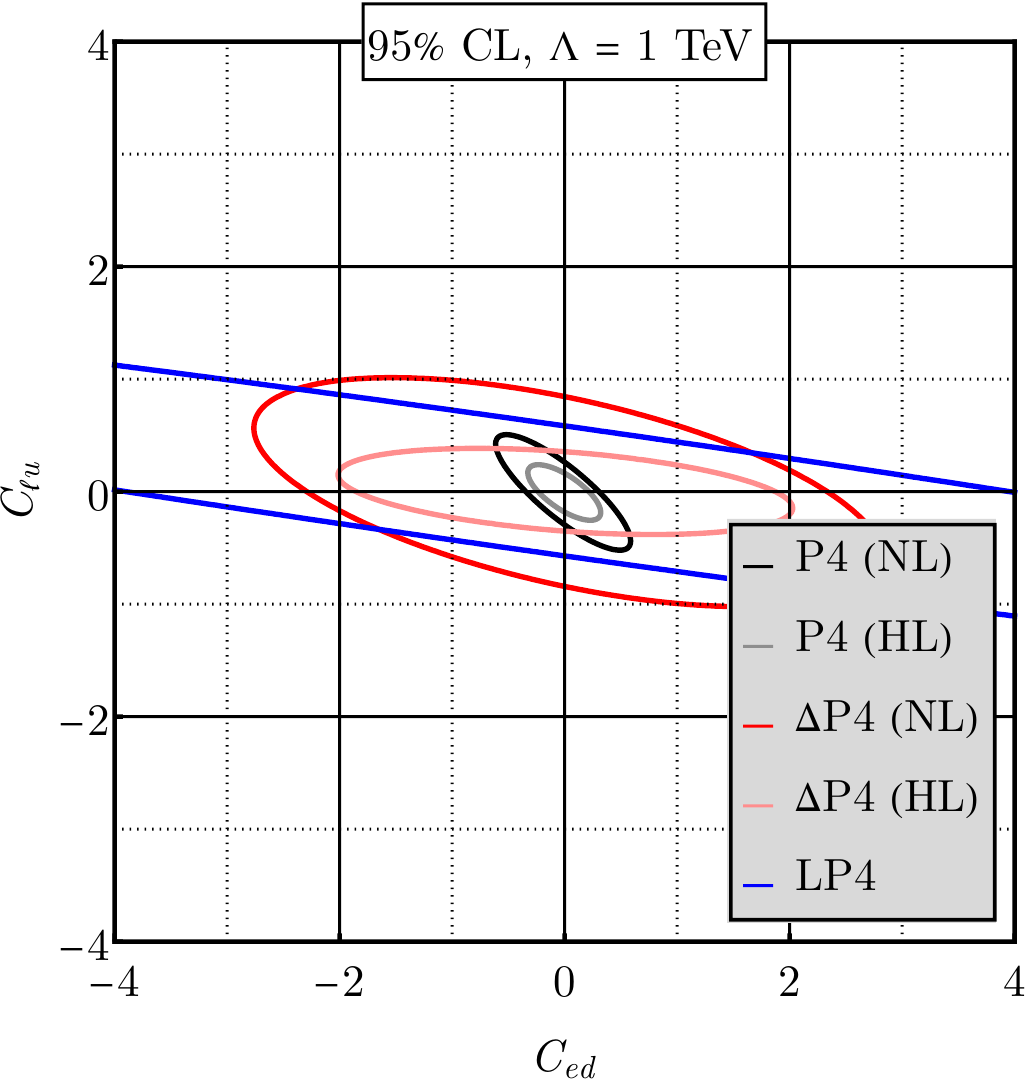}
	\includegraphics[width=\elliwi\textwidth]{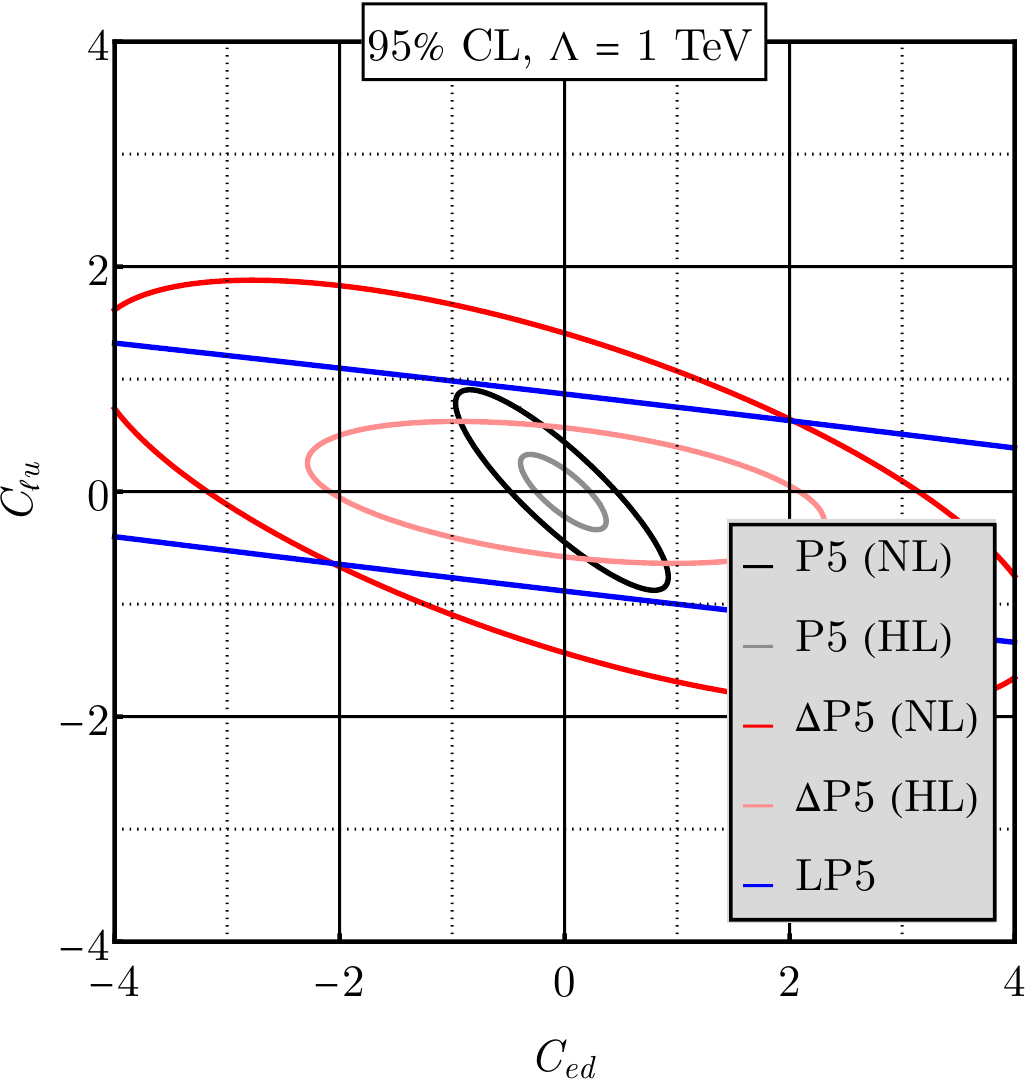}
	\caption{The same as in Fig.~\ref{fig:all-ellipses-Ceu-Ced} but for $\Ced$ and $\Clu$.}
	\label{fig:all-ellipses-Ced-Clu}
\end{figure}

\begin{figure}
	[H]\centering
	\includegraphics[width=\elliwi\textwidth]{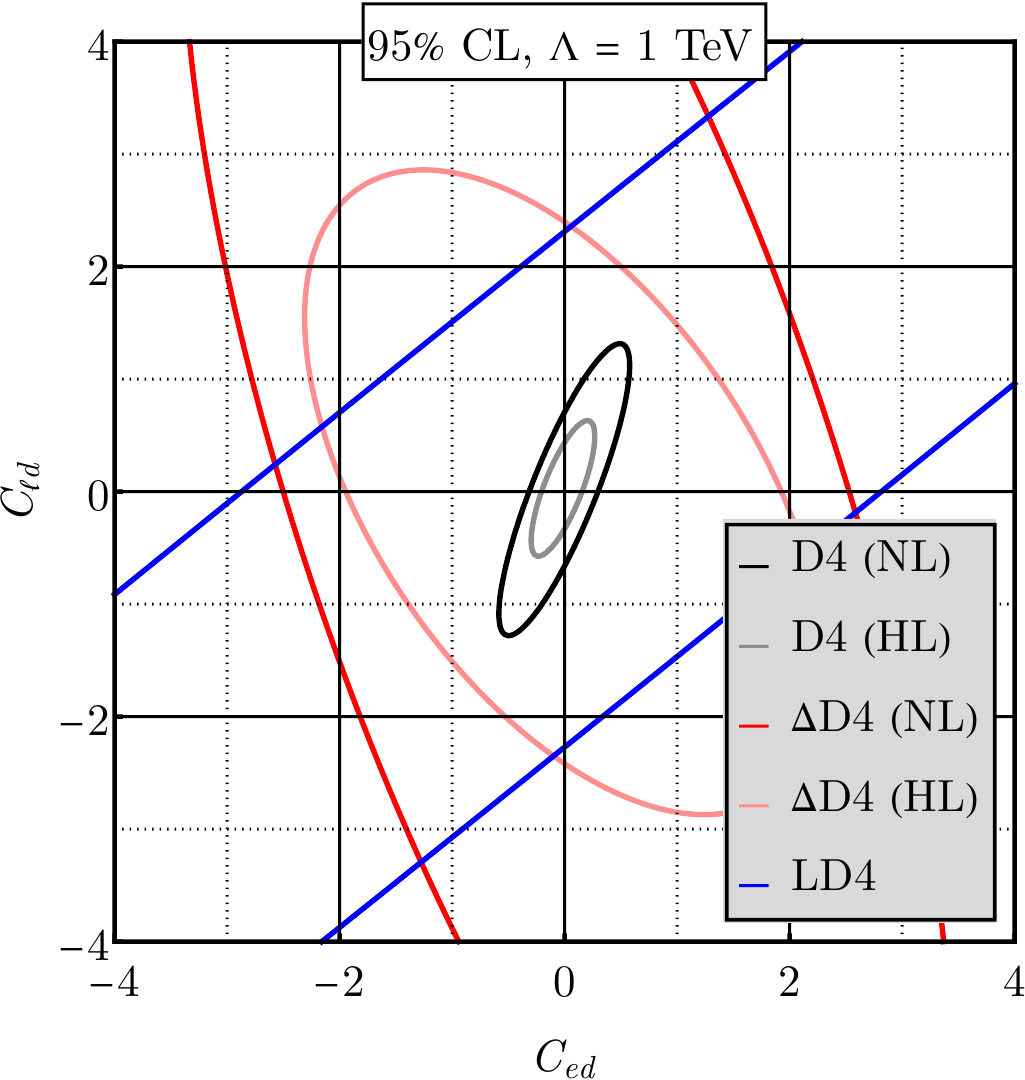}
	\includegraphics[width=\elliwi\textwidth]{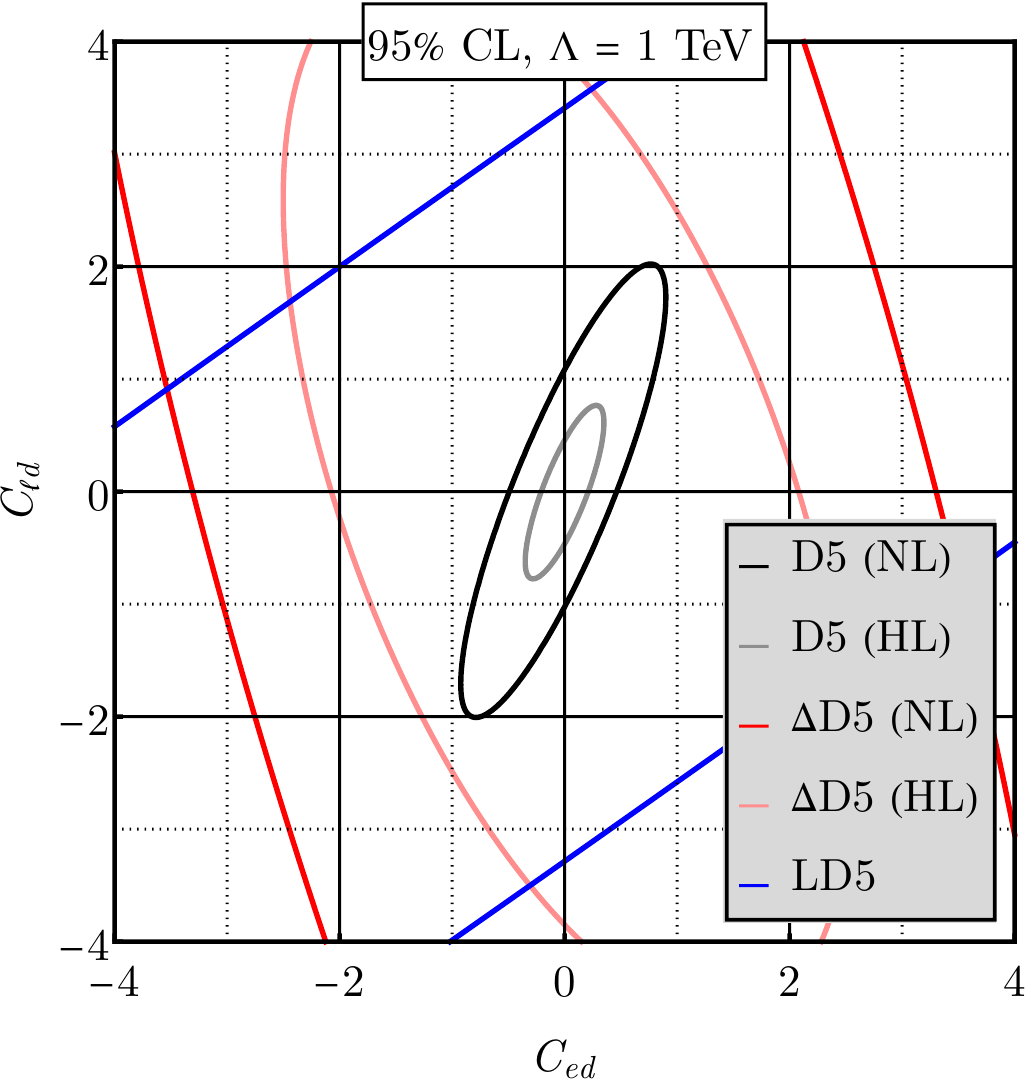}
	\includegraphics[width=\elliwi\textwidth]{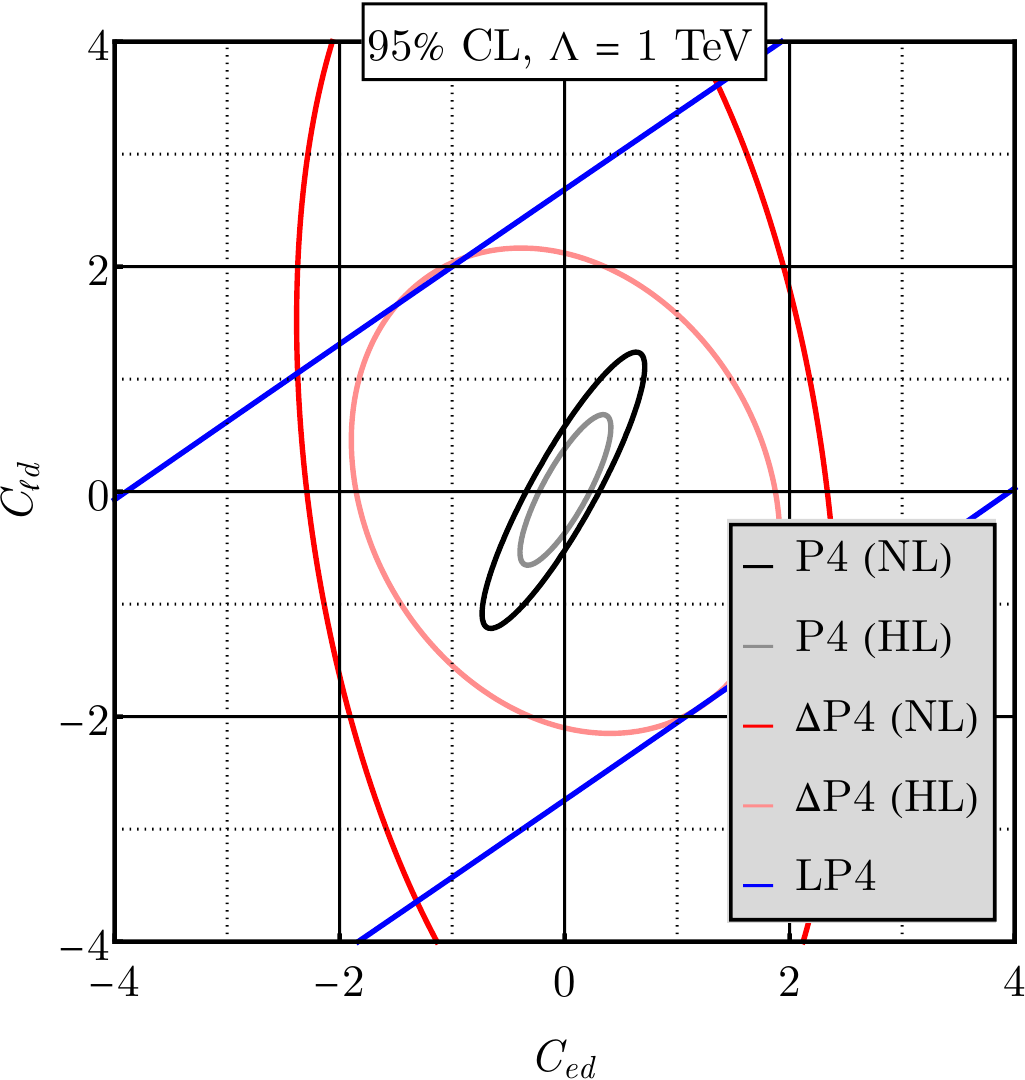}
	\includegraphics[width=\elliwi\textwidth]{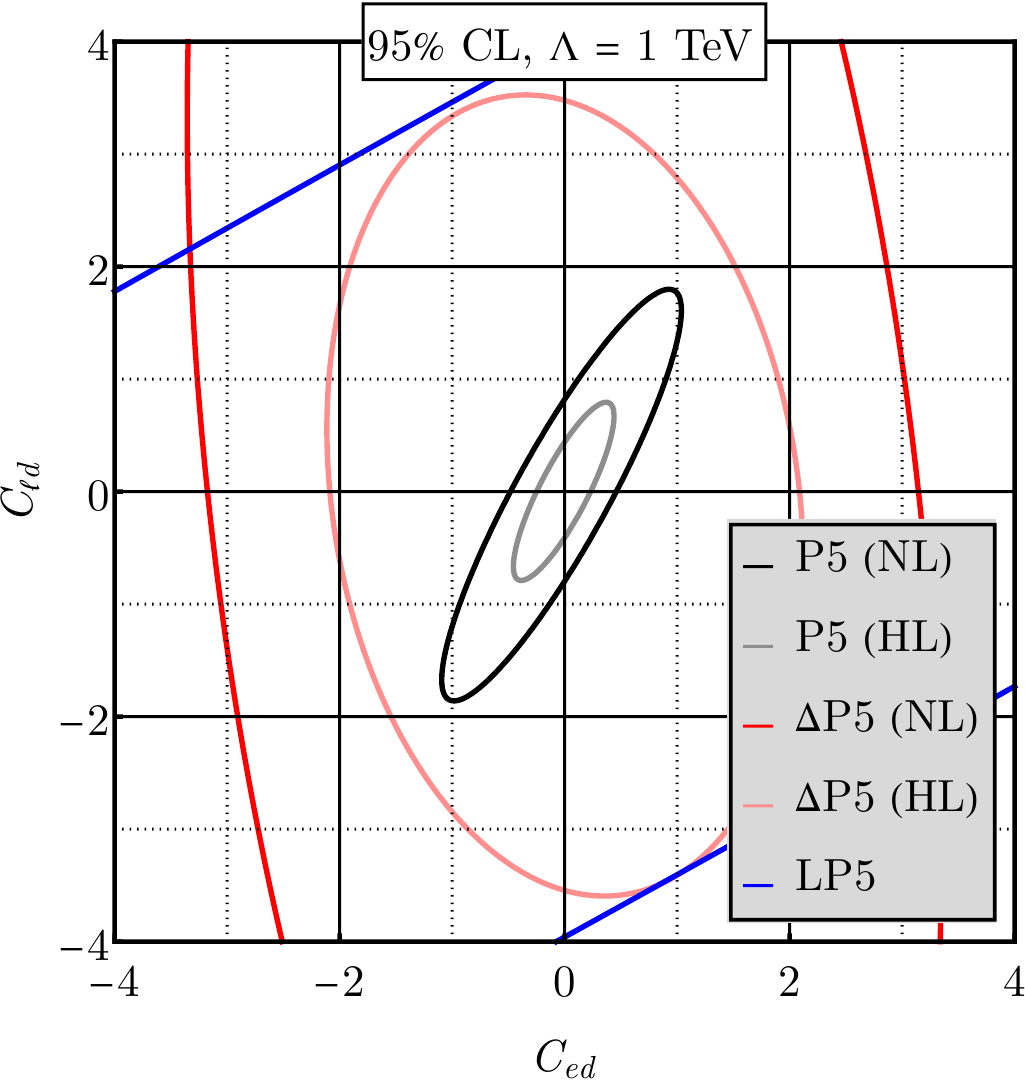}
	\caption{The same as in Fig.~\ref{fig:all-ellipses-Ceu-Ced} but for $\Ced$ and $\Cld$.}
	\label{fig:all-ellipses-Ced-Cld}
\end{figure}

\begin{figure}
	[H]\centering
	\includegraphics[width=\elliwi\textwidth]{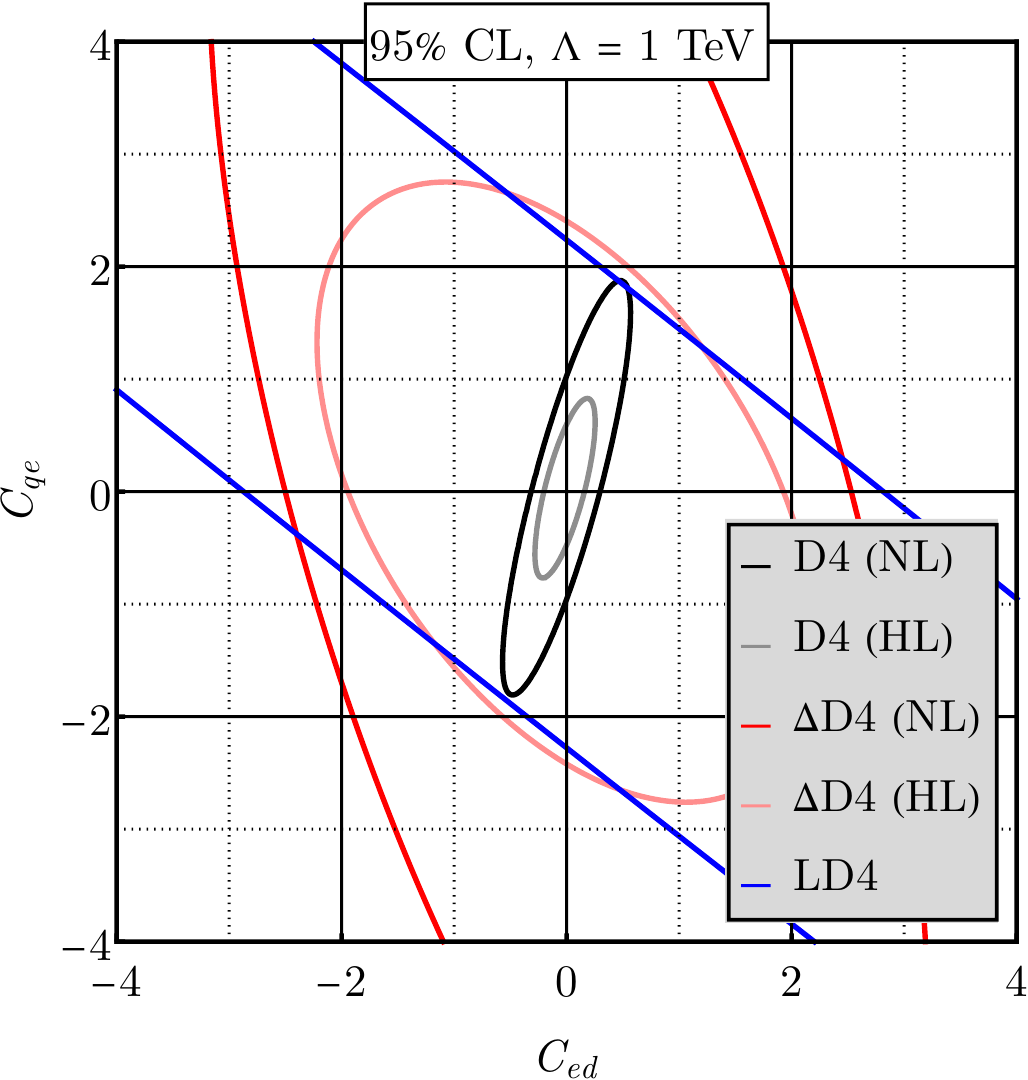}
	\includegraphics[width=\elliwi\textwidth]{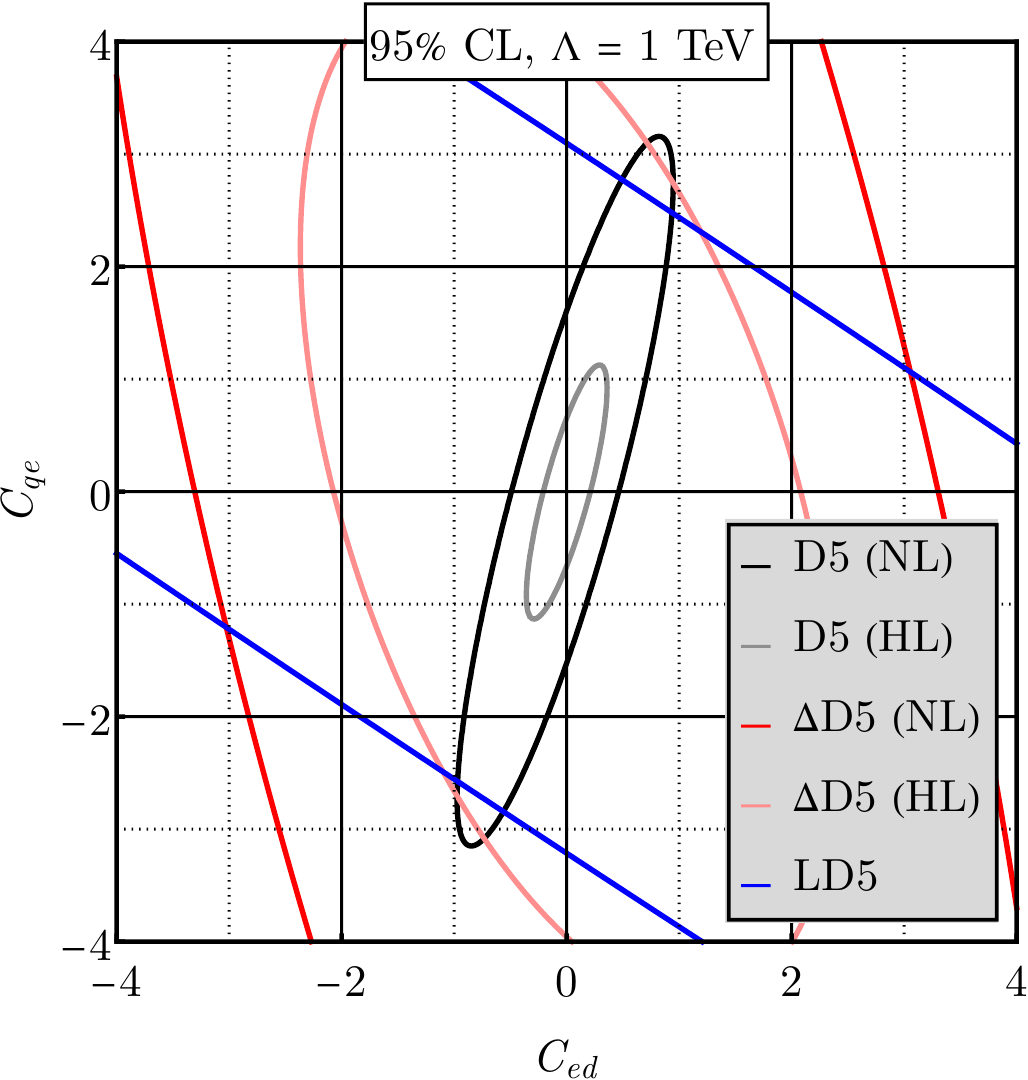}
	\includegraphics[width=\elliwi\textwidth]{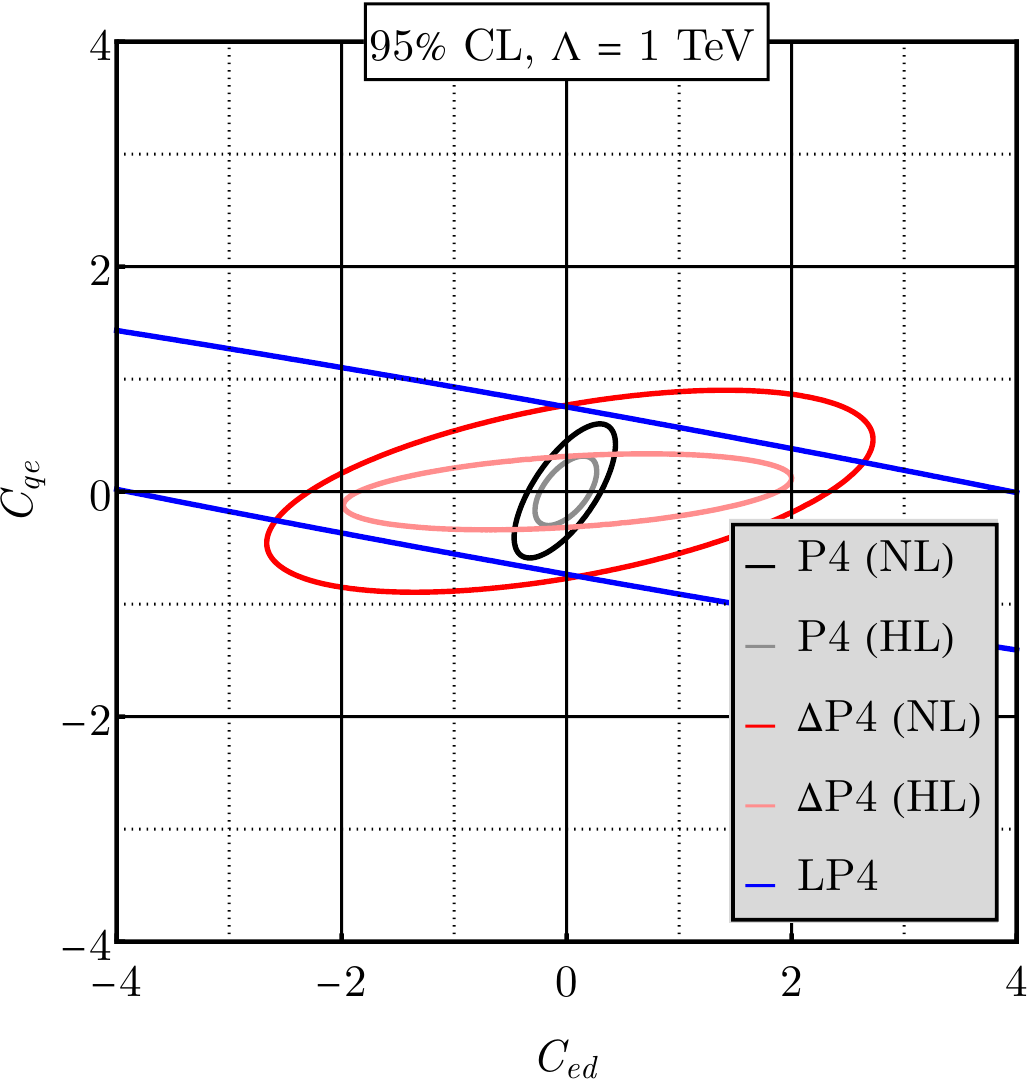}
	\includegraphics[width=\elliwi\textwidth]{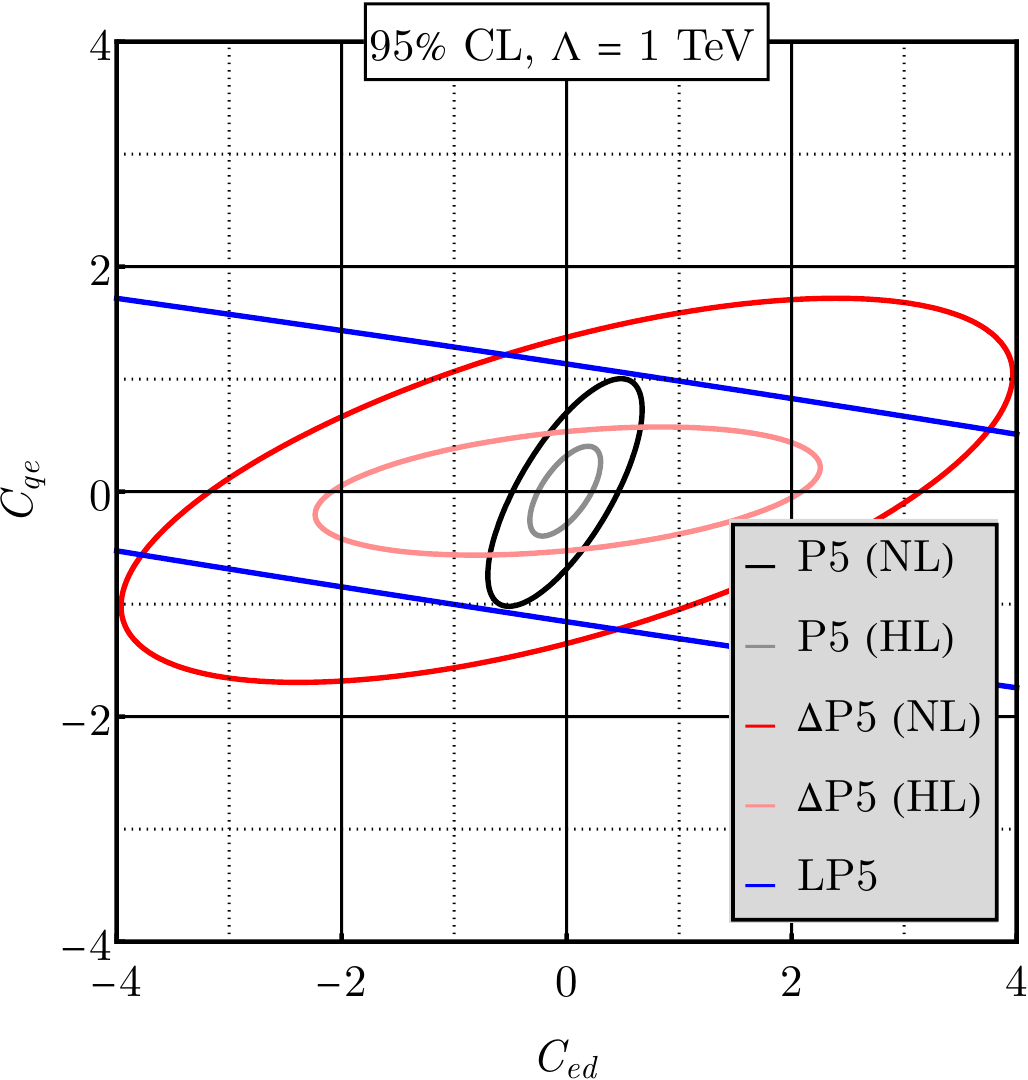}
	\caption{The same as in Fig.~\ref{fig:all-ellipses-Ceu-Ced} but for $\Ced$ and $\Cqe$.}
	\label{fig:all-ellipses-Ced-Cqe}
\end{figure}

\begin{figure}
	[H]\centering
	\includegraphics[width=\elliwi\textwidth]{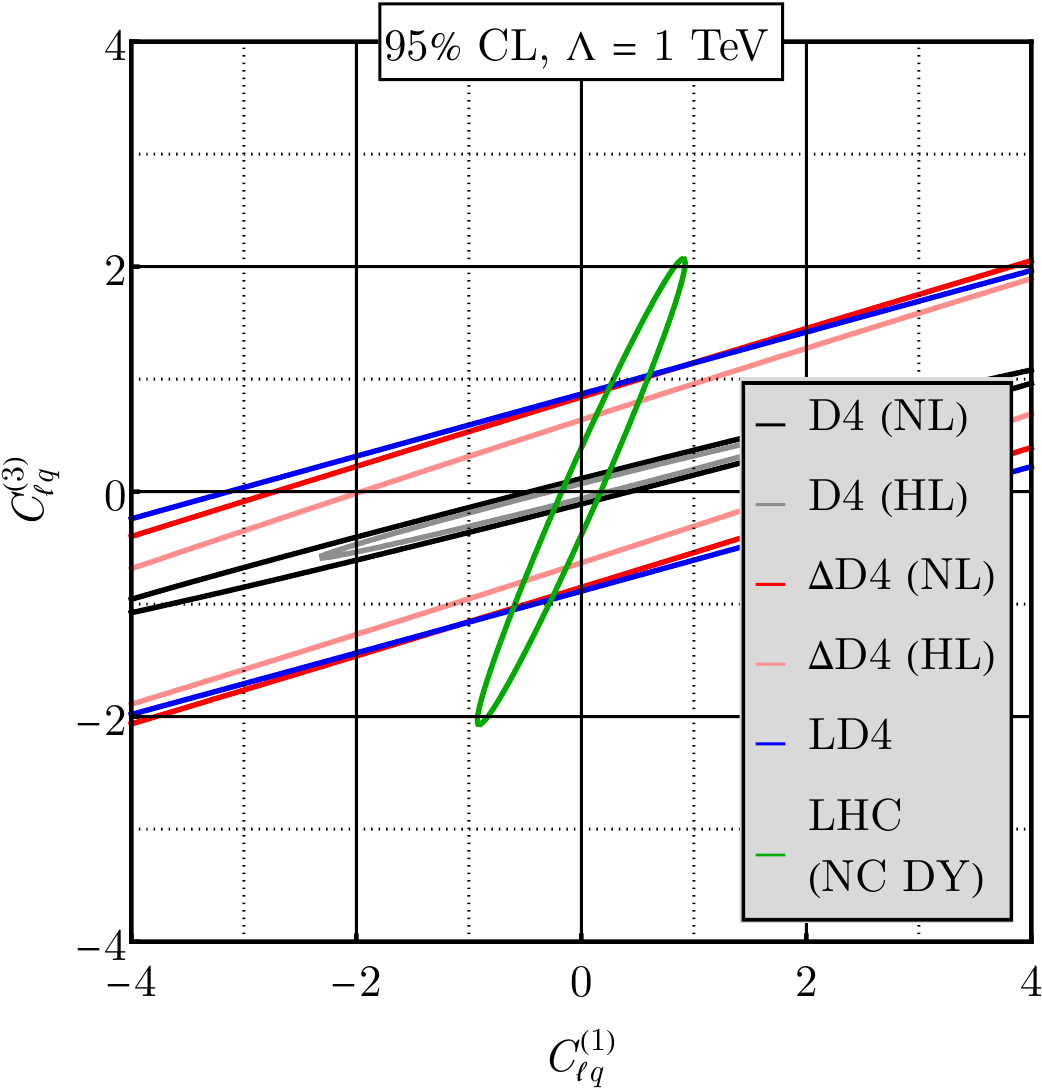}
	\includegraphics[width=\elliwi\textwidth]{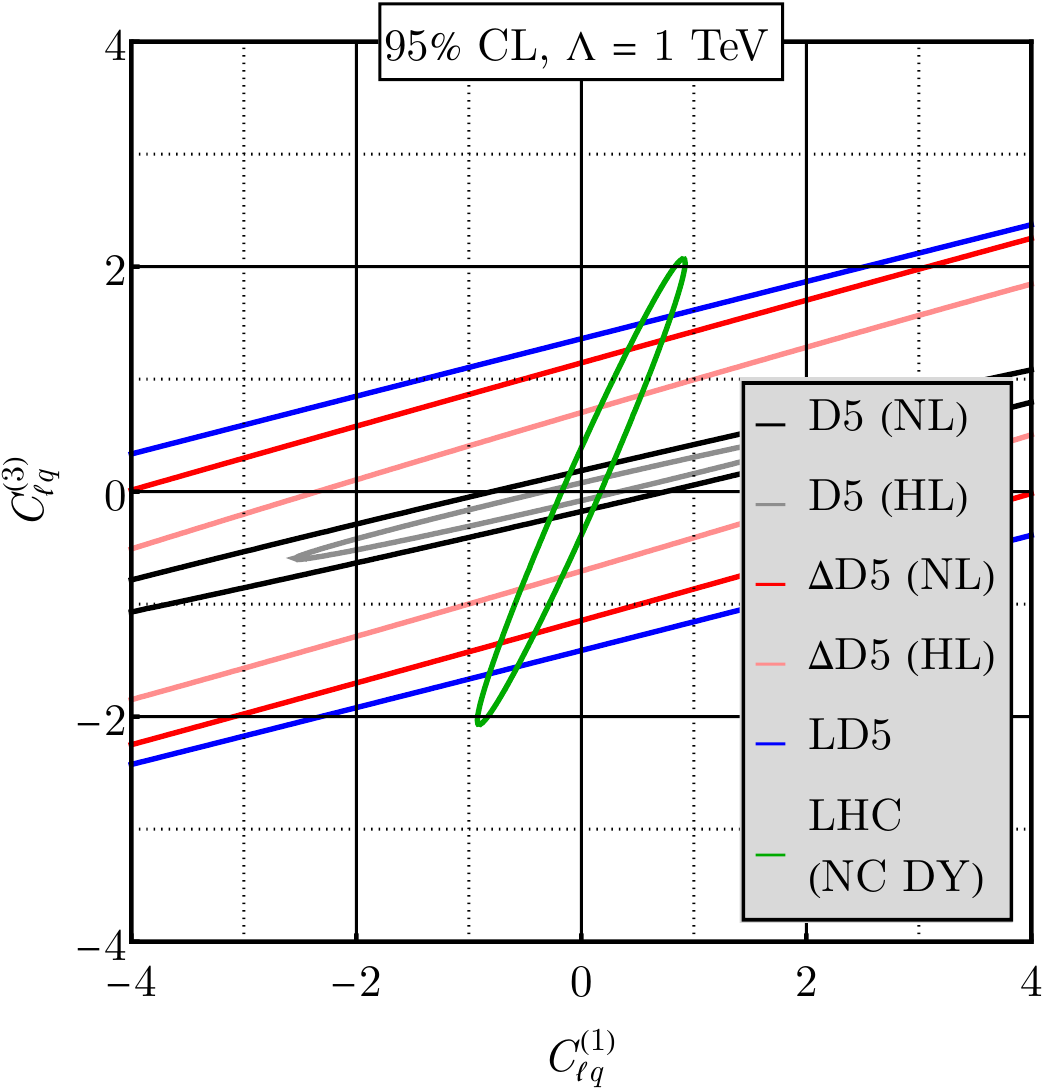}
	\includegraphics[width=\elliwi\textwidth]{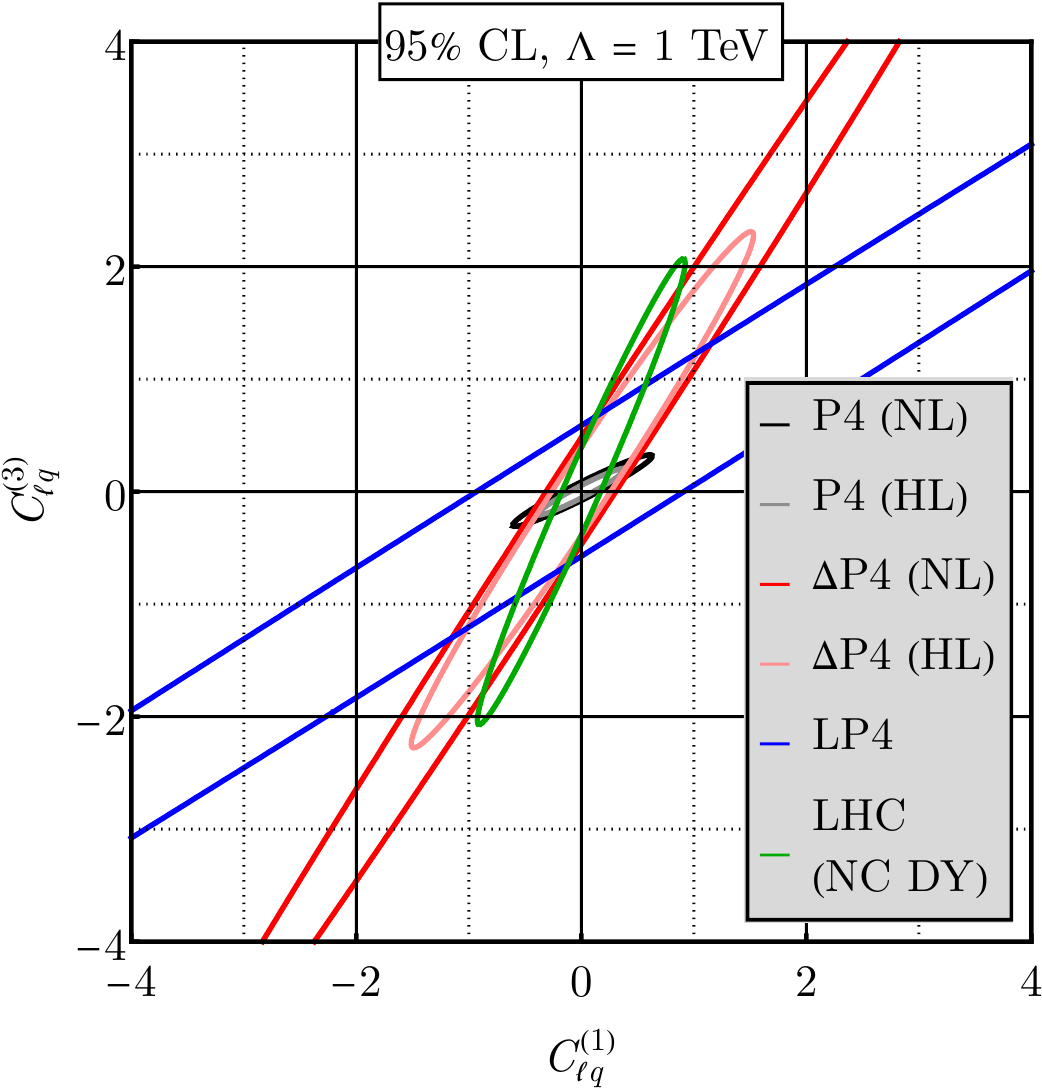}
	\includegraphics[width=\elliwi\textwidth]{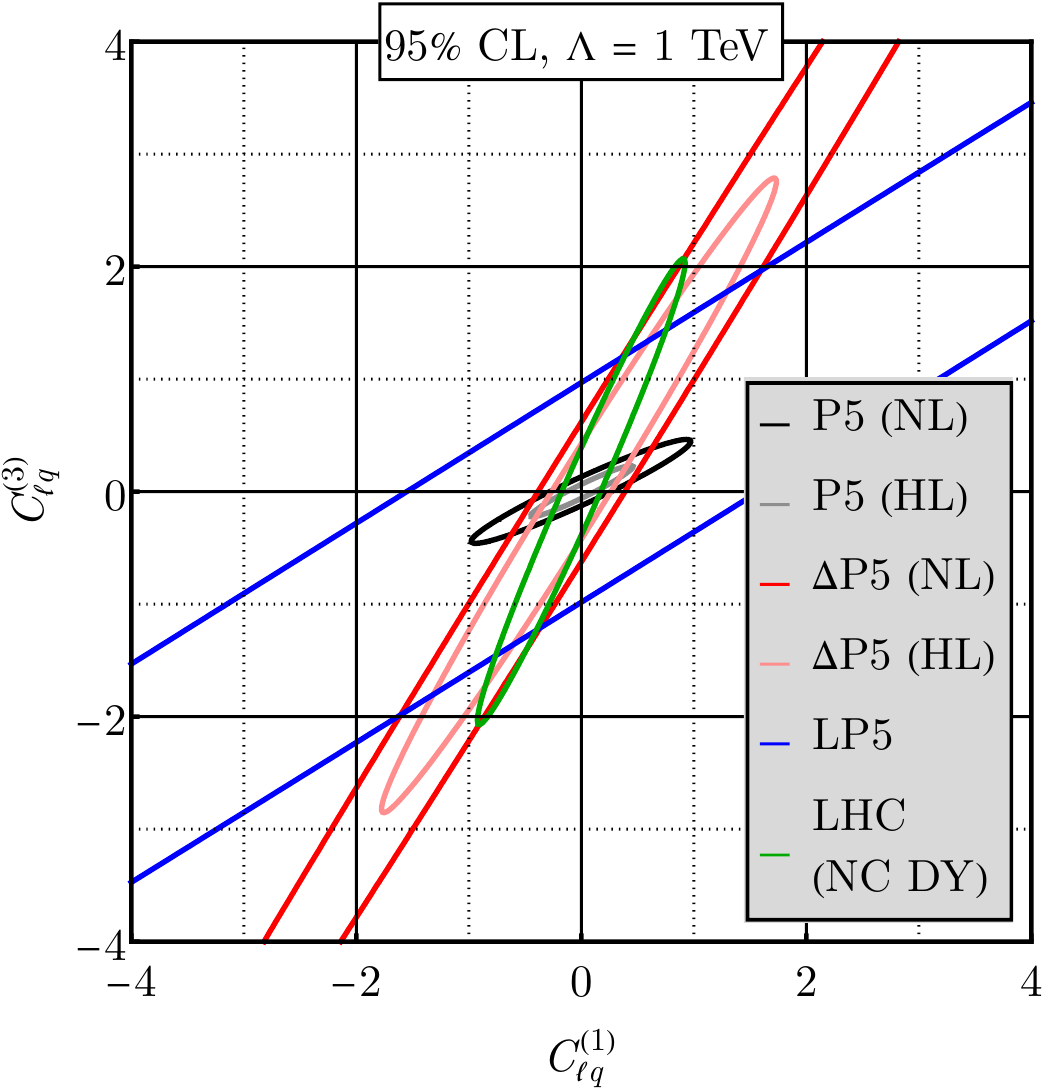}
	\caption{The same as in Fig.~\ref{fig:all-ellipses-Ceu-Ced} but for $\Clqi$ and $\Clqiii$.}
	\label{fig:all-ellipses-Clq1-Clq3}
\end{figure}

\begin{figure}
	[H]\centering
	\includegraphics[width=\elliwi\textwidth]{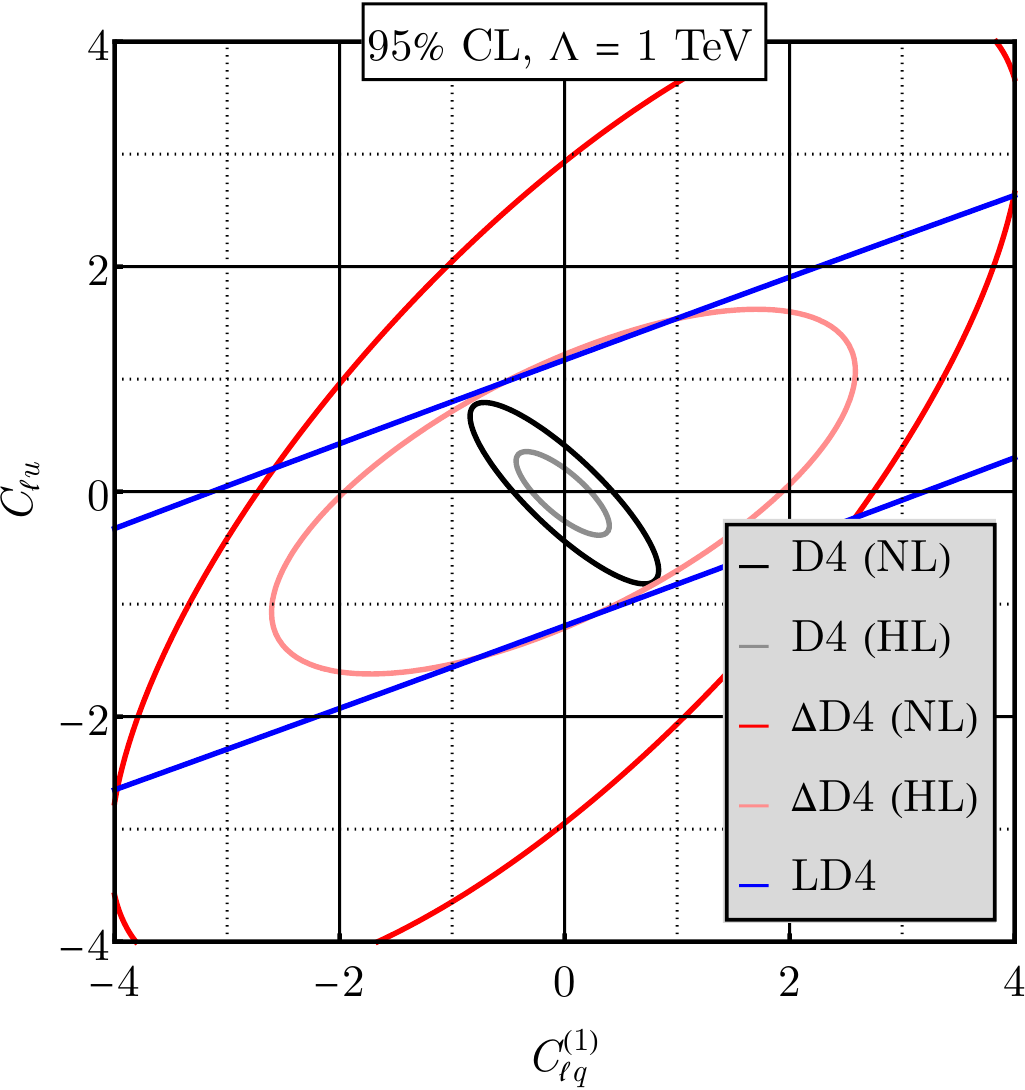}
	\includegraphics[width=\elliwi\textwidth]{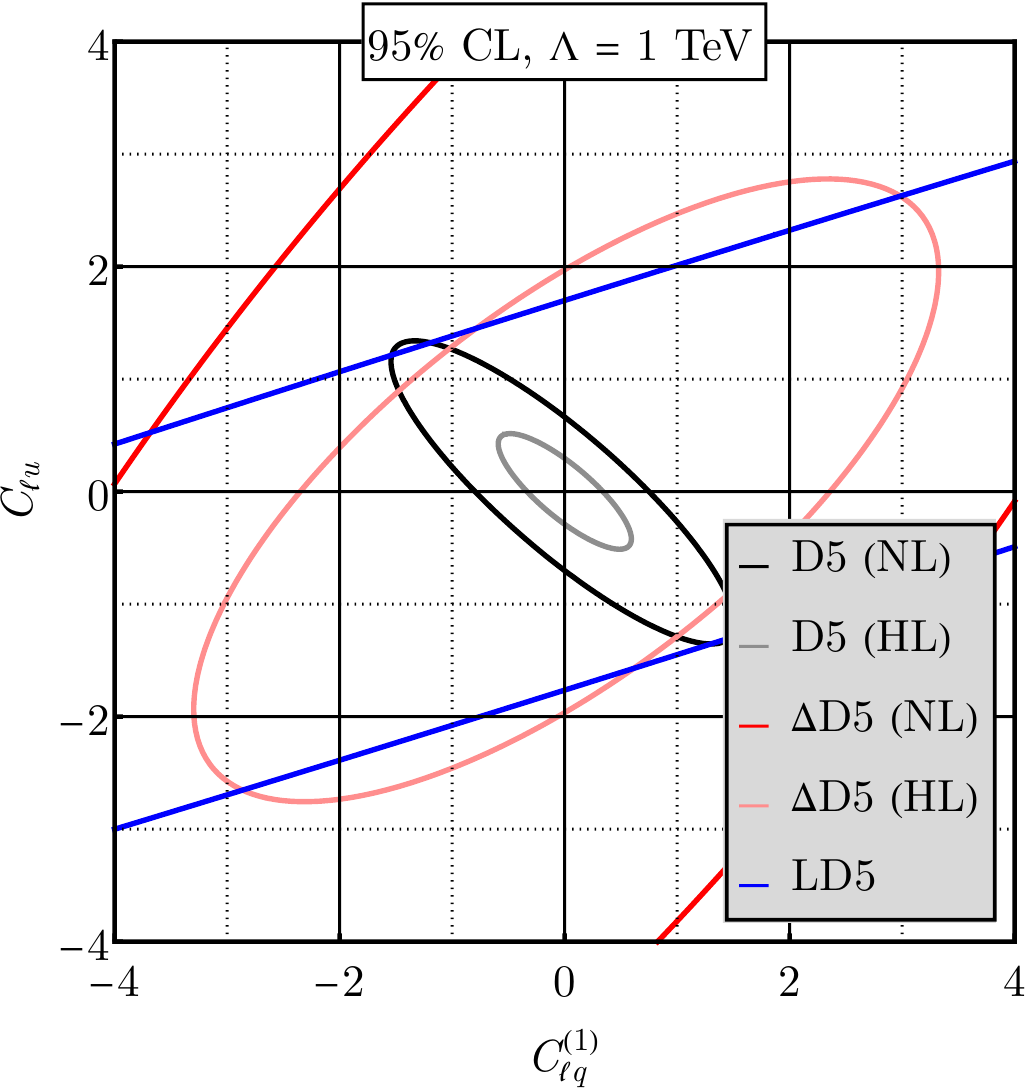}
	\includegraphics[width=\elliwi\textwidth]{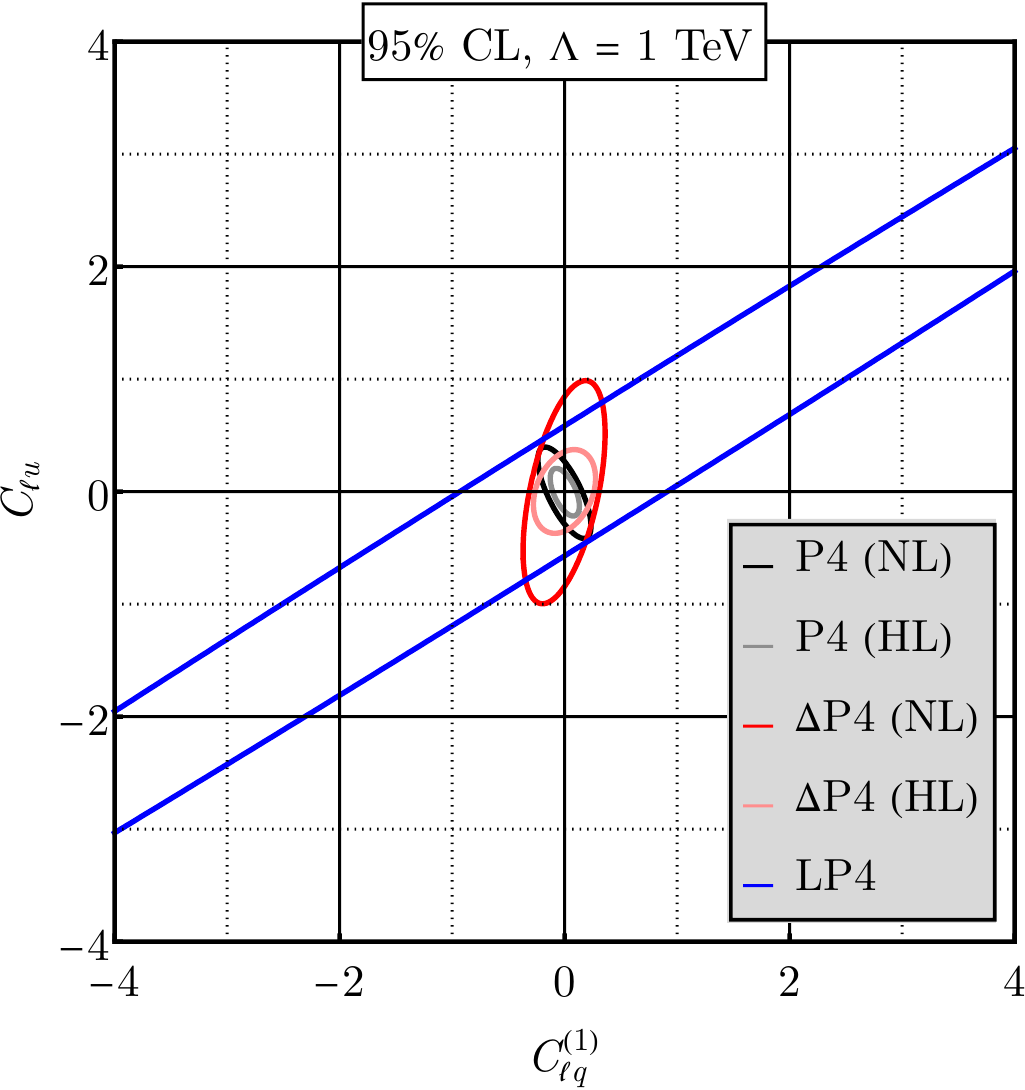}
	\includegraphics[width=\elliwi\textwidth]{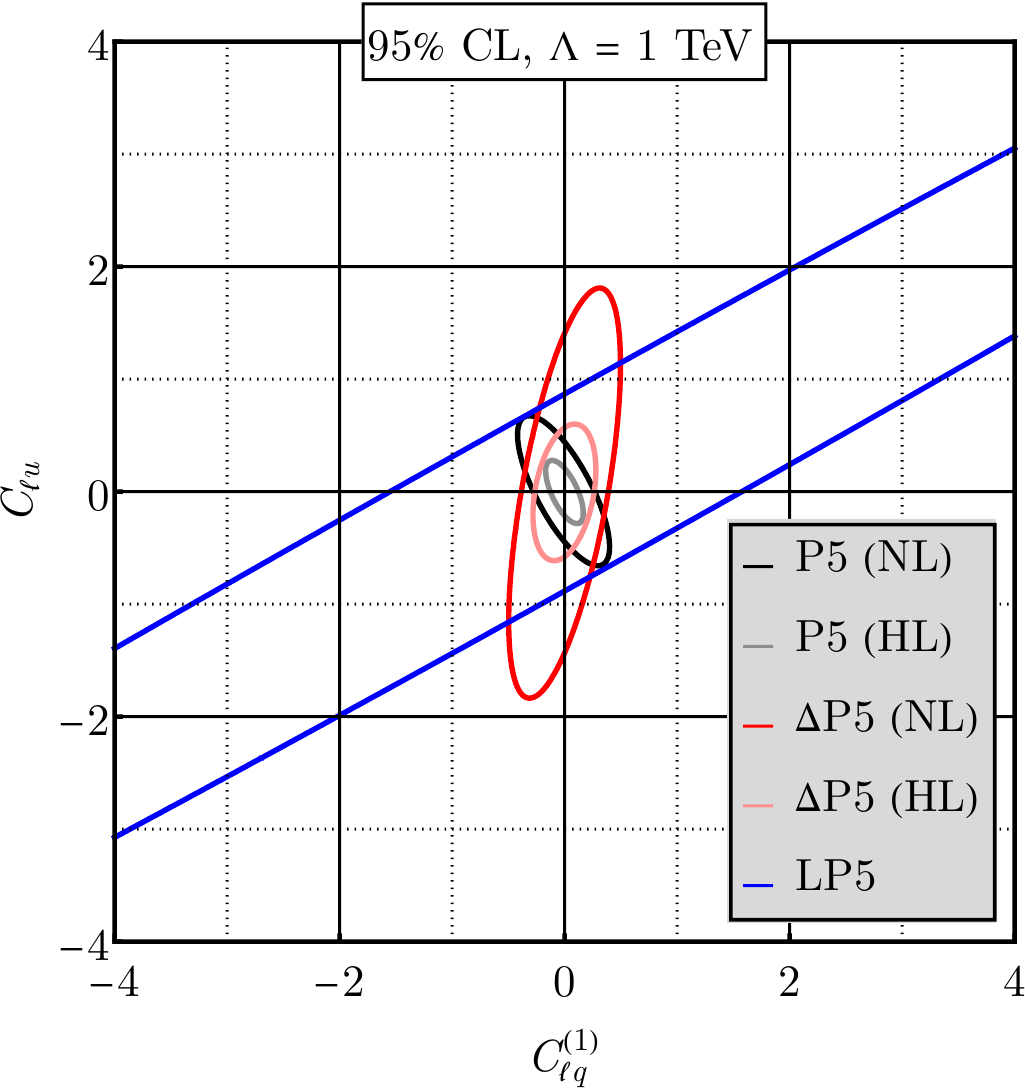}
	\caption{The same as in Fig.~\ref{fig:all-ellipses-Ceu-Ced} but for $\Clqi$ and $\Clu$.}
	\label{fig:all-ellipses-Clq1-Clu}
\end{figure}

\begin{figure}
	[H]\centering
	\includegraphics[width=\elliwi\textwidth]{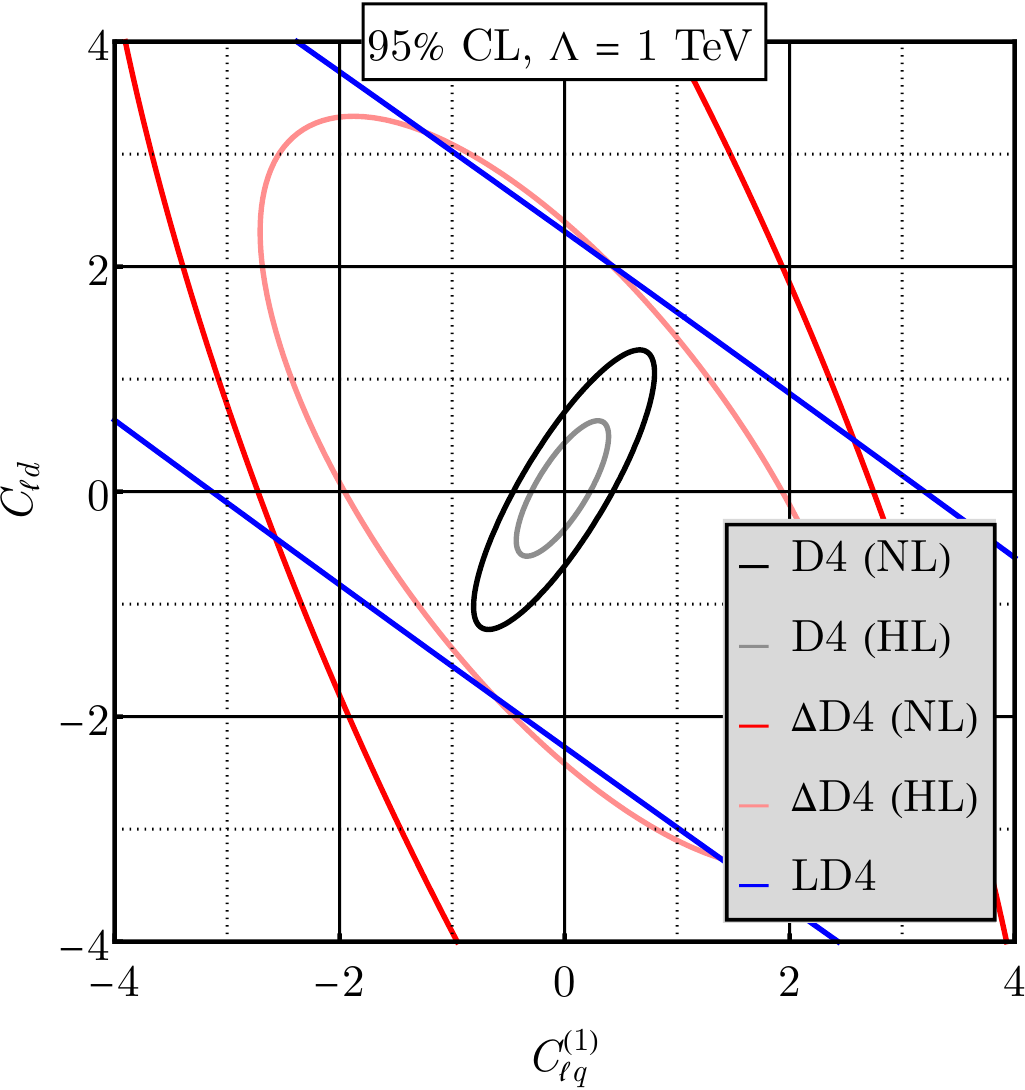}
	\includegraphics[width=\elliwi\textwidth]{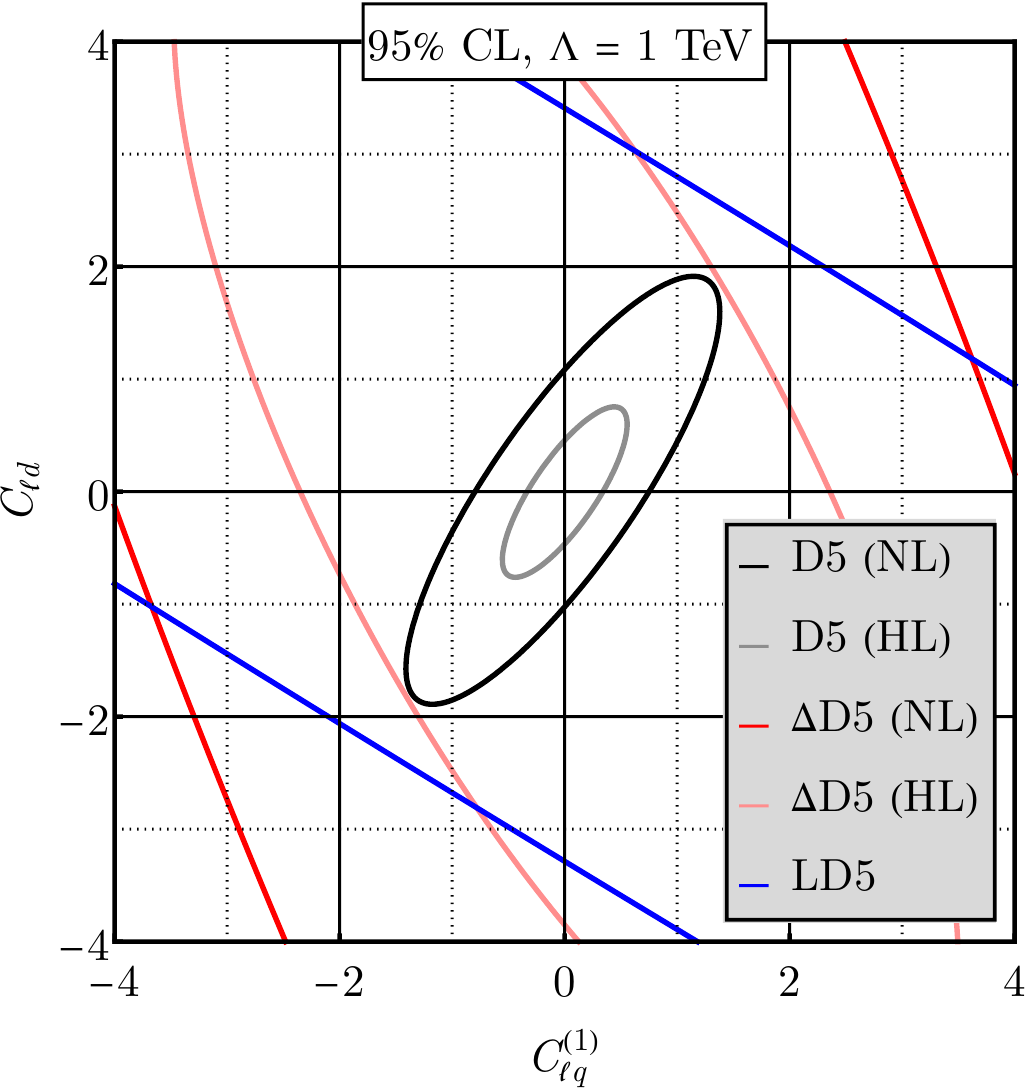}
	\includegraphics[width=\elliwi\textwidth]{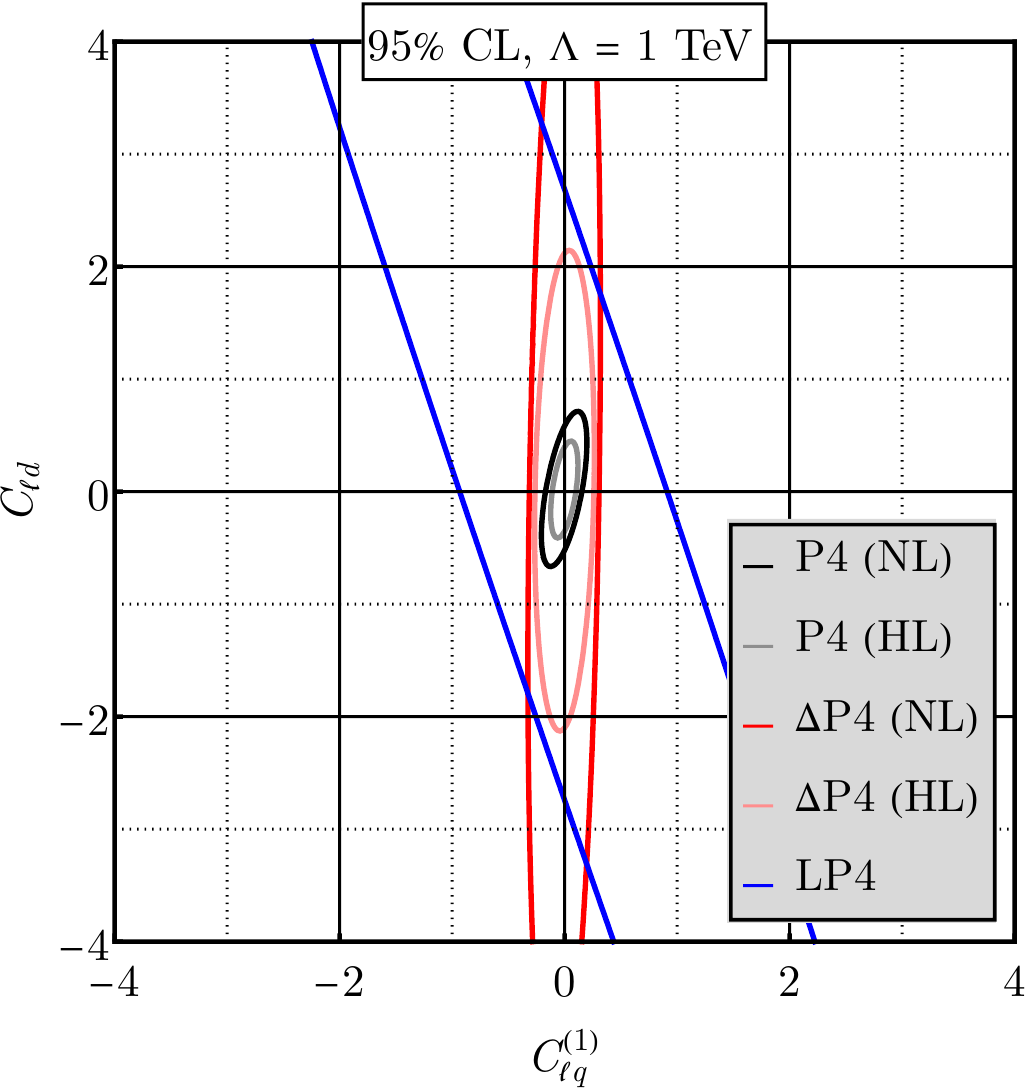}
	\includegraphics[width=\elliwi\textwidth]{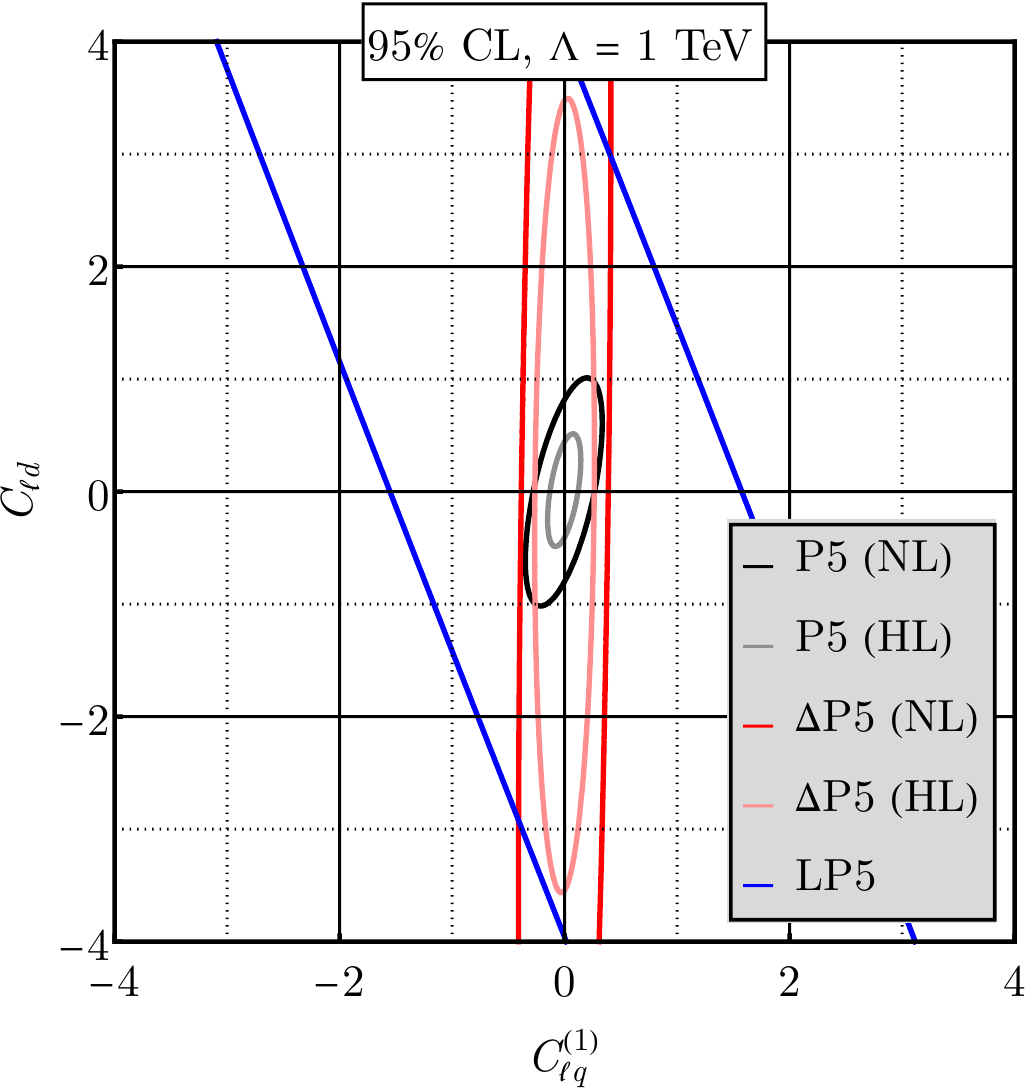}
	\caption{The same as in Fig.~\ref{fig:all-ellipses-Ceu-Ced} but for $\Clqi$ and $\Cld$.}
	\label{fig:all-ellipses-Clq1-Cld}
\end{figure}

\begin{figure}
	[H]\centering
	\includegraphics[width=\elliwi\textwidth]{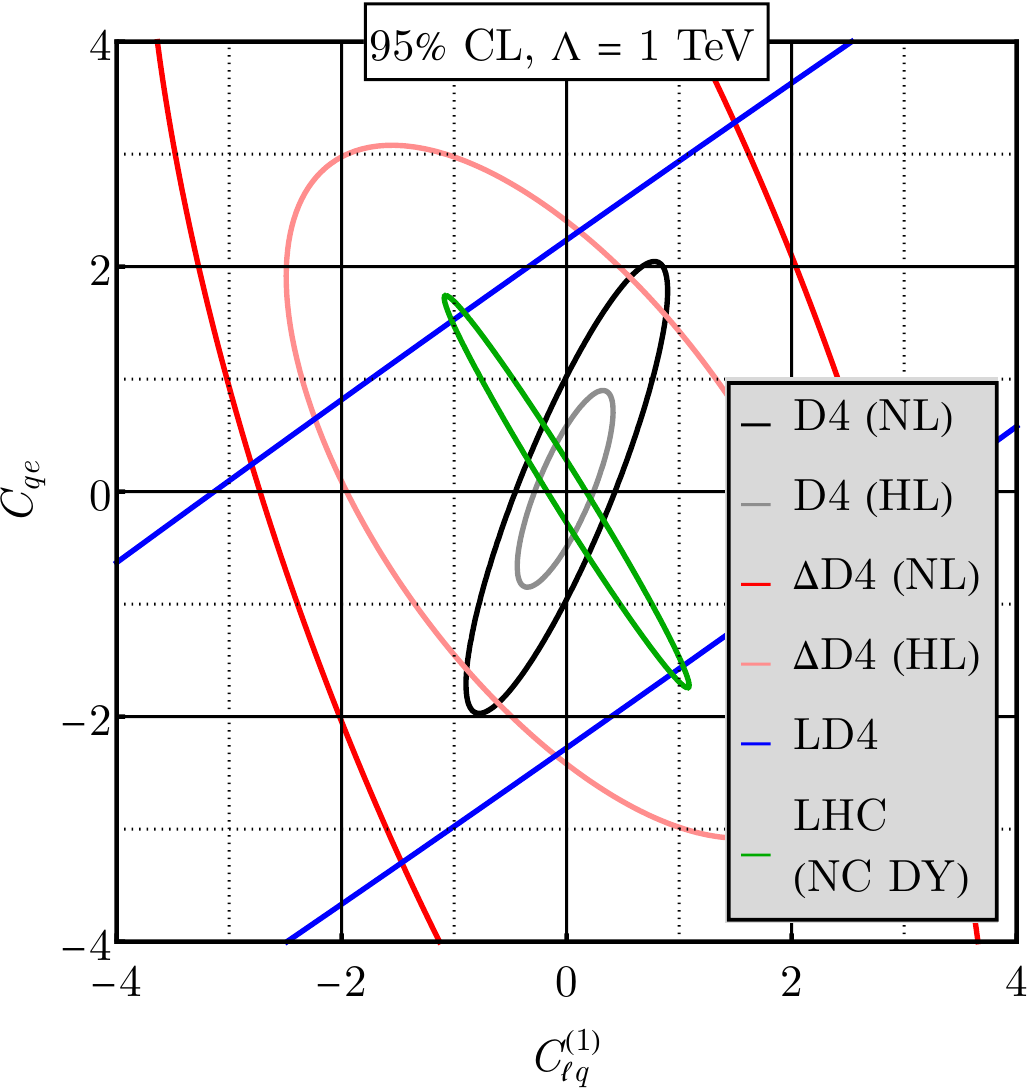}
	\includegraphics[width=\elliwi\textwidth]{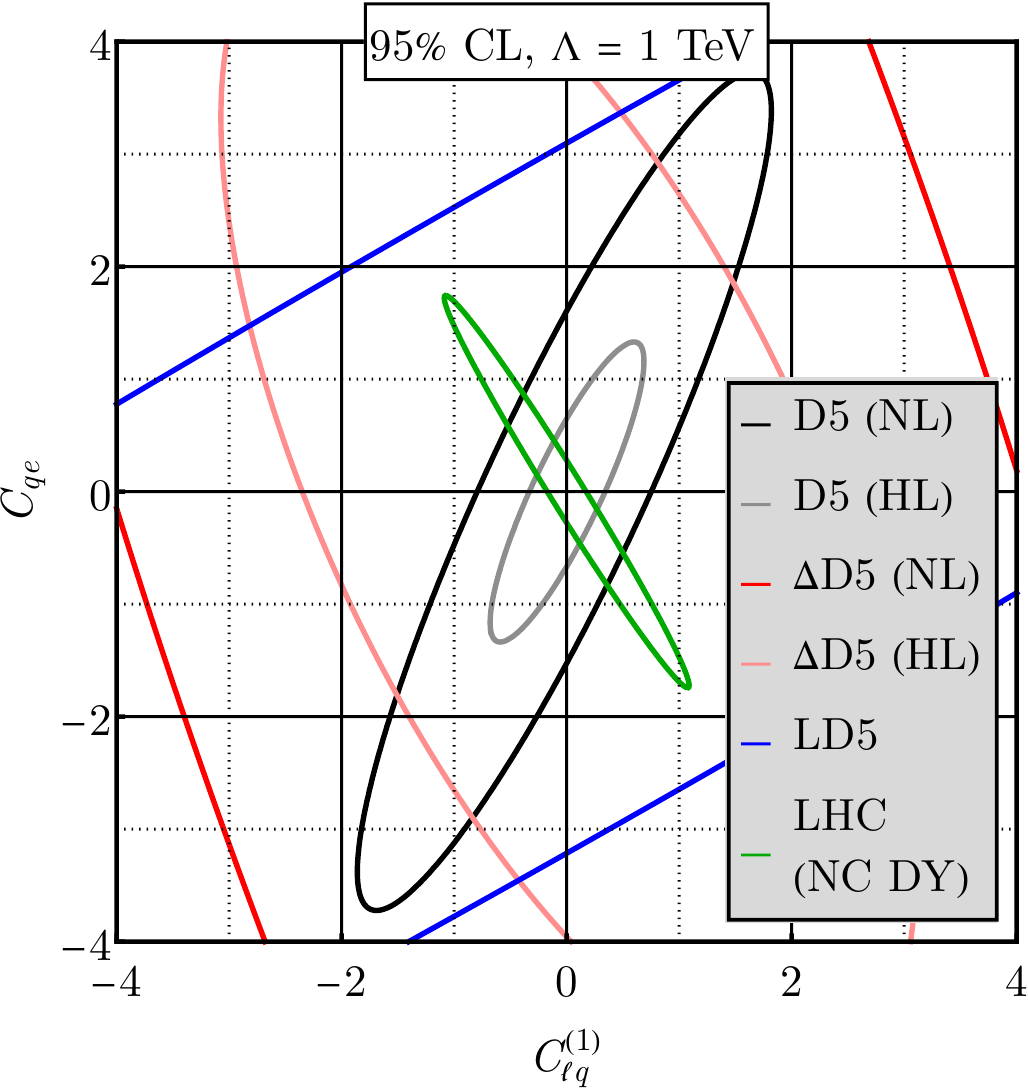}
	\includegraphics[width=\elliwi\textwidth]{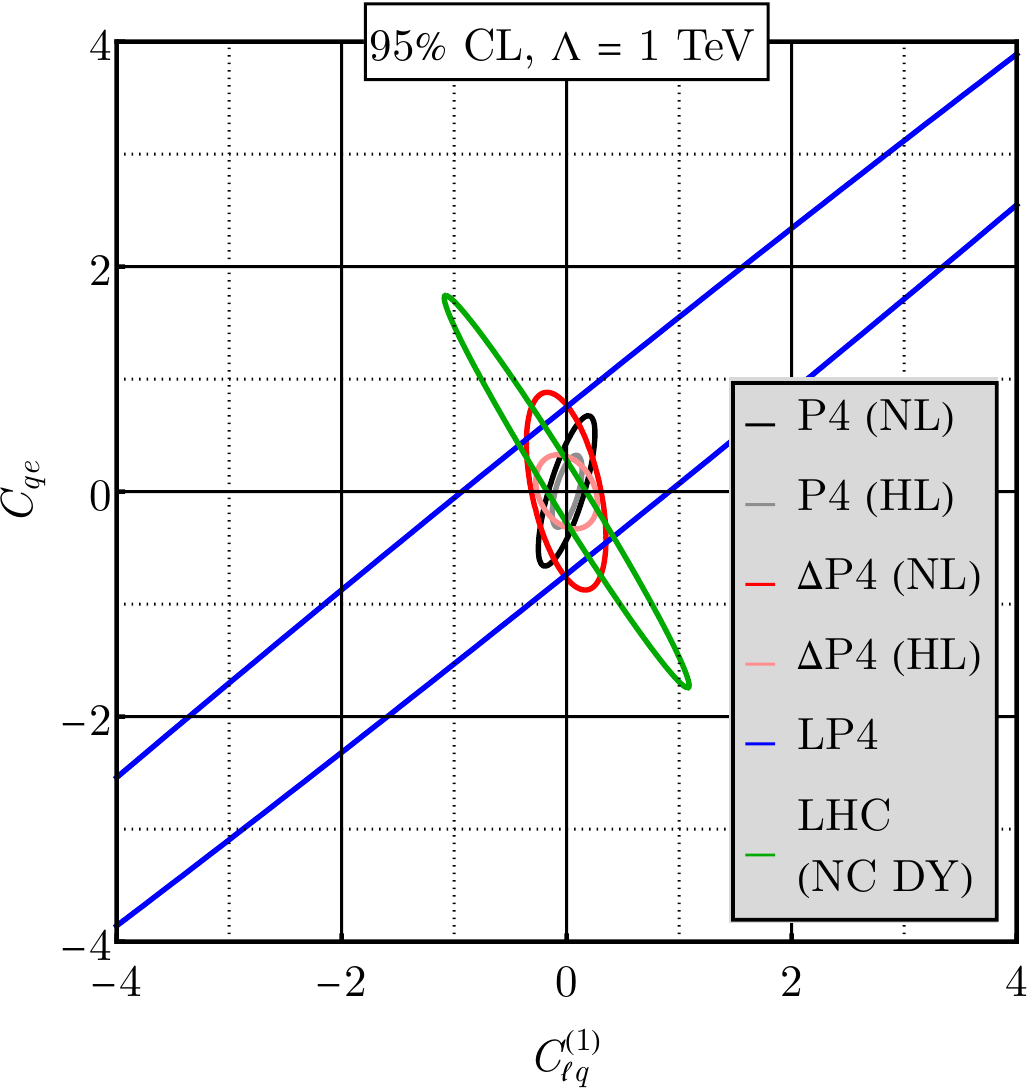}
	\includegraphics[width=\elliwi\textwidth]{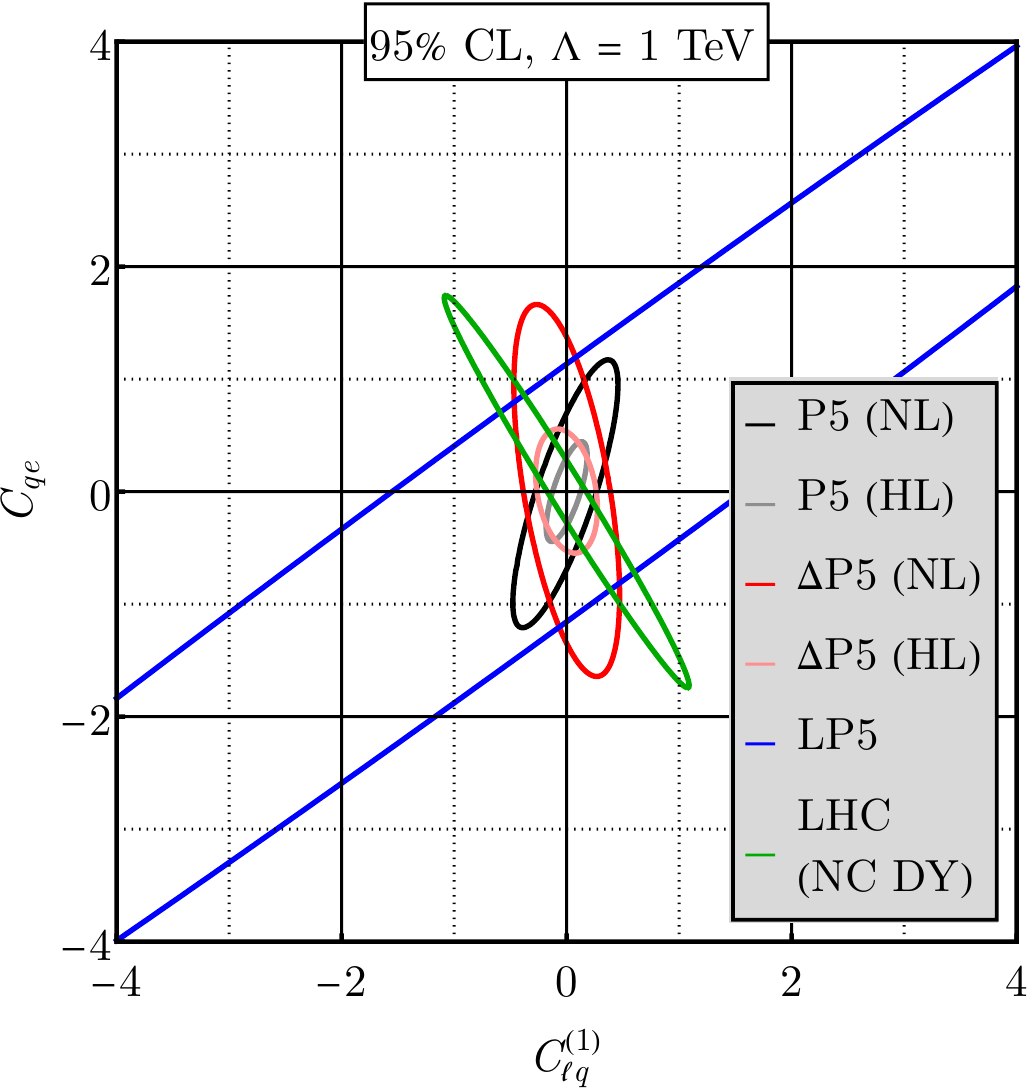}
	\caption{The same as in Fig.~\ref{fig:all-ellipses-Ceu-Ced} but for $\Clqi$ and $\Cqe$.}
	\label{fig:all-ellipses-Clq1-Cqe}
\end{figure}

\begin{figure}
	[H]\centering
	\includegraphics[width=\elliwi\textwidth]{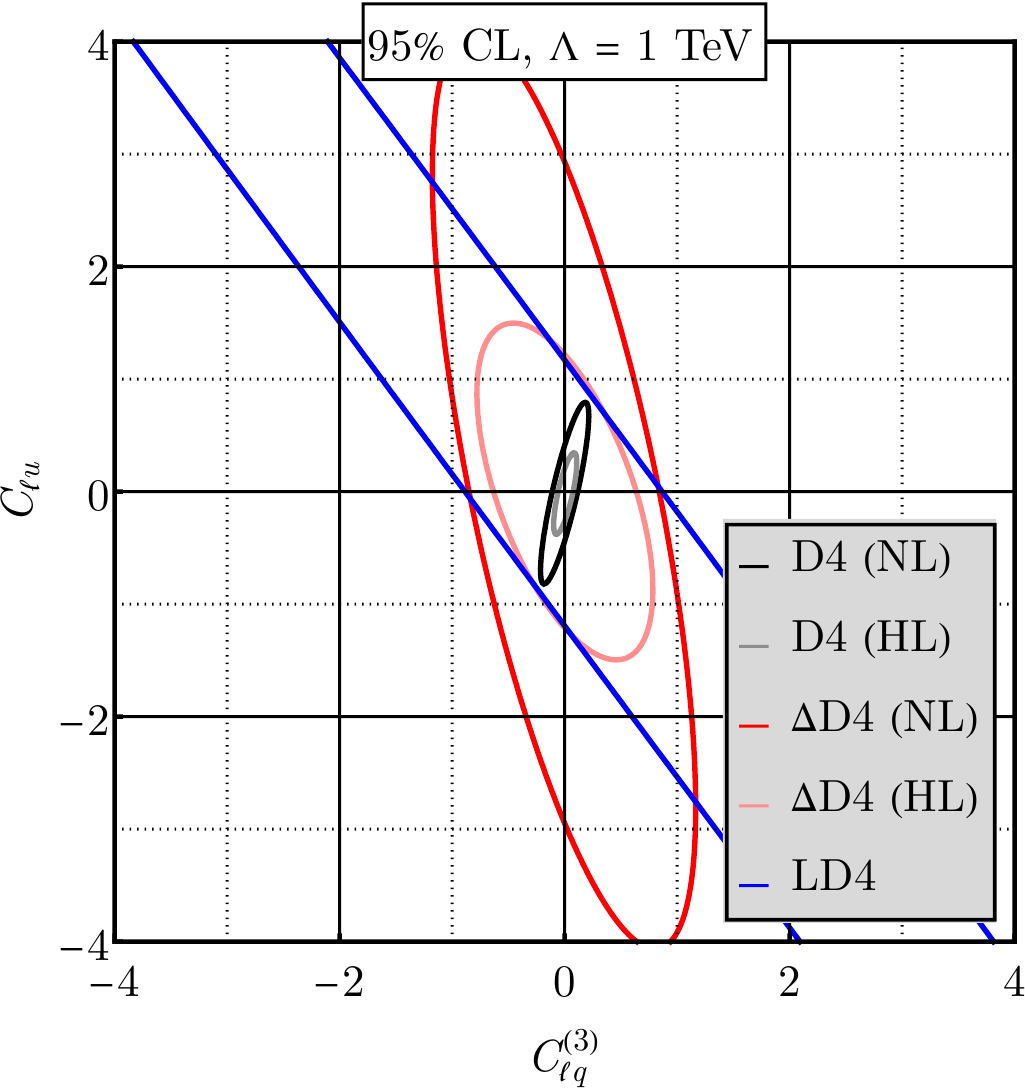}
	\includegraphics[width=\elliwi\textwidth]{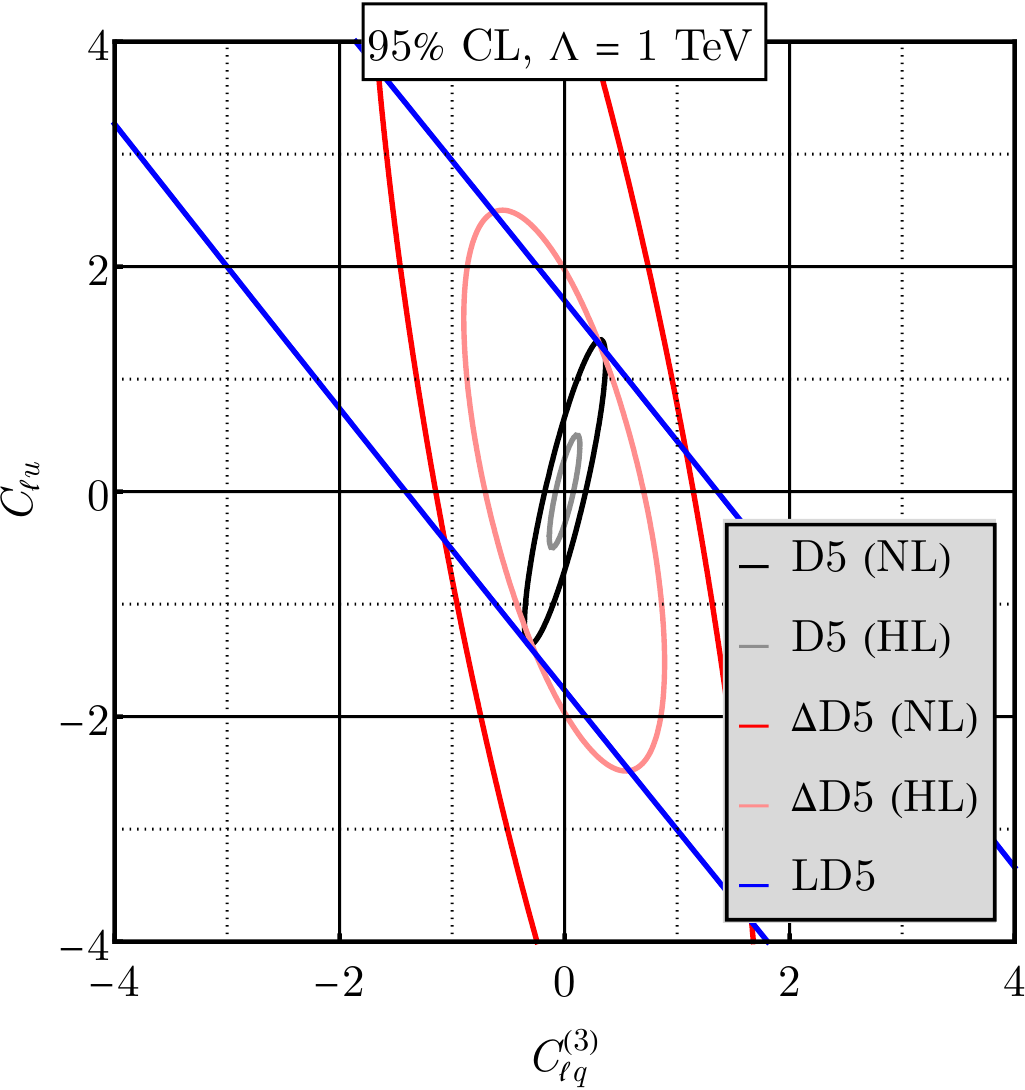}
	\includegraphics[width=\elliwi\textwidth]{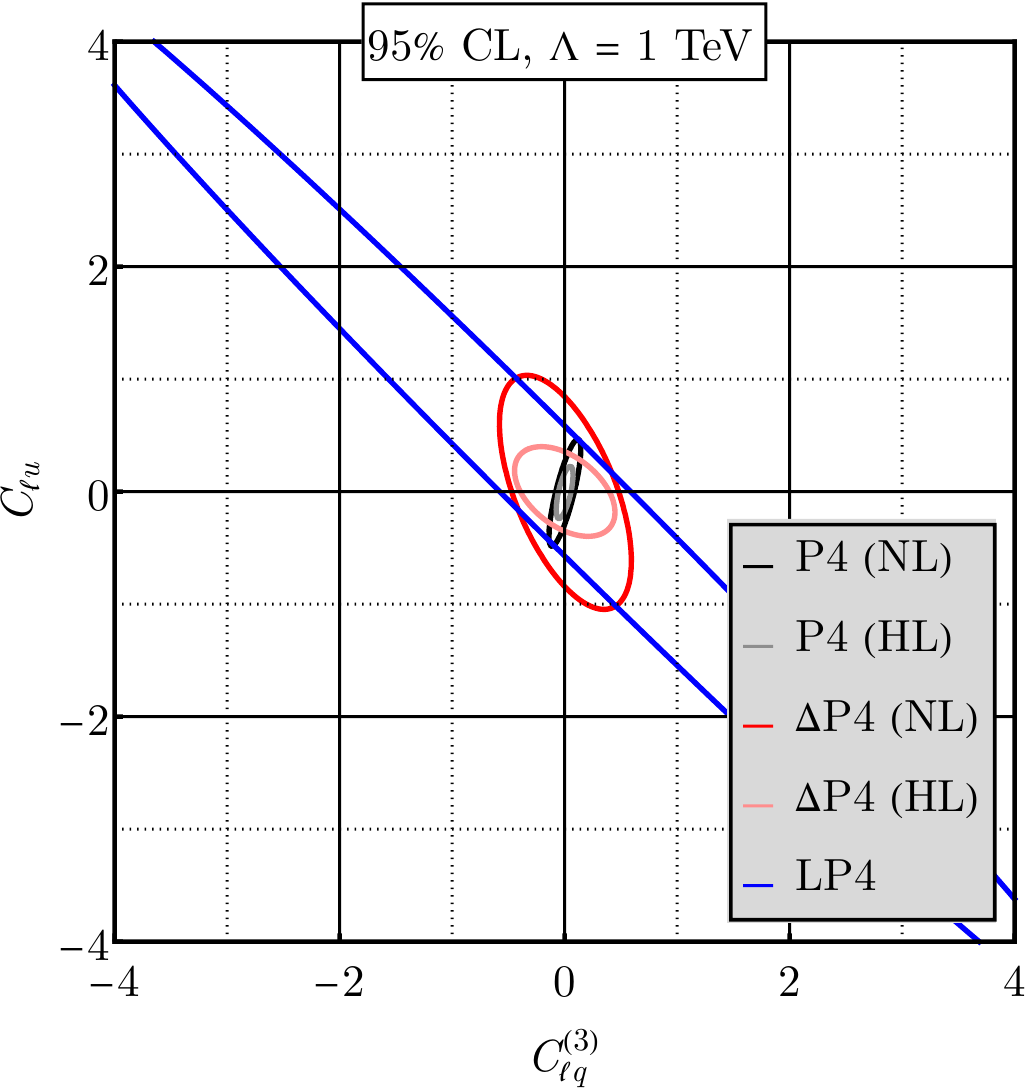}
	\includegraphics[width=\elliwi\textwidth]{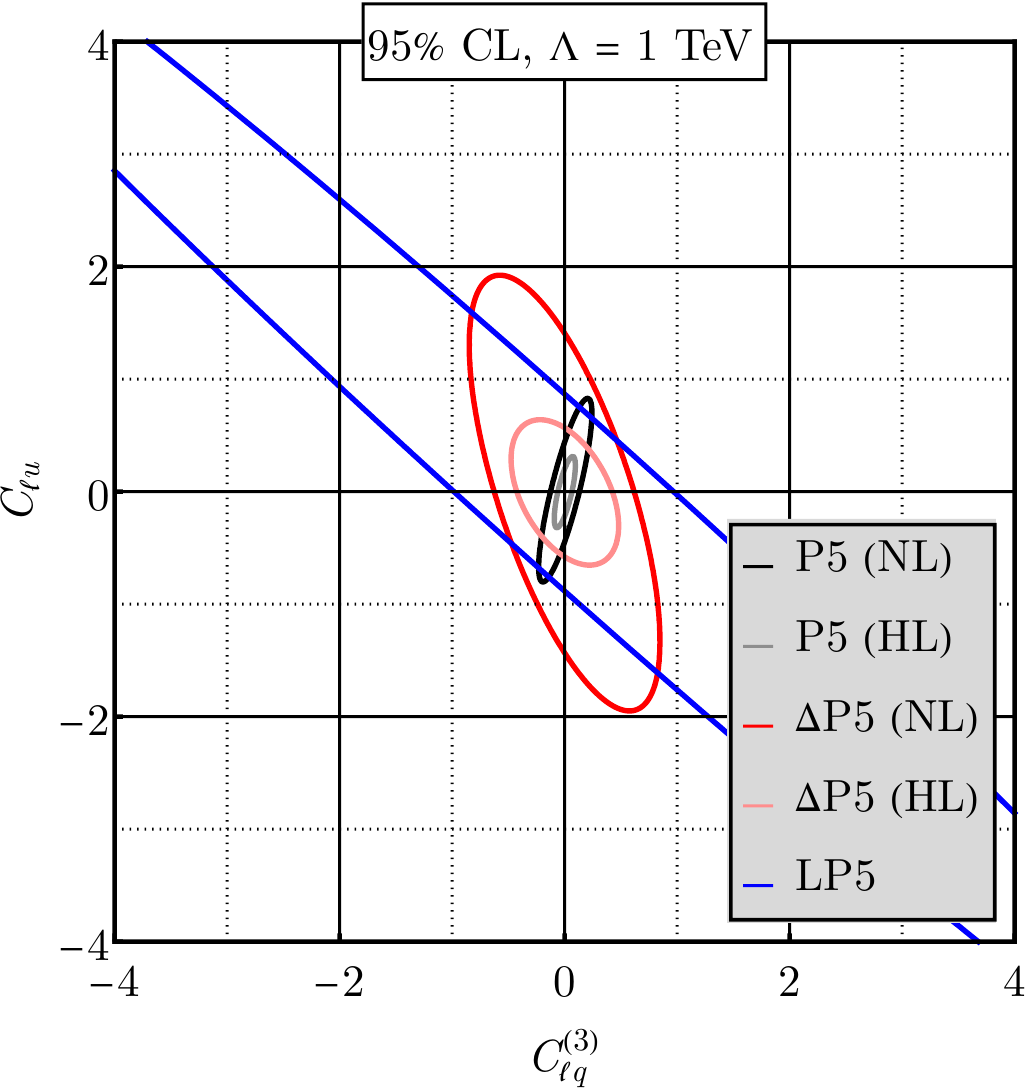}
	\caption{The same as in Fig.~\ref{fig:all-ellipses-Ceu-Ced} but for $\Clqiii$ and $\Clu$.}
	\label{fig:all-ellipses-Clq3-Clu}
\end{figure}

\begin{figure}
	[H]\centering
	\includegraphics[width=\elliwi\textwidth]{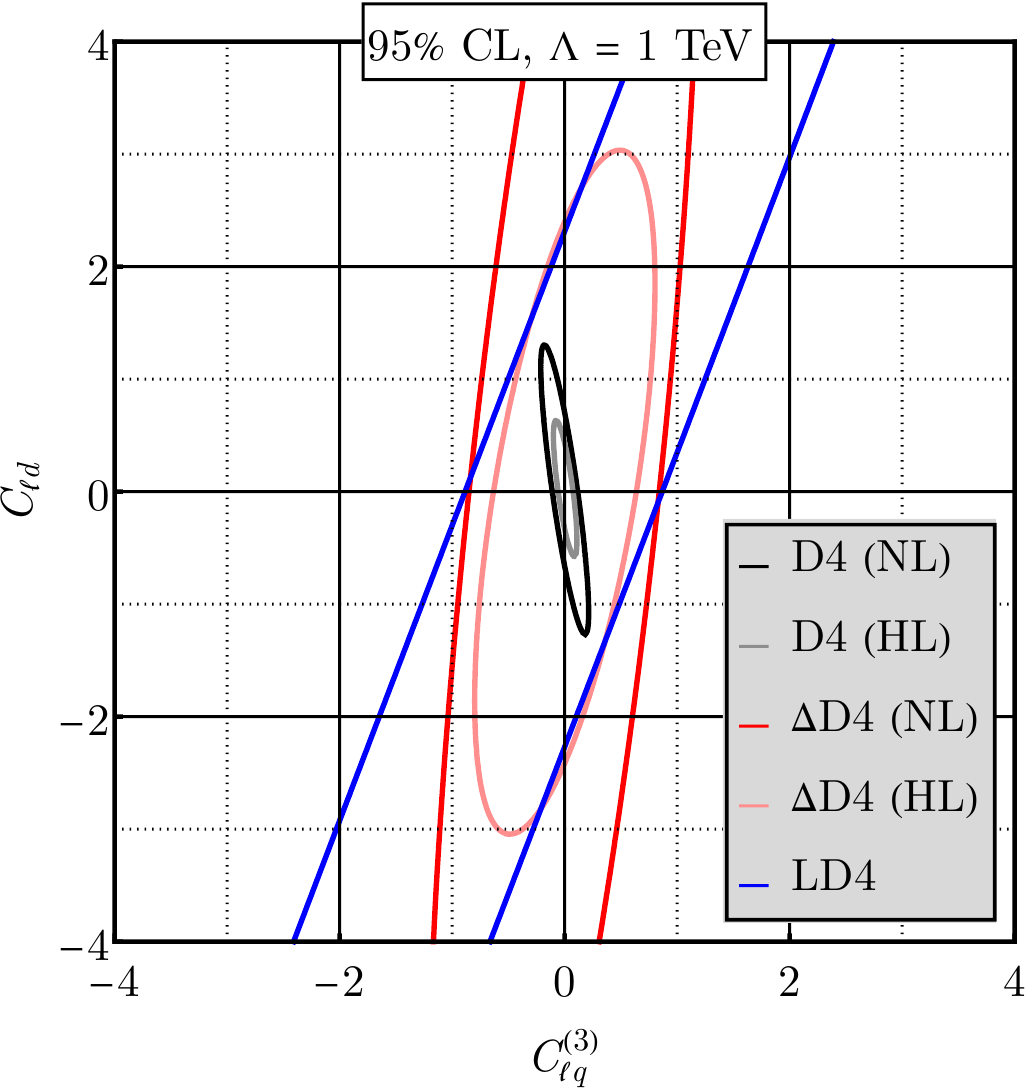}
	\includegraphics[width=\elliwi\textwidth]{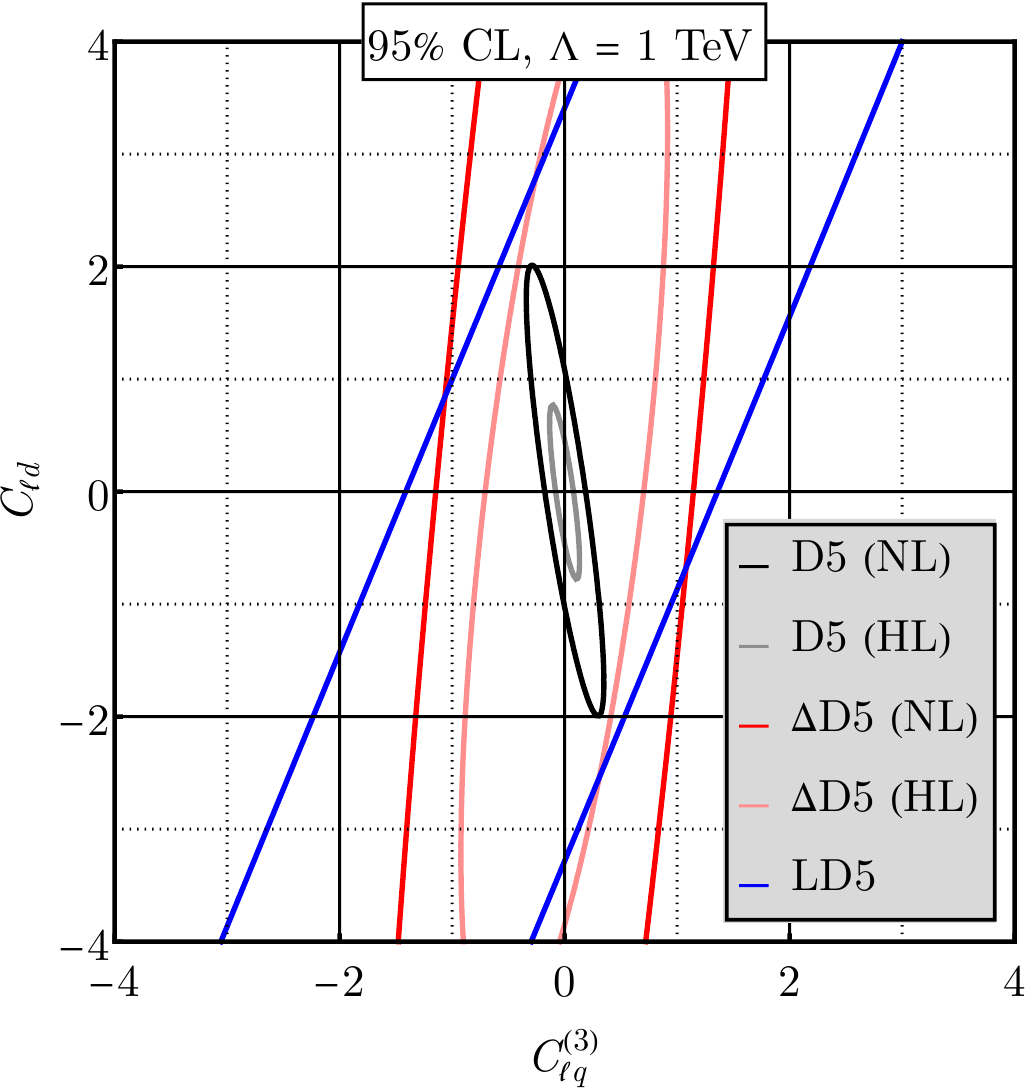}
	\includegraphics[width=\elliwi\textwidth]{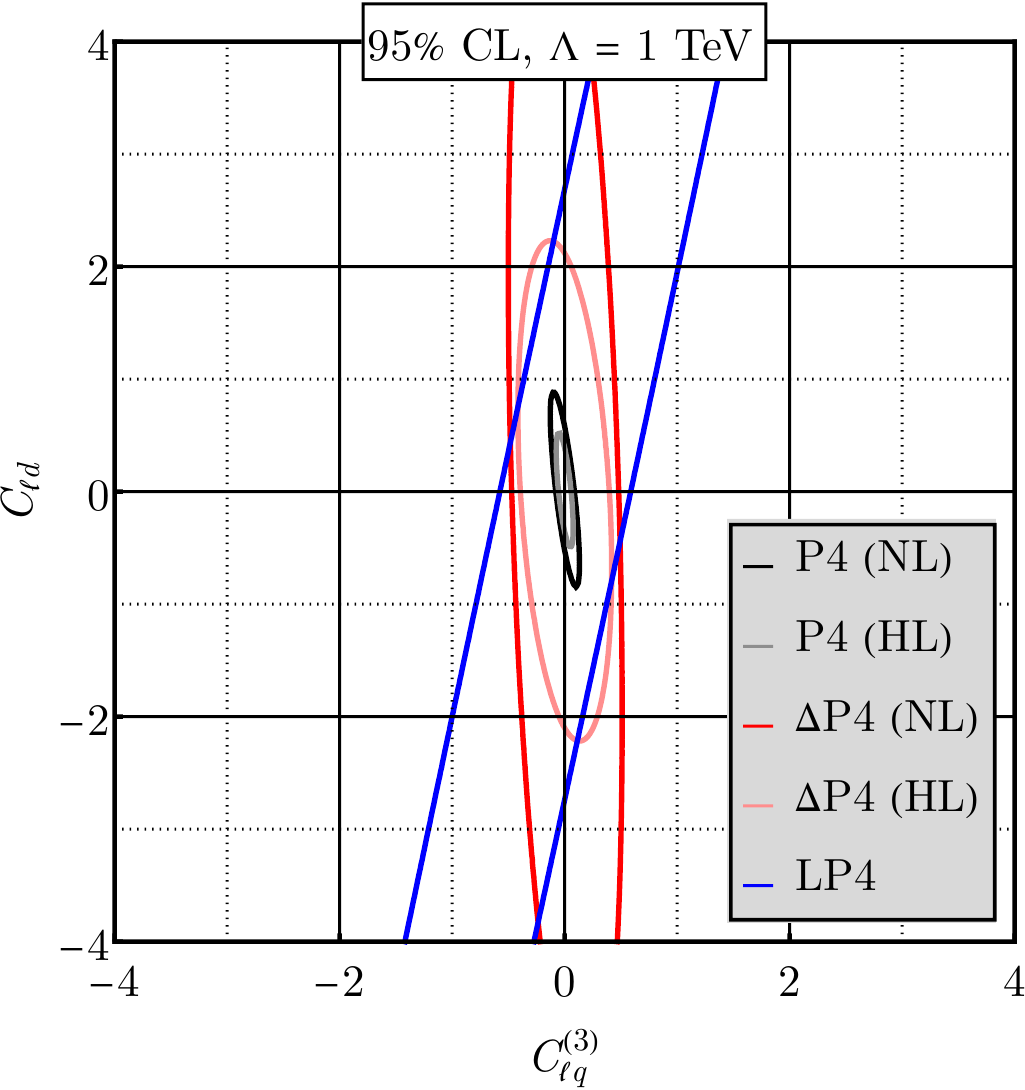}
	\includegraphics[width=\elliwi\textwidth]{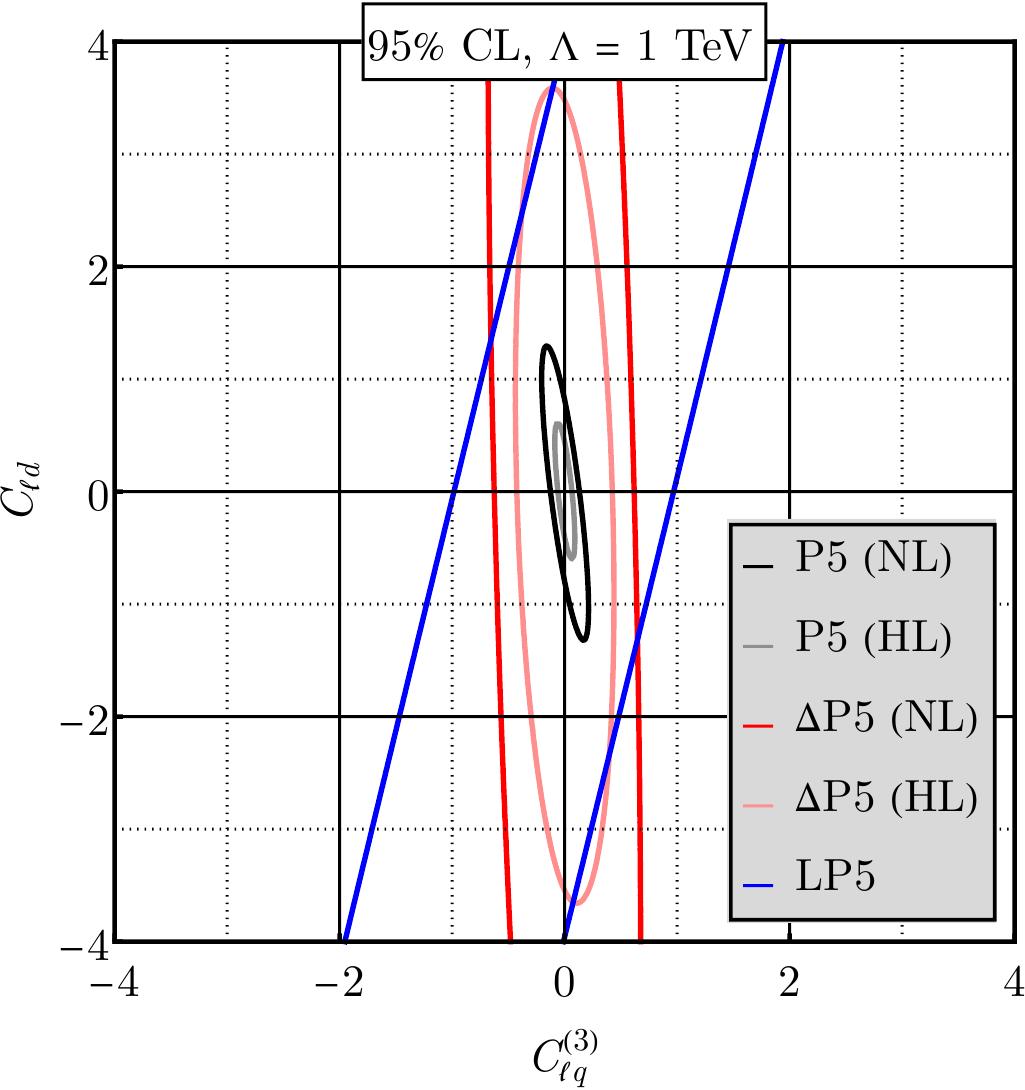}
	\caption{The same as in Fig.~\ref{fig:all-ellipses-Ceu-Ced} but for $\Clqiii$ and $\Cld$.}
	\label{fig:all-ellipses-Clq3-Cld}
\end{figure}

\begin{figure}
	[H]\centering
	\includegraphics[width=\elliwi\textwidth]{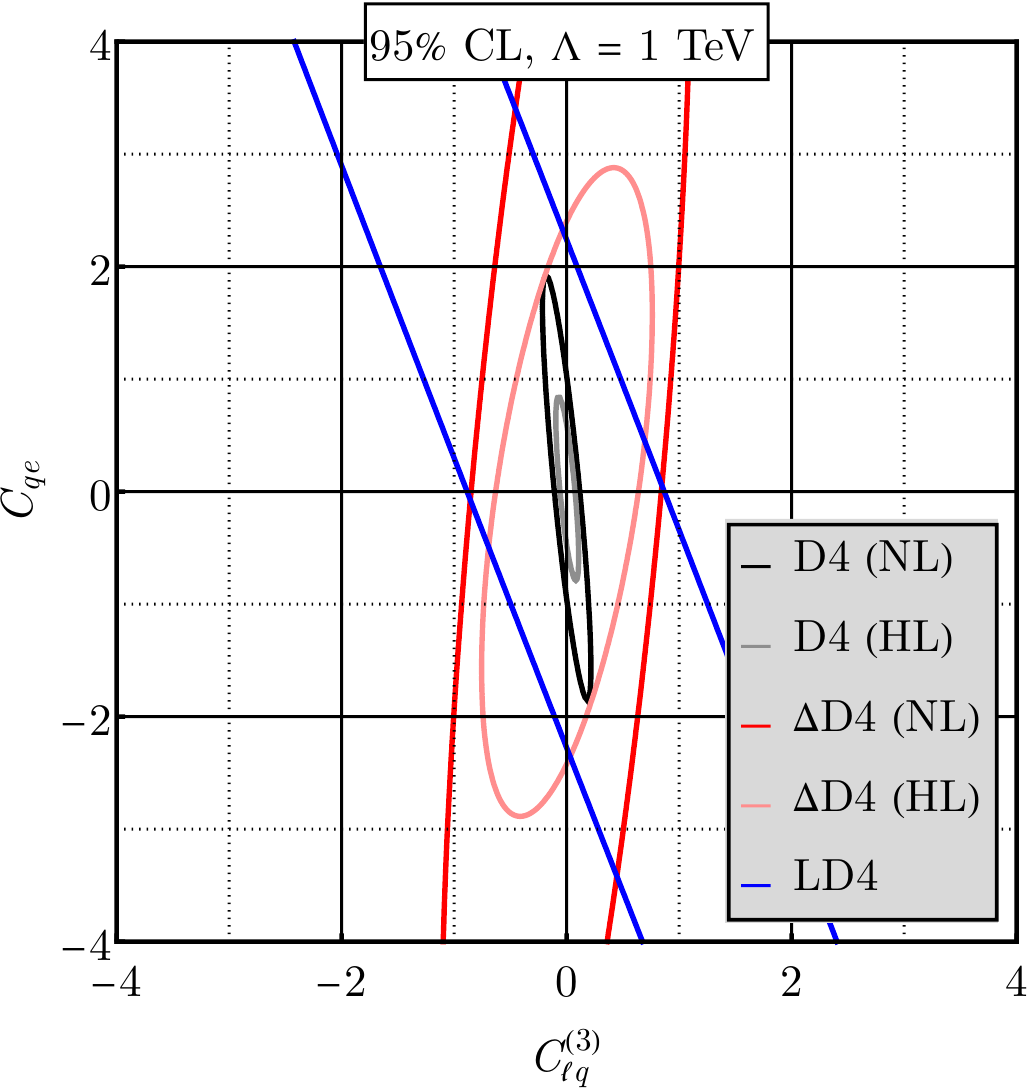}
	\includegraphics[width=\elliwi\textwidth]{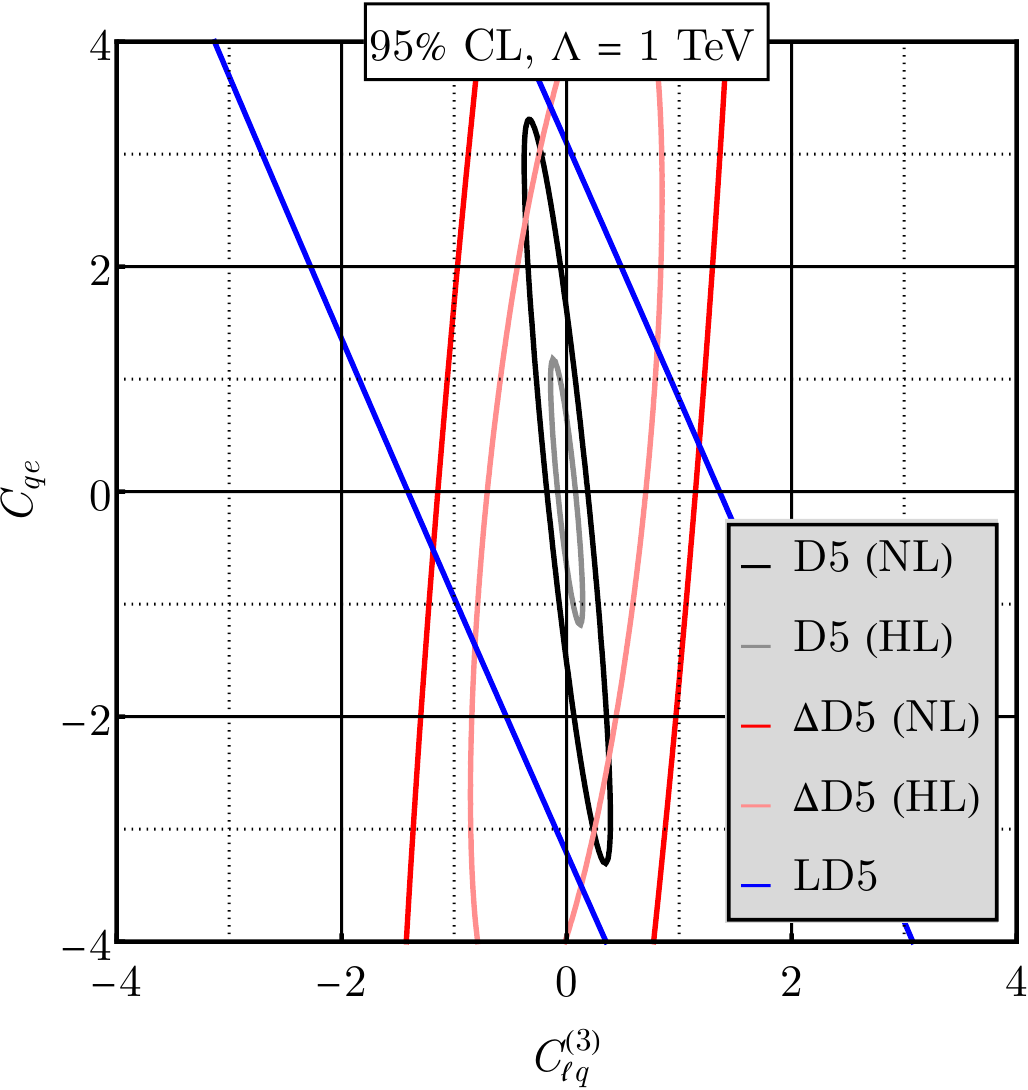}
	\includegraphics[width=\elliwi\textwidth]{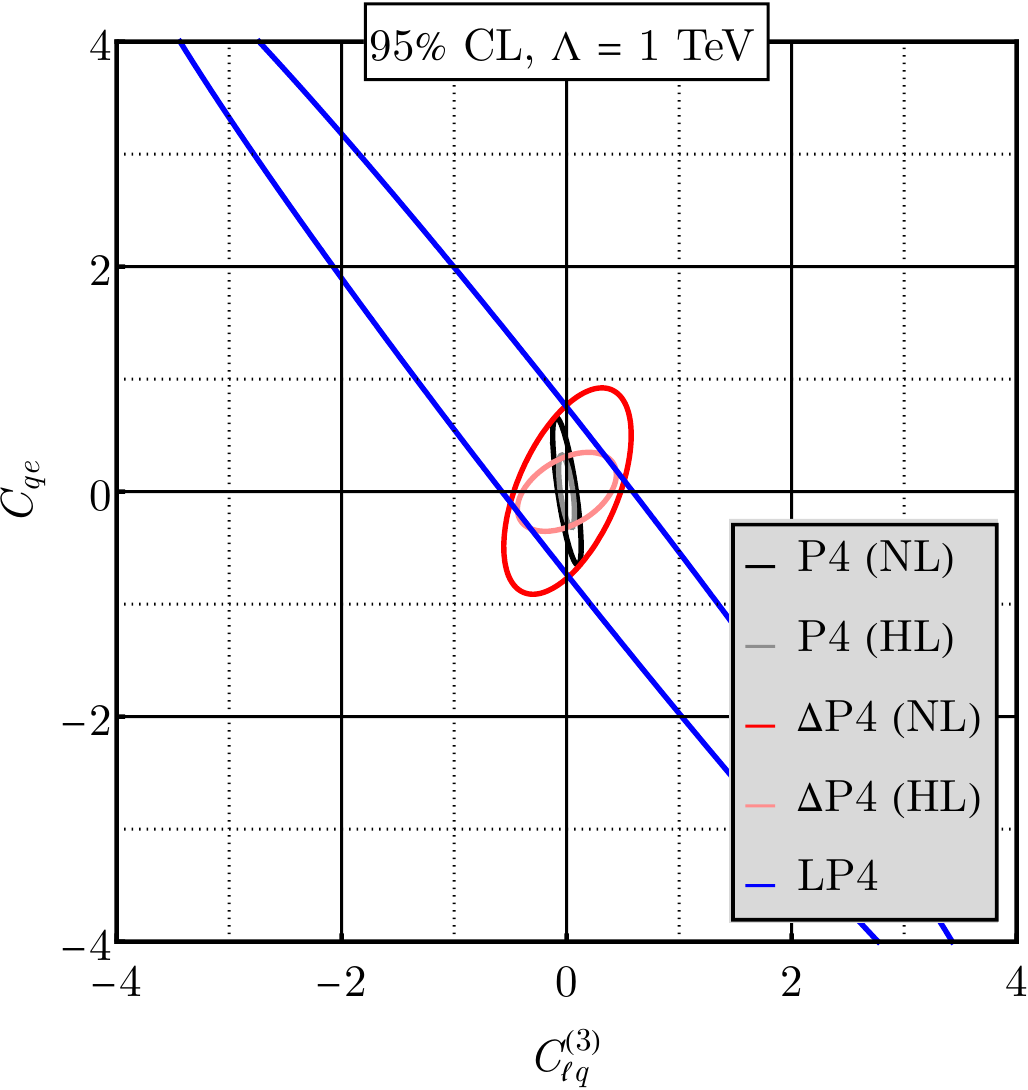}
	\includegraphics[width=\elliwi\textwidth]{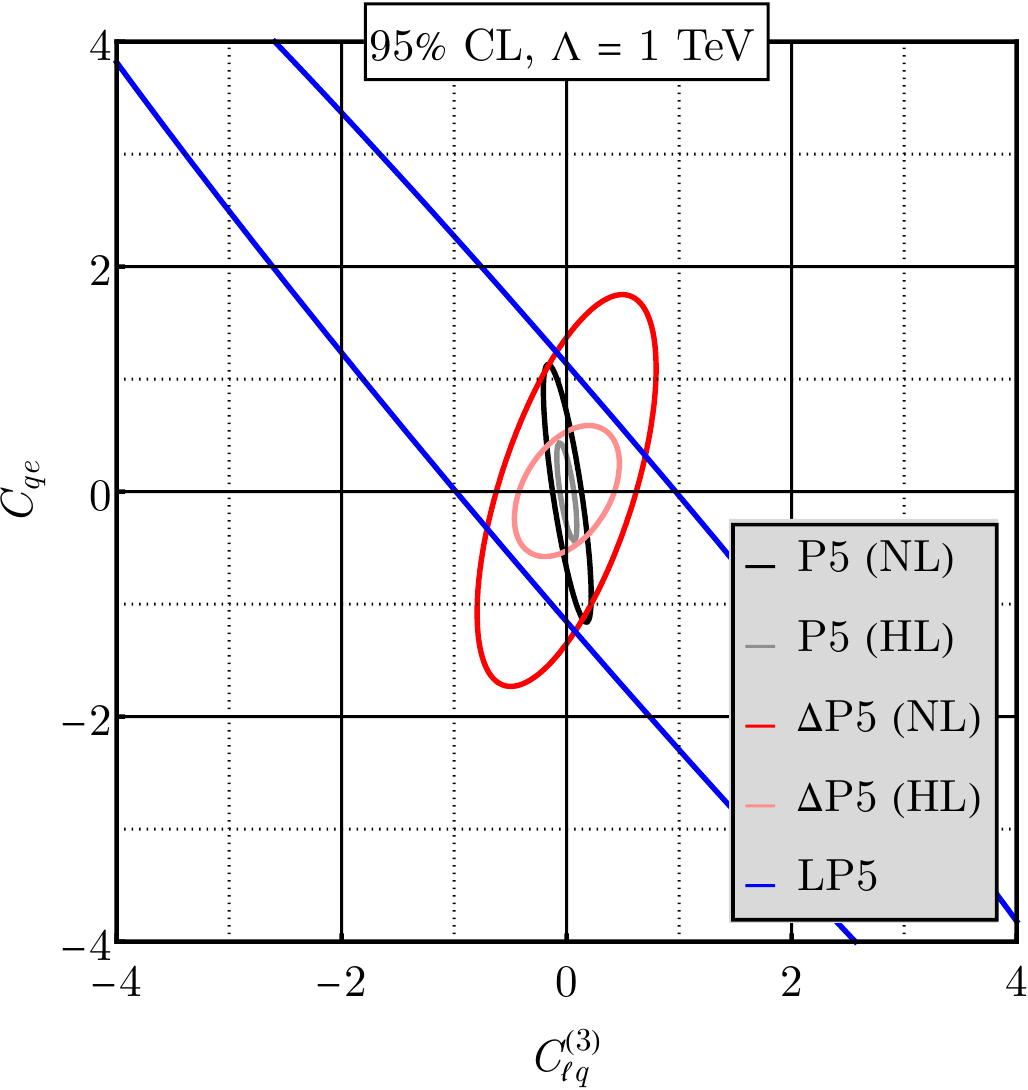}
	\caption{The same as in Fig.~\ref{fig:all-ellipses-Ceu-Ced} but for $\Clqiii$ and $\Cqe$.}
	\label{fig:all-ellipses-Clq3-Cqe}
\end{figure}

\begin{figure}
	[H]\centering
	\includegraphics[width=\elliwi\textwidth]{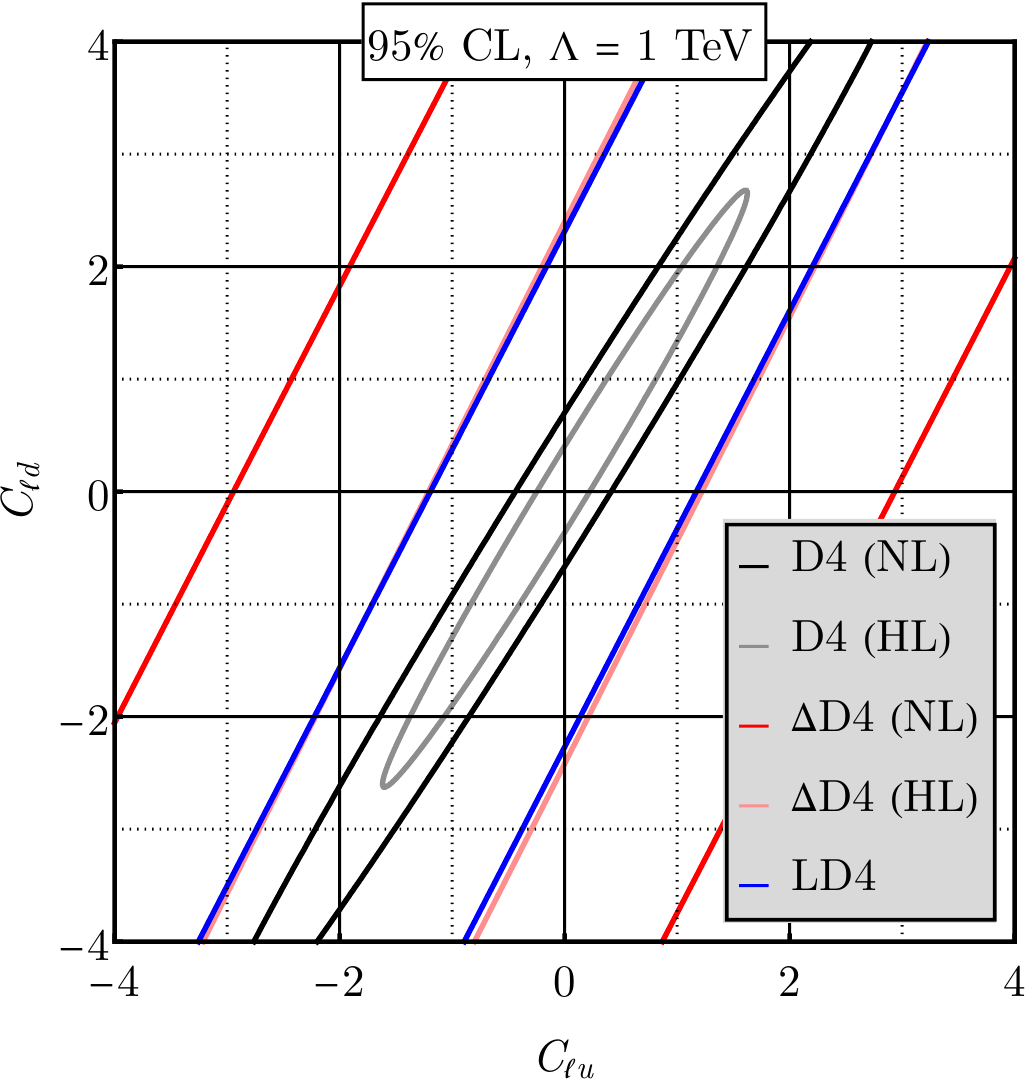}
	\includegraphics[width=\elliwi\textwidth]{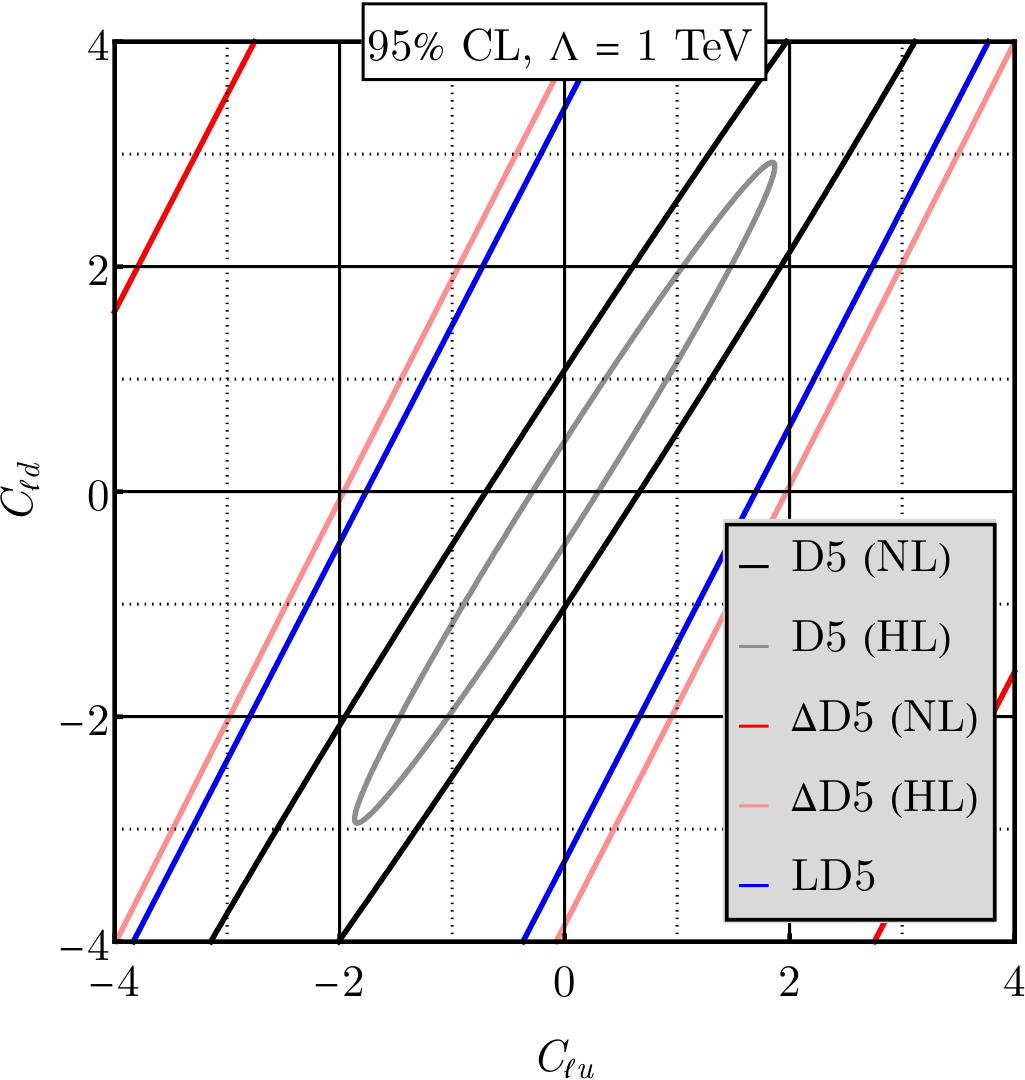}
	\includegraphics[width=\elliwi\textwidth]{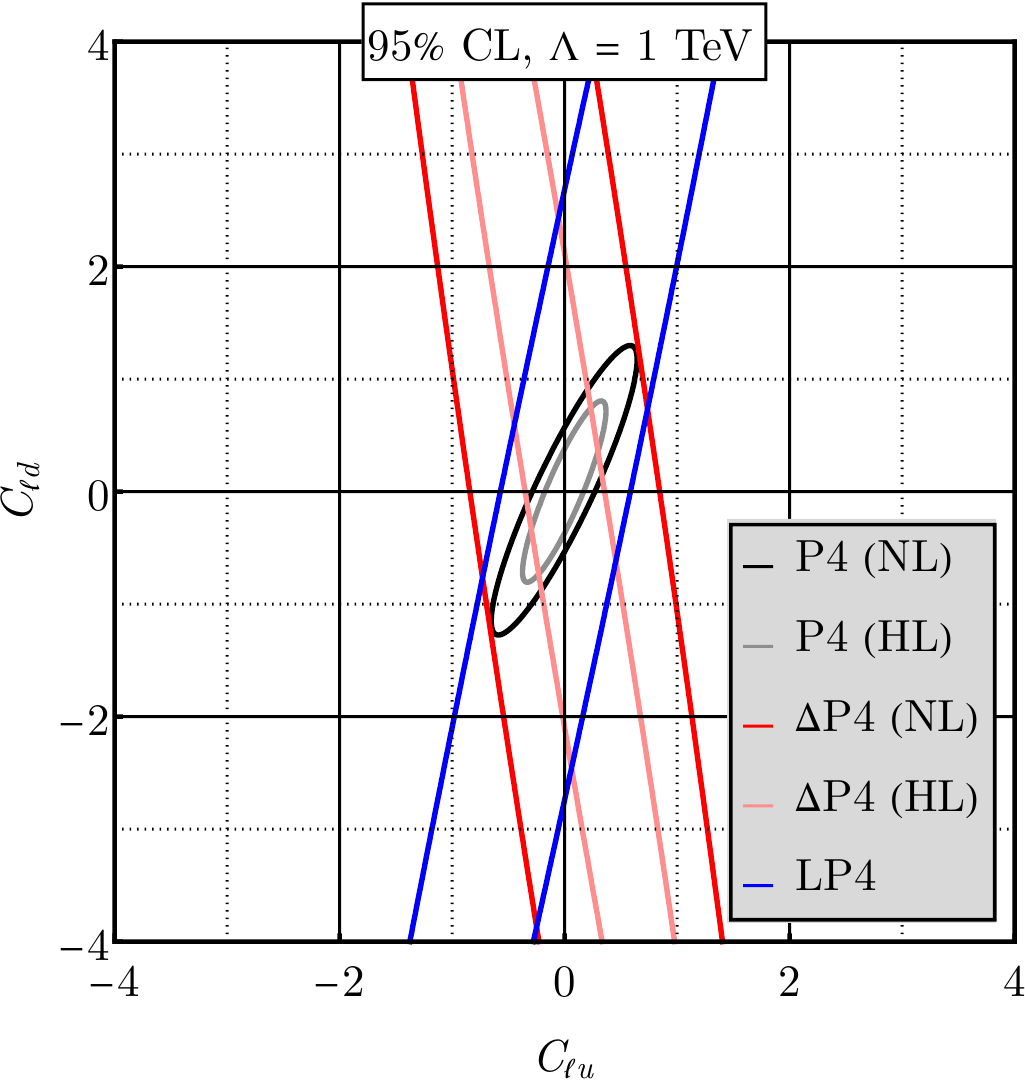}
	\includegraphics[width=\elliwi\textwidth]{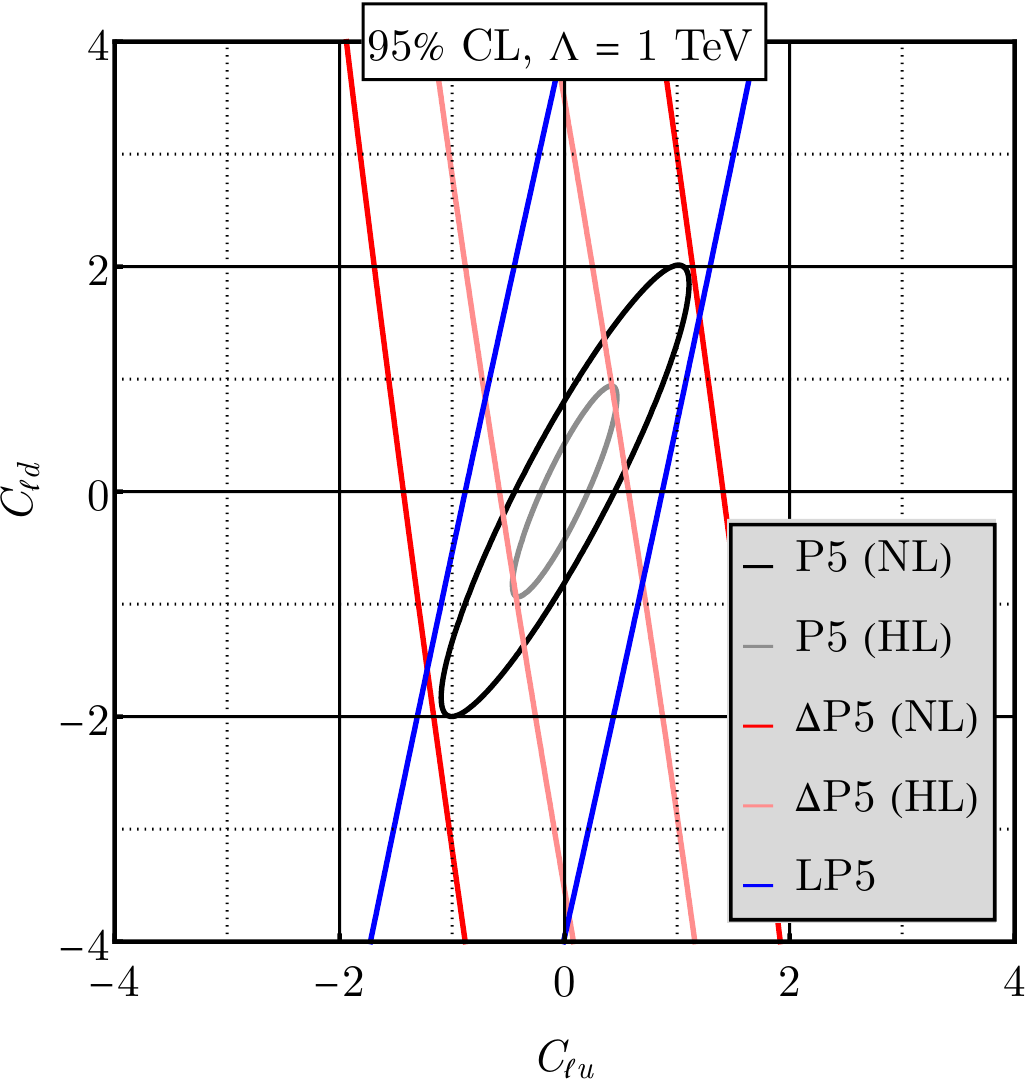}
	\caption{The same as in Fig.~\ref{fig:all-ellipses-Ceu-Ced} but for $\Clu$ and $\Cld$.}
	\label{fig:all-ellipses-Clu-Cld}
\end{figure}

\begin{figure}
	[H]\centering
	\includegraphics[width=\elliwi\textwidth]{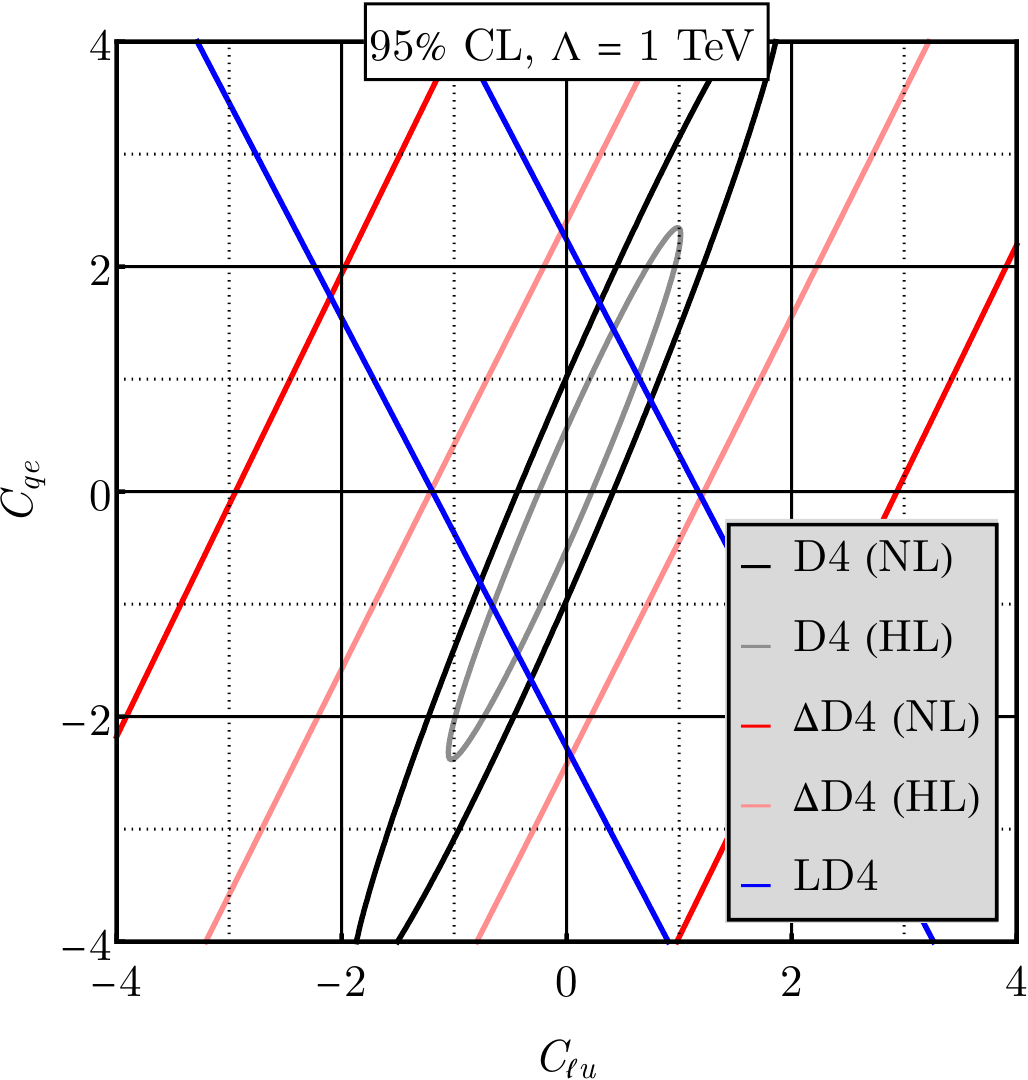}
	\includegraphics[width=\elliwi\textwidth]{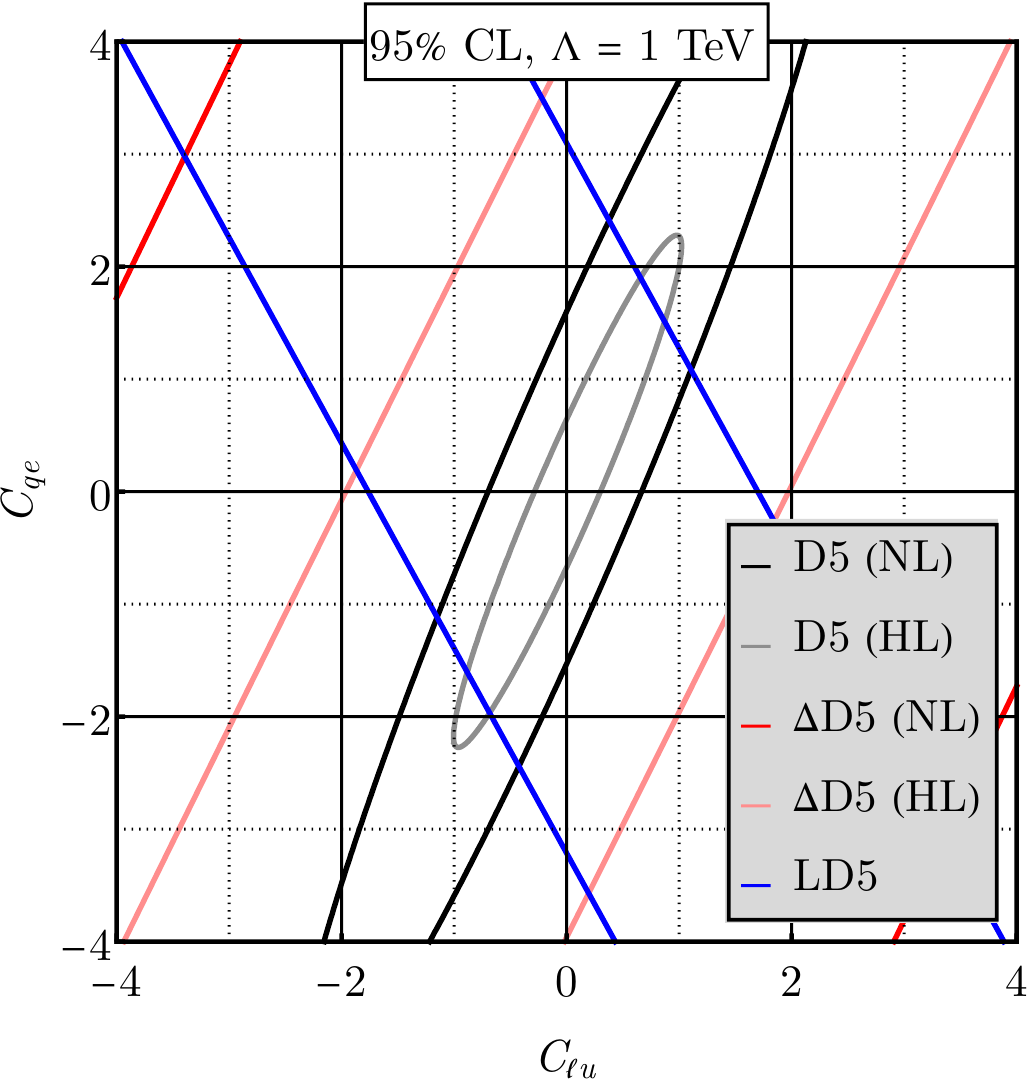}
	\includegraphics[width=\elliwi\textwidth]{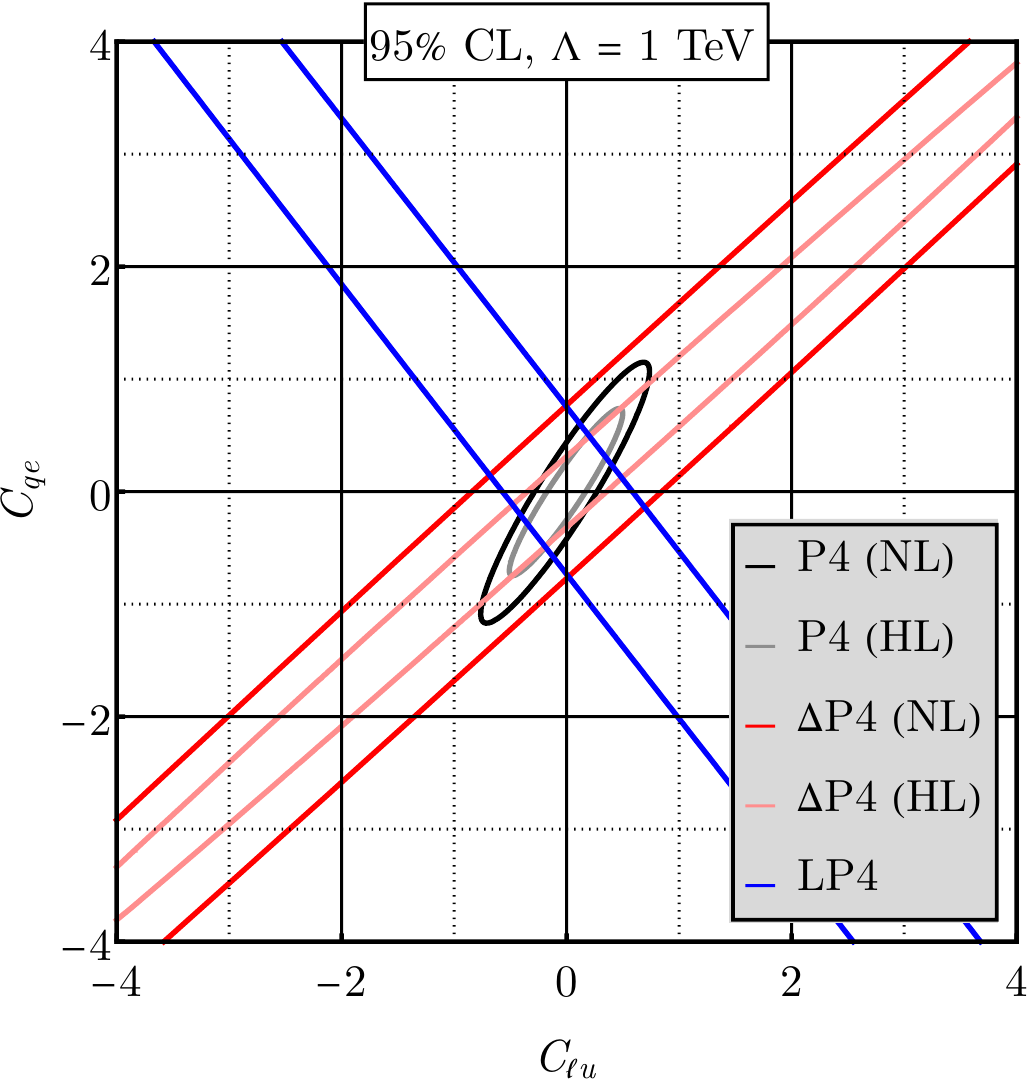}
	\includegraphics[width=\elliwi\textwidth]{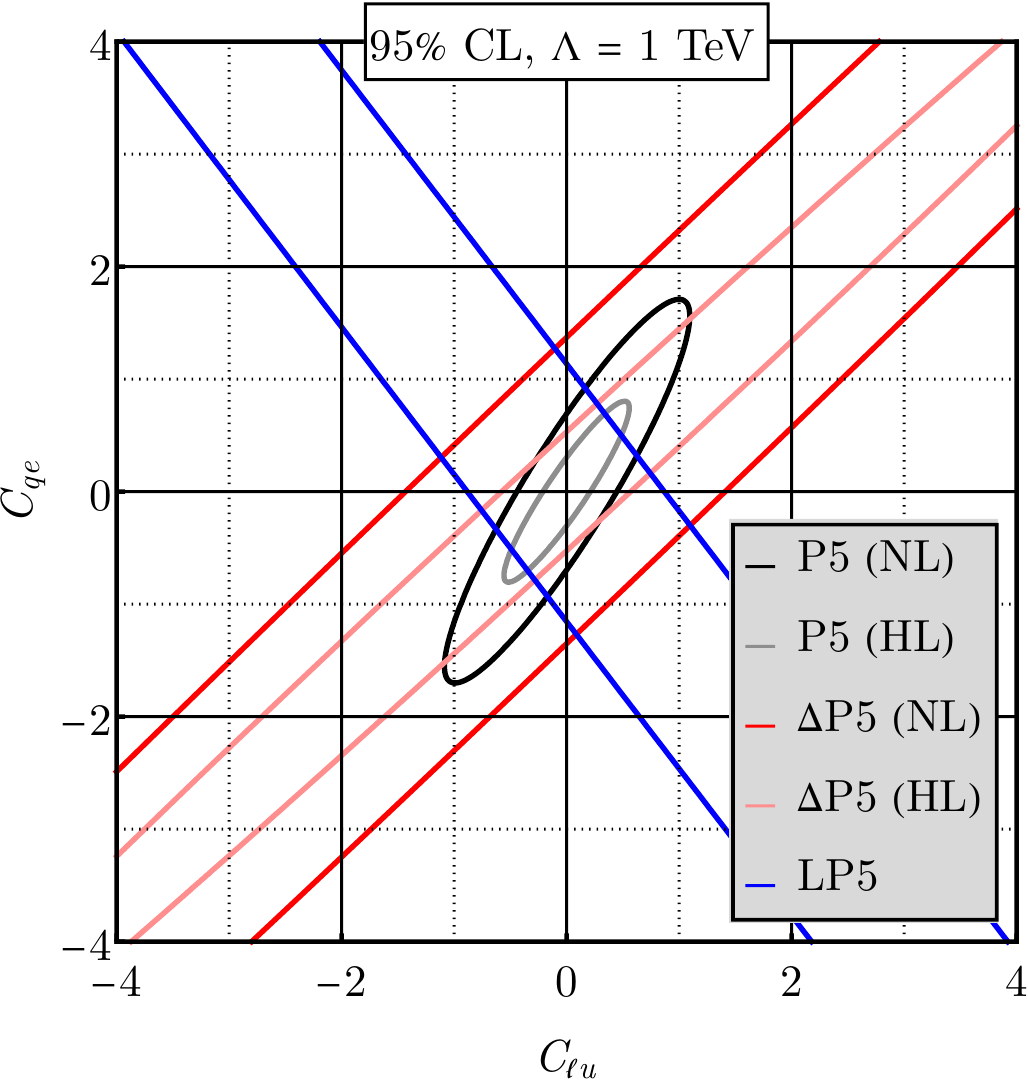}
	\caption{The same as in Fig.~\ref{fig:all-ellipses-Ceu-Ced} but for $\Clu$ and $\Cqe$.}
	\label{fig:all-ellipses-Clu-Cqe}
\end{figure}

\begin{figure}
	[H]\centering
	\includegraphics[width=\elliwi\textwidth]{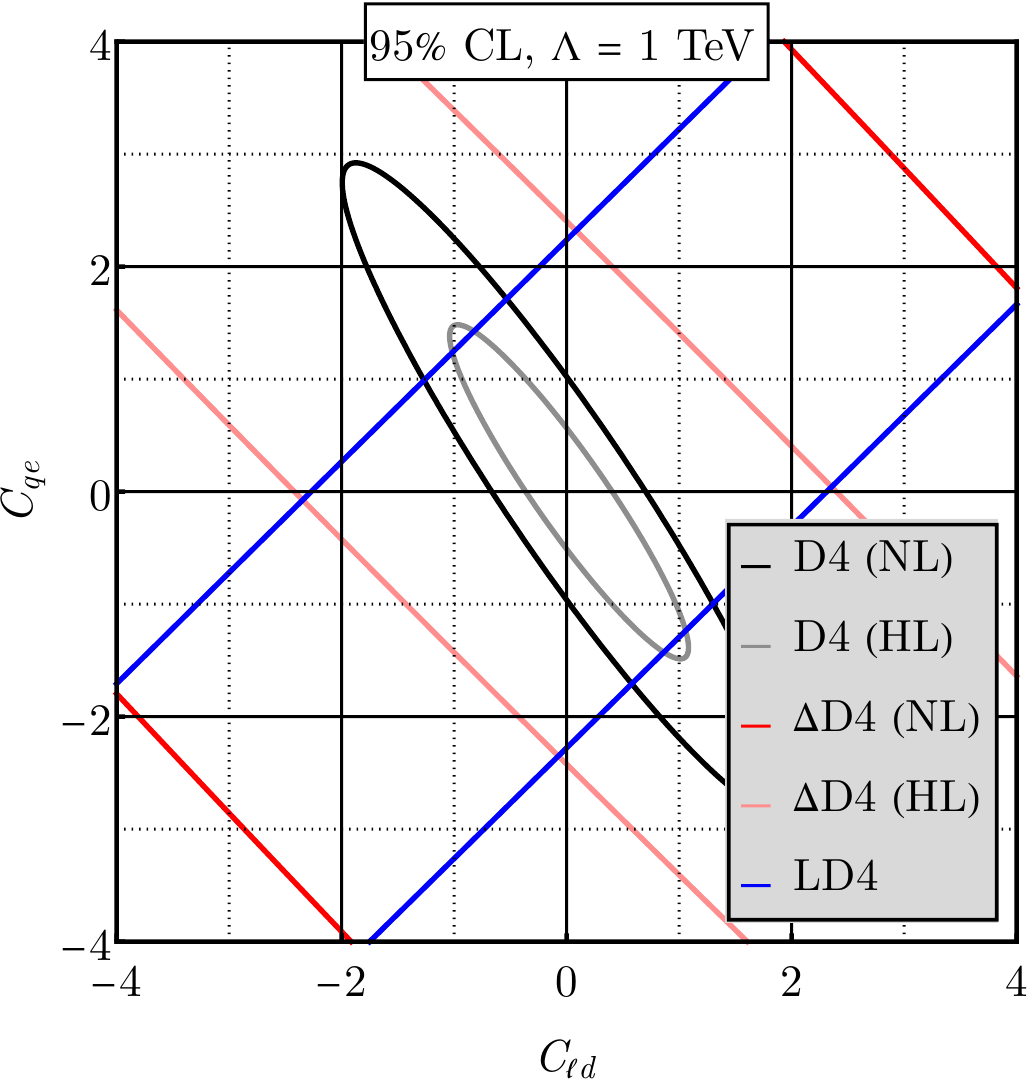}
	\includegraphics[width=\elliwi\textwidth]{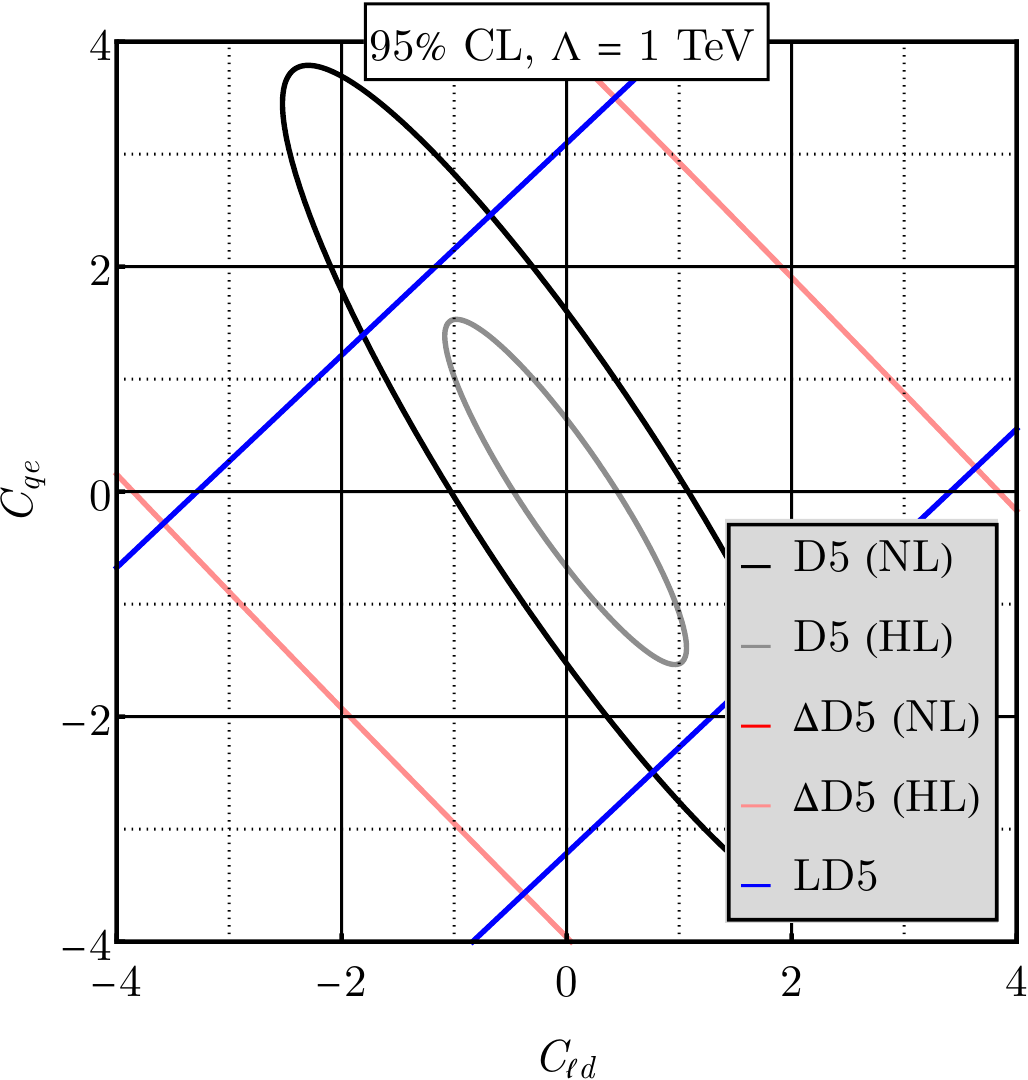}
	\includegraphics[width=\elliwi\textwidth]{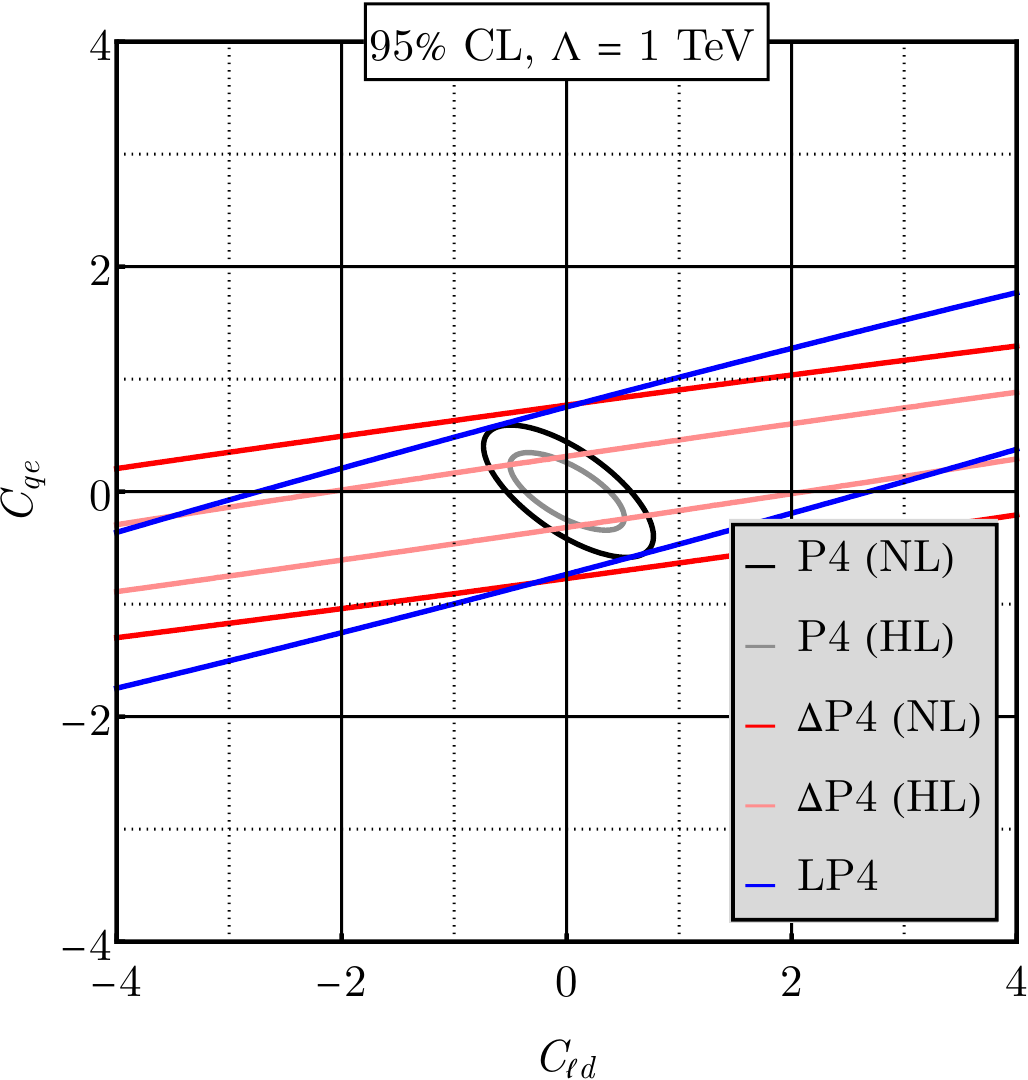}
	\includegraphics[width=\elliwi\textwidth]{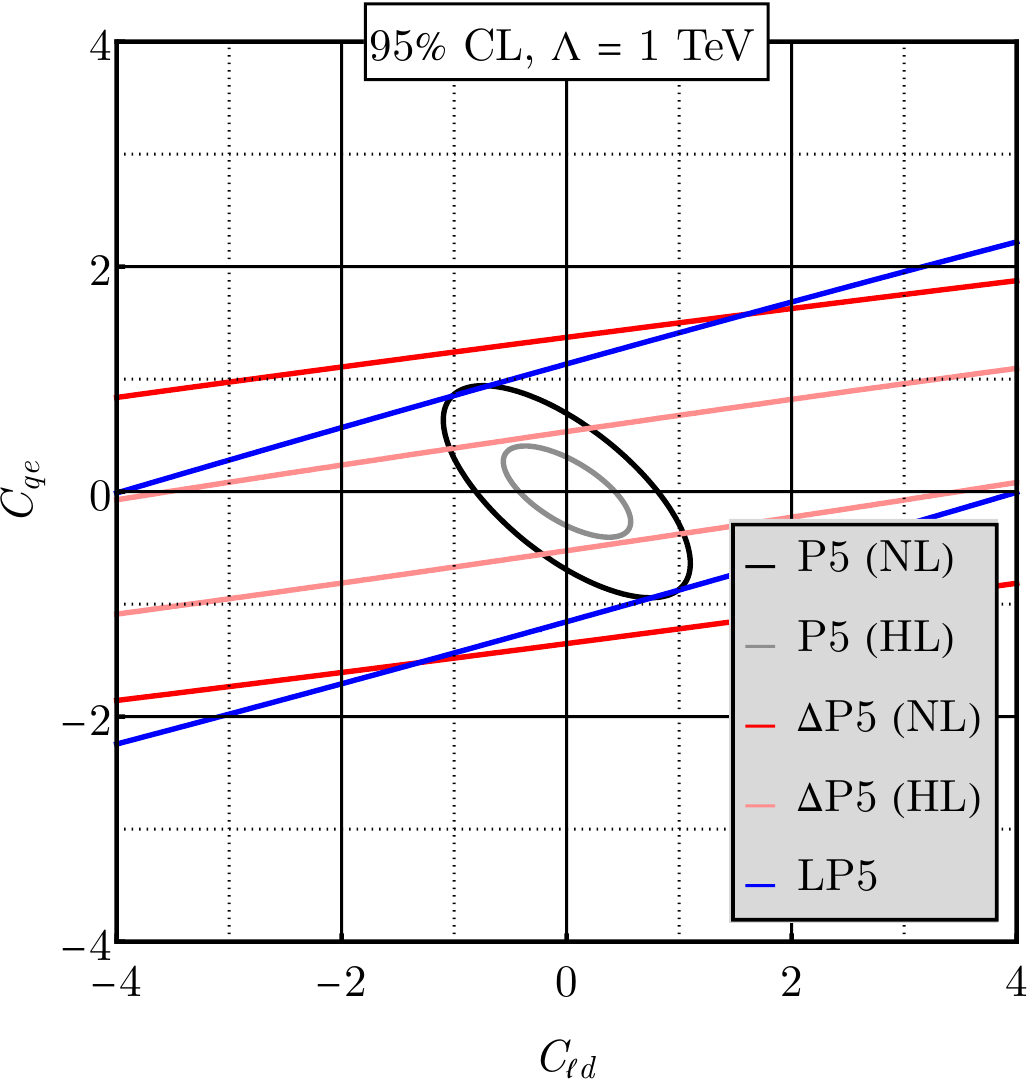}
	\caption{The same as in Fig.~\ref{fig:all-ellipses-Ceu-Ced} but for $\Cld$ and $\Cqe$.}
	\label{fig:all-ellipses-Cld-Cqe}
\end{figure}

\begin{figure}
	[H]\centering
	\includegraphics[height=.9\textheight]{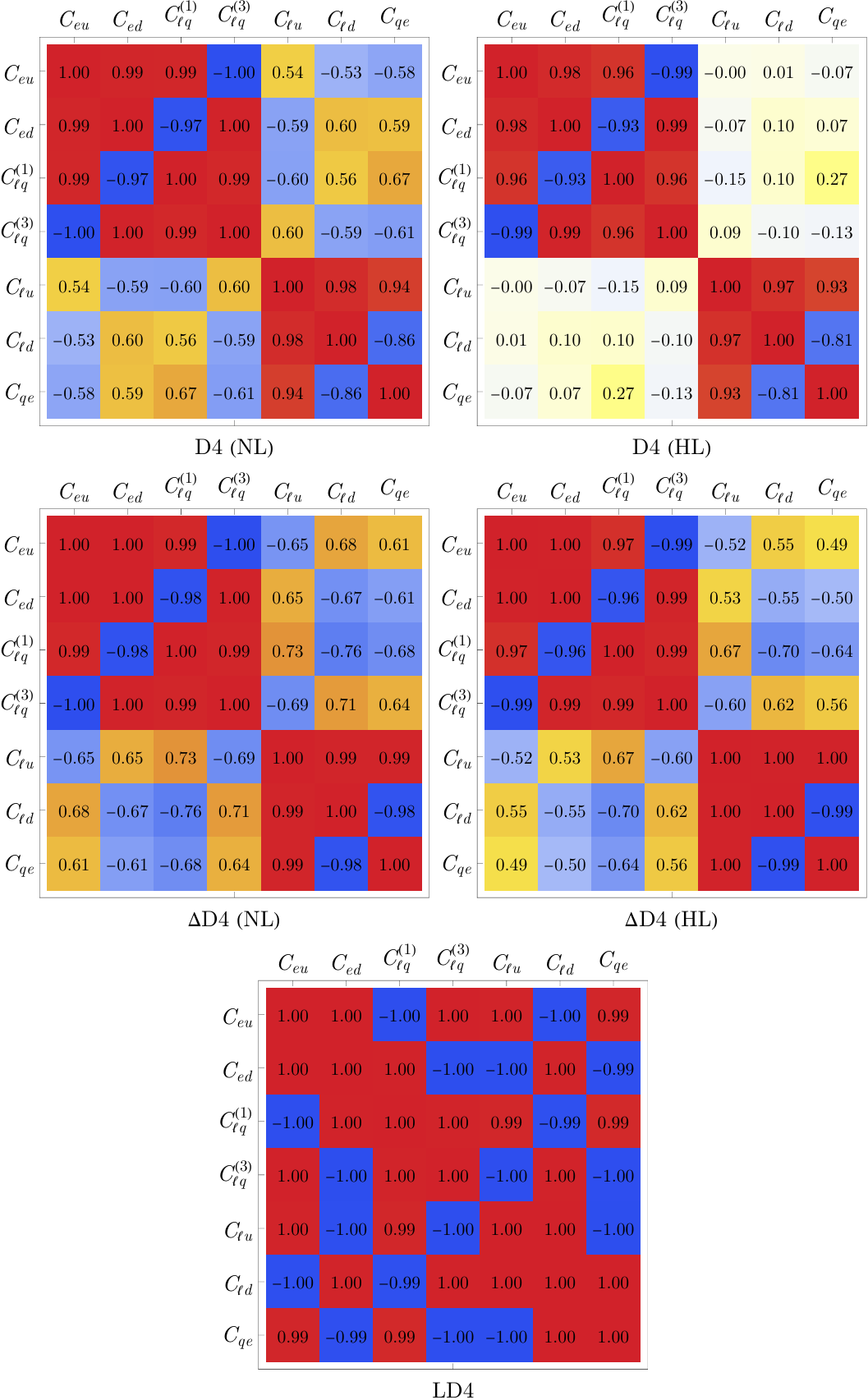}
	\caption{The correlation table of Wilson coefficients in the D4 data family. The off-diagonal entries are from the results of simultaneous fits of the $(2+1)$-parameter fit of two Wilson coefficients plus the beam-polarization parameter $P$.}
	\label{tab:correlation-table-D4}
\end{figure}

\begin{figure}
	[H]\centering
	\includegraphics[height=.9\textheight]{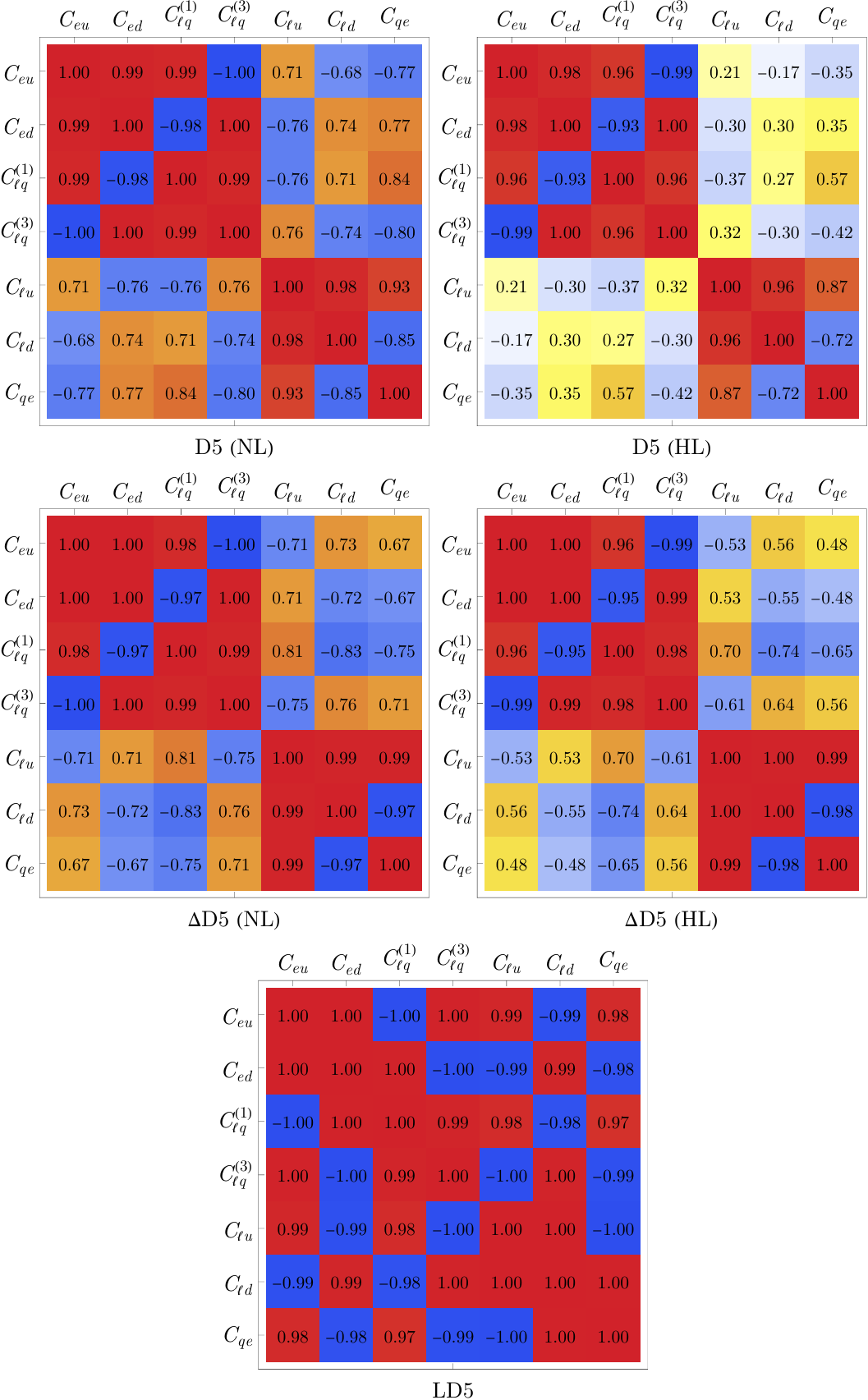}
	\caption{The same as in Fig.~\ref{tab:correlation-table-D4} but for D5.}
	\label{tab:correlation-table-D5}
\end{figure}

\begin{figure}
	[H]\centering
	\includegraphics[height=.9\textheight]{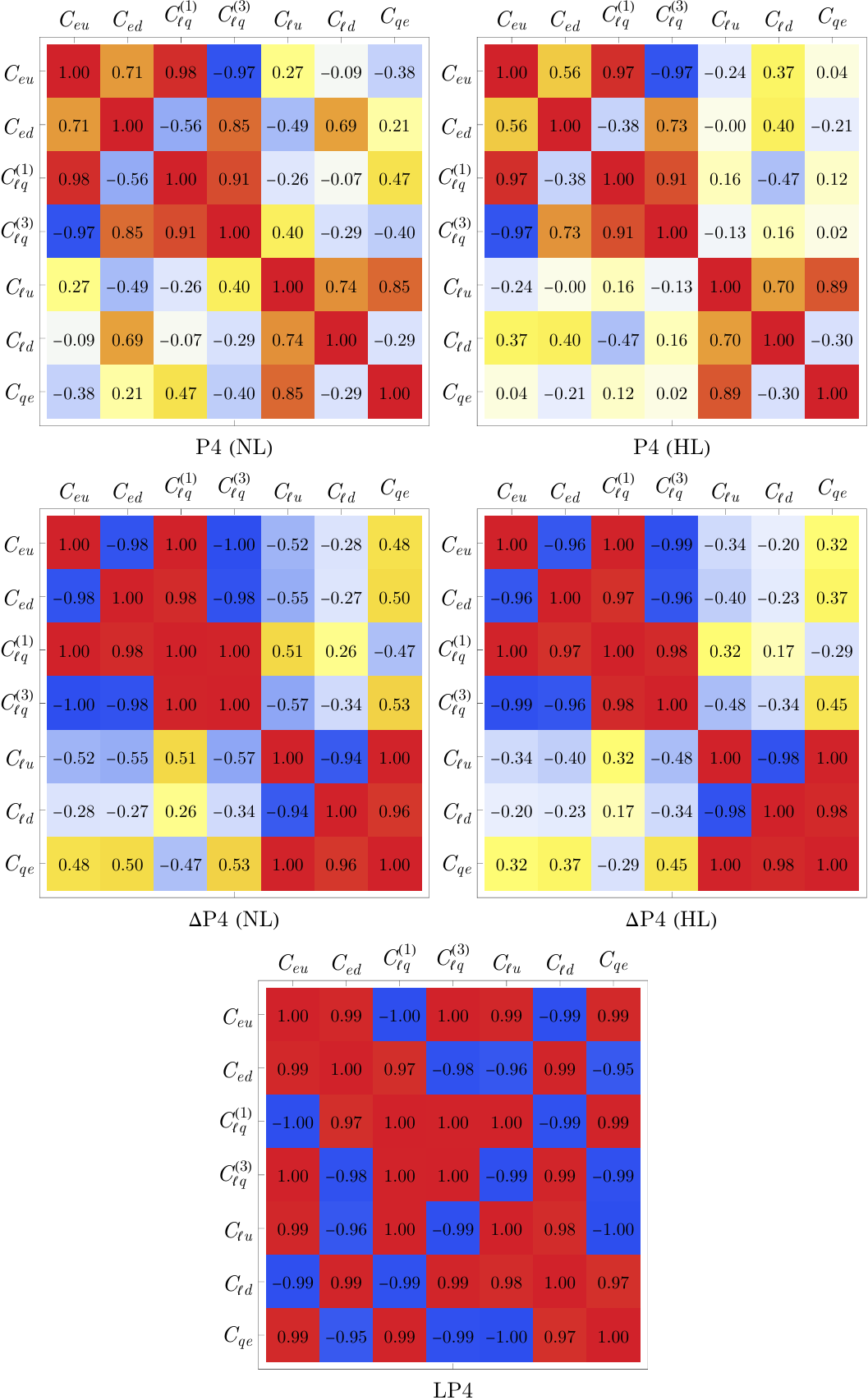}
	\caption{The same as in Fig.~\ref{tab:correlation-table-D4} but for P4.}
	\label{tab:correlation-table-P4}
\end{figure}

\begin{figure}
	[H]\centering
	\includegraphics[height=.9\textheight]{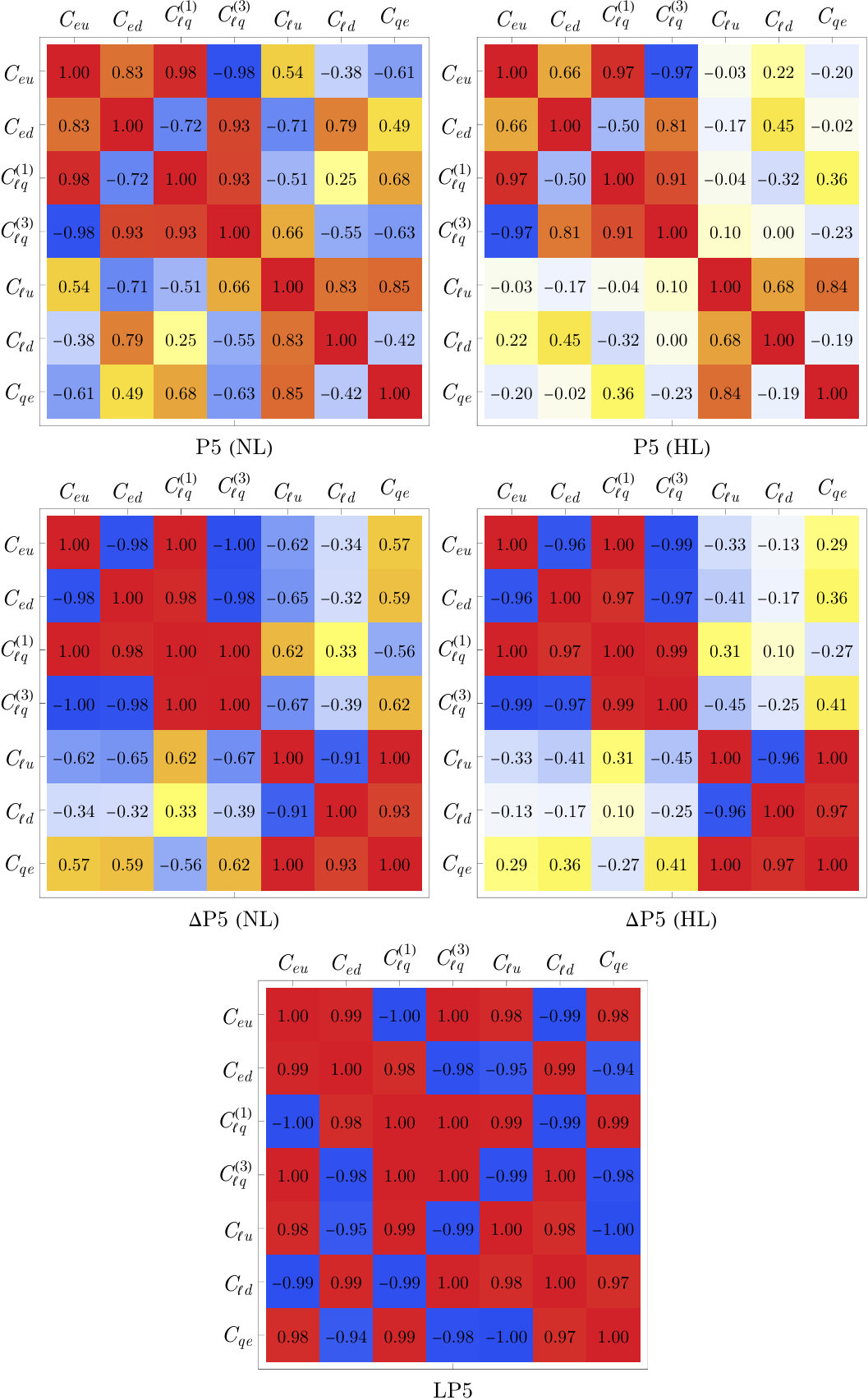}
	\caption{The same as in Fig.~\ref{tab:correlation-table-D4} but for P5.}
	\label{tab:correlation-table-P5}
\end{figure}

For completeness, we show the correlations of Wilson coefficients in Figs. \ref{tab:correlation-table-D4}--\ref{tab:correlation-table-P5}.

%% file: appendices/c1-fits_of_six_wilson_coefficients.tex
\subsection{Six-dimensional fits \label{app:sixd_fits}}

In this section, we discuss projected bounds arising from a fit of six Wilson coefficients. We show projections onto one and two Wilson coefficients from the full 6$d$ hyper-ellipse. The computational power required for higher-dimensional fits increases because we increase the number of pseudoexperiments to reflect the required statistics for the fits. We find that  $N_{\rm exp} = 10^3 $ pseudoexperiments for $2d$ fits lead to stable results, meaning the best-fit values and the corresponding bounds do not change for $N_{\rm exp} \geq 10^3$. This number  becomes $N_{\rm exp} = 10^4$ for $3d$ fits, $N_{\rm exp} = 10^5$ for $4d$ fits, $N_{\rm exp} = 10^6$ for $5d$ fits, and $N_{\rm exp} = 10^7$ for $6d$ fits. We note that starting from $3d$ fits, the size and characteristics of multidimensional fits stabilize as we increase the number of fitted parameters, in the sense that the bounds do not change further. This indicates that the $6d$ fits stand as a useful indicator of what happens in the complete $7d$ fit of the Wilson coefficients.

\begin{figure}
	[H]
	\centering
	\includegraphics[width=.7\textwidth]{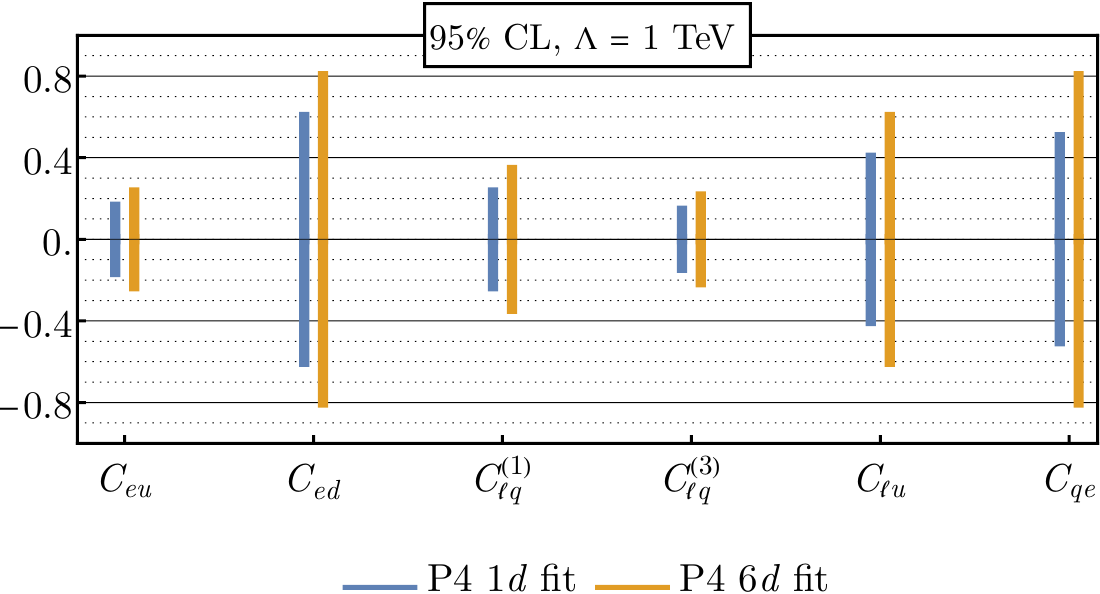}
	\caption{Comparison of the 95\% CL bounds on single Wilson coefficients from one-parameter fits to the projections in the $6d$ fit using the nominal-luminosity data set P4 at $\Lambda = 1~{\rm TeV}$.}
	\label{fig:6d-fit-1d-projections}
\end{figure}

To illustrate our discussion, we pick the representative Wilson coefficients $\Ceu$, $\Ced$, $\Clqi$, $\Clqiii$, $\Clu$, and $\Cqe$ and consider the data from $ep$ collisions in the configuration $10~{\rm GeV} \times 275~{\rm GeV}$ with $100~{\rm fb^{-1}}$, which is the nominal-luminosity P4 data set. In Fig. \ref{fig:6d-fit-1d-projections}, we compare the bounds from the original $1d$ fits to the projected bounds from the $6d$ fit of the Wilson coefficients $\Ceu$, $\Ced$, $\Clqi$, $\Clqiii$, $\Clu$, and $\Cqe$. We observe that the bounds become 25 to 40\% weaker as we increase the number of Wilson coefficients fitted. This is due to an interplay between the increased number of fitted parameters and correlations among them. 
In Figs. \ref{fig:6d-fit-2d-projections_1}--\ref{fig:6d-fit-2d-projections_5}, we compare the confidence ellipses from the initial $2d$ fits to the ones in the two-parameter projections of the $6d$ fit of the aforementioned Wilson coefficients. We find that the bounds become 20 to 30\% weaker as in the case of the comparison of the one-parameter fits and projections, which can be explained by the  same reasoning as mentioned above. The other choices of six Wilson coefficients lead to similar results. We note that no flat directions appear in these fits, indicating that the EIC can fully probe this parameter space without degeneracies. 

\def\elliwi{.32}
\begin{figure}
	[H]\centering
	\includegraphics[width=\elliwi\textwidth]{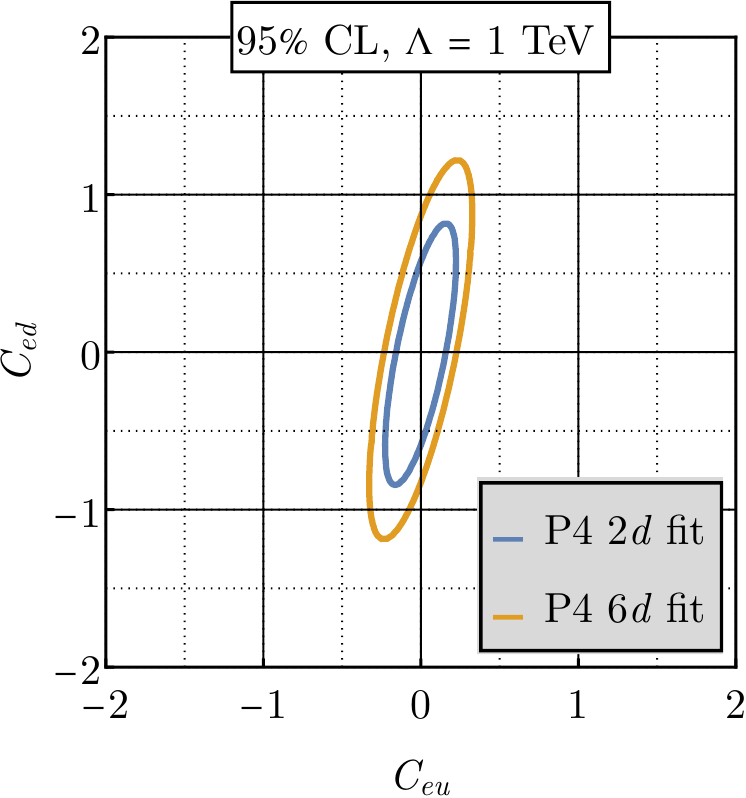}
	\includegraphics[width=\elliwi\textwidth]{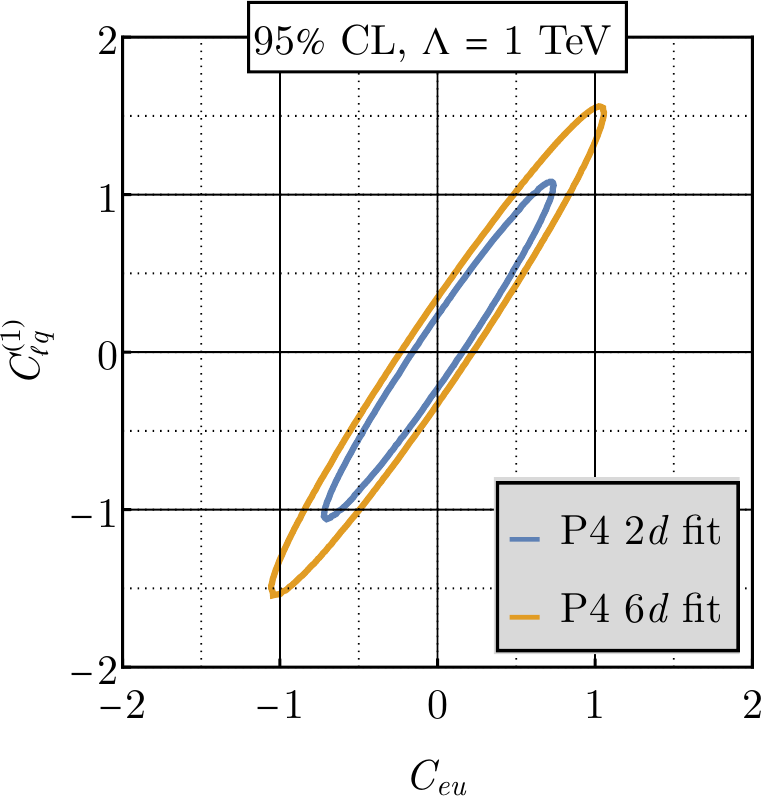}
	\includegraphics[width=\elliwi\textwidth]{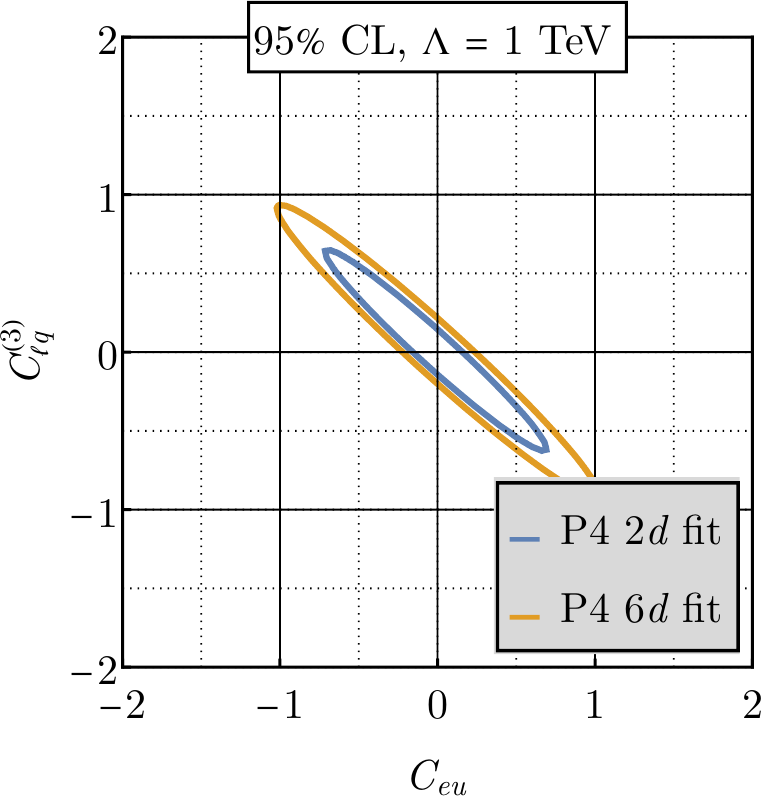}
	\caption{Comparison of the 95\% CL ellipses for the Wilson-coefficient pairs $(\Ceu, \Ced)$, $(\Ceu, \Clqi)$, and $(\Ceu, \Clqiii)$ between the original $2d$ fits and the projections from the simultaneous fit of $\Ceu$, $\Ced$, $\Clqi$, $\Clqiii$, $\Clu$, and $\Cqe$ using the data set P4 at $\Lambda = 1 \ {\rm TeV}$.}
	\label{fig:6d-fit-2d-projections_1}
\end{figure}
\begin{figure}
	[H]\centering
	\includegraphics[width=\elliwi\textwidth]{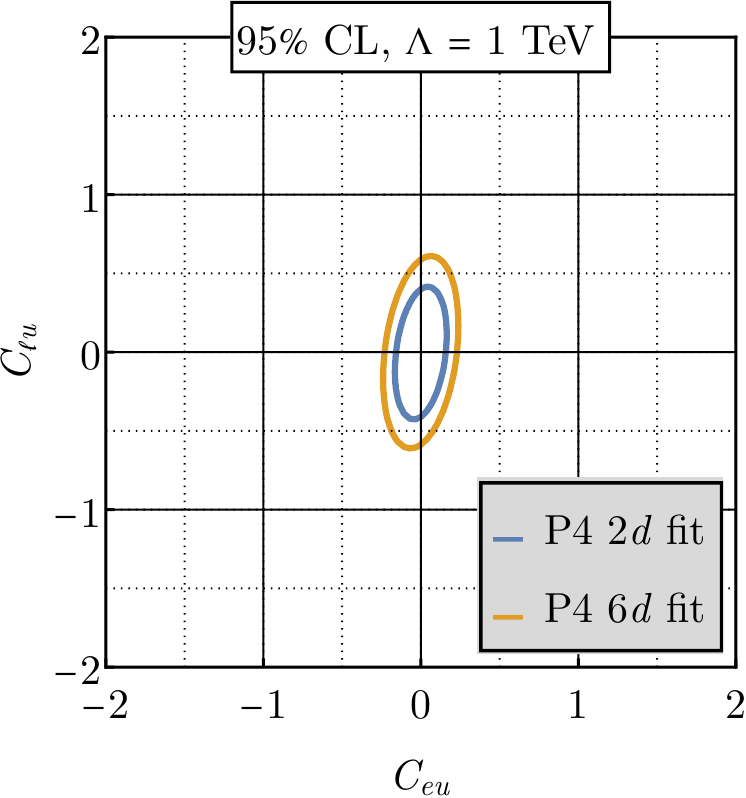}
	\includegraphics[width=\elliwi\textwidth]{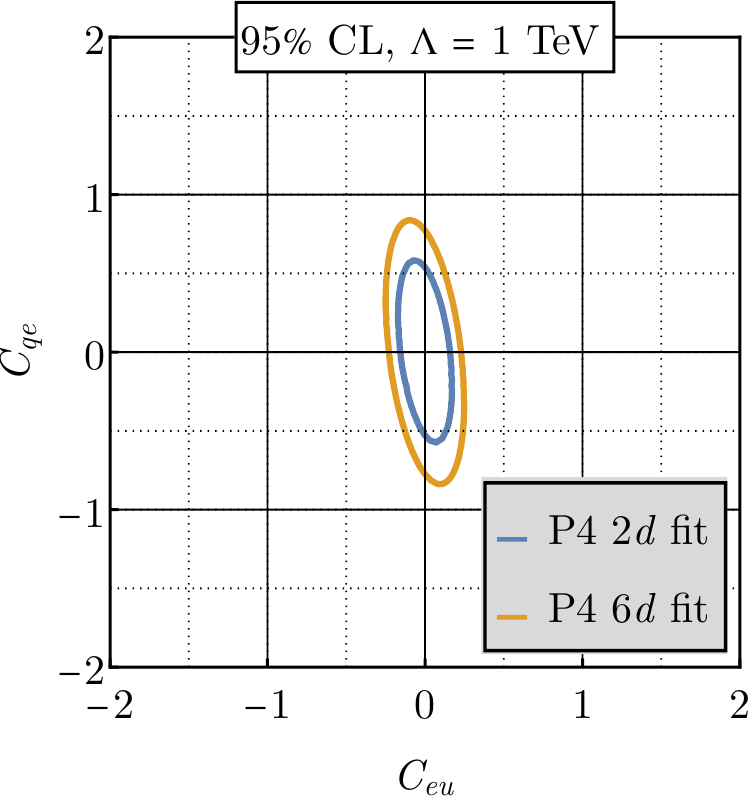}
	\includegraphics[width=\elliwi\textwidth]{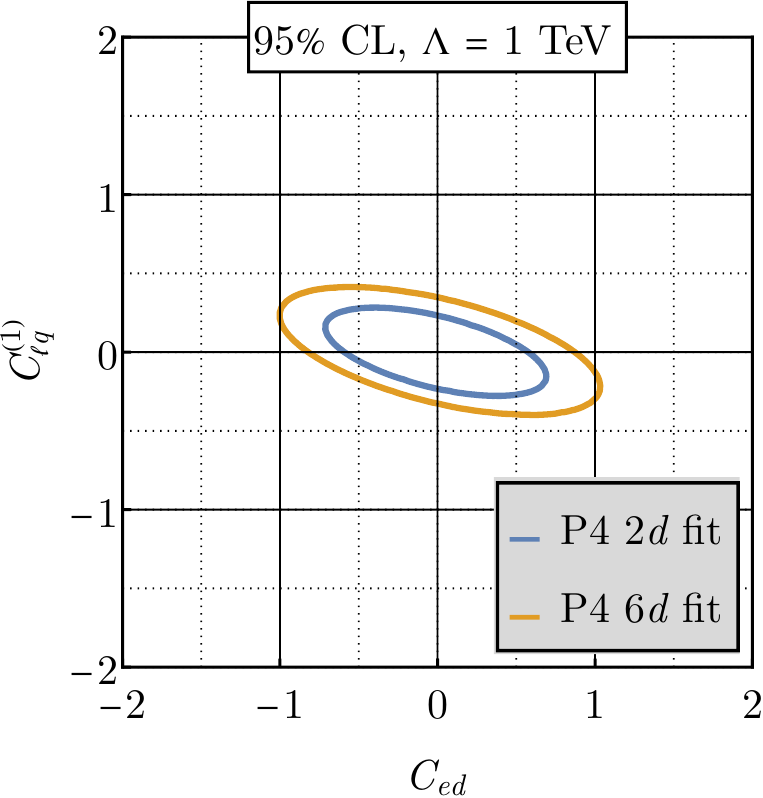}
	\caption{The same as in Fig. \ref{fig:6d-fit-2d-projections_1} but for $(\Ceu, \Clu)$, $(\Ceu, \Cqe)$, and $(\Ced, \Clqi)$.}
	\label{fig:6d-fit-2d-projections_2}
\end{figure}
\begin{figure}
	[H]\centering
	\includegraphics[width=\elliwi\textwidth]{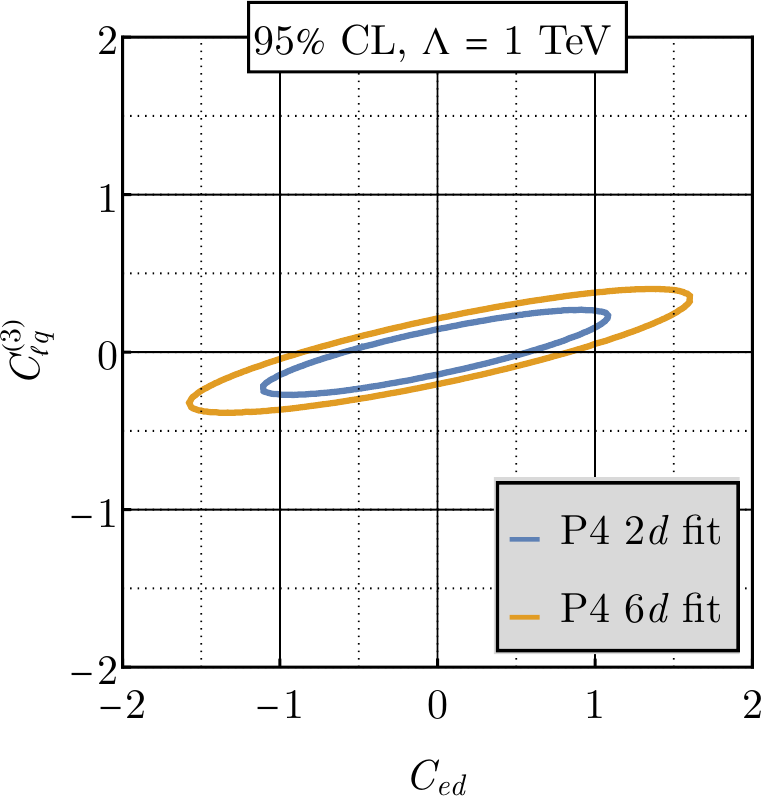}
	\includegraphics[width=\elliwi\textwidth]{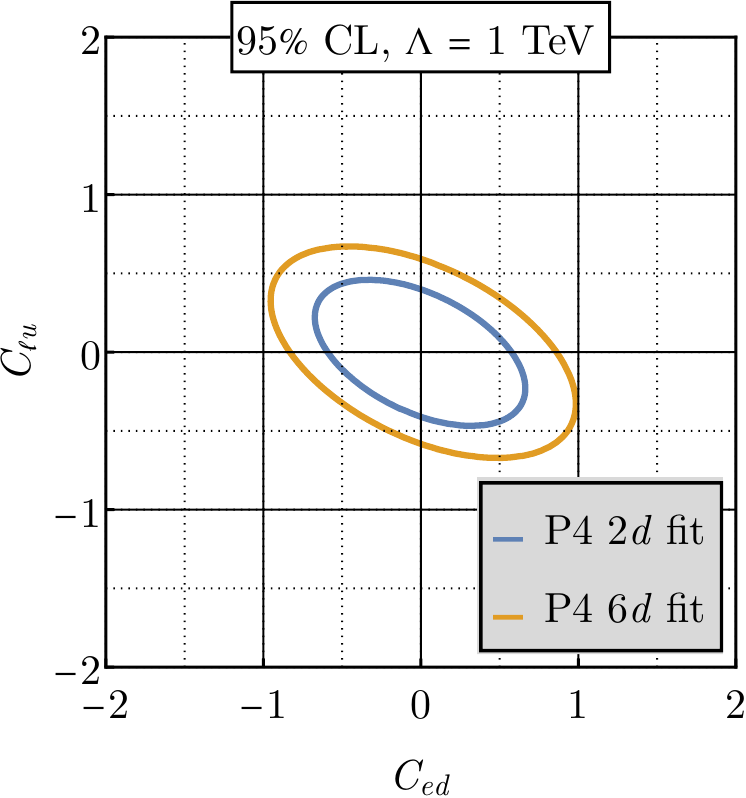}
	\includegraphics[width=\elliwi\textwidth]{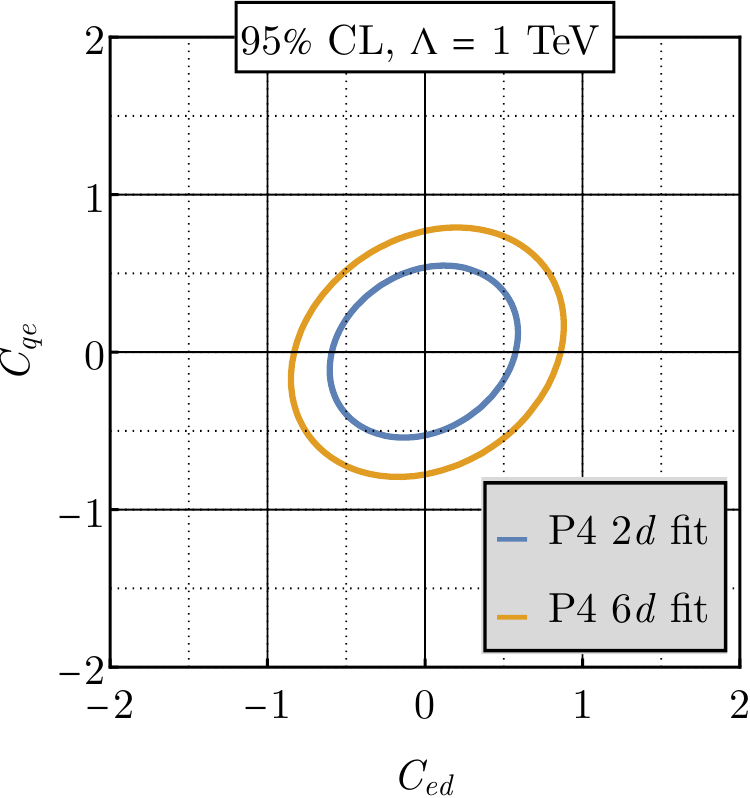}
	\caption{The same as in Fig. \ref{fig:6d-fit-2d-projections_1} but for $(\Ced, \Clqiii)$, $(\Ced, \Clu)$, and $(\Ced, \Cqe)$.}
	\label{fig:6d-fit-2d-projections_3}
\end{figure}
\begin{figure}
	[H]\centering
	\includegraphics[width=\elliwi\textwidth]{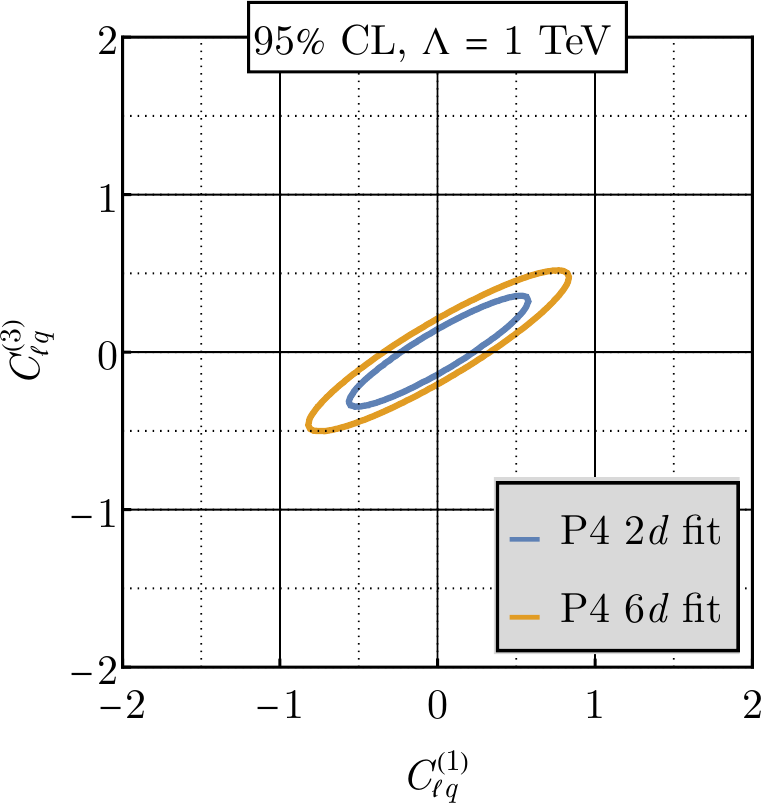}
	\includegraphics[width=\elliwi\textwidth]{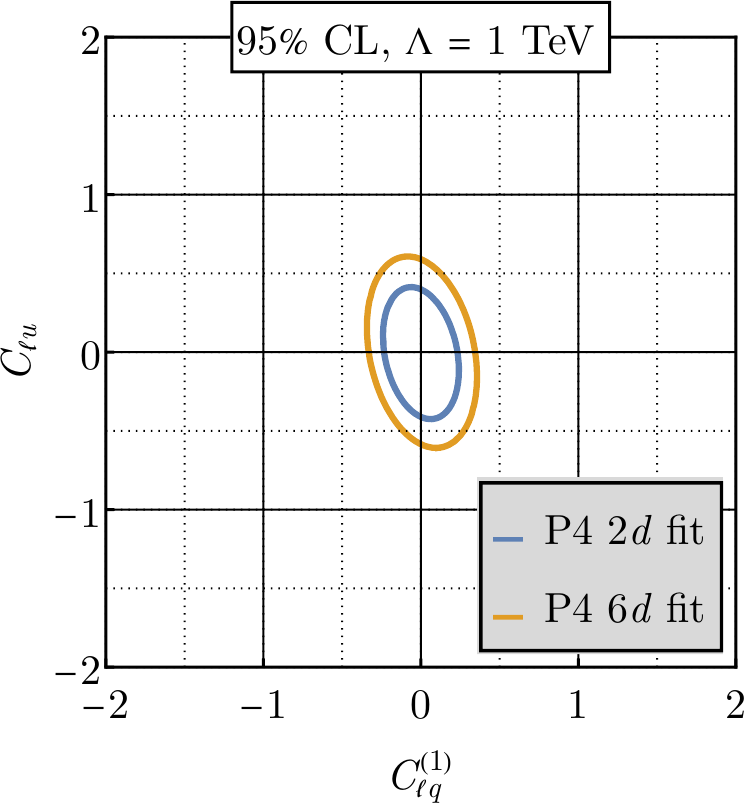}
	\includegraphics[width=\elliwi\textwidth]{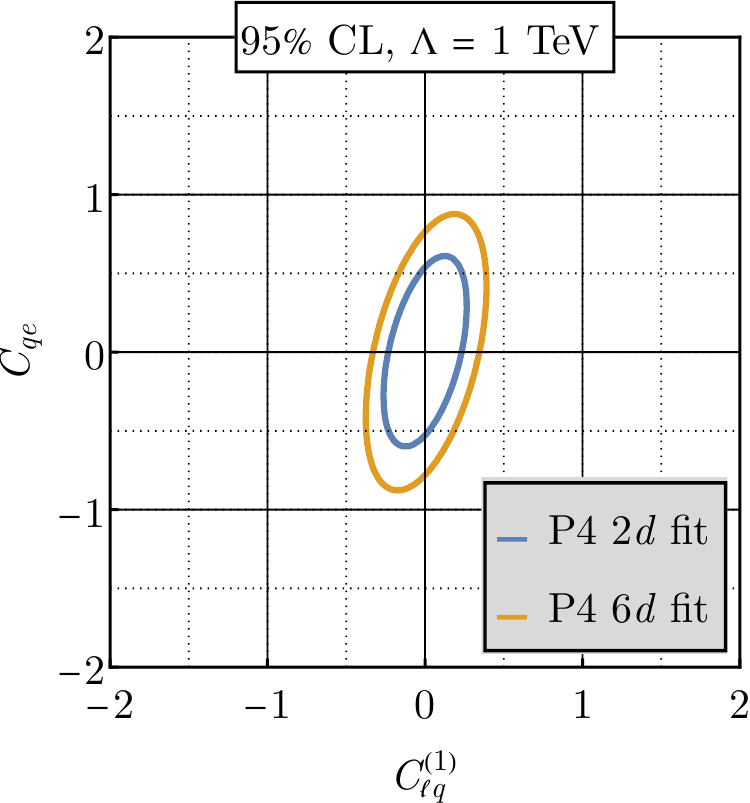}
	\caption{The same as in Fig. \ref{fig:6d-fit-2d-projections_1} but for $(\Clqi, \Clqiii)$, $(\Clqi, \Clu)$, and $(\Clqi, \Cqe)$.}
	\label{fig:6d-fit-2d-projections_4}
\end{figure}
\begin{figure}
	[H]\centering
	\includegraphics[width=\elliwi\textwidth]{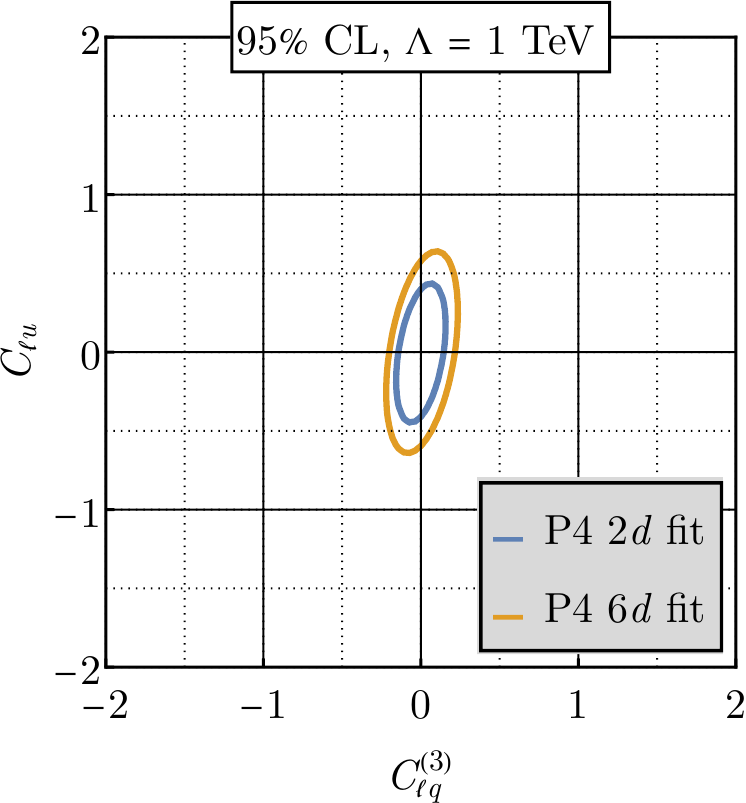}
	\includegraphics[width=\elliwi\textwidth]{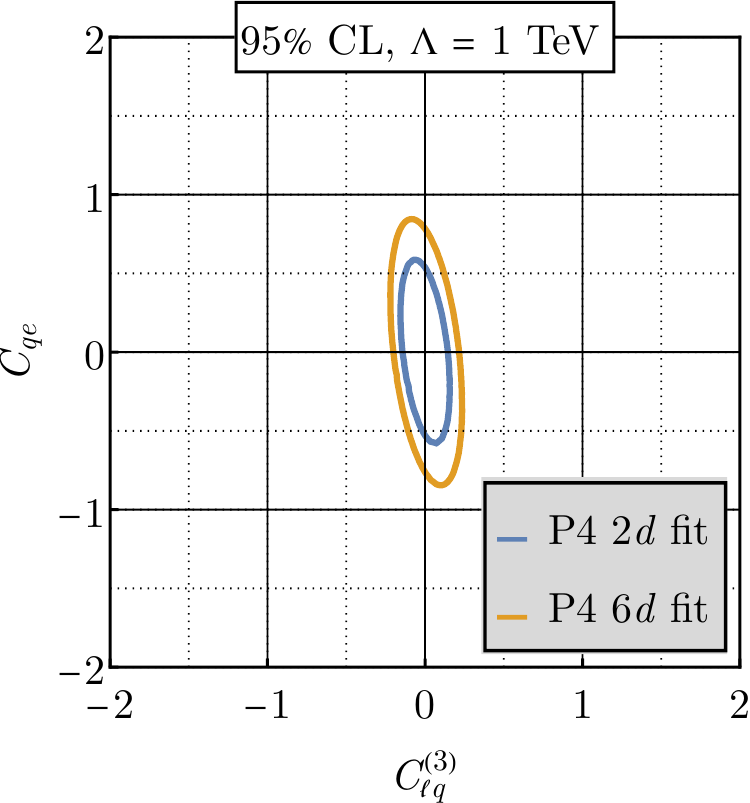}
	\includegraphics[width=\elliwi\textwidth]{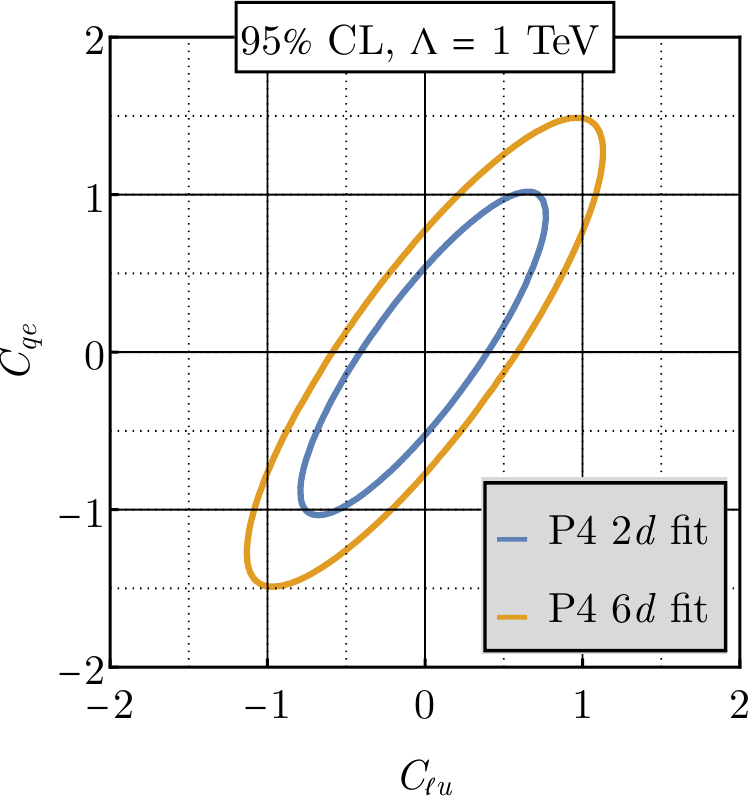}
	\caption{The same as in Fig. \ref{fig:6d-fit-2d-projections_1} but for $(\Clqiii, \Clu)$, $(\Clqiii, \Cqe)$, and $(\Clu, \Cqe)$.}
	\label{fig:6d-fit-2d-projections_5}
\end{figure}